\documentclass{aa}  

\usepackage{graphicx}
\usepackage{txfonts}
\usepackage{color}
\usepackage{ulem}
\begin{document} 

   \title{Seven new triply eclipsing triple star systems}


   \author{S. A. Rappaport\inst{1,2}
          \and
          T. Borkovits\inst{1,3,4,5,6}  
	  \and
	  T. Mitnyan\inst{1}
	  \and
	  R. Gagliano\inst{7}
	  \and
	  N. Eisner\inst{8,9}\fnmsep\thanks{Flatiron Research Fellow}\fnmsep\thanks{Henry Norris Russel Fellow}
	  \and
	  T. Jacobs\inst{10}
	  \and
	  A. Tokovinin\inst{11}
	  \and
	  B. Powell\inst{12}
	  \and
	  V. Kostov\inst{12,13}
	  \and
	  M. Omohundro\inst{14}
	  \and
	  M. H. Kristiansen\inst{15,16}
	  \and
	  R. Jayaraman\inst{2}
	  \and
	  I. Terentev\inst{14}
	  \and
	  H.M. Schwengeler\inst{14}
	  \and
	  D. LaCourse\inst{17}
	  \and
	  Z. Garai\inst{5,6,18}
	  \and
	  T. Pribulla\inst{18}
	  \and
	  P.F.L. Maxted\inst{19}
	  \and
	  I. B. B\'\i r\'o\inst{1,3}
          \and
          I. Cs\'anyi\inst{3}
	  \and
	  A. P\'al\inst{4}
	   \and
	  A. Vanderburg\inst{2}
          }

   \institute{HUN--REN--SZTE Stellar Astrophysics Research Group, H-6500 Baja, Szegedi \'ut, Kt. 766, Hungary 
	\and
	  Department of Physics, Kavli Institute for Astrophysics and Space Research, M.I.T., Cambridge, MA 02139, USA \\
	   \email{sar@mit.edu}
         \and
   	   Baja Astronomical Observatory of University of Szeged, H-6500 Baja, Szegedi \'ut, Kt. 766, Hungary.\\
             \email{borko@electra.bajaobs.hu} 
	\and
	   Konkoly Observatory, Research Centre for Astronomy and Earth Sciences,  H-1121 Budapest, Konkoly Thege Mikl\'os \'ut 15-17, Hungary	   
	\and
	   ELTE E{\"o}tv{\"o}s Lor\'and University, Gothard Astrophysical Observatory, Szent Imre h. u. 112, 9700 Szombathely, Hungary 
	\and
	   HUN--REN--ELTE Exoplanet Research Group, H-9700 Szombathely, Szent Imre h. u. 112, Hungary 
	\and
	   Amateur Astronomer, Glendale, AZ 85308 
	\and
	  Centre For Computation Astrophysics, NYC, USA 
	\and
	   Princeton University, Princeton, USA 
	\and
	   Amateur Astronomer, 12812 SE 69th Place Bellevue, WA 98006, USA 
	\and
	   Cerro Tololo Inter-American Observatory $|$ NSF’s NOIRLab, Casilla 603, La Serena, Chile 
	\and
	   NASA Goddard Space Flight Center, 8800 Greenbelt Road, Greenbelt, MD 20771, USA 
	\and
	   SETI Institute, 189 Bernardo Avenue, Suite 200, Mountain View, CA 94043, USA
	\and
	   Citizen Scientist, c/o Zooniverse, Dept,~of Physics, University of Oxford, Denys Wilkinson Building, Keble Road, Oxford, OX1 3RH, UK 
	\and
	   Brorfelde Observatory, Observator Gyldenkernes Vej 7, DK-4340 T\o ll\o se, Denmark 
	\and 
	   National Space Institute, Technical University of Denmark, Elektrovej 327, DK-2800 Lyngby, Denmark 
	\and
	   Amateur Astronomer, 7507 52nd Place NE Marysville, WA 98270, USA 
	\and
	   Astronomical Institute, Slovak Academy of Sciences, 05960 Tatransk\'a Lomnica, Slovakia 
	 \and
	    Astrophysics Group, Keele University, Staffordshire, ST5 5BG, UK
             }

   \date{Received January .., 2024; accepted ... , ...}

 
  \abstract
  {}
   {We have identified nearly a hundred close triply eclipsing hierarchical triple star systems from data taken with the space telescope TESS. These systems are noteworthy in that we can potentially determine their dynamical and astrophysical parameters with a high precision.  In the present paper, we report the comprehensive study of seven new compact triply eclipsing triple star systems taken from this larger sample: TICs 133771812, 176713425, 185615681, 287756035, 321978218, 323486857, and 650024463.}
   {Most of the data for this study come from TESS observations, but two of them have Gaia measurements of their outer orbits, and we obtained supplemental radial velocity (RV) measurements for three of the systems.   The eclipse timing variation curves extracted from the TESS data, the photometric light curves, the RV points, and the spectral energy distribution (SED) are combined in a complex photodynamical analysis to yield the stellar and orbital parameters of all seven systems.}
   {Four of the systems are quite compact with outer periods in the range of 41-56 days. All of the systems are substantially flat, with mutual inclination angles of $\lesssim 2^\circ$. Including the systems reported in this work, we have now studied in considerable detail some 30 triply eclipsing triples with TESS, and are accumulating a meaningful census of these systems.}
   {}


   \maketitle
%

\nolinenumbers

\section{Introduction}  
\label{sec:intro}

Before the epoch of the {\it Kepler} \citep{borucki10}, and TESS \citep{ricker15} missions there were very few well-studied compact hierarchical triple star systems.  Here we define `compact triple' as a triple star system with an outer period of $\lesssim 1000$ days, down to the currently shortest known outer period of 33.0\,d \citep[$\lambda$~Tau;][]{ebbighausenstruve956}.  Thanks to the precision space-based photometry of these missions, as well as Gaia astrometry and radial velocity measurements \citep{Gaia16,babusiaux23}, and the Optical Gravitational Lensing Experiment (OGLE) ground-based survey \citep{udalski97}, we now know of some thousand compact triple systems \citep[see, e.g.,][]{rappaport13,borkovitsetal15,borkovitsetal16,czavalinga23a,hajdu19}.  Among the best studied of these are about 50 that have years-long eclipse timing variation (ETV) curves and/or third body eclipses where the tertiary star eclipses the inner binary or vice versa \citep{borkovitsetal15,rappaport22,rappaport23}. 

In general, the triple systems with third body eclipses end up providing the most detailed information about their respective triple systems, including a complete understanding of the stellar parameters (masses, radii, $T_{\rm eff}$, and ages) as well as the orbital parameters, including the inclination angles of the two orbital planes, and the mutual inclination angle.  This detailed information can often be obtained without the need for extensive radial velocity (RV) measurements.  The reasons for this success stem from the fact that triply eclipsing triples tend to be quite compact, with orbital periods typically in the range of 40-300 days. In turn, the compactness of the triples can lead to interesting and important dynamical interactions, including so-called `dynamical' delays, driven apsidal motion in the inner binary, possible precession of the orbital planes, as well as some new and fascinating anomalous dynamical interactions, \citep[see, e.g.,][]{rappaport13,borkovitsetal15,borkovitsmitnyan23}. 

The items that allow for robust system solutions involve the following five ingredients: (1) precision space-based photometry of the inner binary eclipses and the third-body eclipses; (2) eclipse timing variation curves (ETV) which are distinct from measuring the shapes and depths of the eclipses; (3) archival ground-based photometry that has the potential to determine the outer orbital period via the third-body eclipses; (4) archival spectral energy distributions (SED); and (5) a comprehensive complex photodynamical  code which puts all this information together in a global analysis.  Of course, supplemental RV data are always useful for further reducing the uncertainties and are much appreciated. 

Knowing the system parameters is important in trying to understand the complex  formation scenarios of triple and multiple star systems \citep[see, e.g.,][]{tokovinin_moe}, as well as for investigating expected and unexpected dynamical interactions in the present-day systems.  The latter category includes dynamical interactions on timescales intermediate between the binary and outer period \citep[see, e.g., Appendix A1 of][]{rappaport22}, and between the outer period and the apse-node long term period \citep{pribulla23}.

In this work we report on seven new triply eclipsing triples.  In Sect.~\ref{sec:7triples} we discuss how these systems are found, and the basic properties of the seven systems presented in this work.  Segments of the  TESS light curves exhibiting third body eclipses are shown in Sect.~\ref{sec:lightcurves} along with model fits.  The folded outer orbital light curves obtained from archival ground-based data are presented in Sec.~\ref{sec:archival_folds}.  In Sect.~\ref{sec:etvs_rvs} we show ETV curves extracted from the TESS light curves, as well as RV curves for a few sources where we were able to acquire them.  The photodynamical analysis code is described briefly in Sect.~\ref{sec:photodynamics}. The individual systems are discussed in Sect.~\ref{sec:results}.  Finally, in Sect.~\ref{sec:discuss} we summarize our work and draw some broader conclusions.
   
\section{The seven triply eclipsing triples}  
\label{sec:7triples}

\subsection{Discovery of the triples}

Most of the triply eclipsing triples reported by us over the past several years \citep[see, e.g.,][]{borkovitsetal20b,borkovitsetal22a,rappaport22,rappaport23} were found by our `Visual Survey Group' (VSG; \citealt{kristiansen22}).  This involved visual inspection of millions of TESS light curves.  The vast majority of the light curves were produced from TESS full-frame images (FFI) of anonymous stars down to TESS magnitude $T \lesssim 15$.  However, most of the discovered triply eclipsing triples were actually found among a much smaller number ($\sim$1 million) of light curves of preselected eclipsing binaries (EBs).  The latter were identified in the TESS data via machine learning (ML) searches\footnote{We use a neural network binary classifier trained to find eclipses in light curves as an initial screening tool.} \citep[see][]{powell21} of more than 100 million TESS FFI light curves from Sectors 1-48.  We made our own light curves from the raw FFIs on the NASA Center for Climate Simulation (NCCS) Discover supercomputer using the {\sc eleanor} \citep{feinstein19} code. The search of these ML-selected binaries was therefore the more productive of the two approaches, but such searches are less likely to discover new classes of objects because we are specifically selecting binaries to look at. 

The visual surveying of the light curves is done with Allan Schmitt's {\tt LcTools} and {\tt LcViewer} software \citep{schmitt19}, which facilitates the inspection of a typical light curve in just a matter of a few seconds.  Our searches for triply eclipsing triples involve looking for an eclipsing binary light curve with an additional extra eclipse that is typically strangely shaped and of longer duration than the regular EB eclipses, or a rapid succession of isolated eclipses.  Once such `extra' eclipses, or `third body' events are found, there is usually very little doubt that they are due to a tertiary star eclipsing the EB or vice versa.  

From searches through light curves obtained from the first three full years of TESS observations, we have found more than 70 of these triply eclipsing triples. Of these 70 triply eclipsing systems, we have been able to determine the outer orbital period for about half of them via a combination of TESS and archival ground-based photometry, and, occasionally, from Gaia-determined orbits. We have previously reported five of these 30  systems in \citet{borkovitsetal20b}, \citet{mitnyanetal20} and \citet{borkovitsetal22a}; six more in \citet{rappaport22}; and an additional nine systems in \citet{rappaport23}.  Here we present a detailed analysis and study of seven of the remaining triply eclipsing triples from among this set. In particular, we have chosen those triples which are not scheduled for additional TESS observations, at least not until the end of the ongoing second extended mission. Hence, it is not surprising that six of our seven targets are located in the southern hemisphere.

\subsection{Summary of available data}

Fortunately, there exist excellent archival data for all of these sources, in addition to the TESS data.  In particular, we made use of archival data from (1) Gaia Data Release 3 \citep{gaia23}; (2) the Mikulski Archive for Space Telescopes (MAST)\footnote{\url{https://mast.stsci.edu/portal/Mashup/Clients/Mast/Portal.html}}, in particular the TESS Input Catalog v8.2 \citep{TIC8}; (3) the All-Sky Automated Survey for SuperNovae (ASAS-SN; \citealt{shappee14}; \citealt{kochanek17}); (4) Asteroid Terrestrial-impact Last Alert System (ATLAS\footnote{\url{https://fallingstar-data.com/forcedphot/}}; \citealt{tonry18}; \citealt{smith20}); and (5) the online VizieR SED viewer (\citealt{ochsenbein00})\footnote{\url{http://vizier.cds.unistra.fr/vizier/sed/}}. Last but not least, (6) for one target, TIC~321978218, we also used SuperWASP data \citep{pollacco06}. This relatively bright system was observed with SuperWASP cameras during five consecutive seasons between 2010 and 2014, and these observations partly cover 11 third-body eclipses, which we have included in our analysis (see below). We note that SWASP observations are  also available for another target, TIC~176713425, but in this latter case, we did not find these data to be useful for our analysis and, hence, neglected them.  

The Gaia and MAST data were used primarily to construct a table of the basic properties of the seven systems reported in this work (Table \ref{tbl:mags7} with, e.g., coordinates, magnitudes, distances, proper motions, and so forth).  Amongst the Gaia data we also found the outer orbits for two of our systems.  We used the ASAS-SN and ATLAS data primarily to obtain an independent and accurate determination of the outer orbital period for 6 of the 7 systems via the third-body eclipses over a baseline of $\sim$10\,yr.  The SED data were obtained from VizieR; these points were used in conjunction with other data to make some of the initial fits to the system parameters.

The sectors in which the sources were observed by TESS, and the sectors during which third-body events were actually seen, are summarized in Table \ref{tbl:sectors}.  In Table \ref{tbl:noticed} we point out the two sources whose outer orbits were found with Gaia\footnote{\url{https://vizier.cds.unistra.fr/viz-bin/VizieR-3?-source=I/357&-out.max=50&-out.form=HTML}} and references to another source that had previous reports in the literature, and we list the outer orbital parameters of period, eccentricity, and argument of periastron we find from this work. 

In addition to the above archival data, we were able to acquire new RV data  for two of the sources with the CHIRON spectrometer mounted on the 1.5-m SMARTS telescope at CTIO in Chile \citep{tokovinin13}. Spectra with a resolution of 25,000 were taken in the service mode and processed by the standard pipeline \citep{paredes21}. We note that a set of archive RV data were also collected for a third target -- TIC~185615681. 

Finally, in Table \ref{tbl:input} we summarize the information we have available for all of the seven sources.  This includes eclipse timing variation (ETV) curves that exhibit non-linear behavior, RV data, and spectroscopic outer orbits from Gaia NSS catalog (in two cases).

\begin{table*}
\hspace{-20px}
\footnotesize
\centering
\caption{Main properties of the seven triple systems from different catalogs}
\tiny
\begin{tabular}{lccccccc}
\hline
\hline
Parameter			              & 133771812 	      &  176713425 	      & 185615681	        &  287756035	           &  321978218            &   323486857 	         &  650024463  	 \\  
\hline
RA J2000 			               & 08:05:16.560        &  23:35:35.42	      & 08:41:08.26              & 13:14:51.86               &  00:04:12.35          &  11:45:48.28	           &  02:41:20.17            \\
Dec J2000 		               & $-$41:49:57.97     &  42:22:17.57	      &  $-$32:12:03.06        & $-$46:44:29.21	    &   $-$50:55:00.91     &  $-$65:02:48.99 	  &  $-$82:25:16.73        \\
$T^a$                                       & $13.30 \pm 0.01$ & $13.75 \pm 0.01$   & $10.52 \pm 0.01$     & $12.86 \pm 0.01$      & $11.35 \pm 0.01$   & $11.61 \pm 0.02$        &  $14.4 \pm 0.6$    \\
G$^b$ 				       & $13.76 \pm 0.00$  & $14.25 \pm 0.00$   & $10.65 \pm 0.00$    & $13.47 \pm 0.01$     & $11.73 \pm 0.00$    & $12.31 \pm 0.00$        &  $14.85 \pm 0.00$   \\ 
G$_{\rm BP}$$^b$ 		       & $14.12 \pm 0.00$  & $14.62 \pm 0.00$   & $10.75 \pm 0.00$     & $14.00 \pm 0.01$    & $12.00 \pm 0.00$   & $12.93 \pm 0.01$        & $15.20 \pm 0.02$    \\
G$_{\rm RP}$$^b$ 		       & $13.19 \pm 0.00$  & $13.68 \pm 0.00$  & $10.46 \pm 0.00$       & $12.79 \pm 0.00$    & $11.28 \pm 0.00$   & $11.54 \pm 0.01$        & $14.17 \pm 0.01$     \\
B$^a$                                       & $14.62 \pm 0.01$ & $15.40 \pm 0.46$   & $10.97 \pm 0.16$       & $14.77 \pm 0.08$    & $12.42 \pm 0.06$   & $13.68 \pm 0.02$        &  ...                            \\
V$^c$                                       & $14.00 \pm 0.12$ & $14.60 \pm 0.11$   & $10.68 \pm 0.03$       & $13.75 \pm 0.13$     &  $11.92 \pm 0.05$  & $12.68 \pm 0.07$        &  ...                            \\       
J$^d$				       & $12.49 \pm 0.03$  & $13.06 \pm 0.02$  & $10.56 \pm 0.03$       & $11.92 \pm 0.02$    & $10.77 \pm 0.03$    & $10.53 \pm 0.02$          & ...                          \\
H$^d$ 				       & $12.23 \pm 0.02$  & $12.73 \pm 0.03$    & $10.50 \pm 0.02$      & $11.35 \pm 0.02$    & $10.50 \pm 0.02$    & $9.97 \pm 0.02$          & ...                           \\
K$^d$ 			                & $12.17 \pm 0.02$  & $12.66 \pm 0.03$   & $10.48 \pm 0.02$       & $11.28 \pm 0.02$    & $10.44 \pm 0.03$    & $9.83 \pm 0.02$          & ...                           \\
W1$^e$ 				        & $12.09 \pm 0.02$  & $12.65 \pm 0.02$  & $10.16 \pm 0.02$        & $11.22 \pm 0.02$    & $10.40 \pm 0.02$    & $9.70 \pm 0.02$          & ...                           \\
W2$^e$ 				        & $12.09 \pm 0.02$  & $12.67 \pm 0.02$  & $10.21 \pm 0.02$        & $11.27 \pm 0.02$    & $10.44 \pm 0.02$    & $9.75 \pm 0.02$           & ...                           \\
W3$^e$ 				        & $12.08 \pm 0.28$  & $12.02 \pm 0.22 $   & $10.17 \pm 0.07$       & $11.29 \pm 0.11$    & $10.41 \pm 0.08$    & $9.57 \pm 0.04$           & ...                           \\
$T_{\rm eff}$ [K]$^b$ 	        & $5342 \pm 300$  & $5355 \pm 115$     & $8325 \pm 950$        & $4973 \pm 230$       & $6506 \pm 800$        &  $4518 \pm 150$          & $4362 \pm 125$      \\  
$T_{\rm eff}$ [K]$^a$ 	        & $8357 \pm 123$  & $5937 \pm 125$    & $8696 \pm 155$        & $5090 \pm 122$       & $5949 \pm 130$        &  $7197 \pm 130$	       &  ...                            \\  
Radius [$R_\odot$]$^b$ 	        & $3.41 \pm 0.41$   & $3.48 \pm 0.11$     & ...                              &  $5.07 \pm 0.47$       &  $1.97 \pm 0.50$      & $11.28 \pm 1.2$	       &  $14.3 \pm 0.8$        \\
Radius [$R_\odot$]$^a$ 	        & $2.82 \pm NA$     & $3.08 \pm NA$      & $3.11 \pm 0.12 $       & $5.20 \pm NA$        & $2.35 \pm 0.13$        & $9.47 \pm NA$ 	      &  ...                              \\
Distance [pc]$^f$			& $1803 \pm 46$     & $2144\pm 86$       &  $833 \pm 13$          & $2139 \pm 83$ 	     &  $570 \pm 7$            & $1818 \pm 38$              & $1126 \pm 26$       \\  
$E(B-V)$$^a$                            & $0.59 \pm 0.01$    & $0.15 \pm 0.01$       & $0.13 \pm 0.01$       & $0.12 \pm 0.00$      & $0.01 \pm 0.01$     & $0.75 \pm 0.02$           & $0.09 \pm 0.00$     \\ 
$\mu_\alpha$ [mas/yr]$^b$	& $-2.56 \pm 0.01$  & $-3.82 \pm 0.01$    & $-3.51 \pm 0.01$       & $-15.07 \pm 0.02$   & $24.38 \pm 0.01$    & $10.61 \pm 0.01$ 	     & $-2.43 \pm 0.02$    \\   
$\mu_\delta$ [mas/yr]$^b$  	& $+5.63 \pm 0.01$  & $-6.43 \pm 0.01$   & $-3.58 \pm 0.02$       &  $4.73 \pm 0.01$     & $-0.83 \pm 0.02$     & $1.18 \pm 0.01$ 	     & $-5.01 \pm 0.02$    \\  
RUWE$^{b,g}$                          &  0.975                      & 1.00                      & 1.11                           & 1.15                         & 1.79                         & 0.957                            & 1.07                        \\
astr\_ex\_noise [mas]$^{b,h}$   &    0                           &  0                          &  0.13                          &   0.066                     &  0.23                        &   0.043                          &   0.065                    \\
astr\_ex\_noise\_sig$^{b,h}$    &    0                            &  0                          &   29                             &    3.3                        &  117                         &   2.1                              &   1.2                       \\
$P_{\rm binary}$$^i$ [d]          & 12.370                      & 1.8987                  & 2.3185                      & 2.0814                     & 0.5698                     & 0.8840                          & 7.1971                     \\
$P_{\rm triple}$$^i$ [d]            & 242.3                        & 52.55                    & 55.84                        & 367.6                       & 57.51                       & 41.36                            & 107.95                     \\
\hline
\label{tbl:mags7}
\end{tabular}  

\small
\textit{Notes.}  General: ``NA" and ellipses in this table indicate that the value is not available. (a) TESS Input Catalog (TIC v8.2) \citep{TIC8}. (b) Gaia EDR3 \citep{GaiaEDR3}; the uncertainty in $T_{\rm eff}$ and $R$ listed here is 1.5 times the geometric mean of the upper and lower error bars cited in DR2. Magnitude uncertainties listed as 0.00 are $\lesssim 0.005$. (c) AAVSO Photometric All Sky Survey (APASS) DR9, \citep{APASS}, \url{http://vizier.u-strasbg.fr/viz-bin/VizieR?-source=II/336/apass9}. (d) 2MASS catalog \citep{2MASS}.  (e) WISE point source catalog \citep{WISE}. (f) \citet{bailer-jonesetal21}, geometric distances. (g) The Gaia renormalized unit weight error (RUWE) is the square root of the normalized $\chi^2$ of the astrometric fit to the along-scan observations. Values in excess of about unity are sometimes taken to be a sign of stellar multiplicity. (h) Abbreviations for ``astrometric\_excess\_noise'' and ``astrometric\_excess\_noise\_significance'' (\citealt{lindegren21}; \url{https://gea.esac.esa.int/archive/documentation/GDR2/Gaia_archive/chap_datamodel/sec_dm_main_tables/ssec_dm_gaia_source.html)}; these are a measure of ``the disagreement, expressed as an angle, between the observations of a source and the best-fitting standard astrometric model.'' Values of astr\_ex\_noise\_sig $\gtrsim 2$ are considered significant. (i) Binary and outer orbital periods from this work; given here for reference purposes.\\  

\end{table*}

\begin{table*}
\centering
\caption{TESS observation sectors for the triples$^a$}
\small
\begin{tabular}{lcc}
\hline
\hline
Object & Sectors Observed & Third Body Events  \\
\hline
TIC 133771812 & S7,S8,S34,S35,S61,S62  & S7,S34,S61  \\
TIC 176713425 & S16,S17,S57 & S16 \\ 
TIC 185615681 & S8,S34,S35,S61,S62 & S8,S34,S35,S61,S62 \\
TIC 287756035 & S11,S37,S38,S64  & S37,S64 \\
TIC 321978218 & S2,S28,S29,S68,S69 & S28,S29,S68,S69 \\
TIC 323486857 & S10,S11,S37,S38,S64,S65 & S10$^b$,S11,S37,S38$^b$,S64$^b$,S65  \\
TIC 650024463 & S1,S12,S13,S27,S28,S39,S66,S67,S68 & S12,S27,S28,S39,S66,S67 \\
\hline
\label{tbl:sectors}  
\end{tabular}

\textit{Notes.}  (a) None of these sources will be observed in further TESS sectors until the end of Cycle 6 observations; (b) both primary and secondary third-body eclipses observed in these sectors.

\end{table*}

\begin{table*}
\centering
\caption{Other detections of the triples$^a$}
\small
\begin{tabular}{lccc}
\hline
\hline
Object & This Work$^b$ & Gaia DR3 Orbits$^{b,c}$ & Other \\
        & ($P_{\rm out}$, $e_{\rm out}$, $\omega_{\rm out}$)  &   ($P_{\rm out}$, $e_{\rm out}$, $\omega_{\rm out}$) &    \\
\hline
TIC 133771812 & 243.87 d; 0.217; 195$^\circ$  & ... & ...\\
TIC 176713425 &  52.57 d; 0.412; 95.2$^\circ$ & ... & ...\\ 
TIC 185615681 & 55.86 d; 0.098; 309$^\circ$ & ... & \citet{rowan23} \\
TIC 287756035 & 368 d; 0.235; $53.7^\circ$ &  ... & ... \\
TIC 321978218 & 57.51 d; 0.258; $123\degr$ & 57.52 d; 0.28; 116$^\circ$ & ... \\
TIC 323486857 &  41.36 d; 0.007; $241^\circ$ & 41.35 d; 0.045; 180$^\circ$ & ... \\
TIC 650024463 & 107.88 d; 0.323; 351$^\circ$ & ... & ... \\
\hline
\label{tbl:noticed}  
\end{tabular}

\textit{Notes.}  (a) Three of the seven triply eclipsing systems were spotted in prior broad surveys, but no quantitative analysis of the system parameters (especially of the constituent stars) was undertaken.  Two of them have outer orbits reported by Gaia, but no third body eclipses were reported by Gaia.  One of the systems, TIC 185615681, was included in the \citet{rowan23} list of mutlistellar systems tabulated from ASAS-SN data, and studied for the apsidal motion of its inner binary by \citet{zasche12} and \citet{kimetal18}, as well as for RVs by \citet{duerbeckrucinski07}. (b) Where available, we show for each source, the period, eccentricity, and argument of periastron of the outer orbit, separated by semicolons. (c) The Gaia orbital solutions are both spectroscopic; \citet{babusiaux23}; \citet{gaia23}.  
\end{table*}

\begin{table*}
\centering
\caption{Input information for the system analysis}
\small
\begin{tabular}{lccccccc}
\hline
\hline
Object & TESS 3rd-Body & TESS EB & Archival Outer & SED Points$^c$ & ETV Curve$^d$ & RV Data$^e$ & Gaia Orbit$^f$ \\
 & Eclipse(s)$^a$ & Light curve$^a$ & Eclipses$^b$ &  &  & &  \\
\hline
TIC 133771812 & $\checkmark$  &$\checkmark$ &  & $\checkmark$ & $\checkmark$ &   &  \\
TIC 176713425 & $\checkmark$ & $\checkmark$ & $\checkmark$ & $\checkmark$ & $\checkmark$ &  &  \\ 
TIC 185615681 & $\checkmark$ & $\checkmark$ & $\checkmark$ & $\checkmark$ & $\checkmark$ & $\checkmark$ &  \\
TIC 287756035 & $\checkmark$ &  $\checkmark$ & $\checkmark$ & $\checkmark$ &  & $\checkmark$ & \\
TIC 321978218 & $\checkmark$ & $\checkmark$ & $\checkmark$ & $\checkmark$ & $\checkmark$ & $\checkmark$ & $\checkmark$ \\
TIC 323486857 & $\checkmark$ & $\checkmark$ & $\checkmark$ & $\checkmark$ &$\checkmark$ &  & $\checkmark$ \\
TIC 650024463 & $\checkmark$ & $\checkmark$ & $\checkmark$ & $\checkmark$ & $\checkmark$ & &  \\
\hline
\label{tbl:input}  
\end{tabular}

\textit{Notes.}  (a) See Figs.~\ref{fig:133771812lcs}--\ref{fig:650024463lcs}. (b) Fig.\,\ref{fig:outer_orbit_folds}. (c) Obtained from {\url{http://vizier.cds.unistra.fr/vizier/sed/}}. (d) See Figs.~\ref{fig:133771812ETV}--\ref{fig:650024463ETV}. (e) See Figs. \ref{fig:185615681rv}--\ref{fig:321978218rv}. (f) \citet{babusiaux23}; \citet{gaia23}
\end{table*}

\section{Light curve and model Fits}  
\label{sec:lightcurves}

\begin{figure*}
   \centering
   \includegraphics[width=0.32\textwidth]{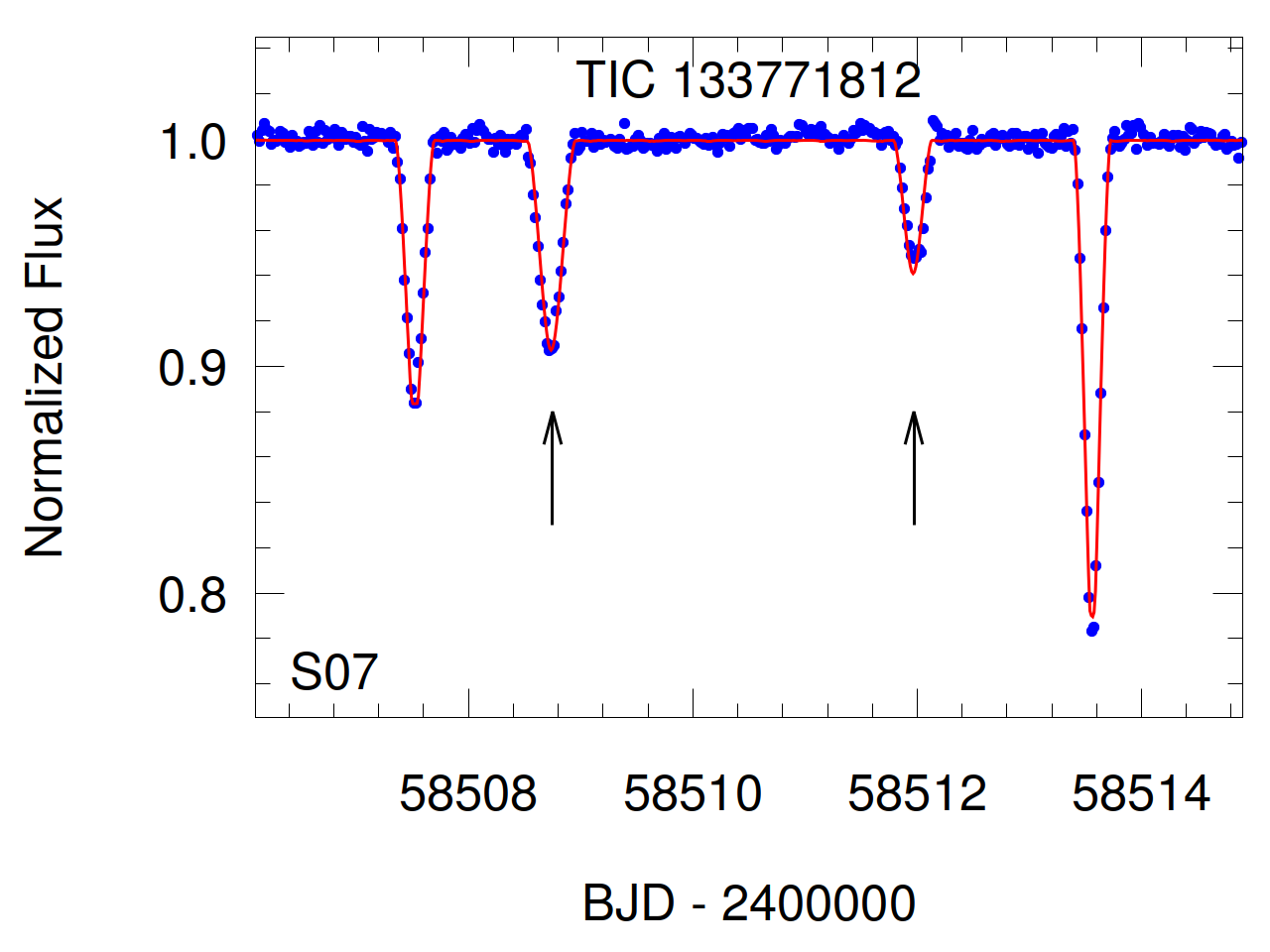}
    \includegraphics[width=0.32\textwidth]{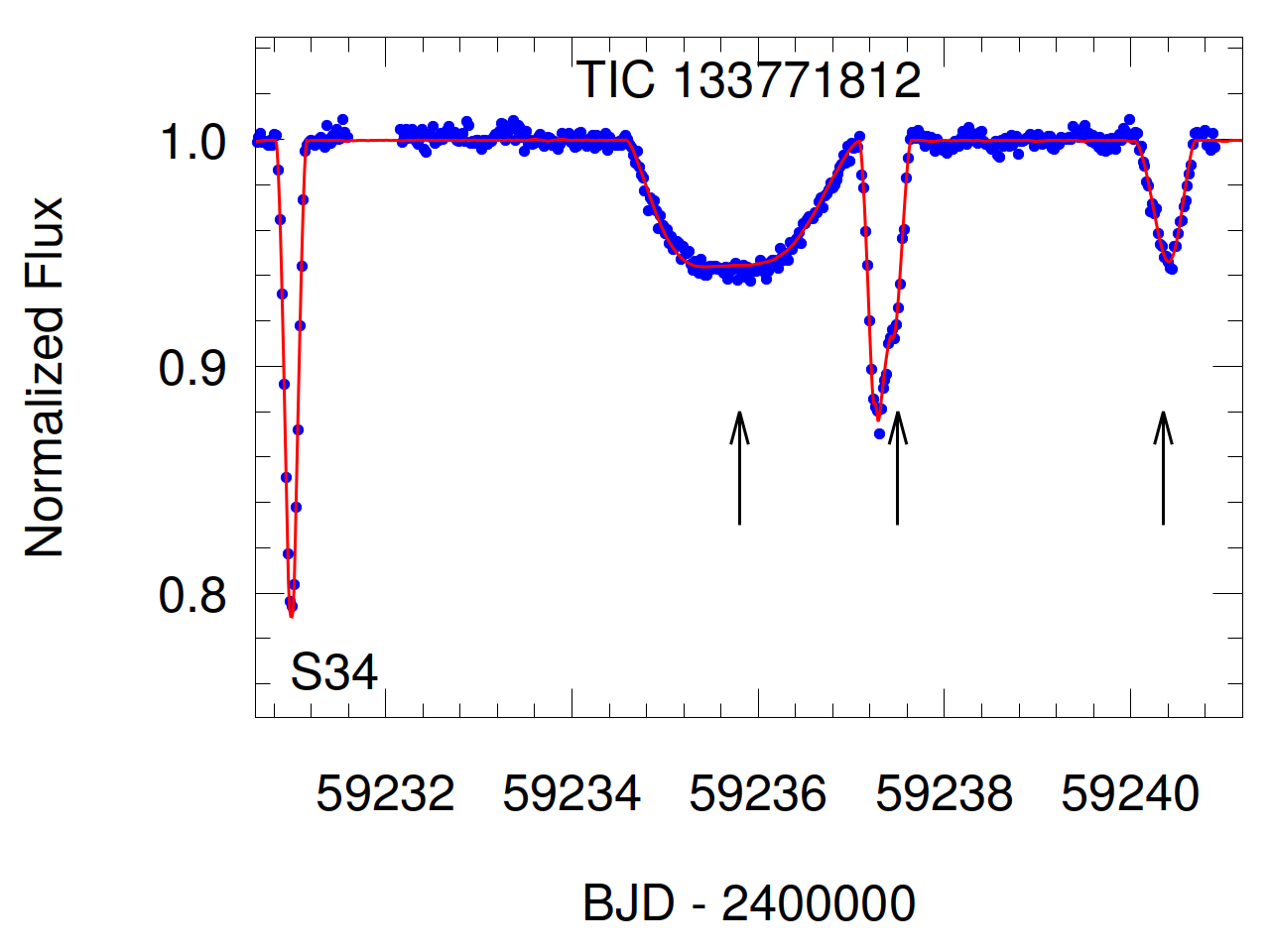}
      \includegraphics[width=0.32\textwidth]{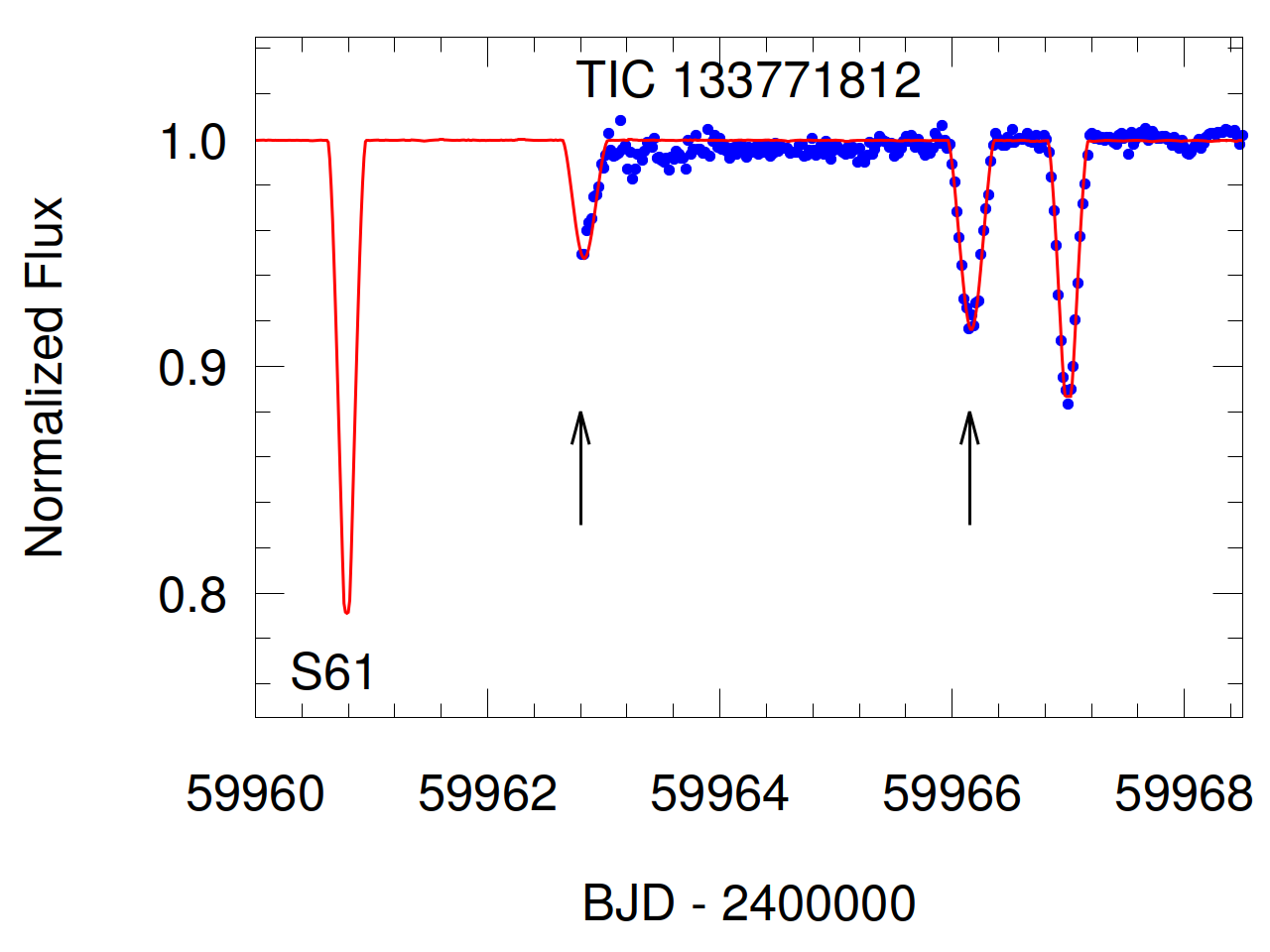}
      \caption{Light curves (blue points) and model fits (smooth red curves) for the third-body eclipses of TIC 133771812.  The third body eclipses are marked with vertical arrows. The sector numbers are indicated in the lower left corner of each panel. Calculation of the model curves will be discussed in Sect.~\ref{sec:photodynamics}. }
         \label{fig:133771812lcs}
   \end{figure*}

\begin{figure}
   \centering
   \includegraphics[width=1.0\columnwidth]{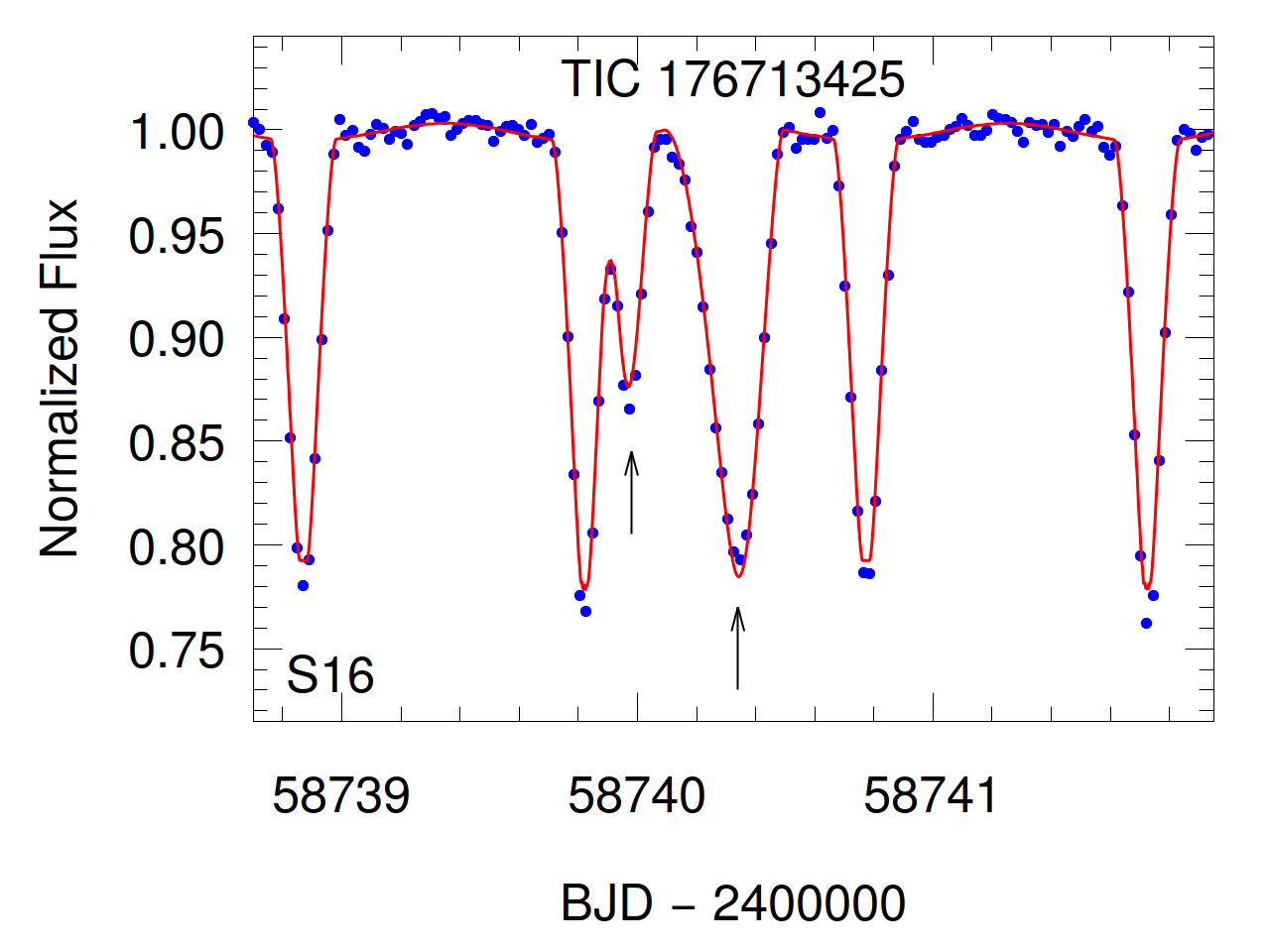}
      \caption{Light curve (blue points) and model fit (smooth red curves) for the only third-body eclipse of TIC 176713425 observed with TESS.  The third body eclipses are marked with vertical arrows. The other notation is the same as in Fig.\,\ref{fig:133771812lcs}.}
         \label{fig:176713425lcs}
   \end{figure}

\begin{figure*}
   \centering
   \includegraphics[width=0.3\textwidth]{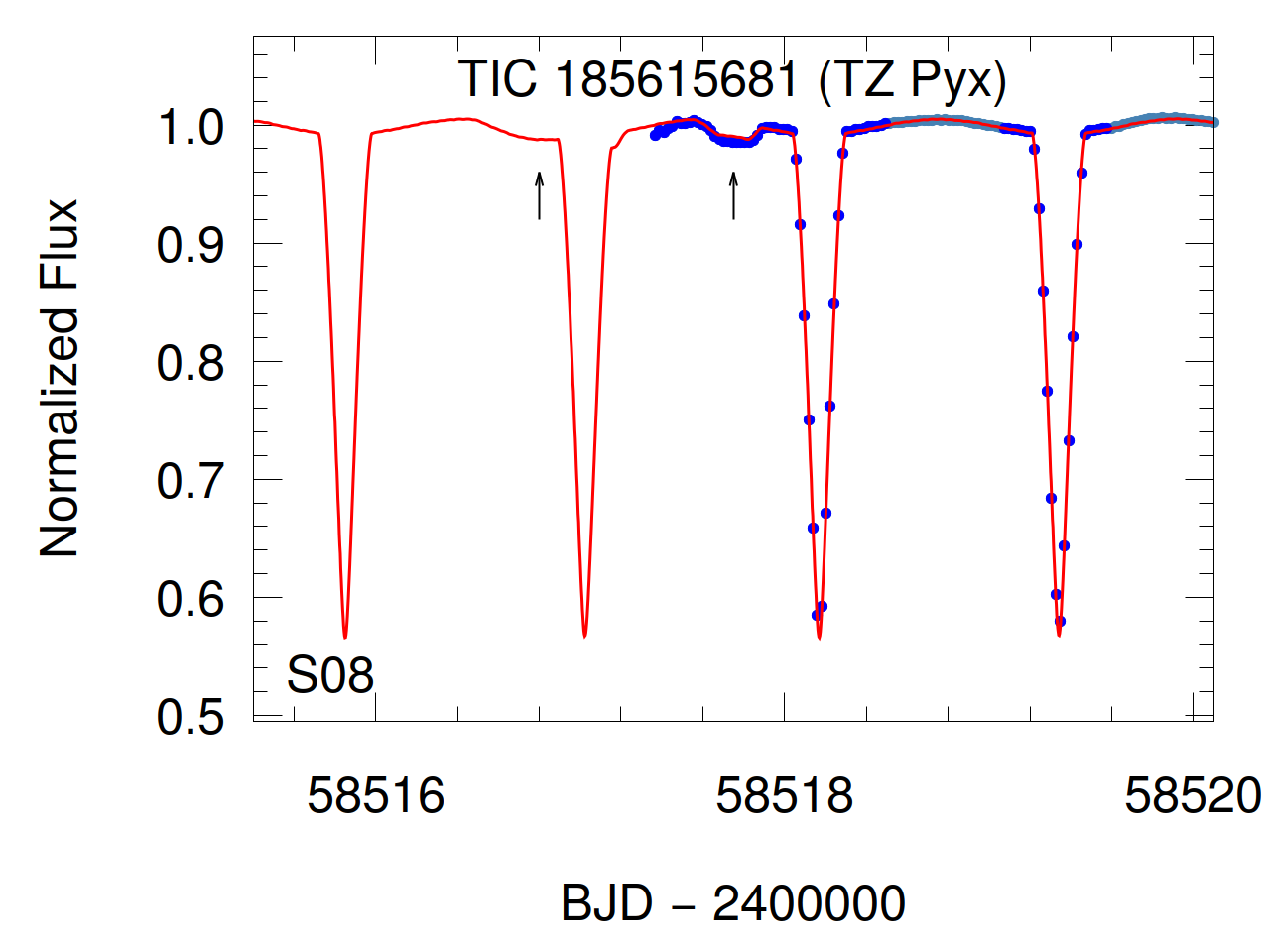}
    \includegraphics[width=0.3\textwidth]{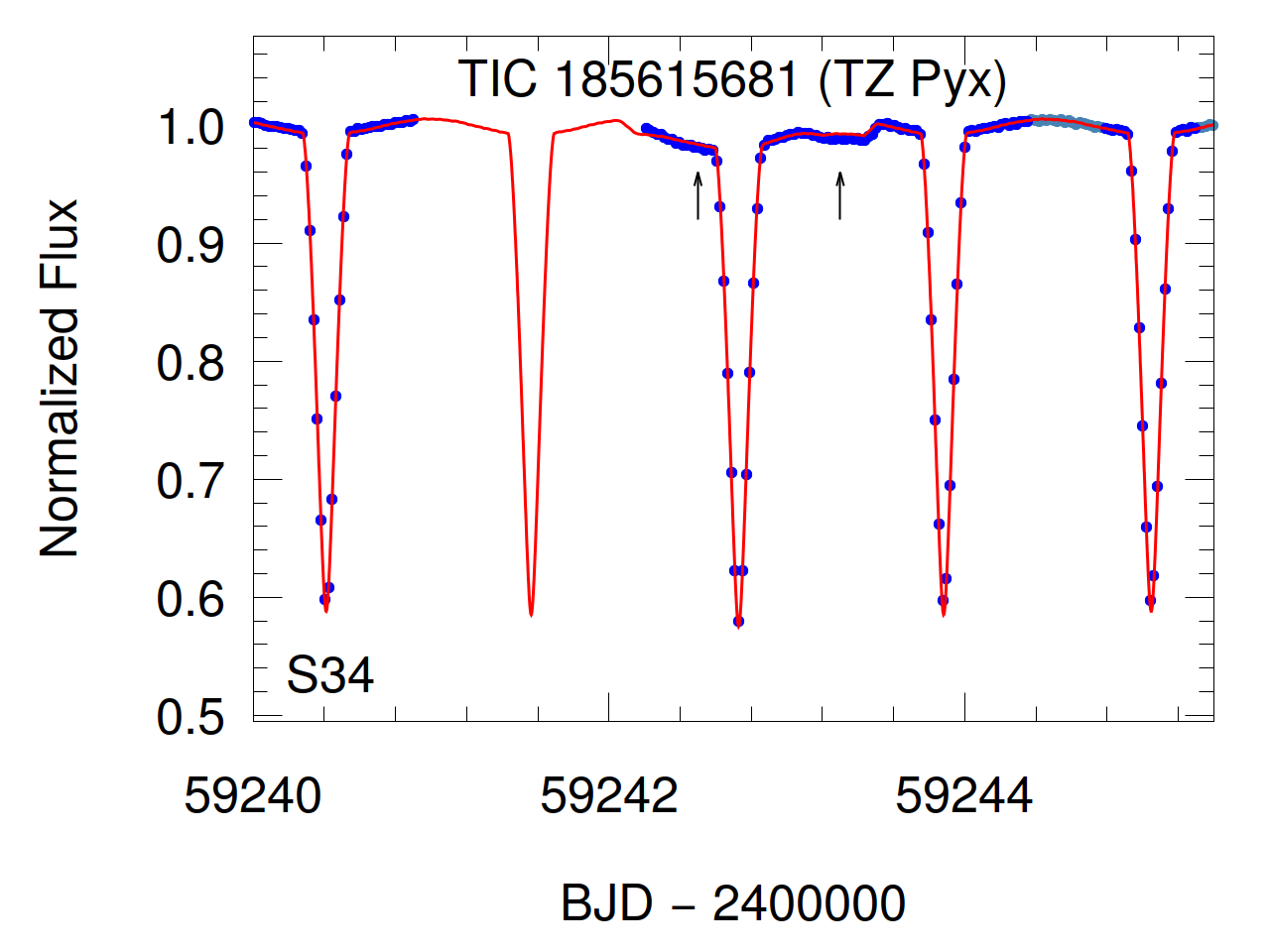}
     \includegraphics[width=0.3\textwidth]{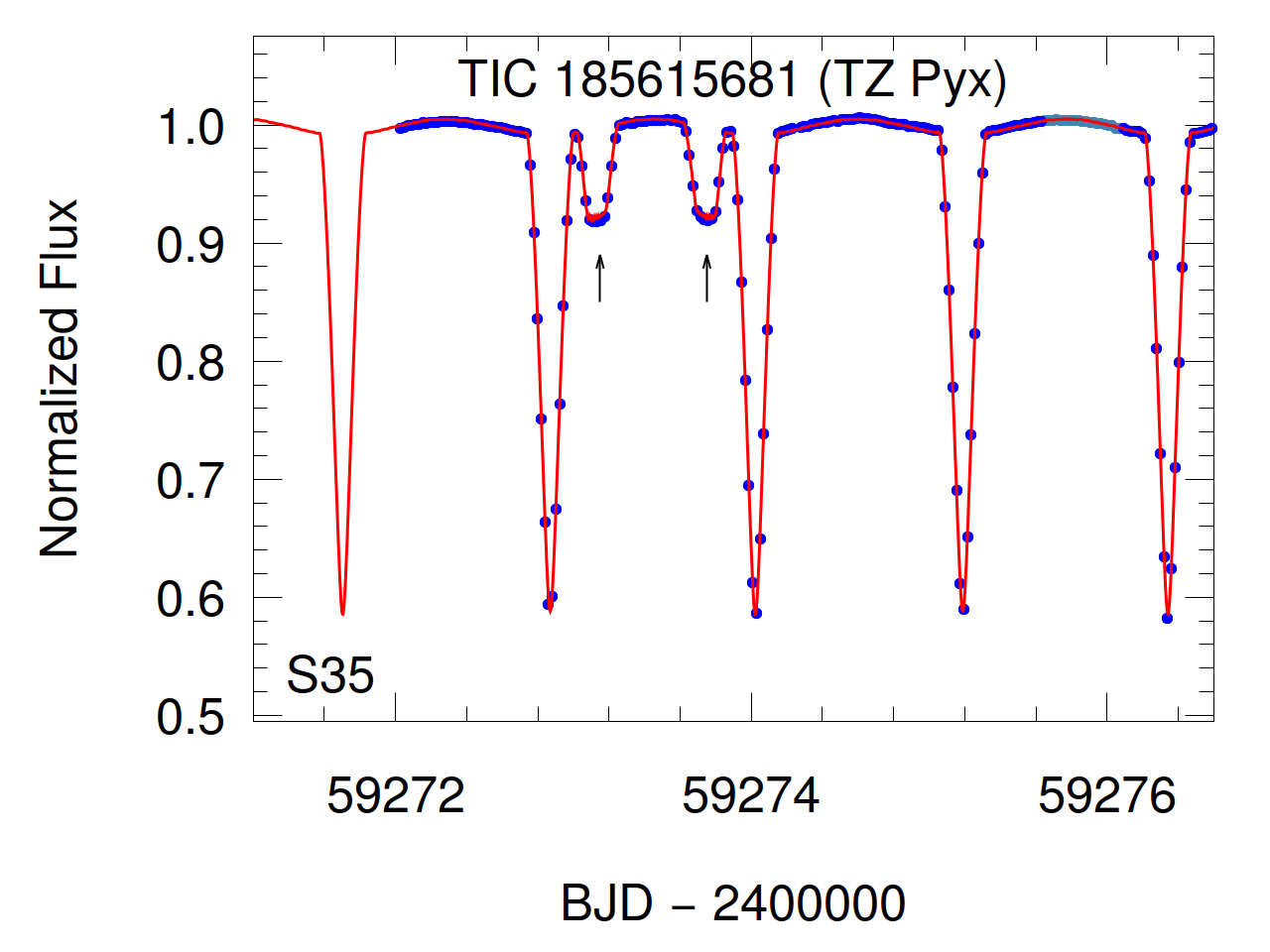}
      \includegraphics[width=0.3\textwidth]{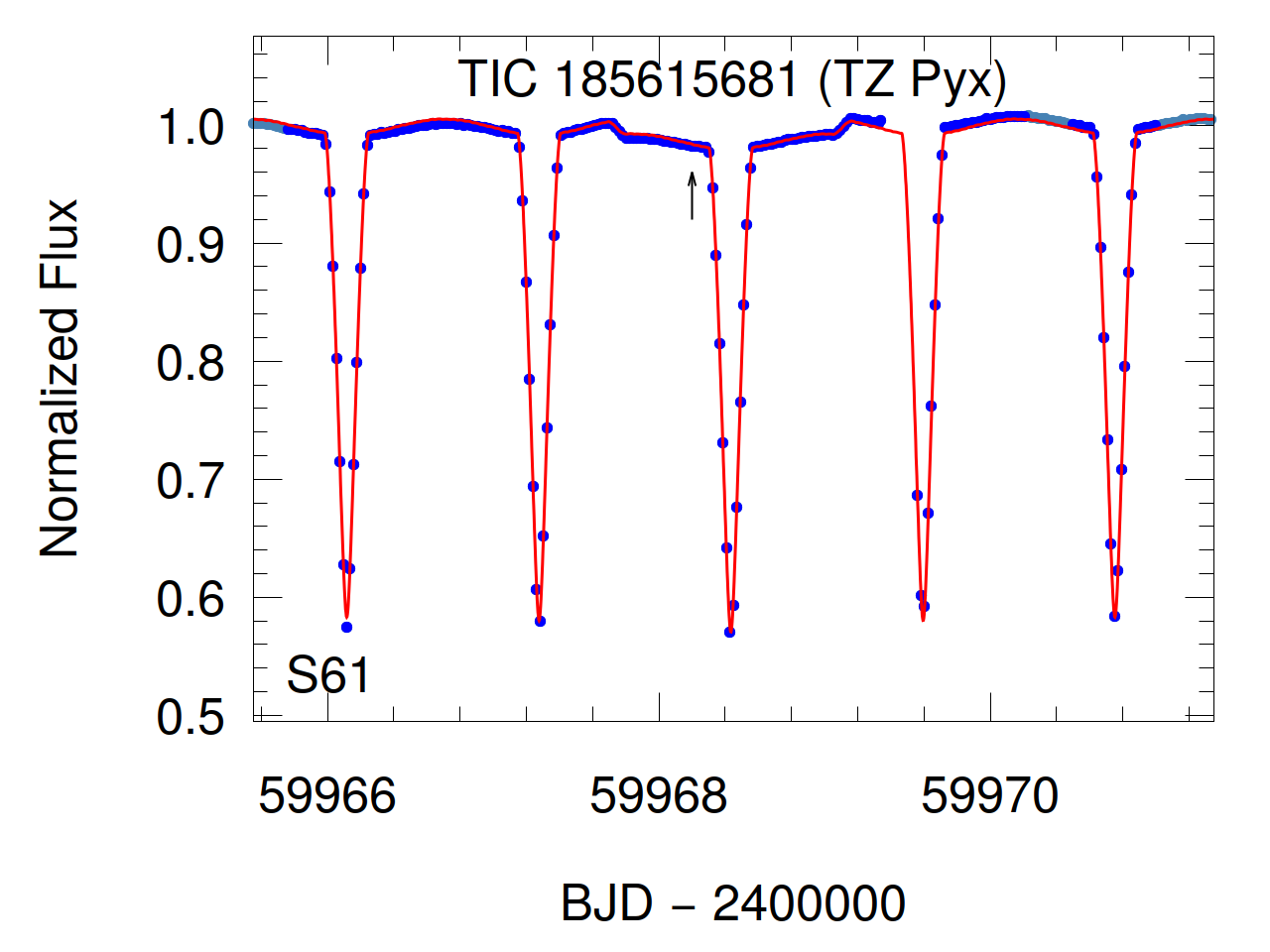}
       \includegraphics[width=0.3\textwidth]{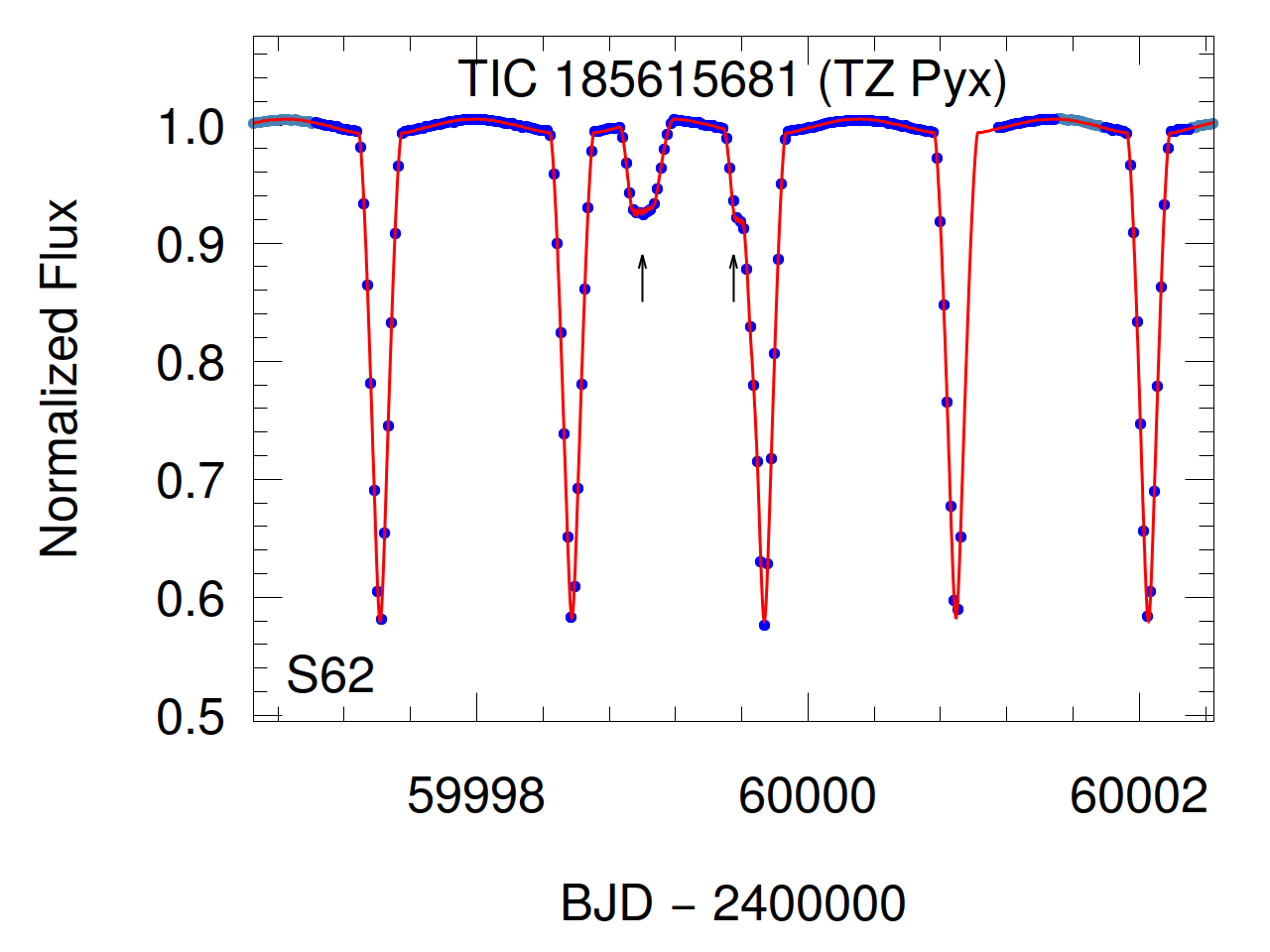}
      \caption{Light curves (blue points) and model fits (smooth red curves) for the third-body eclipses of TIC 185615681 (marked with vertical arrows).  The other notation is the same as in Fig.\,\ref{fig:133771812lcs}.}
         \label{fig:185615681lcs}
   \end{figure*}

\begin{figure*}
   \centering
    \includegraphics[width=0.35\textwidth]{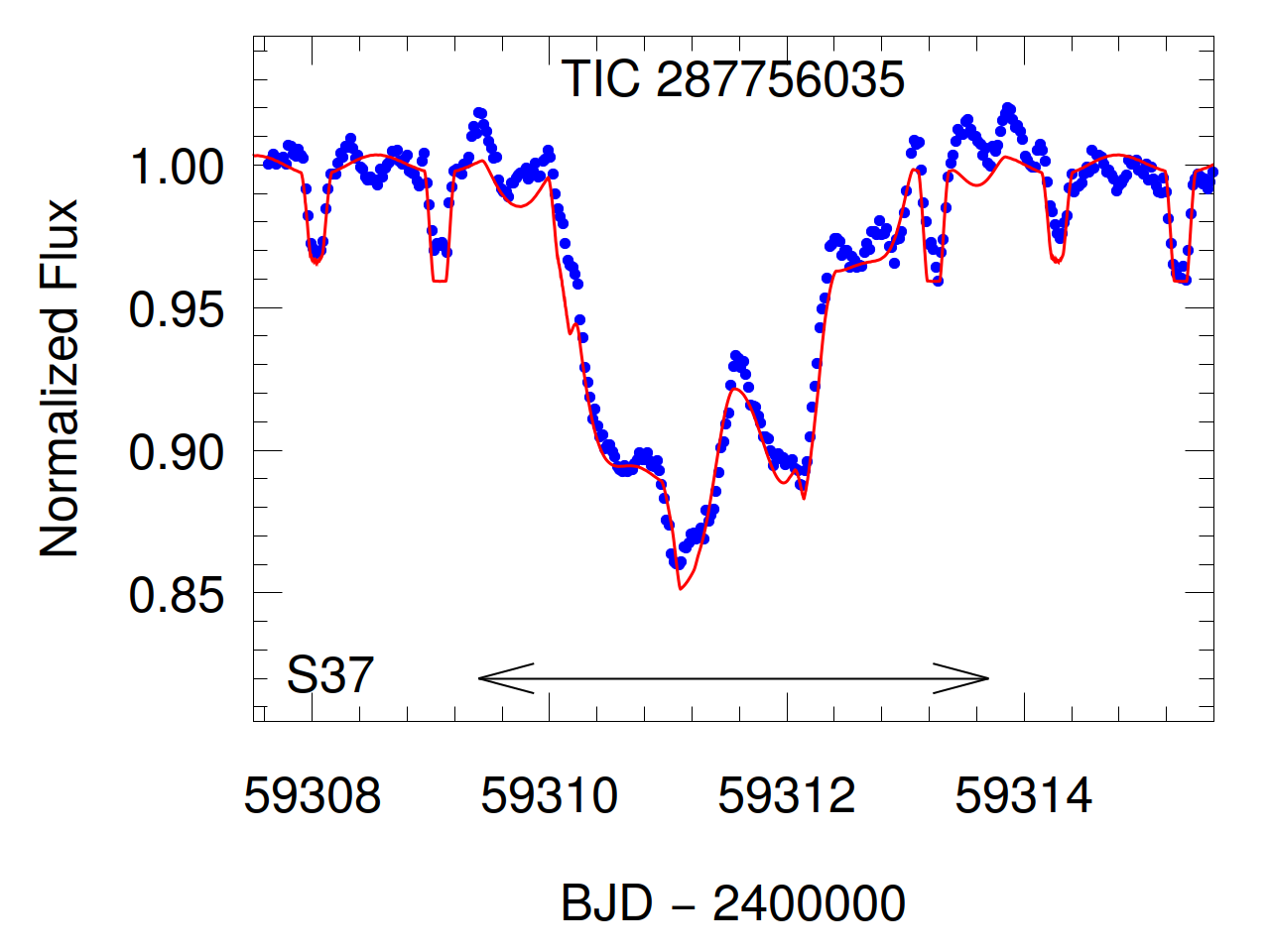}
      \includegraphics[width=0.35\textwidth]{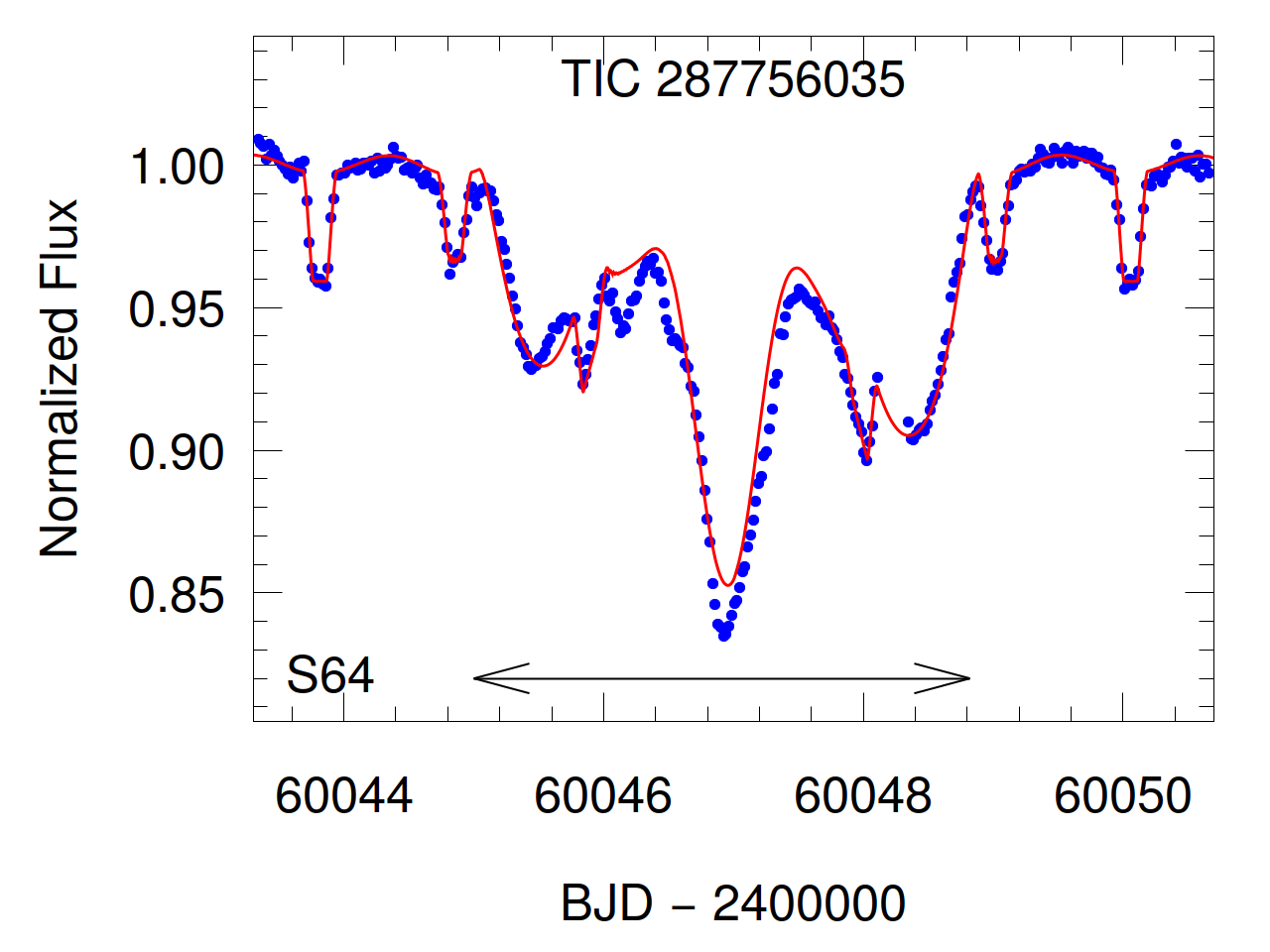}
      \caption{Light curves (blue points) and model fits (smooth red curves) for the third-body eclipses of TIC 287756035 (marked by horizontal arrows). The other notation is the same as in Fig.\,\ref{fig:133771812lcs}.}
         \label{fig:287756035lcs}
   \end{figure*}

\begin{figure*}
   \centering
    \includegraphics[width=0.35\textwidth]{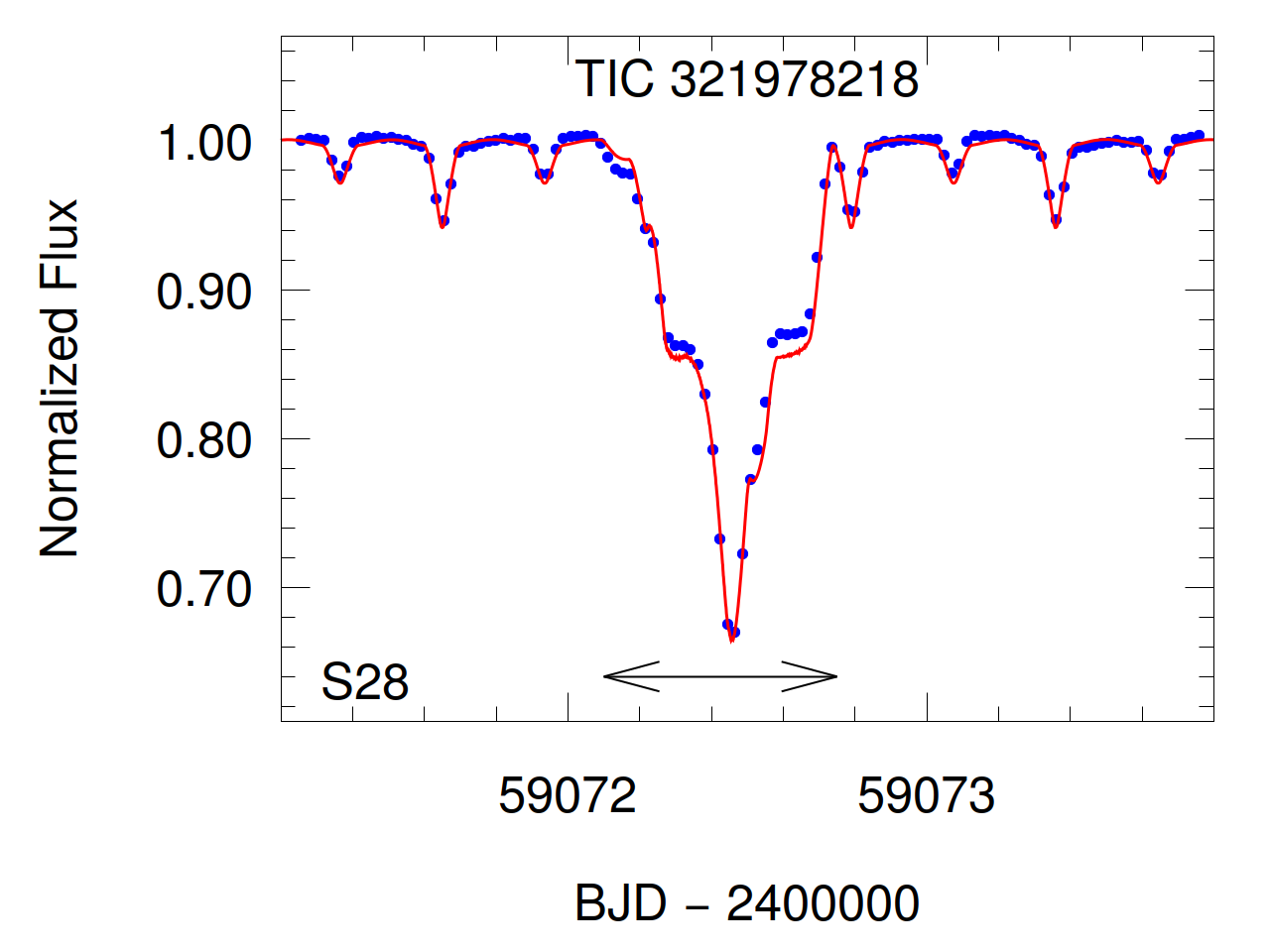}
      \includegraphics[width=0.35\textwidth]{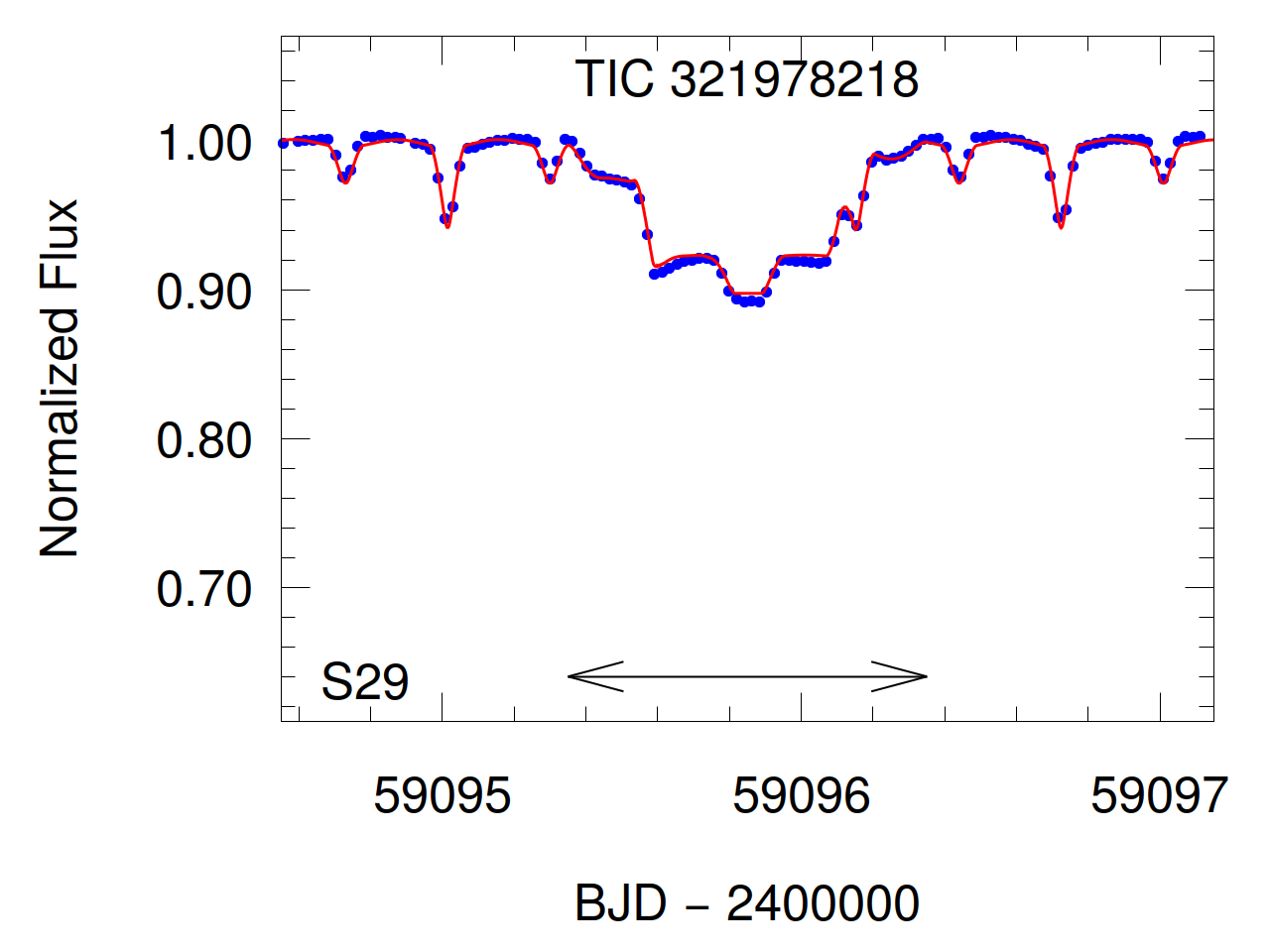}
      \includegraphics[width=0.35\textwidth]{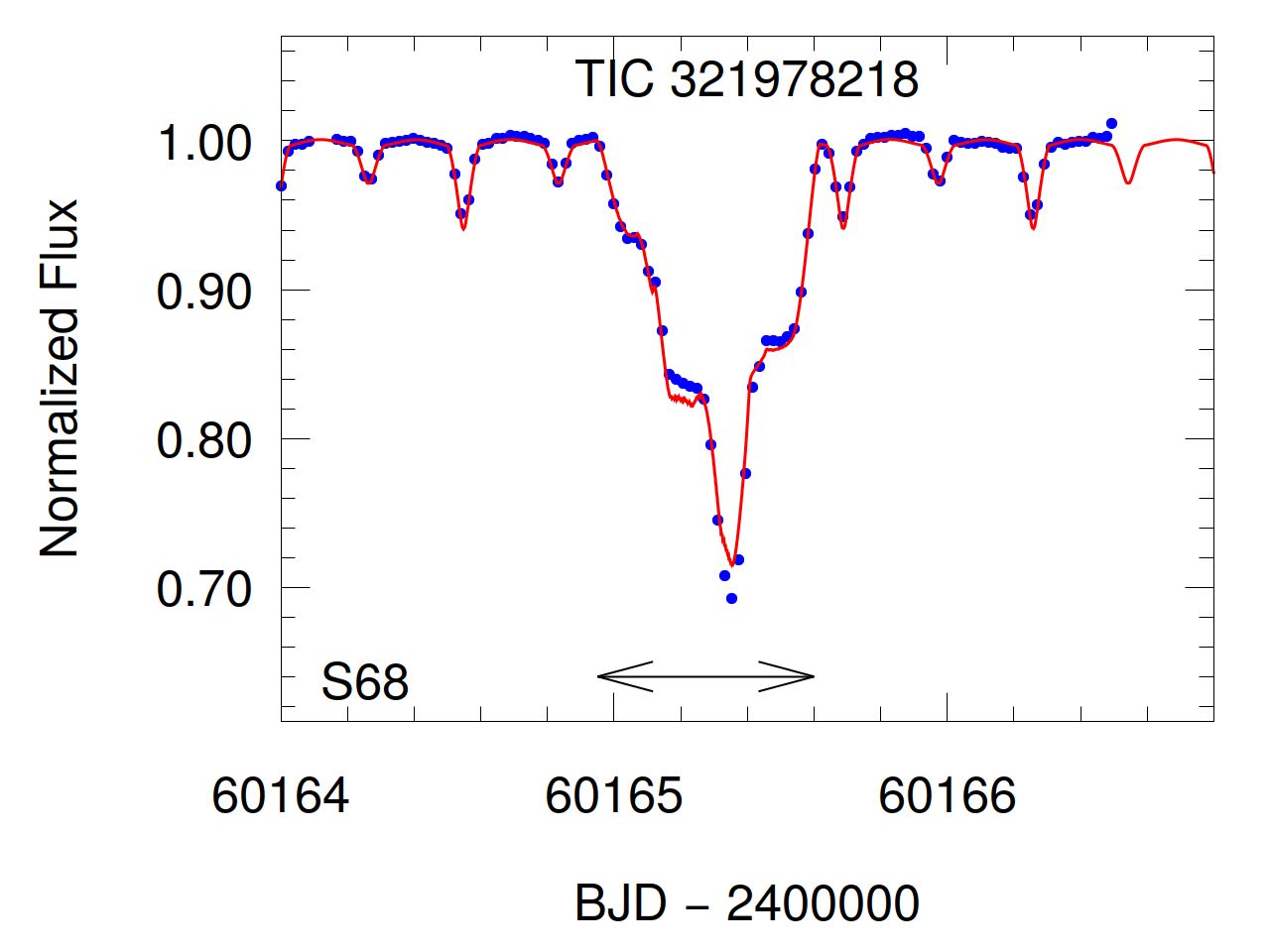}
      \includegraphics[width=0.35\textwidth]{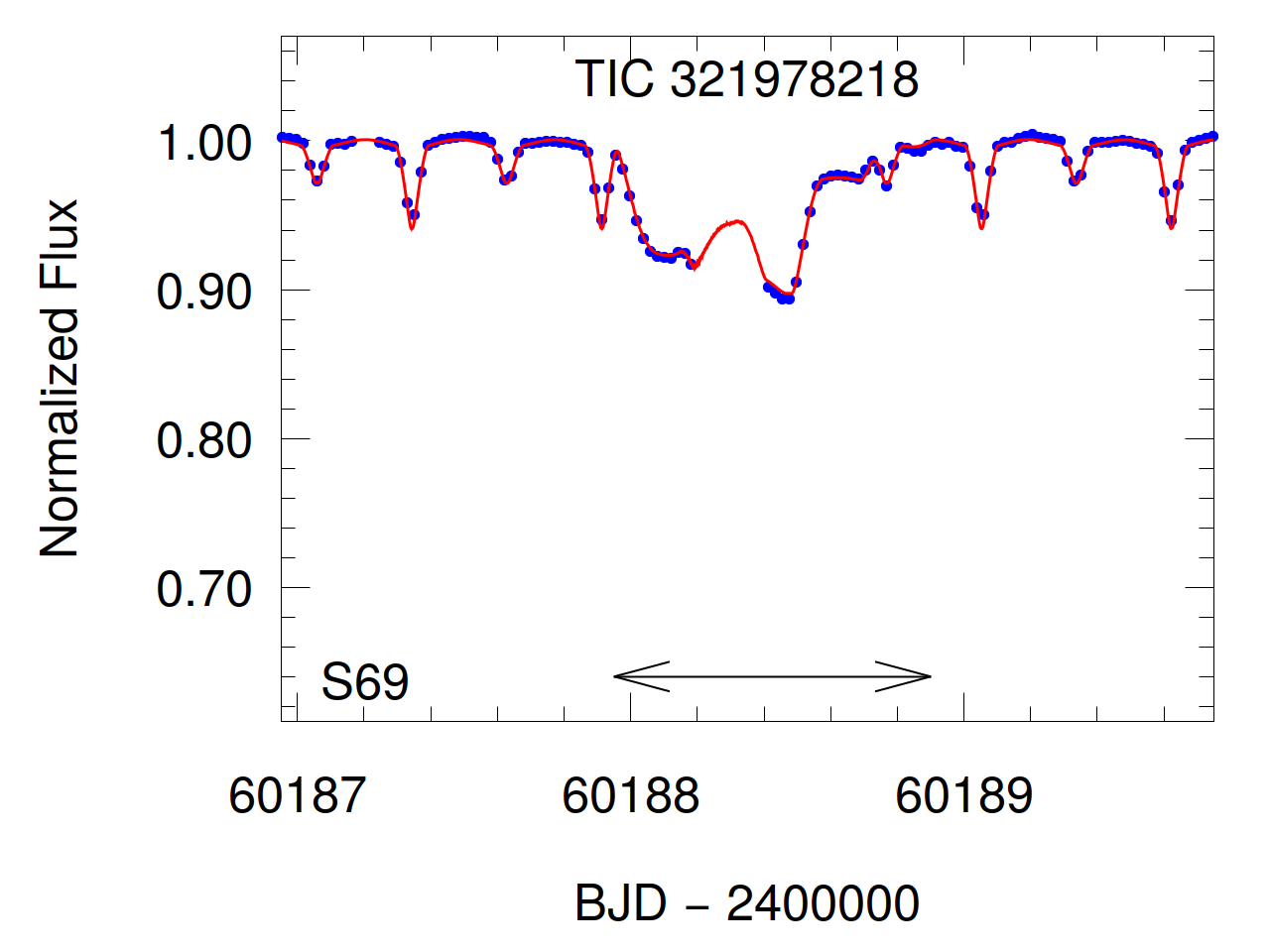}
      \caption{Light curves (blue points) and model fits (smooth red curves) for the TESS-observed third-body eclipses of TIC 321978218 (marked by horizontal arrows). The other notation is the same as in Fig.\,\ref{fig:133771812lcs}.}
         \label{fig:321978218lcs}
   \end{figure*}

\begin{figure*}
   \centering
   \includegraphics[width=0.3\textwidth]{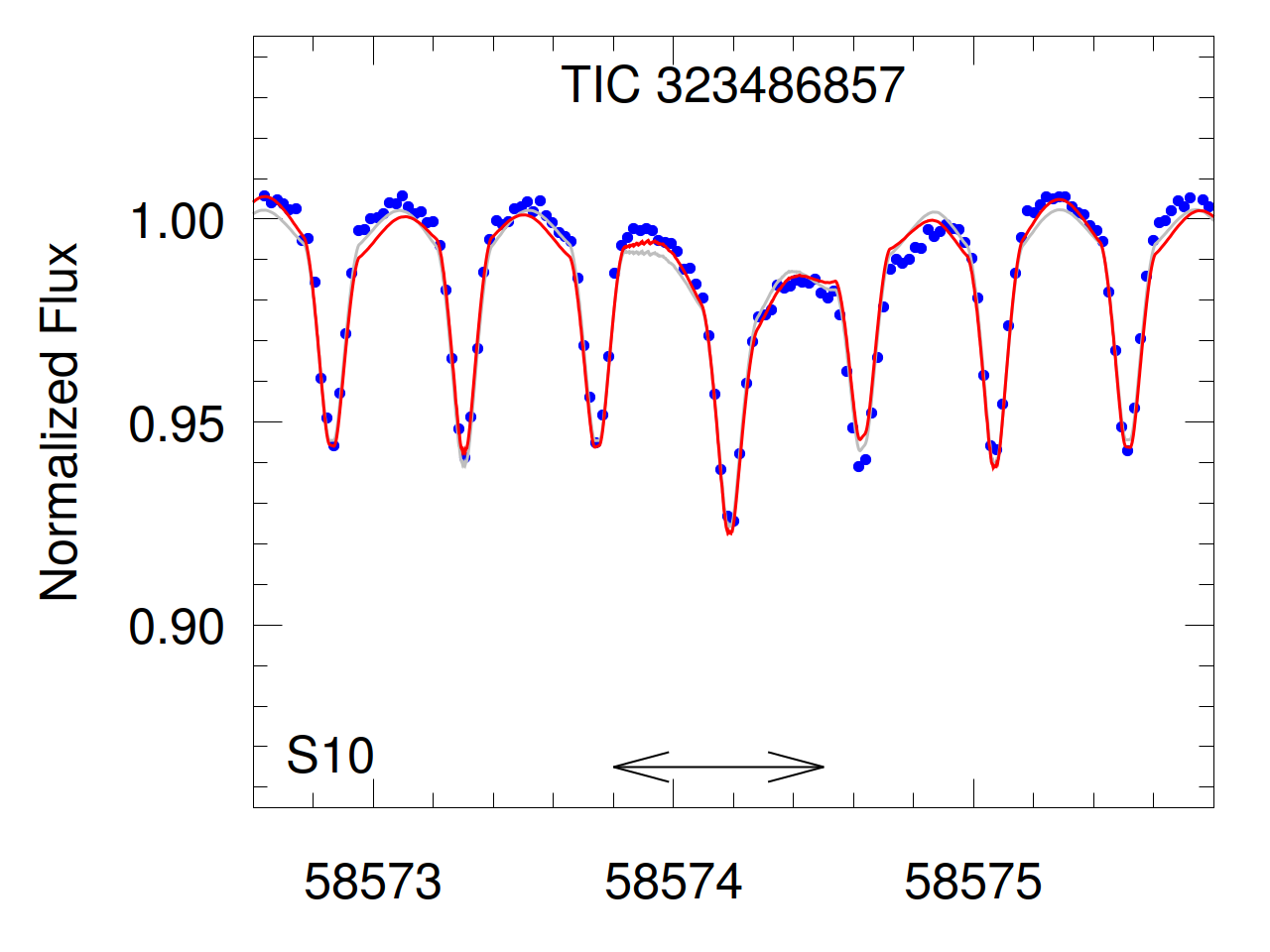}
    \includegraphics[width=0.3\textwidth]{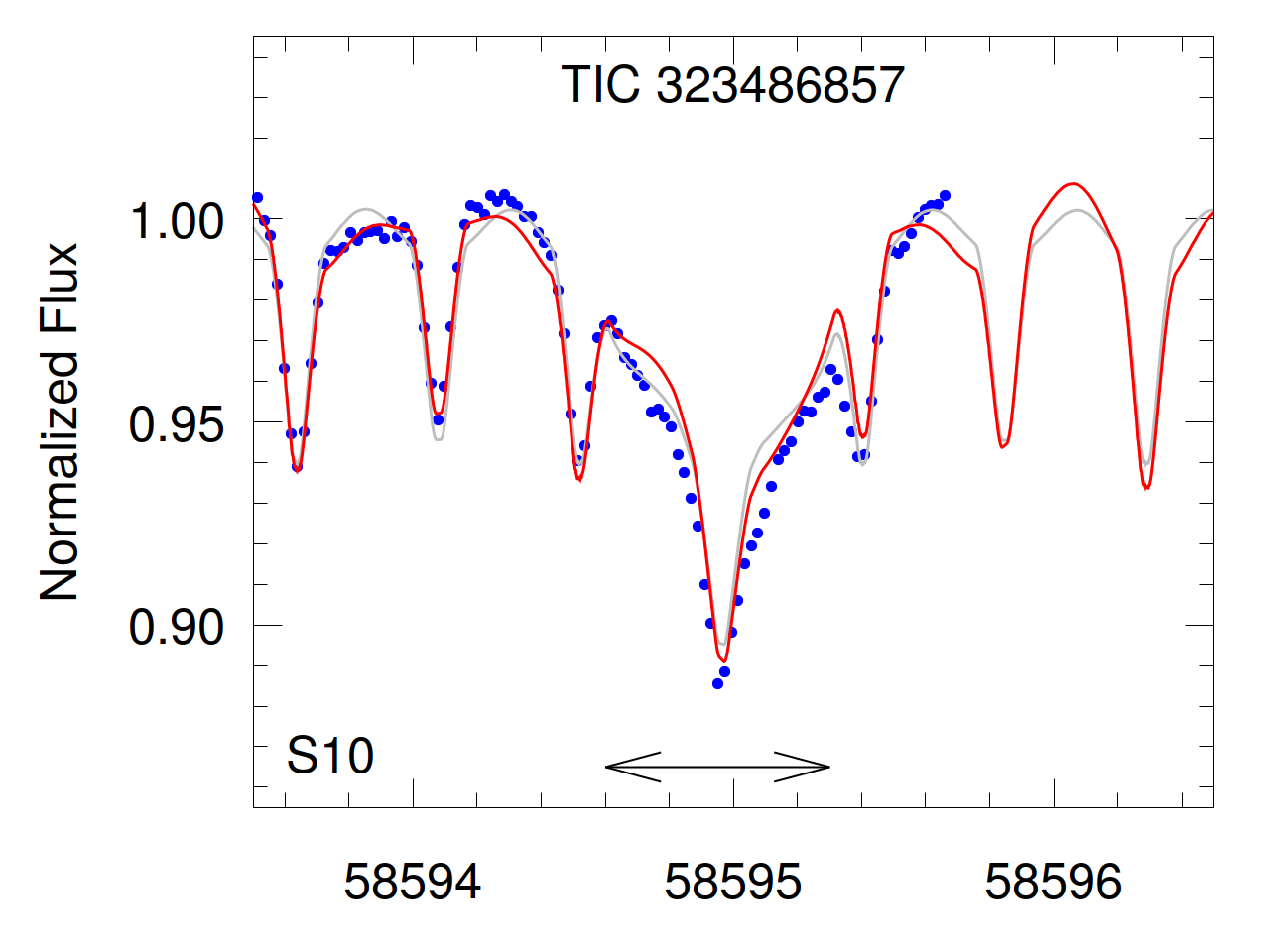}
     \includegraphics[width=0.3\textwidth]{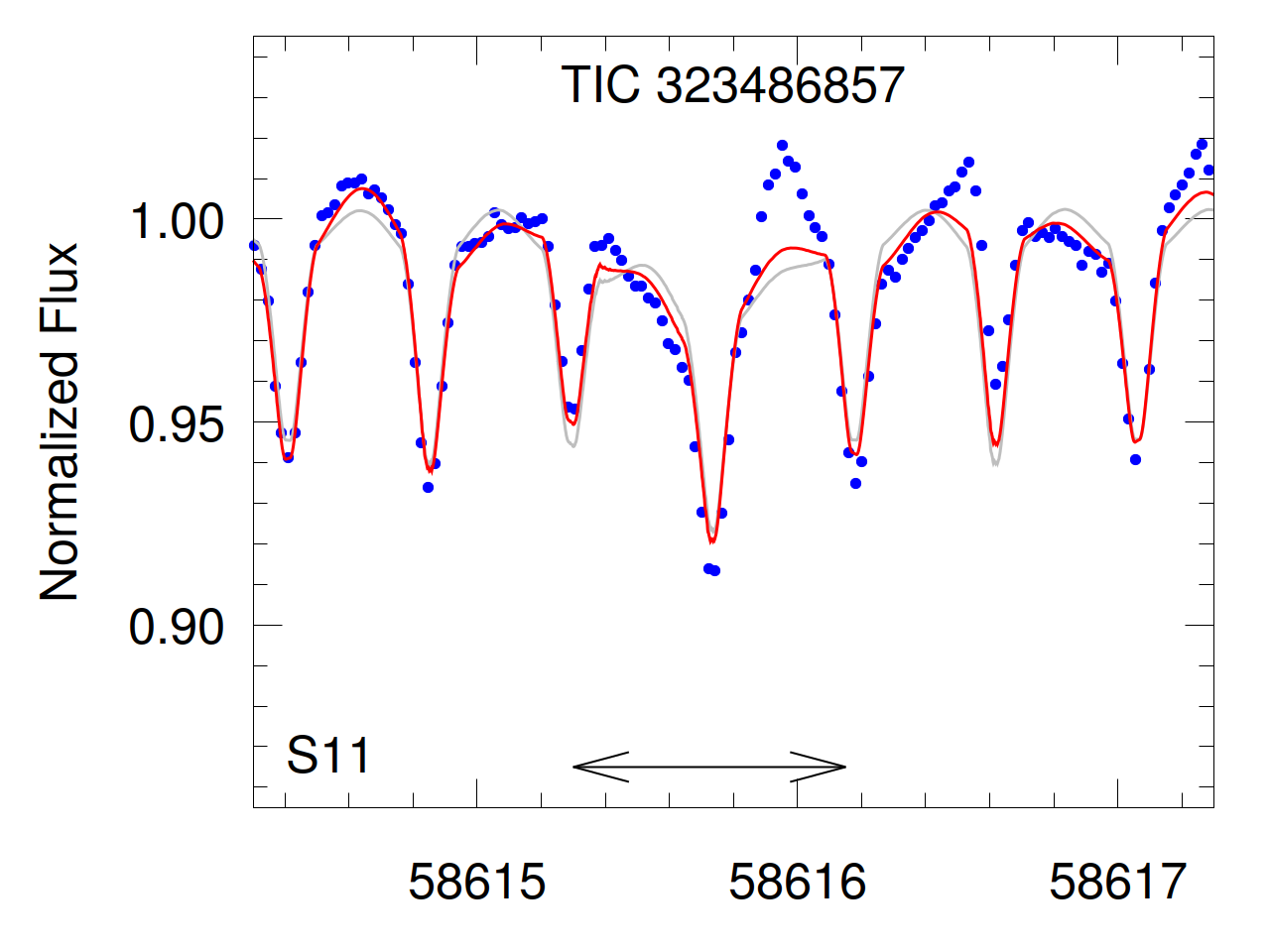}
      \includegraphics[width=0.3\textwidth]{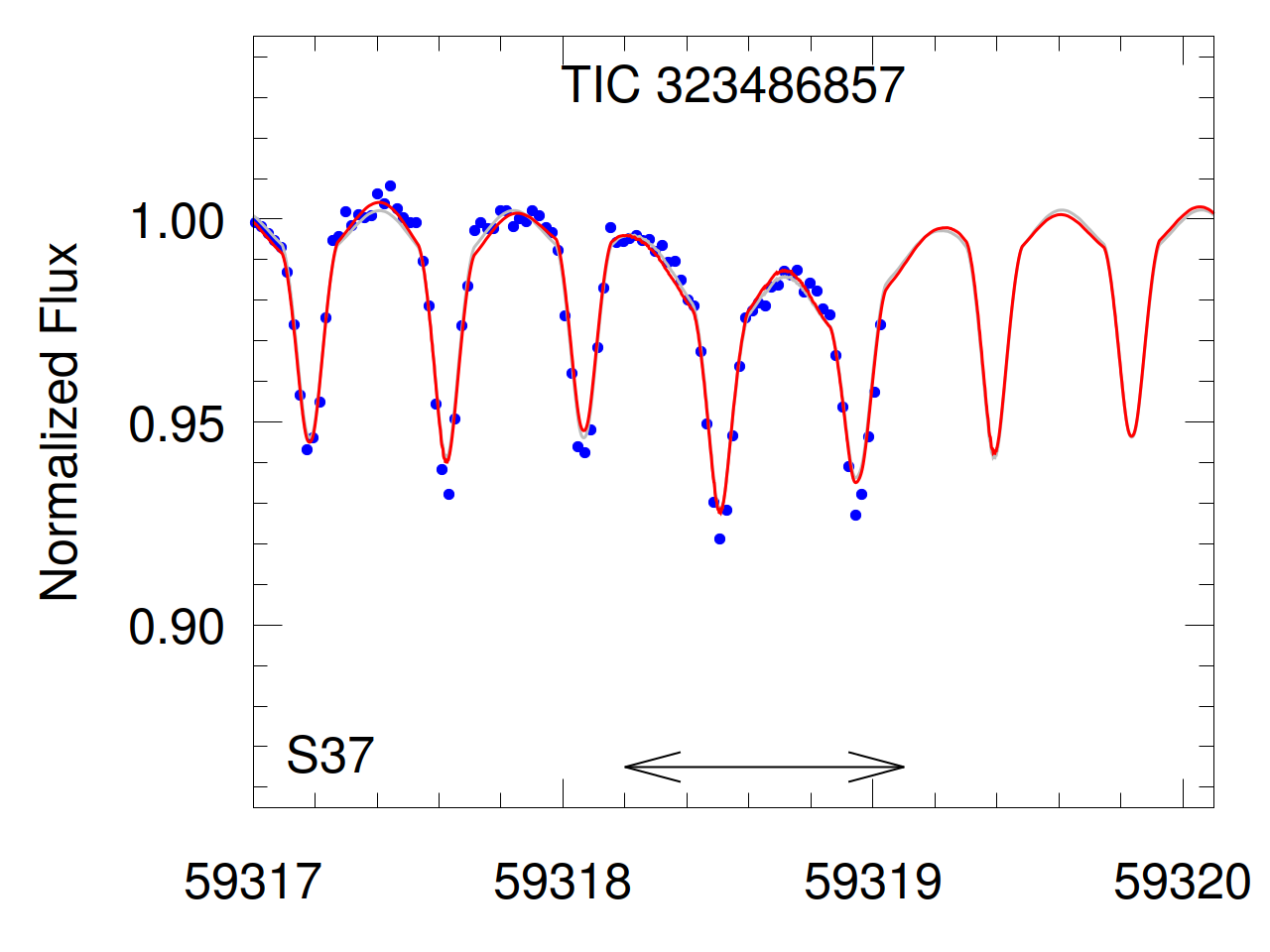}
       \includegraphics[width=0.3\textwidth]{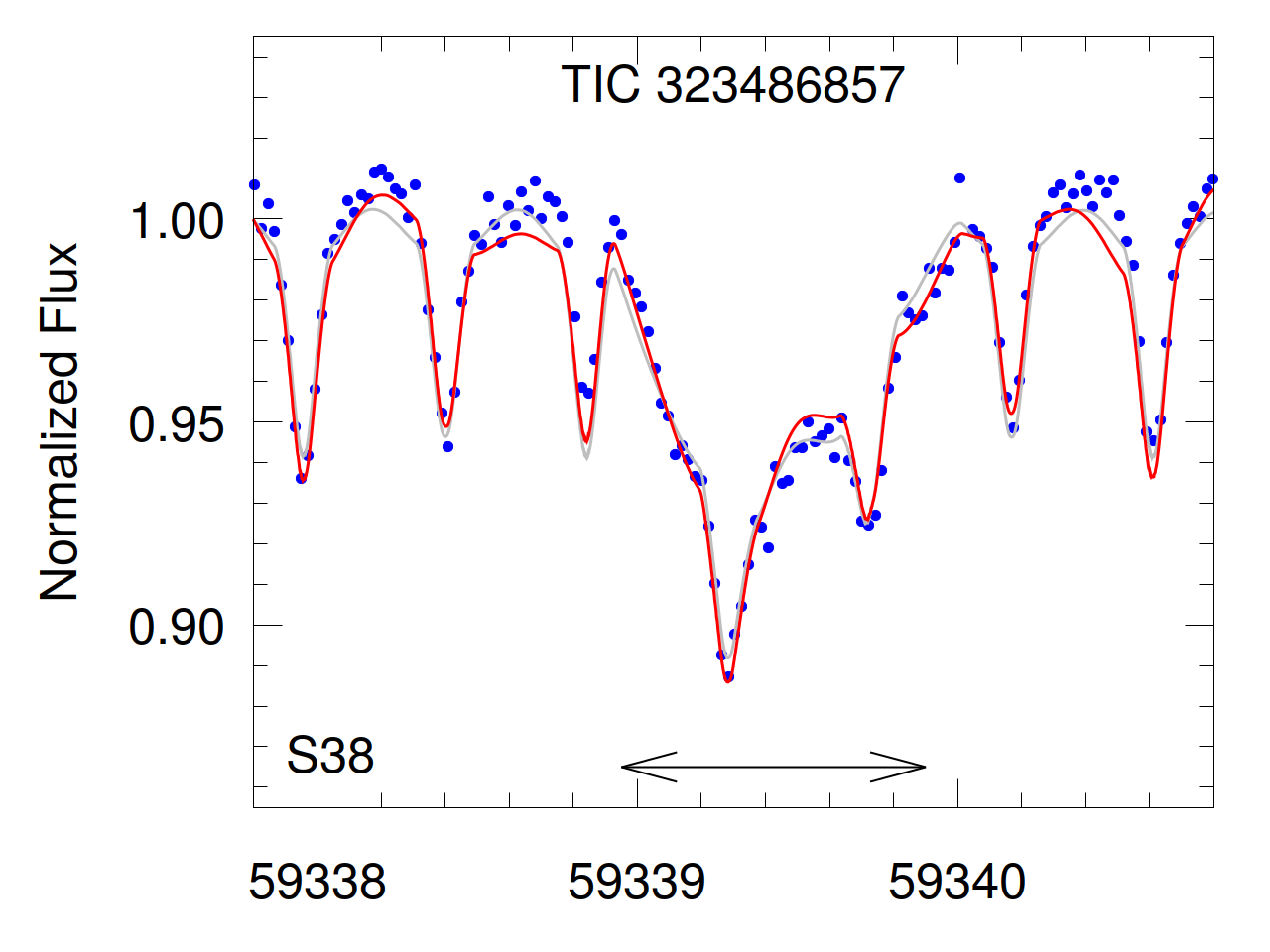}
        \includegraphics[width=0.3\textwidth]{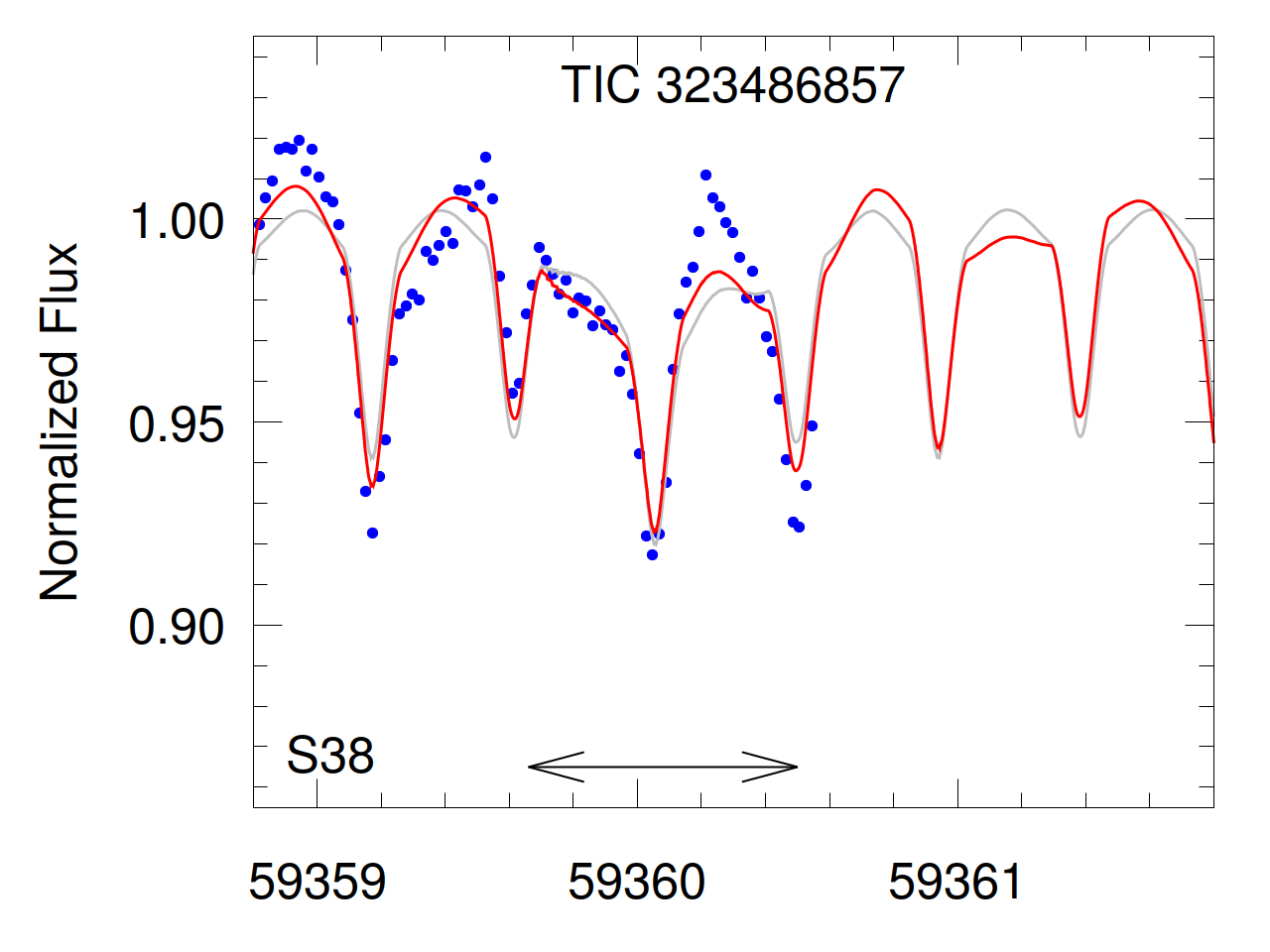}
       \includegraphics[width=0.3\textwidth]{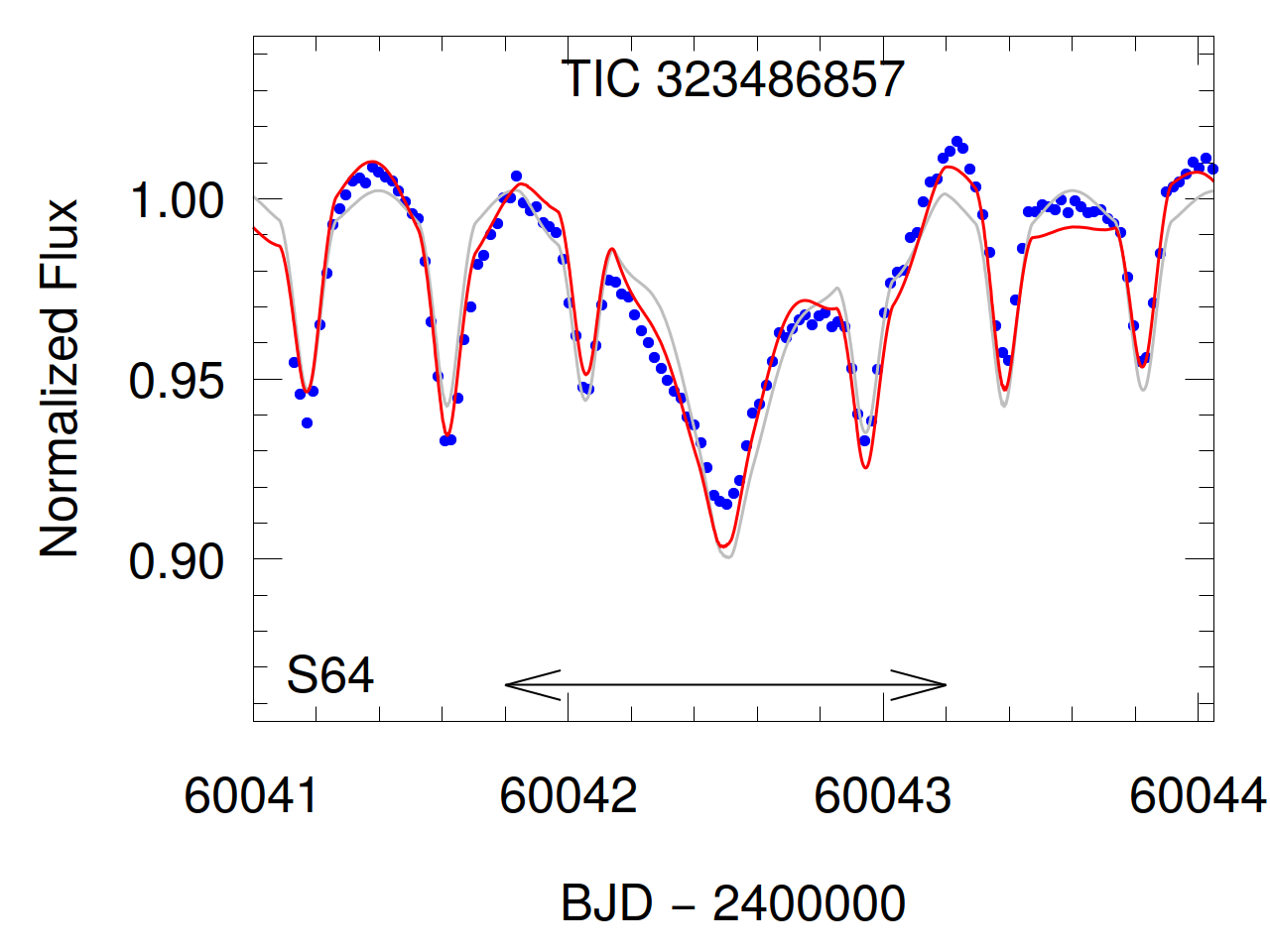}
       \includegraphics[width=0.3\textwidth]{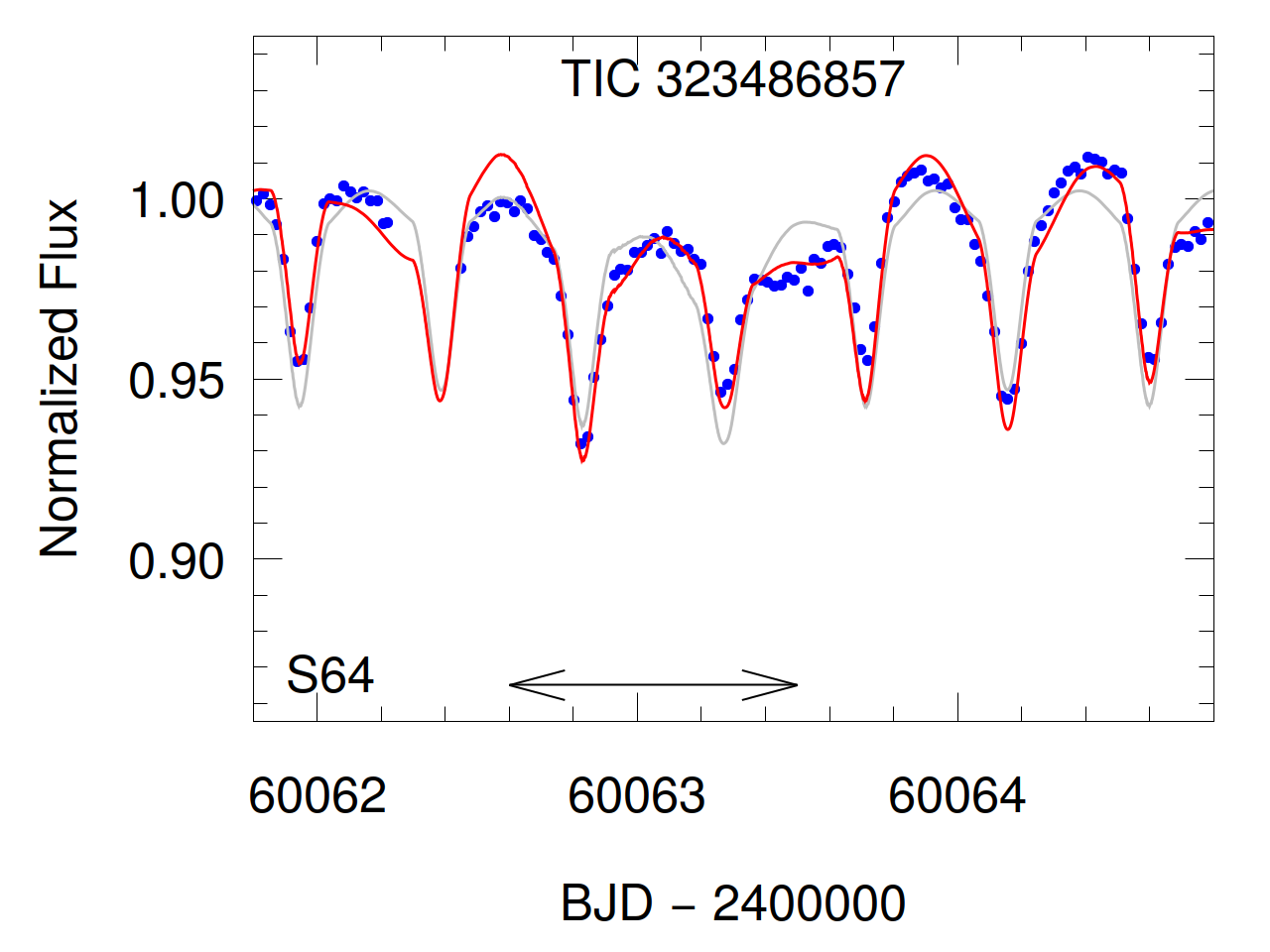}
        \includegraphics[width=0.3\textwidth]{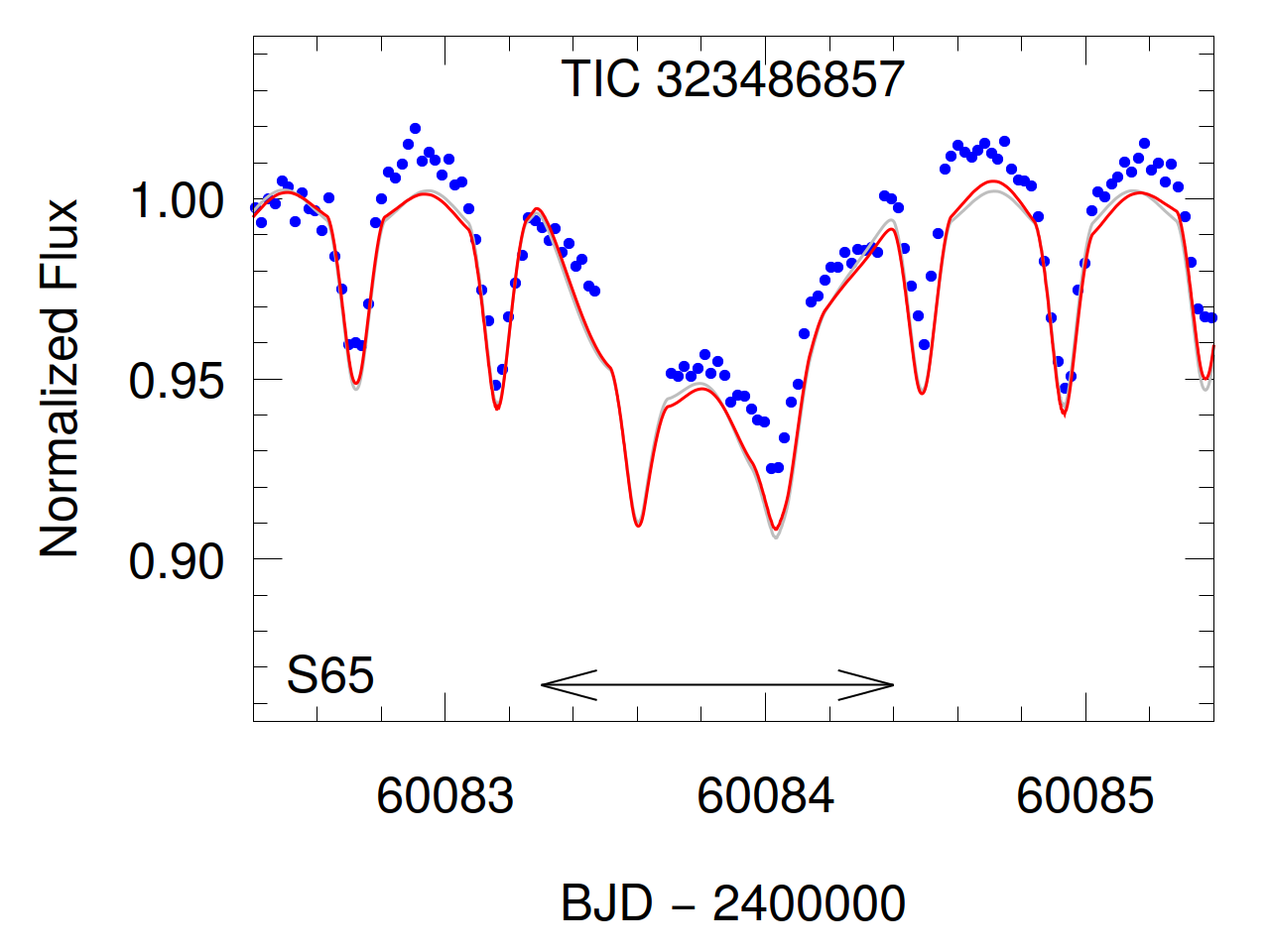}

      \caption{Light curves (blue points) and two kinds of model fits for TIC 323486857. The light-gray fit represents the pure three-body model light curve, while the red fit is the net fit of the triple star model and a four-frequency Fourier model of the light curve distortions (see text for details). The region of the third-body eclipses is indicated with the horizontal arrows. The other notation is the same as in Fig.\,\ref{fig:133771812lcs}.}
         \label{fig:323486857lcs}
   \end{figure*}

\begin{figure*}
   \centering
   \includegraphics[width=0.3\textwidth]{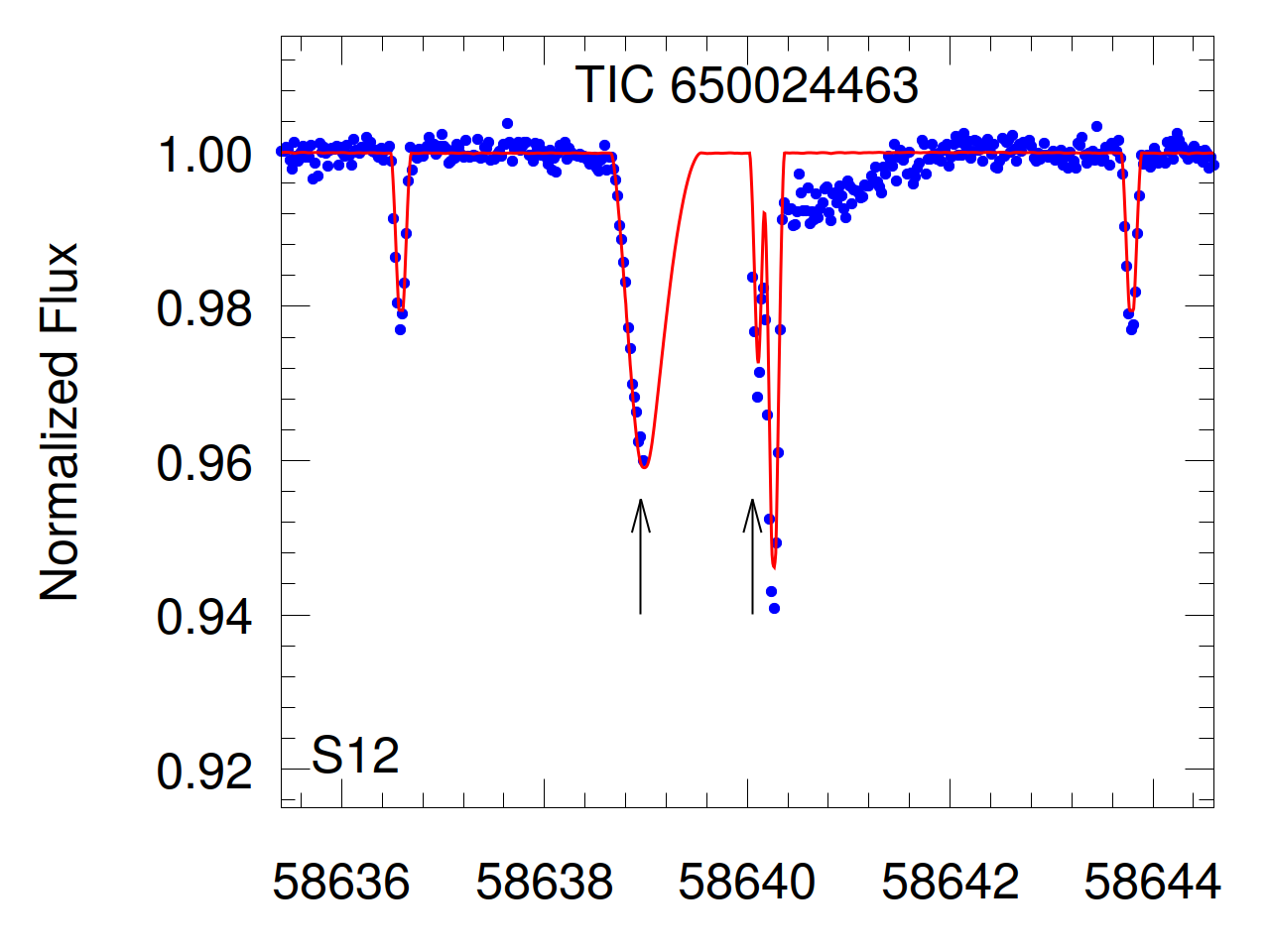}
    \includegraphics[width=0.3\textwidth]{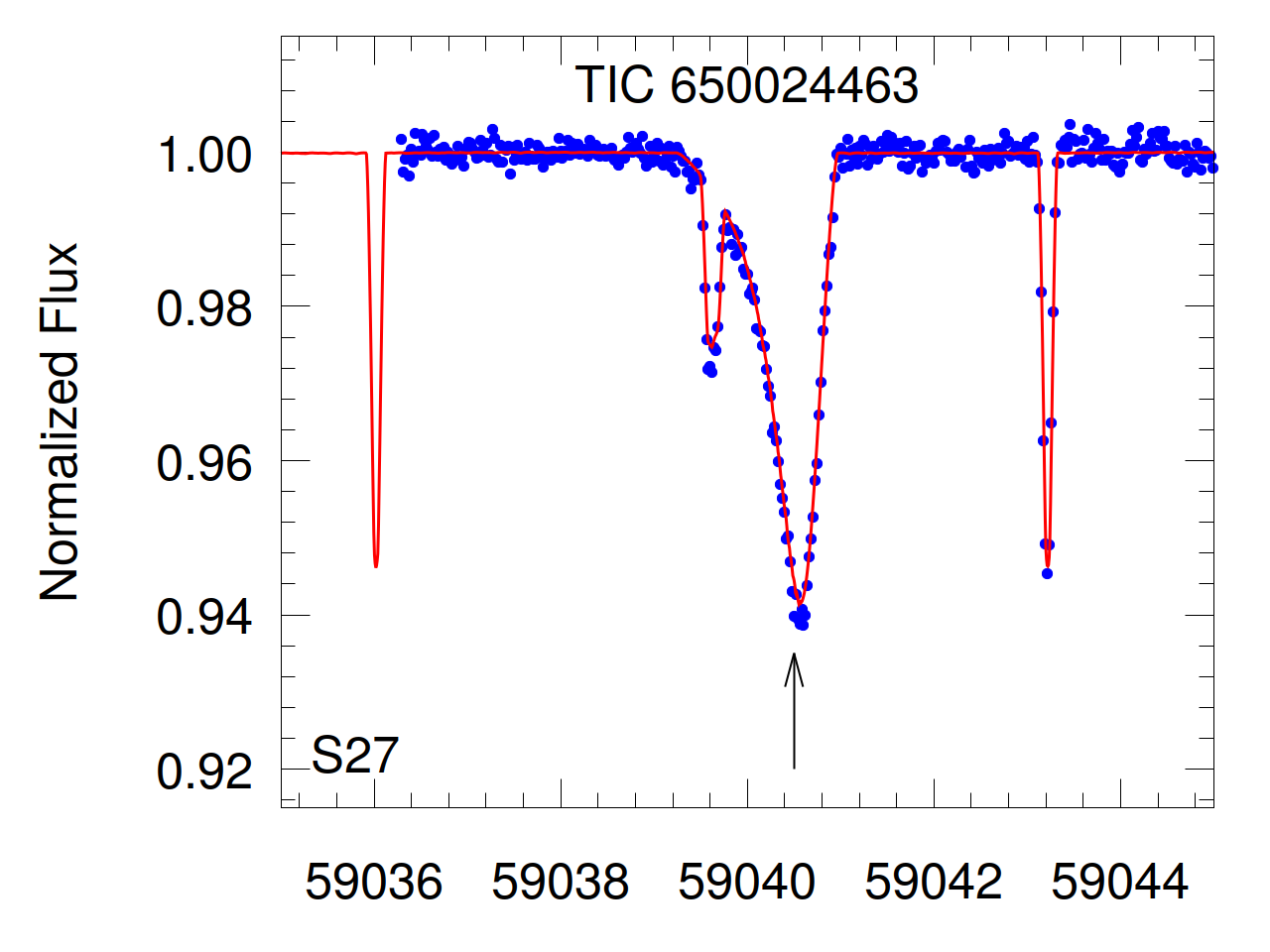}
     \includegraphics[width=0.3\textwidth]{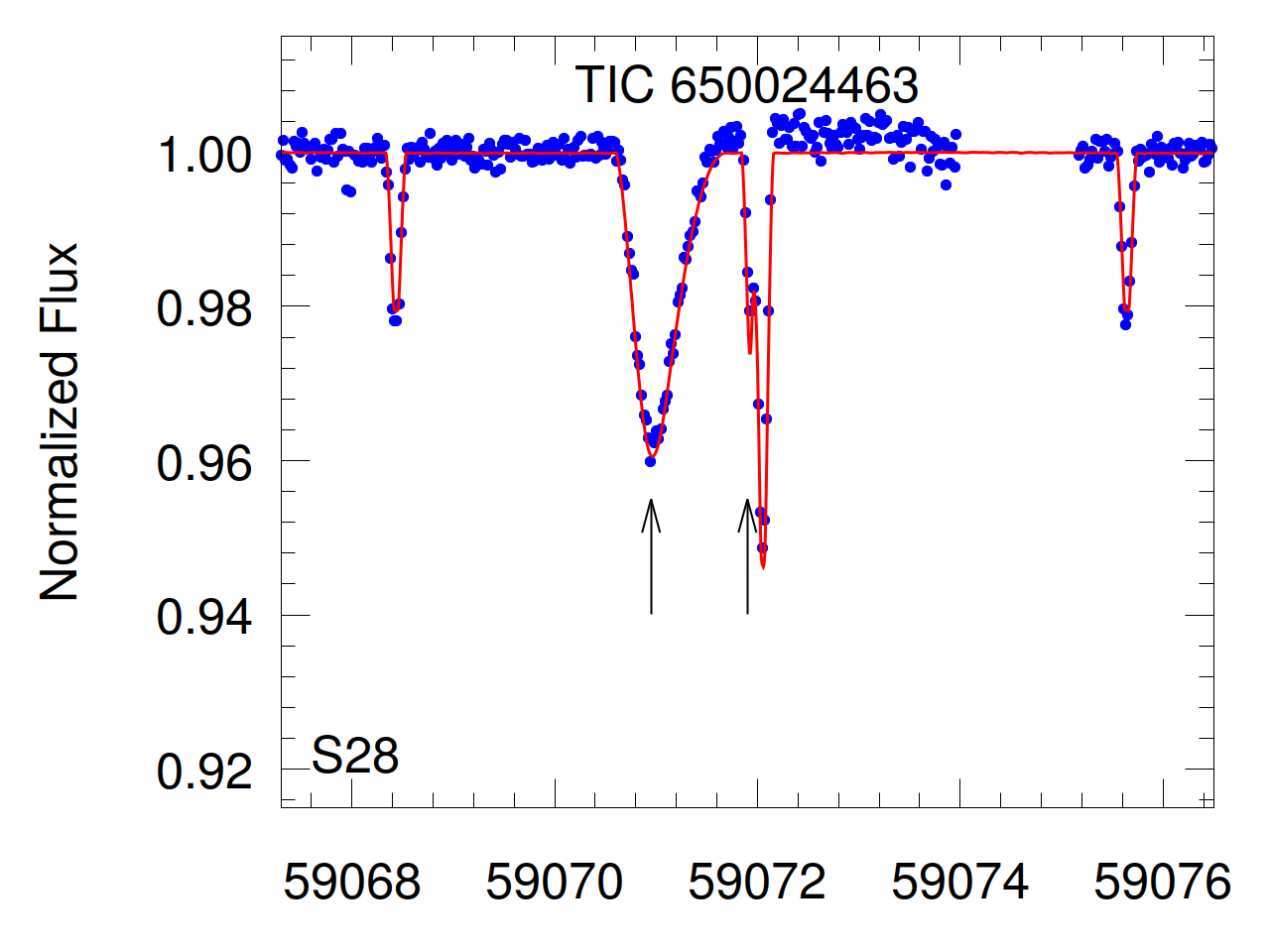}
      \includegraphics[width=0.3\textwidth]{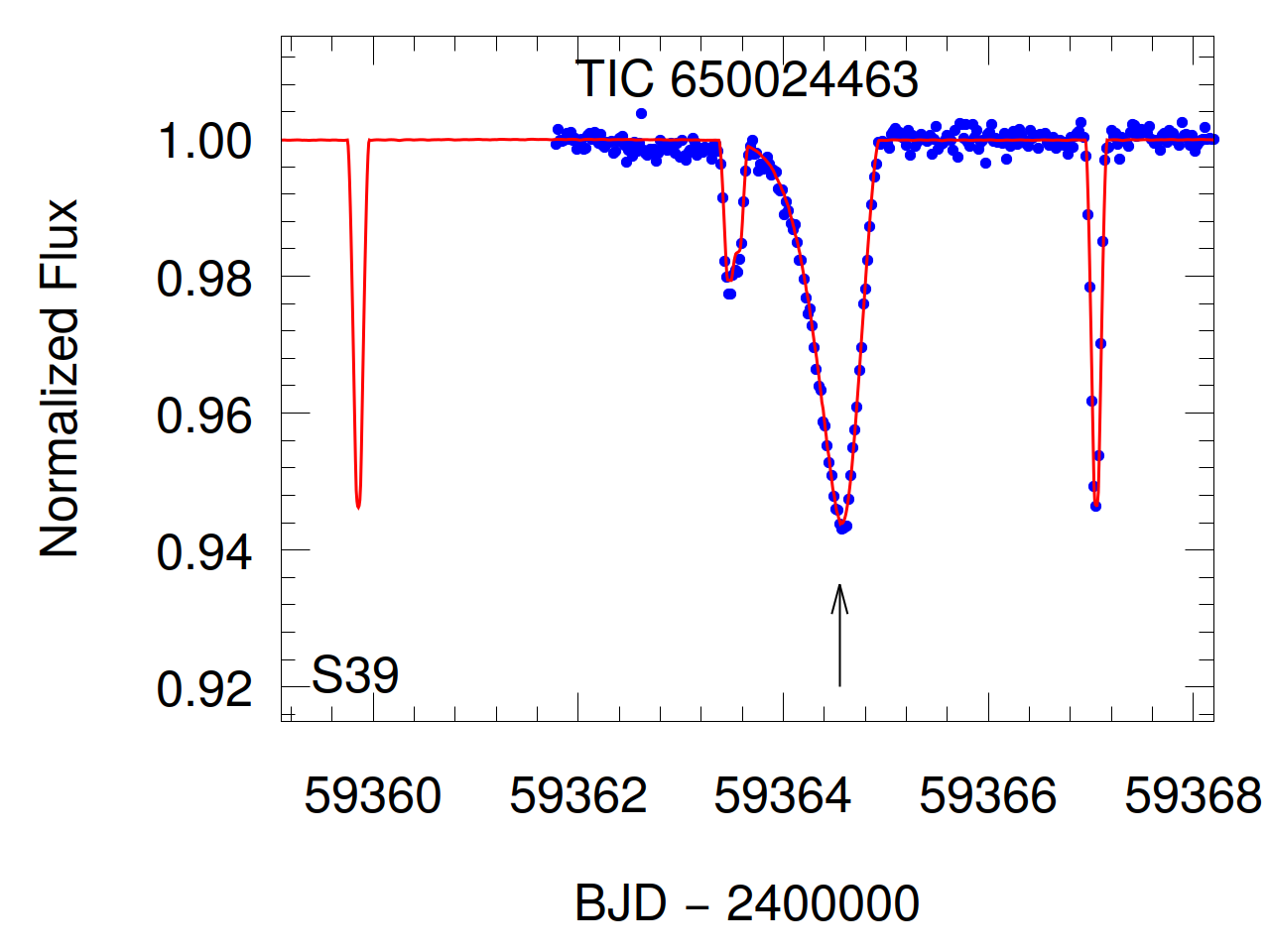}
       \includegraphics[width=0.3\textwidth]{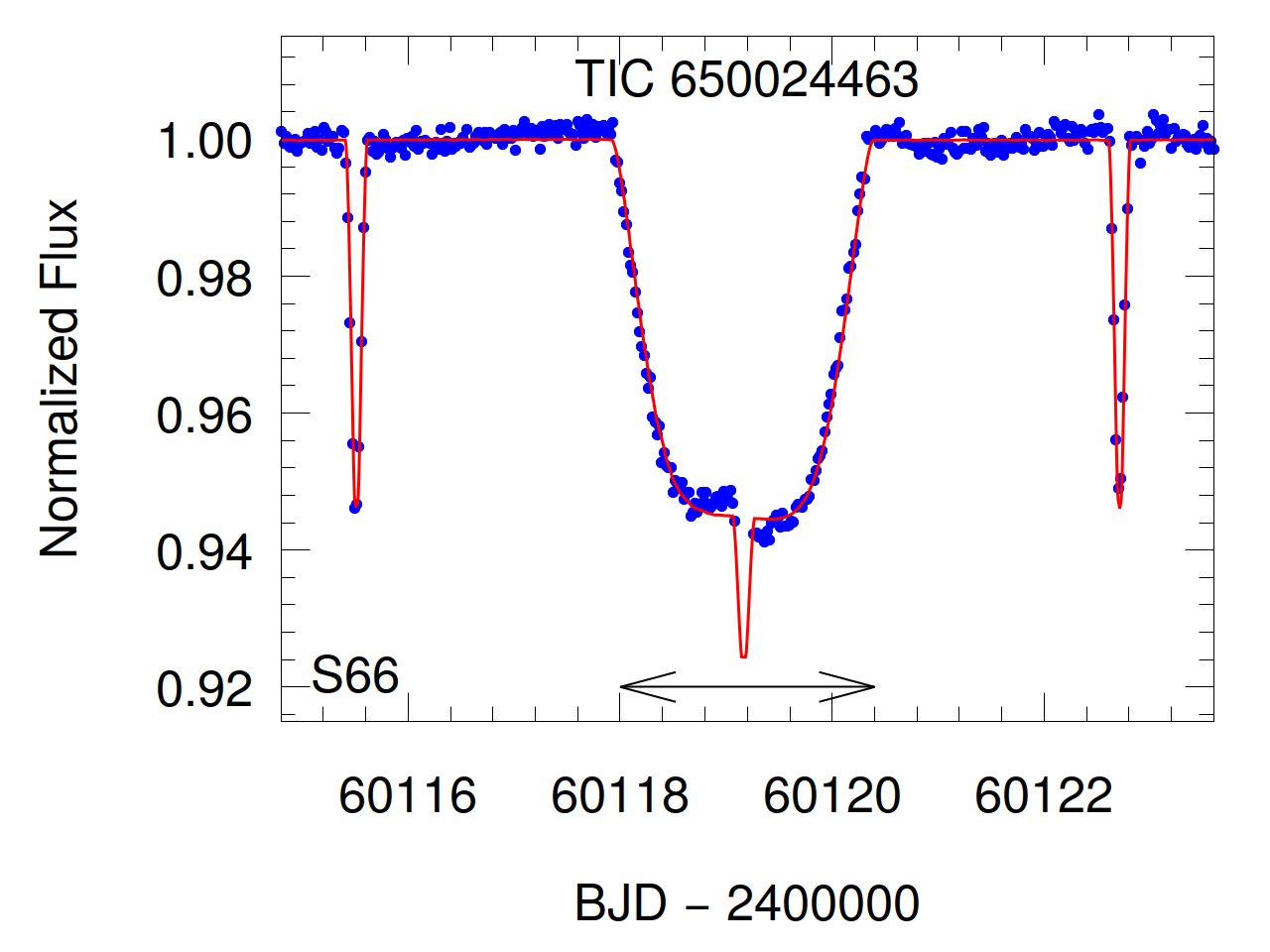}
        \includegraphics[width=0.3\textwidth]{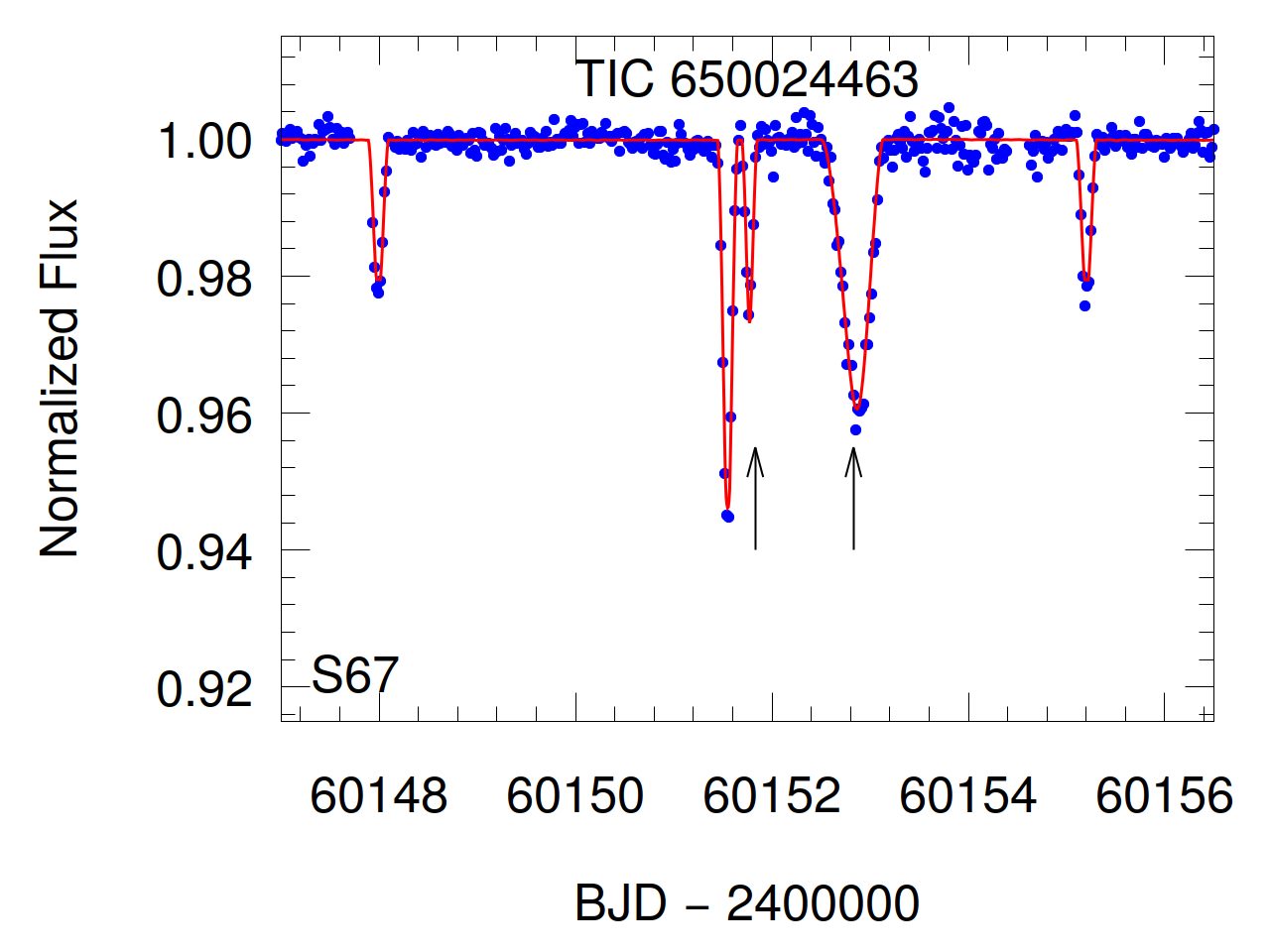}
      \caption{Light curves (blue points) and model fits (smooth red curves) for TIC 650024463. The vertical and horizontal arrows mark the times of third body eclipses. The other notation is the same as in Fig.\,\ref{fig:133771812lcs}.}
         \label{fig:650024463lcs}
   \end{figure*}

\begin{figure*}
   \centering
   \includegraphics[width=0.3\textwidth]{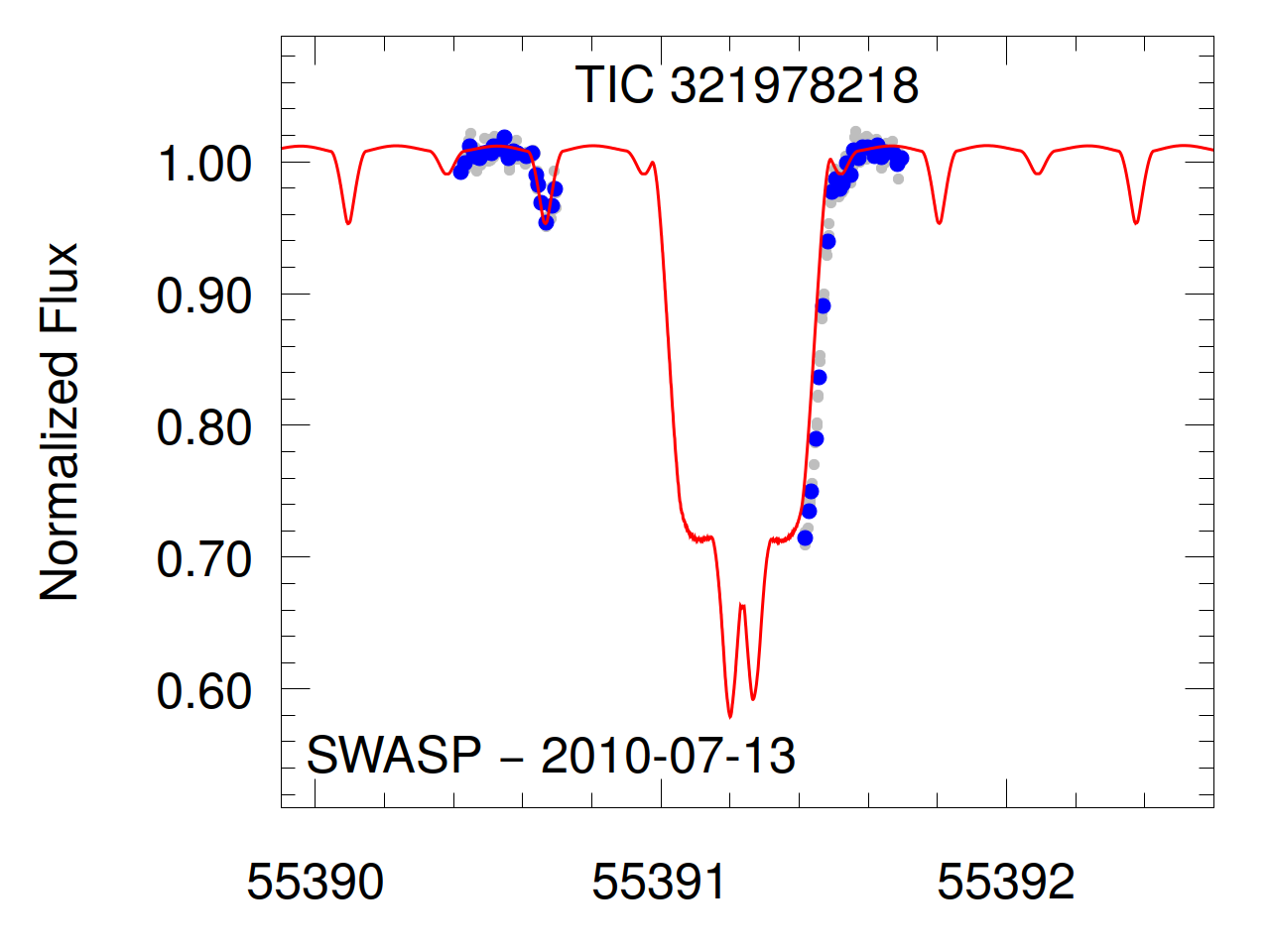}
    \includegraphics[width=0.3\textwidth]{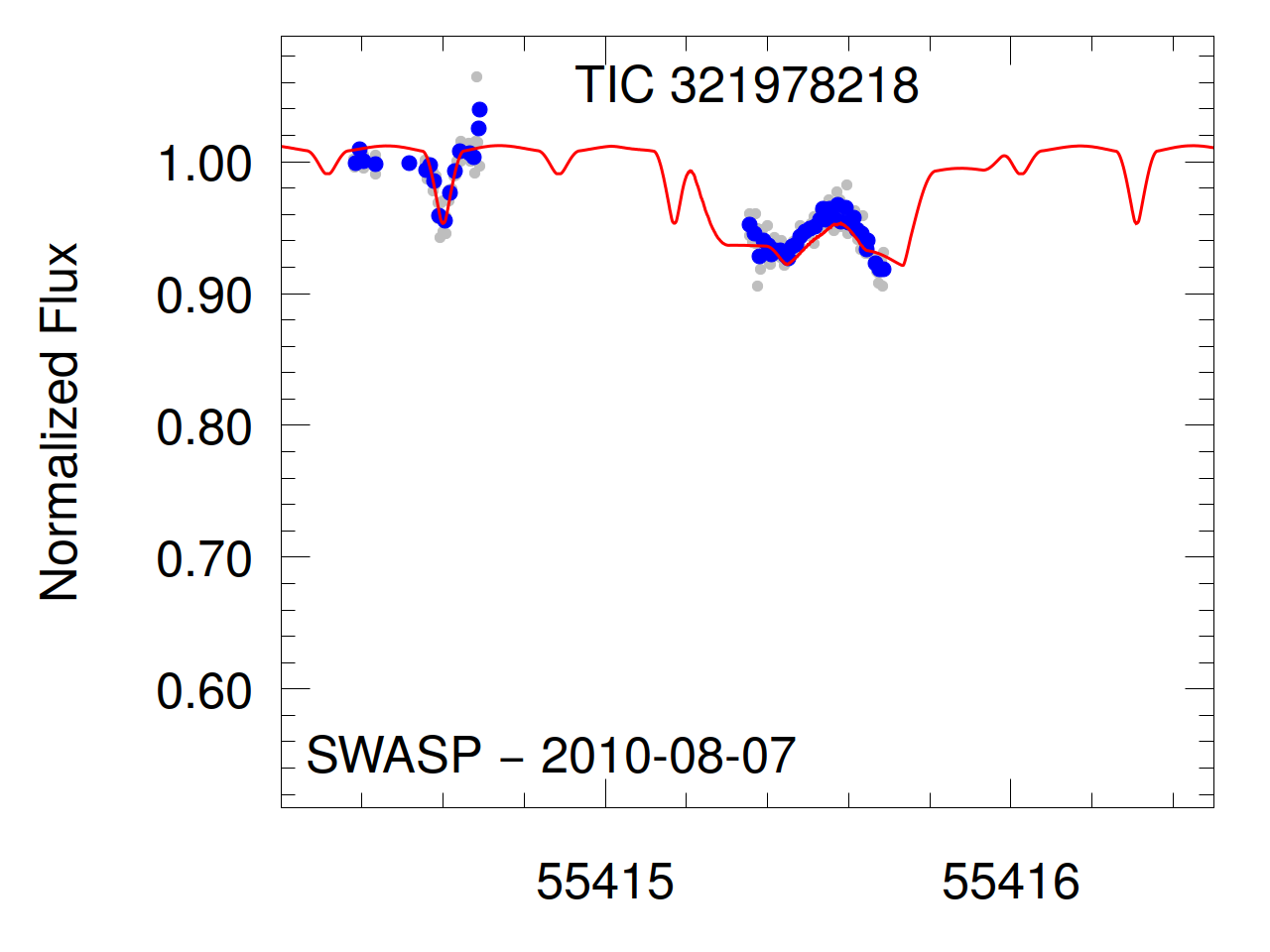}
     \includegraphics[width=0.3\textwidth]{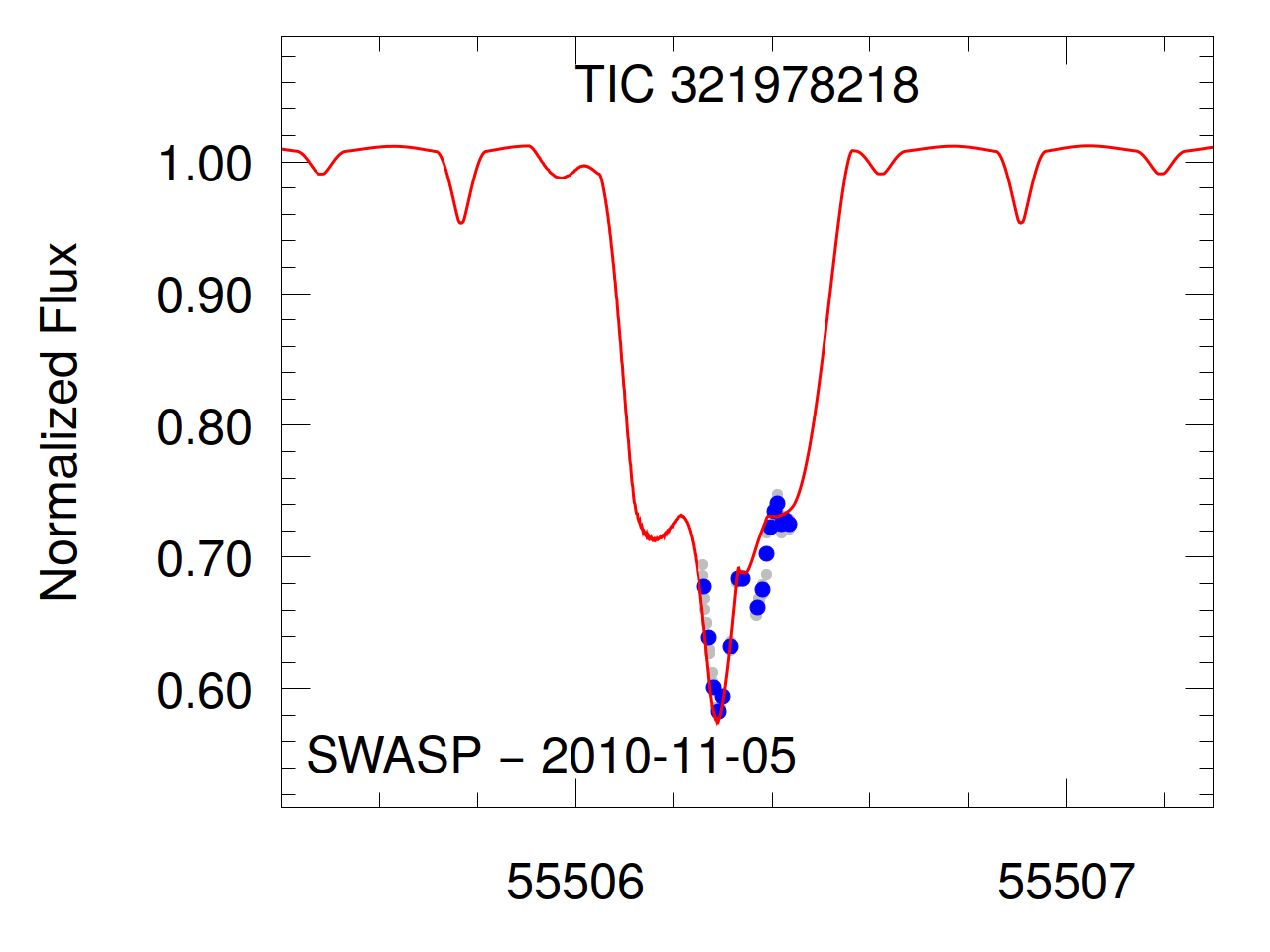}
      \includegraphics[width=0.3\textwidth]{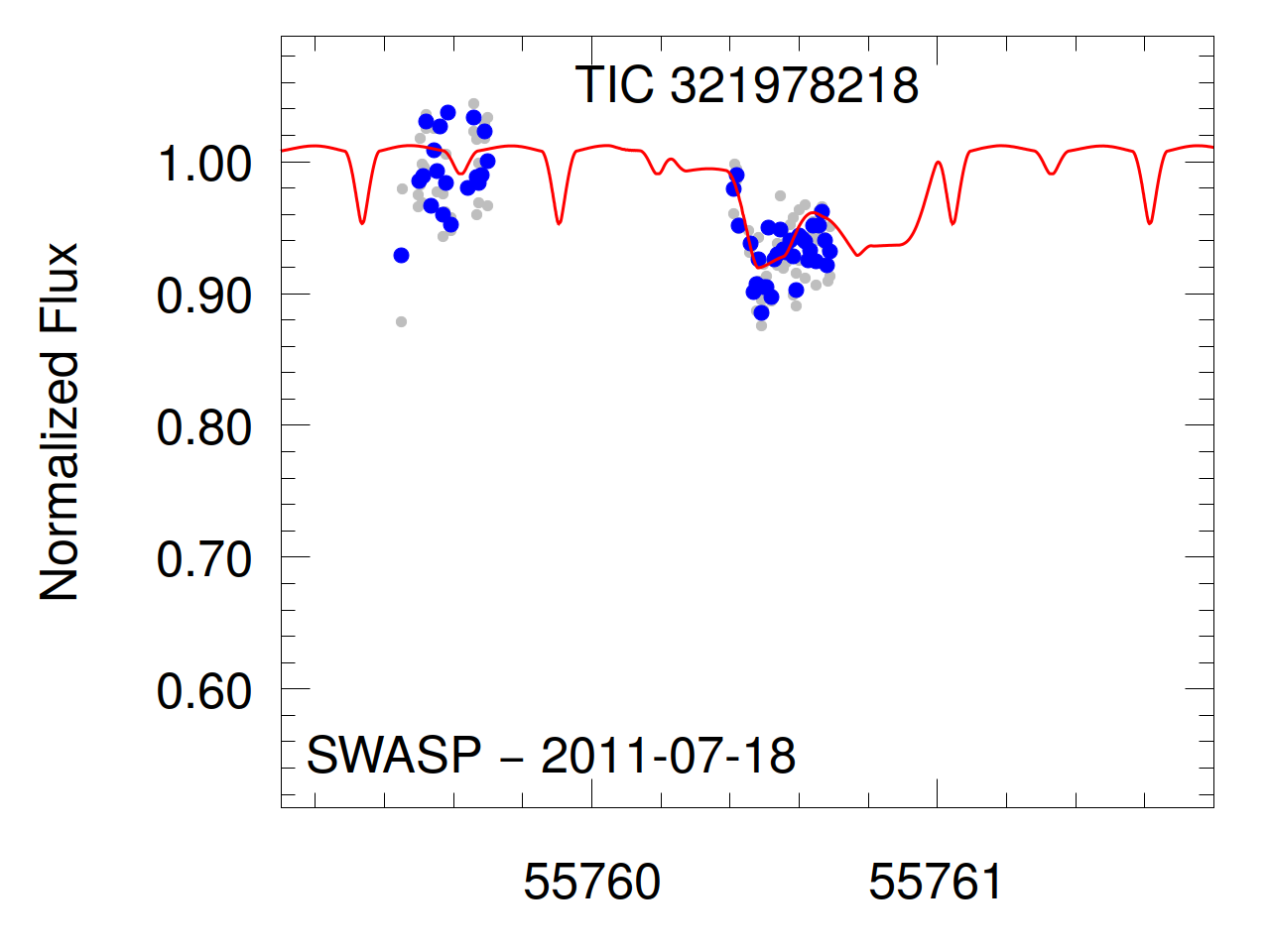}
       \includegraphics[width=0.3\textwidth]{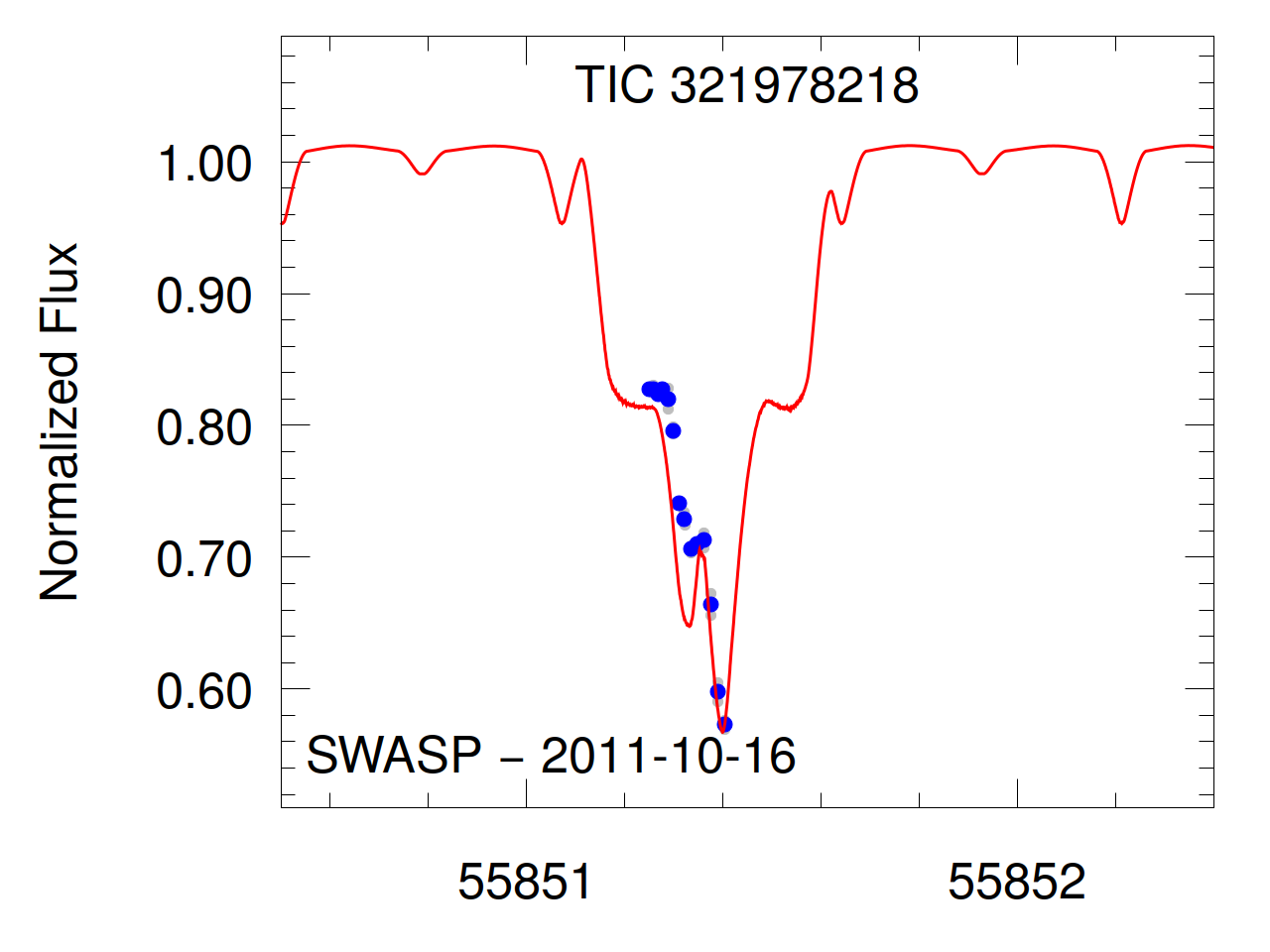}
        \includegraphics[width=0.3\textwidth]{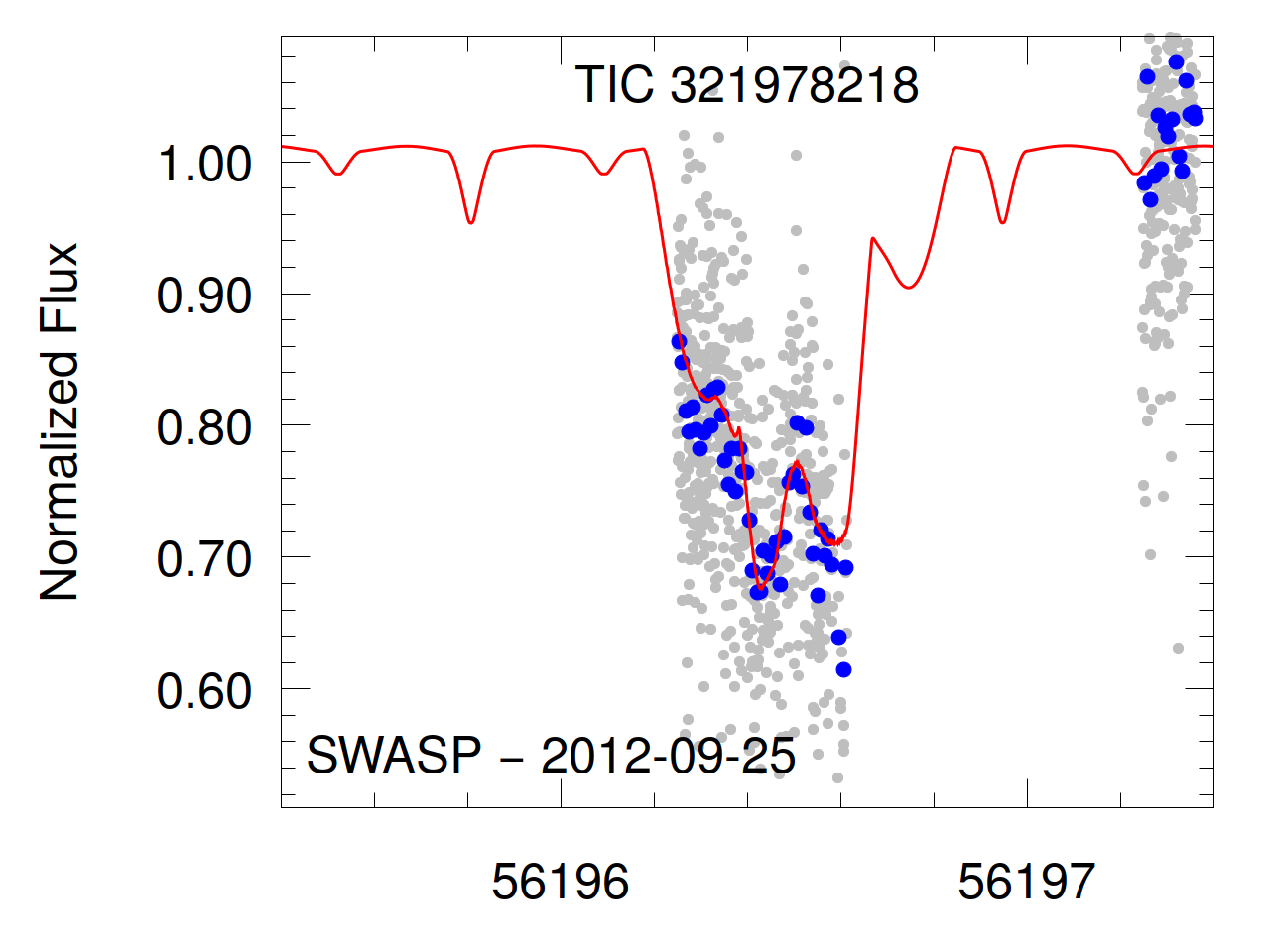}
   \includegraphics[width=0.3\textwidth]{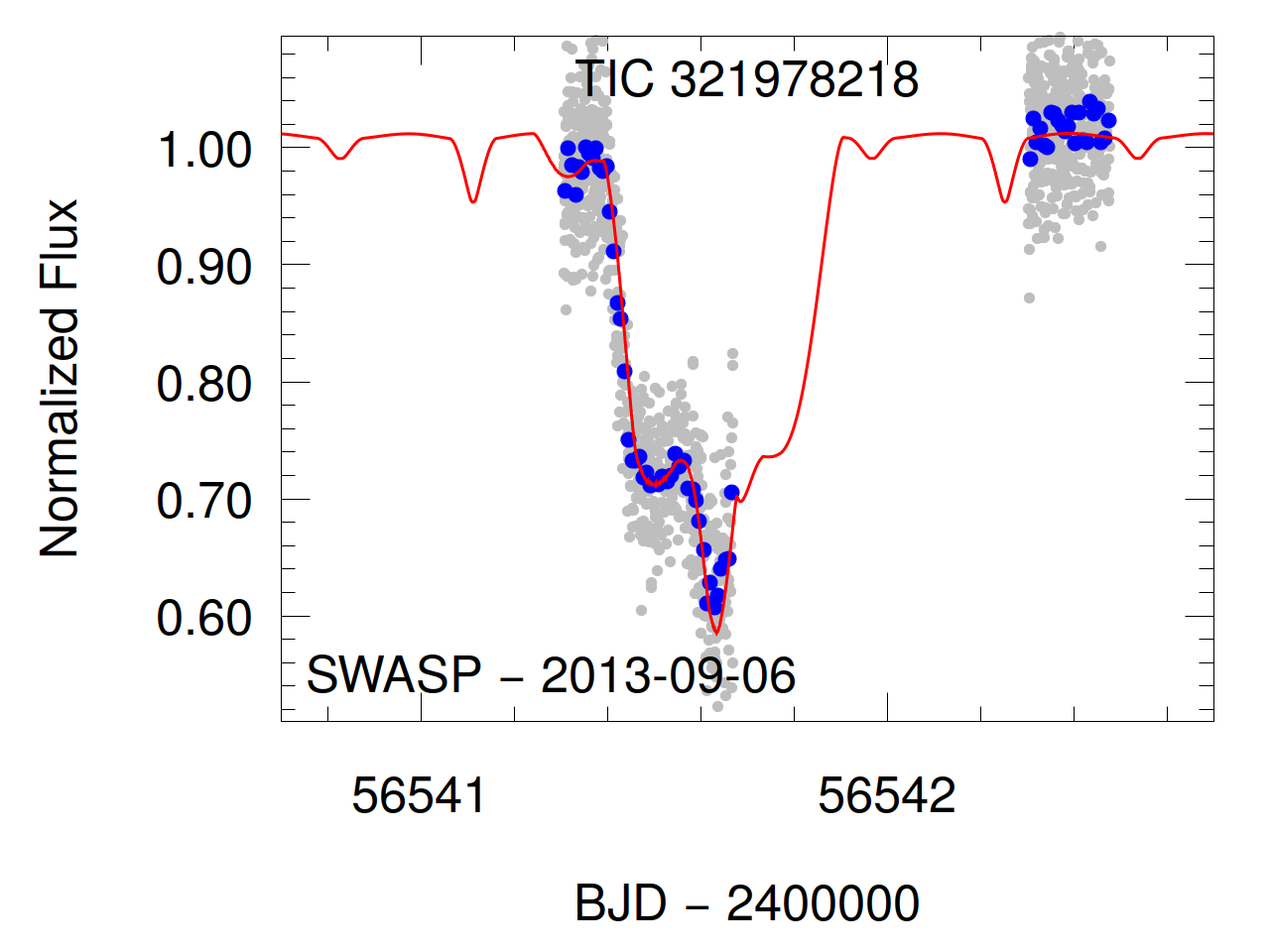}
    \includegraphics[width=0.3\textwidth]{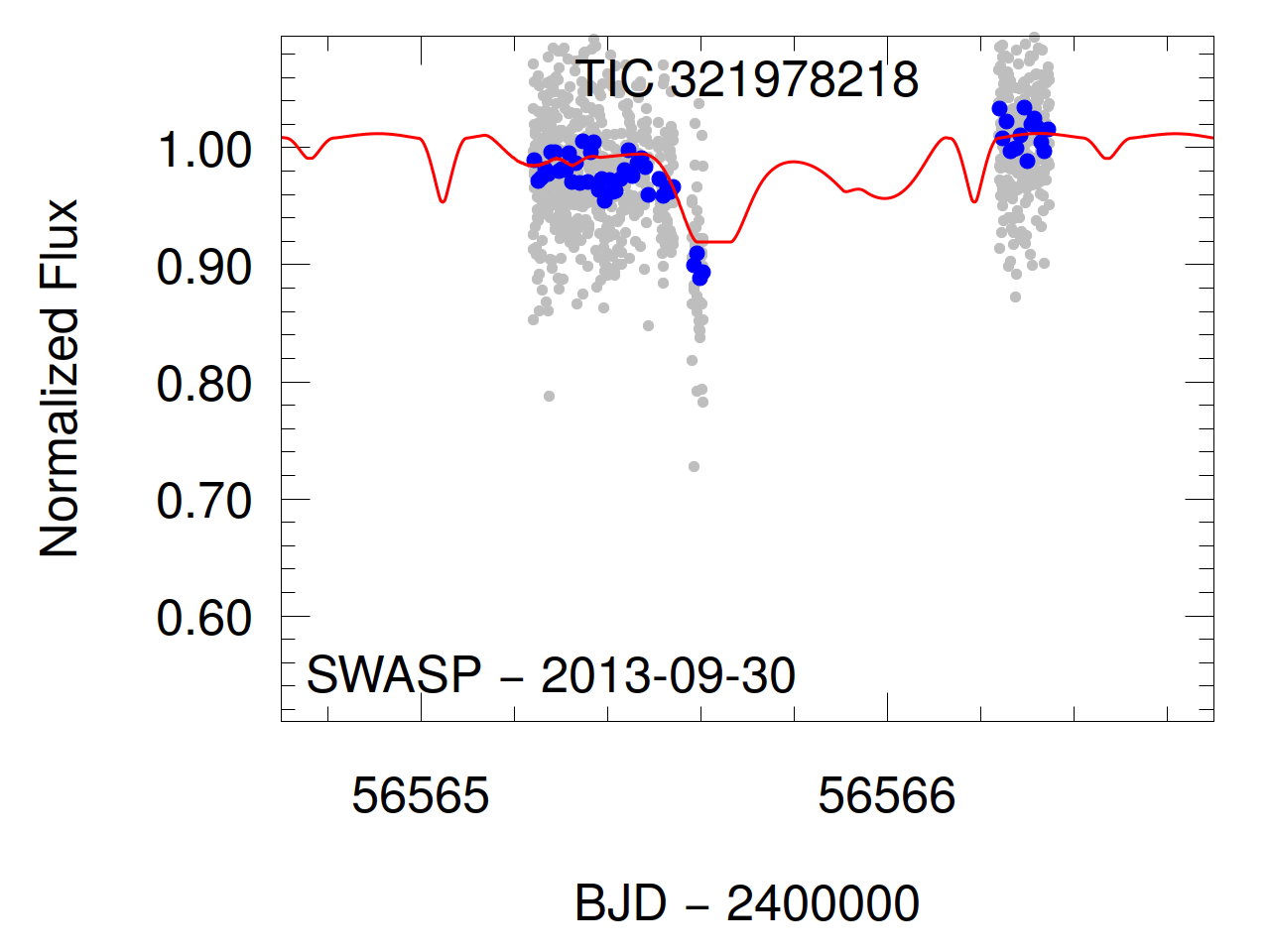}
     \includegraphics[width=0.3\textwidth]{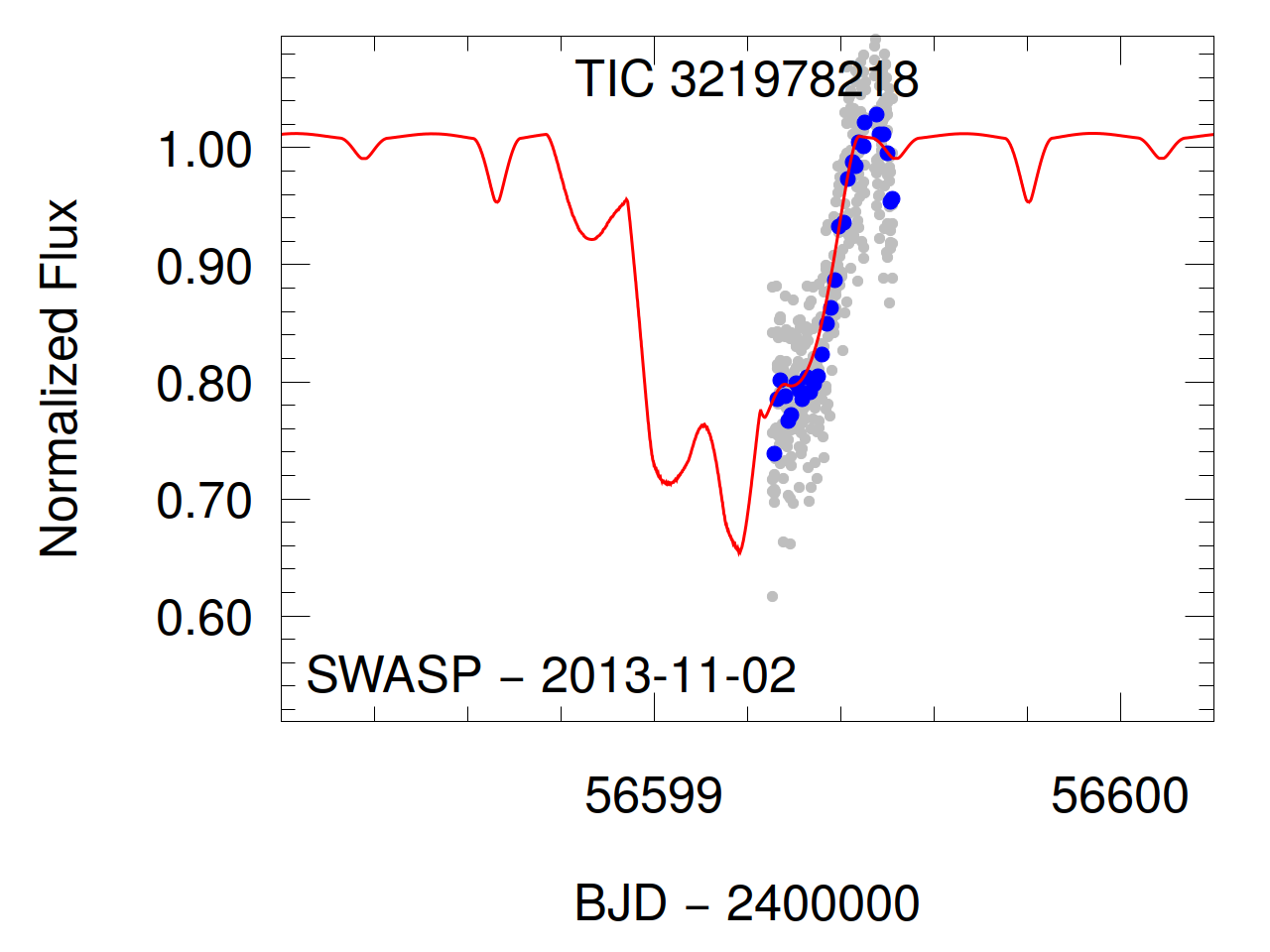}
      \includegraphics[width=0.3\textwidth]{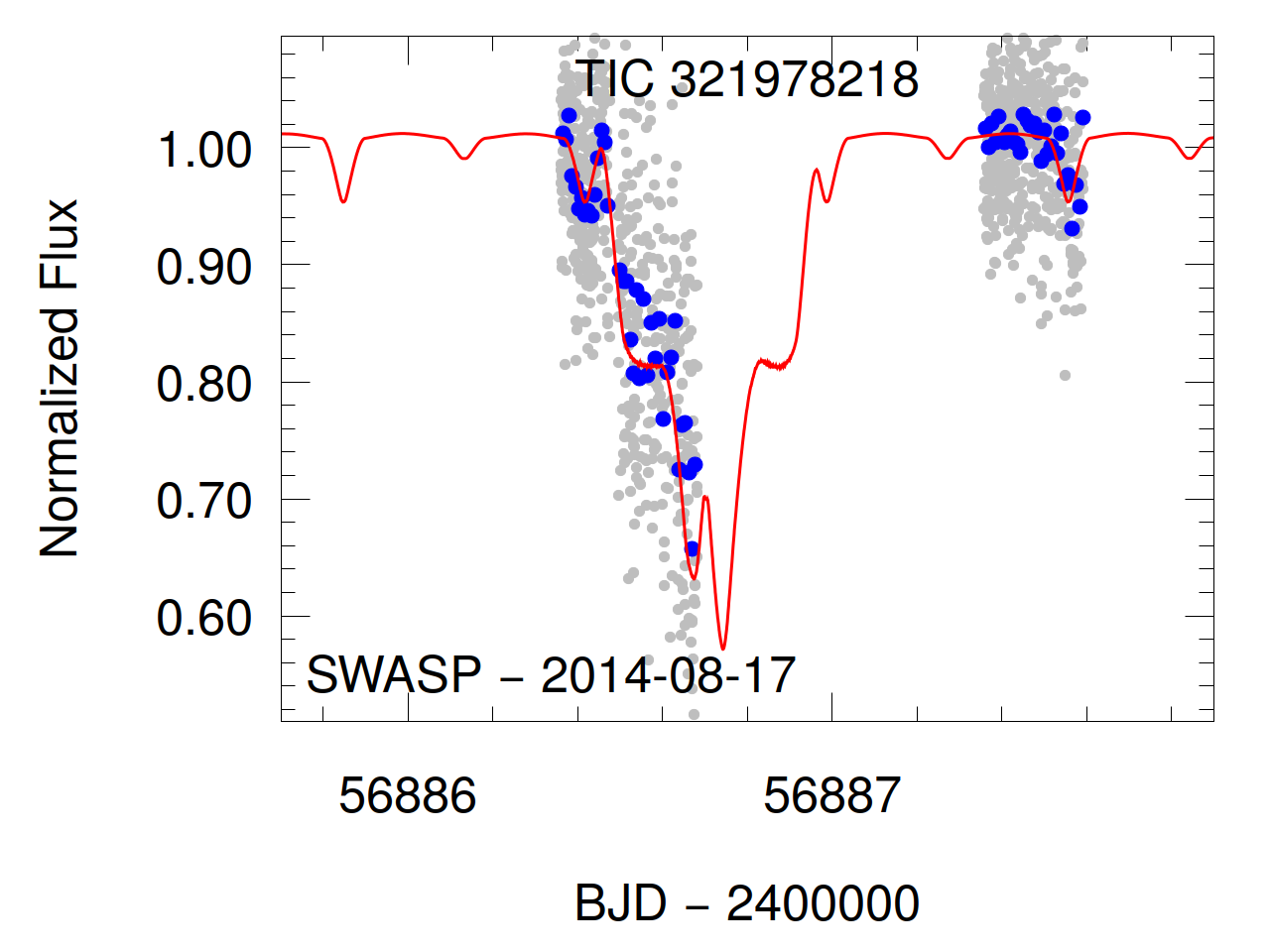}
       \includegraphics[width=0.3\textwidth]{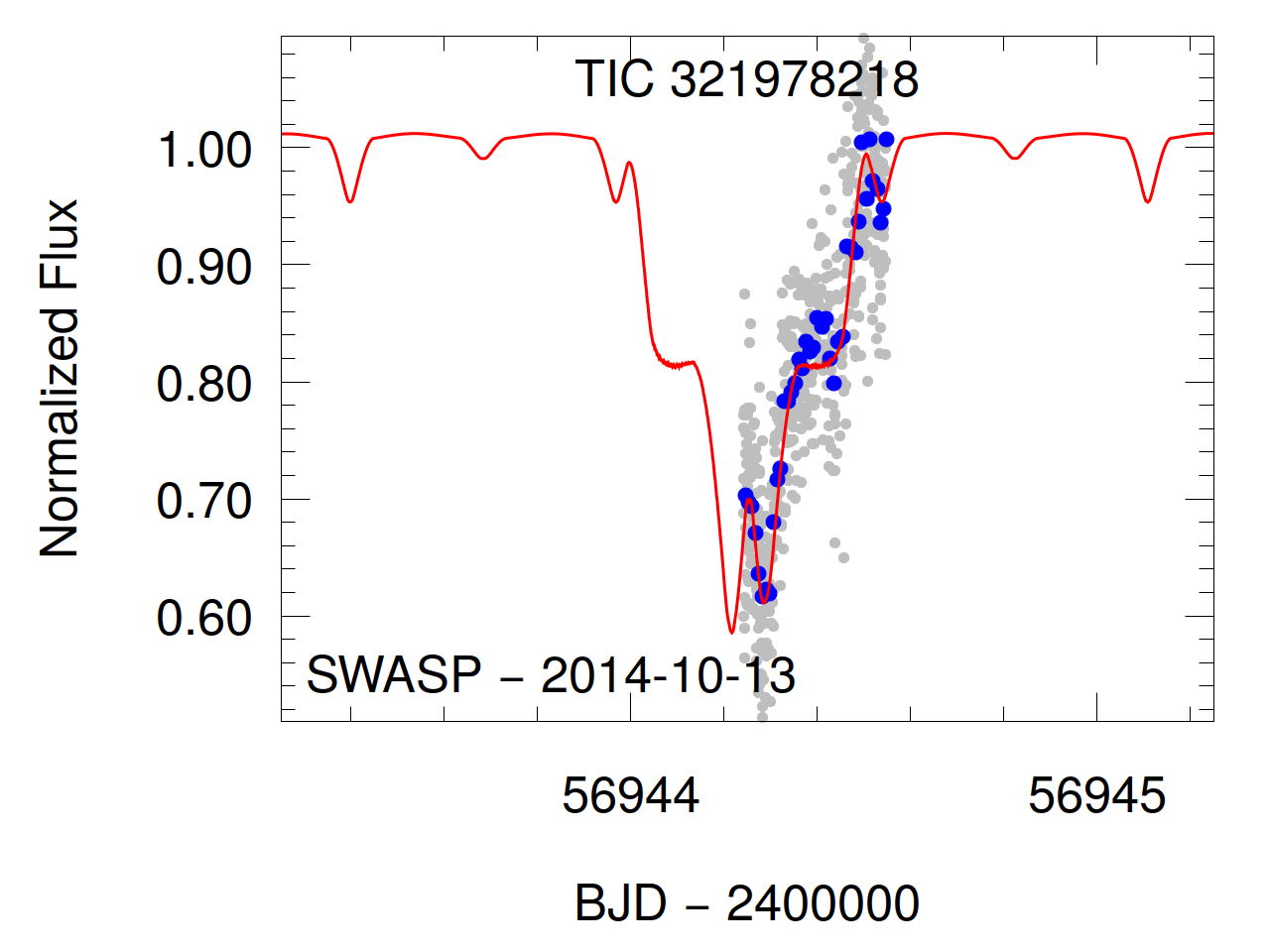}
 \caption{Eleven partially observed third-body eclipses of TIC~321978218 from the five seasons of the SWASP project. Gray dots represent the individual observations, while larger blue circles denote their 10-min. averages which were used for the joint photodynamical analysis. Red curves stand for the best-fit model.}
         \label{fig:321978218E3SWAPSlcs}
   \end{figure*}

In Figures \ref{fig:133771812lcs}--\ref{fig:650024463lcs} we show short segments of the TESS light curves that exhibit third body outer eclipses where the tertiary star eclipses the inner eclipsing binary or vice versa.  The blue points are the TESS photometric measurements.  For the light curve modeling we used 30-min cadence data even in those sectors where shorter cadence data were available. Hence, all these blue points are at a cadence of 30-min.  The regular binary eclipses are either self-evident (or called out in the respective figure captions), while the `extra' eclipses are, for the most part, all the dips in flux that cannot be ascribed to the regular EB eclipses.  In some cases the third-body eclipses are quite irregular and anomalous looking, while in other cases, the eclipses look almost `normal' but occur in rapid succession and/or are clearly out of place.  Moreover, in the eleven panels of Fig.~\ref{fig:321978218E3SWAPSlcs} we plot those third-body eclipses of TIC~321978218 which were observed, at least in part, during the SuperWASP survey, and were used in our analysis.

The smooth red curves are models taken from the full photodynamical analysis.  These will be discussed in Sect.~\ref{sec:photodynamics}.  

These third-body eclipses contain crucial information about both the orbits (in particular, the mutual inclination angle and outer orbital period) as well as the properties of the stars themselves (e.g., the relative sizes and effective temperatures).

\section{Archival outer orbit folds}  
\label{sec:archival_folds}

For each of the seven triply eclipsing triple systems, we looked up the archival ASAS-SN (\citealt{shappee14}; \citealt{kochanek17}) and ATLAS (\citealt{tonry18}; \citealt{smith20}) data sets that were available.  The goal here is to find additional third body eclipses in the archival record and thereby possibly determine the outer orbital period of these systems.  In general the ASAS-SN data typically yield between 3000 and 6000 photometric points spanning the past decade (over the whole sky). The ATLAS data have only a somewhat smaller number of points than ASAS-SN in the northern hemisphere, but often $\lesssim 1000$ points in the south.  The ATLAS photometry is superior for the fainter targets (e.g., $\gtrsim 15$th magnitude). 

We median normalize the $V$ and $g$ bands of ASAS-SN to each other, and likewise the $o$ and $c$ bands of ATLAS.  Finally, we median normalize the ASAS-SN and ATLAS data to each other to form a data set typically consisting of between 4000 and 8000 archival points.  Next, we remove the EB orbital light curve by subtracting out typically 100 harmonics of the EB period from the raw data (see, e.g., \citealt{powell21}). We then use a Box Least Squares (BLS) algorithm\footnote{\url{https://exoplanetarchive.ipac.caltech.edu/cgi-bin/Pgram/nph-pgram}} \citep{kovacs02} to do a blind search for the third-body eclipses of the outer orbit covering periods from 20-1000 days.  These third body events usually occur over an interval of about a day.  The outer periods in this sample of seven sources range from 41 days to a year.  Therefore, the `duty cycle' for the third body events ranges from $\sim$0.02 to $\sim$0.003.  For an archival sample of, for example, 5000 points, we can then expect somewhere between 15-100 points to land within the outer orbital phase range of the third body events, and thereby measure an average drop in flux at that phase.  Thus, for systems with longer outer periods, the task of uniquely identifying the outer orbital period from the archival data becomes ever more challenging. 

\begin{figure*}
   \centering
   \includegraphics[width=0.45\textwidth]{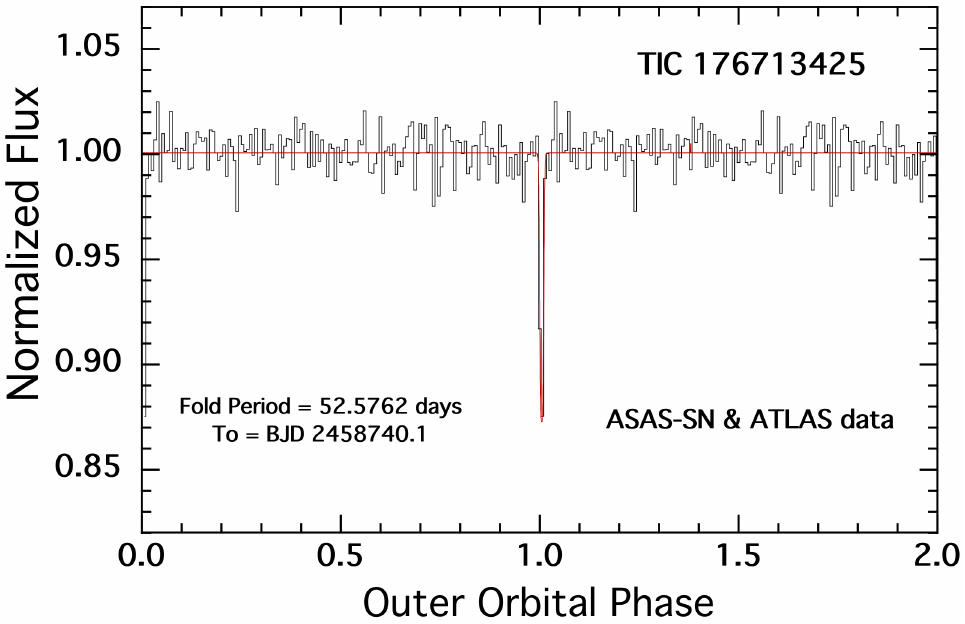}  
    \includegraphics[width=0.45\textwidth]{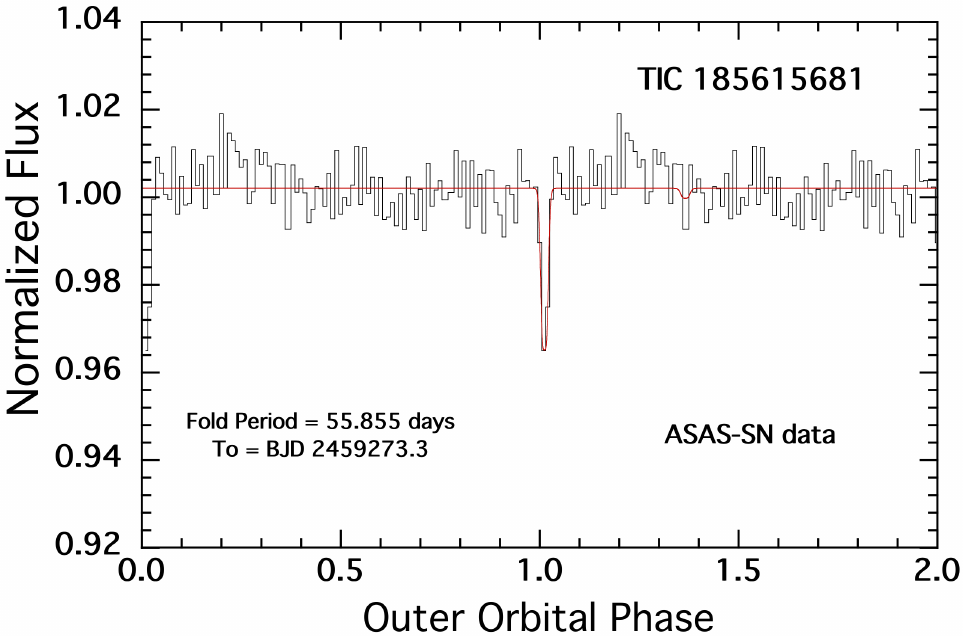}  \vglue0.3cm
    \includegraphics[width=0.43\textwidth]{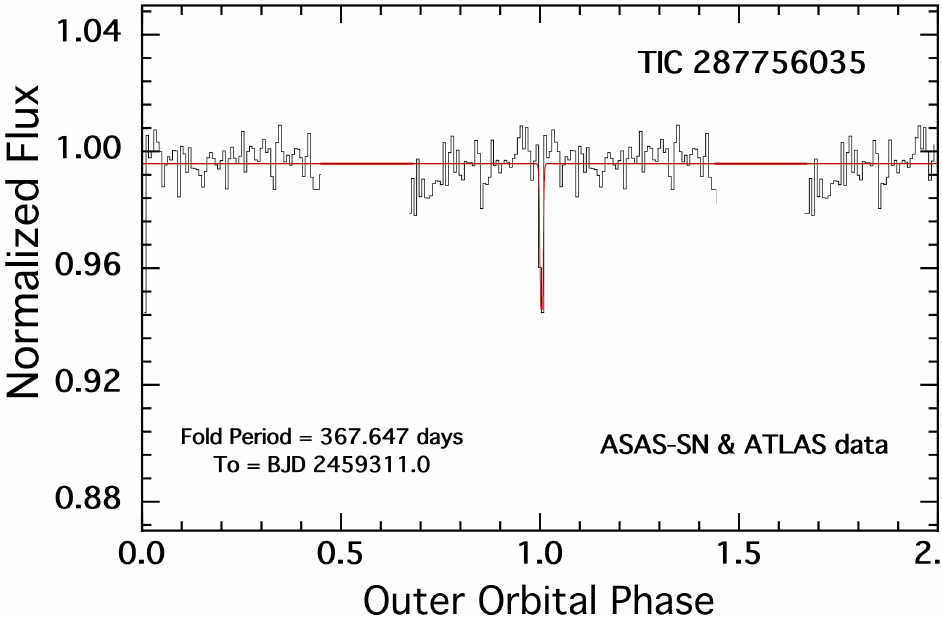}  \hglue0.25cm
    \includegraphics[width=0.45\textwidth]{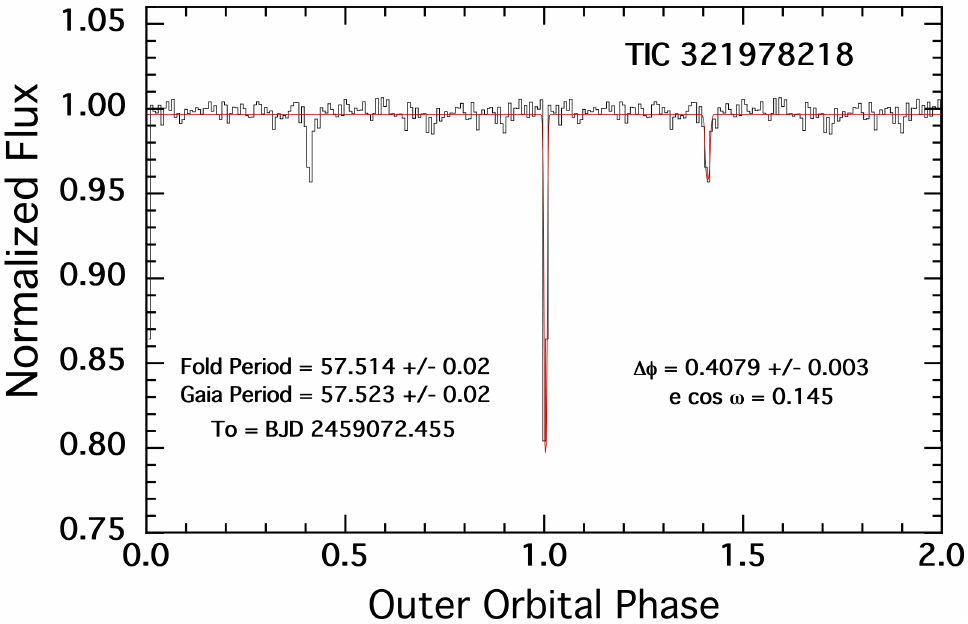} 
    \includegraphics[width=0.45\textwidth]{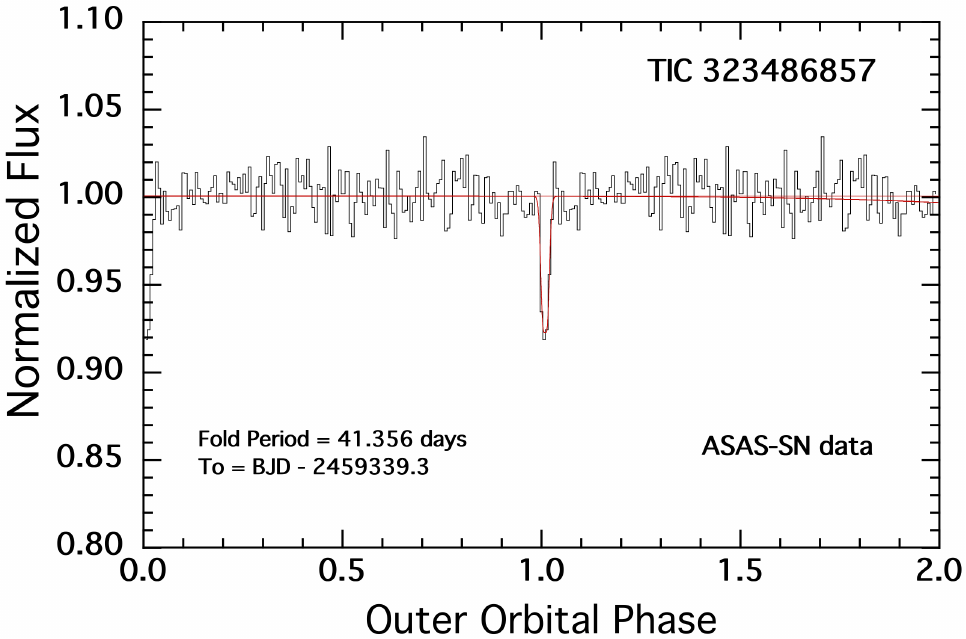}  
    \includegraphics[width=0.45\textwidth]{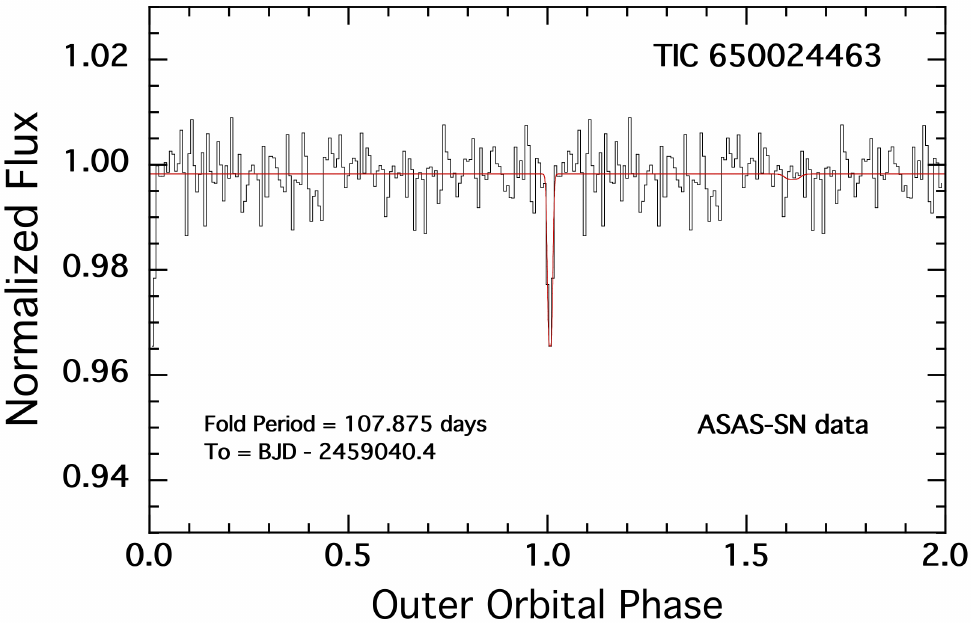}
       \caption{Archival outer orbit folds. The ASAS-SN and ATLAS archival data were used to find the outer period of these triple systems independently of what we were able to learn from ETV  or RV curves, or third-body eclipses observed with TESS. Typically, there are 3000-6000 archival photometric points spanning a decade.  In the case of TIC 287756035, there is about 1/4 of the outer orbital phase which is missing due to the fact that the outer period nearly exactly matches that of the Earth year, and the consequential observing seasons. The light red curves are fits to a non-physical function consisting of a modified hyperbolic secant (e.g., Eqn.~1 of \citealt{rappaport16}).  The fit covers only one of the two cycles for clarity in viewing the eclipses, and is used only to find the phases of the eclipses.}
        \label{fig:outer_orbit_folds}
   \end{figure*}

The results of our search for the outer orbital periods are shown in Fig.~\ref{fig:outer_orbit_folds}.  For six of the seven sources (all except TIC 133771812) there is a clear BLS signal at the outer period.  For TIC 133771812, with an outer period of 244 days, and third-body eclipse depths of only $\sim$6\%, we were not able to detect the outer eclipses at a significant level.  On the other hand, we were able to robustly detect the outer eclipses in TIC 287756035 with a year-long outer period.  The only difficulty with that one is that due to the annual observing schedule of the ground-based observations, we cannot tell whether the outer period is 368 days or half that.  However, other information that is available allows us to break this degeneracy and select the 368 day period.  

We note that four of the seven triples studied in this work (TICs 185615681, 321978218, 323486857, and 650024463) exhibit both types of outer eclipses (i.e., secondary vs.~primary) in the TESS data -- thanks to our nearly edge-on view of their outer orbits.  Additionally, in the cases of TICs~176713425 and 287756035, although TESS has observed only one type of outer eclipse, our photodynamical analyses (see later in Sect.~\ref{sec:photodynamics}) predict the existence of the other type of outer eclipse, as well. In contrast to these, only TIC 321978218 exhibits detectable third body primary and secondary eclipses in our archival folded light curves (Fig.~\ref{fig:outer_orbit_folds}).

In general, these outer eclipses we see in the archival folds are detected at the $\sim$8-20 $\sigma$ confidence level. Therefore, if the secondary outer eclipses have several times less area under their normalized light curve, then they can become very difficult to detect.  In fact, we do not detect the secondary eclipse in five of the six above-mentioned sources.  In four of the six cases, it is obvious that the shallow secondary eclipses would be undetectable in the archival data. In the specific case of TIC 650024463, where the secondary outer eclipses have $\sim$2/3 the depths of the primary eclipses, one might wonder whether these should have been detected in the archival fold for that source.  However, when we take into account that the primary outer eclipses have a duration that is $\sim$2-3 times longer than the secondary outer eclipses, then the expected detection significance of the secondary eclipses will drop by about a factor of 3-4.  This is a sufficient decrease so we cannot expect to detect it robustly--and we do not.  Finally, in the case of TIC~287756035, the above mentioned seasonal gaps in the ground-based observations would explain the absence of the archival fold detection of the other type of third-body eclipse\footnote{We return to this question in Sect.~\ref{Sect:discussion_TIC287756035}.}.

  \section{ETV and RV curves}
  \label{sec:etvs_rvs}
  
  The eclipse timing variation curves form a crucial piece of input information for the comprehensive photodynamical analysis we discuss in Sect.~\ref{sec:photodynamics}.  These are based on eclipse midtimes for both the primary and secondary eclipses of the inner EB of the system.  We extract these mid-eclipse times in the manner discussed previously in \citet{borkovitsetal15,borkovitsetal16}.  The non-linear behavior we see in the ETV curves comes predominantly from three basic effects, as follows.  First, there is the classical light-travel-time effect (LTTE; \citealt{roemer1677}) due to the changing distance to the EB as it is pulled around in its outer orbit by the tertiary star.  Second is the `dynamical' or `physical' delays which, in nearly coplanar orbits, are caused largely by the lengthening of the EB orbit due to the presence of the tertiary star (see, e.g., \citealt{rappaport13}; \citealt{borkovitsetal15}).  This effect depends on the instantaneous separation between the EB and the tertiary star, and therefore varies with the phase of the outer orbit if that orbit is eccentric. (3) Finally, there is the well-known apsidal motion exhibited by eccentric binaries. This is a long timescale effect which has three different origins: (i) non-spherical mass distributions of the tidally distorted binary components; (ii) effects of general relativity and; (iii) perturbations forced by the tertiary star. The timescale of this latter effect, dynamically driven apsidal motion, is on the order of $P_{\rm out}^2/P_{\rm in}$ \citep{soderhjelm75}, where $P_{\rm out}$ and $P_{\rm in}$ are the outer and inner binary periods, respectively. The two dominant drivers of apsidal motion in the systems reported here are due to the mutual tidal deformations of the two EB stars and forced apsidal motion by the third body.

\begin{figure}
\centering
     \includegraphics[width=\hsize]{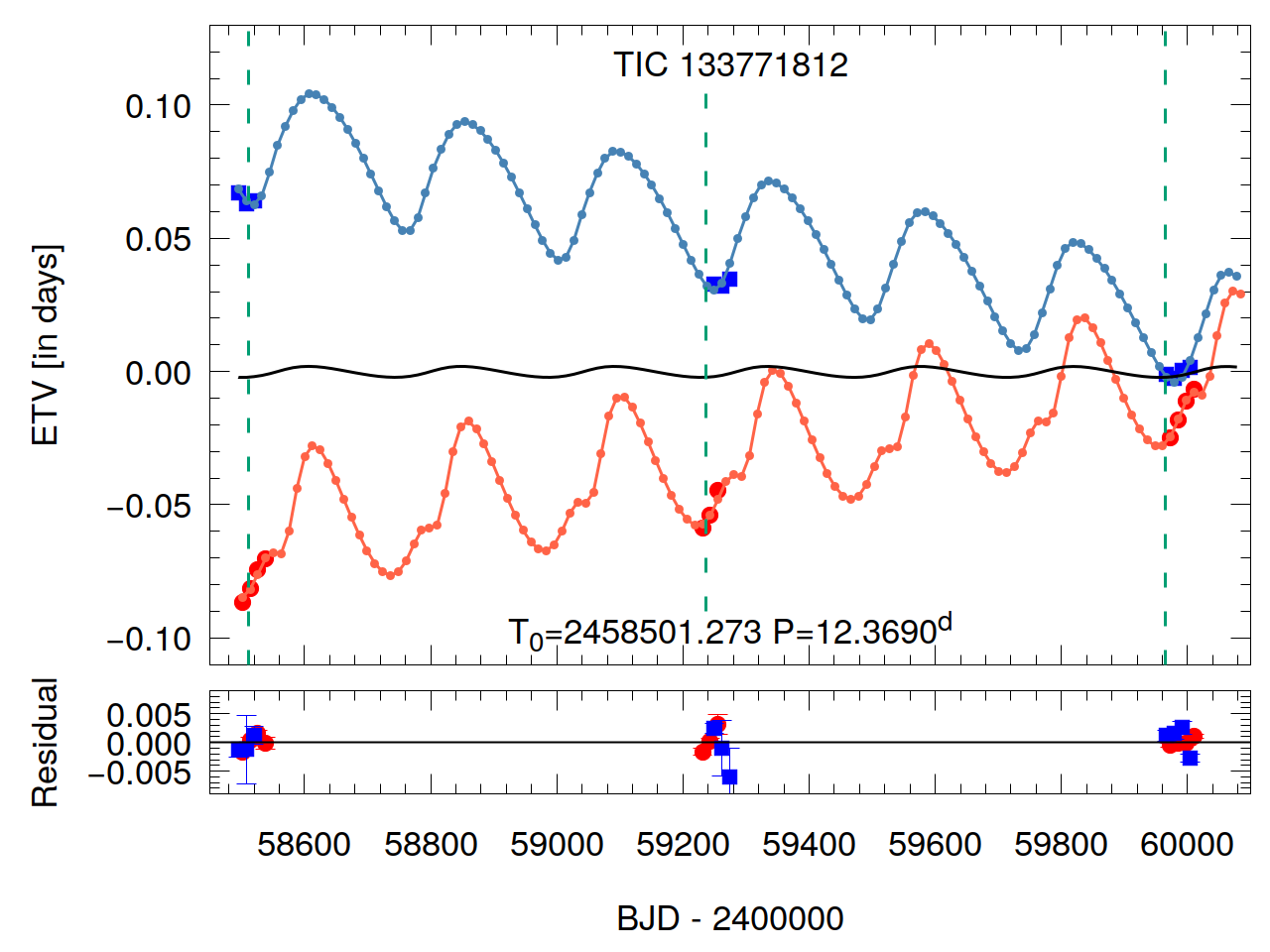}
     \caption{TESS primary and secondary ETV curves (red and blue circles, respectively), with the best-fit photodynamical solution for TIC 133771812 (see Sect.~\ref{sec:photodynamics}). The horizontally centered black curve represents the pure LTTE contribution. Vertical lines mark the times of the observed outer eclipses (dashed green--the binary occulting the tertiary star).}
\label{fig:133771812ETV}
\end{figure}  

\begin{figure}[ht]
\begin{center}
     \includegraphics[width=\hsize]{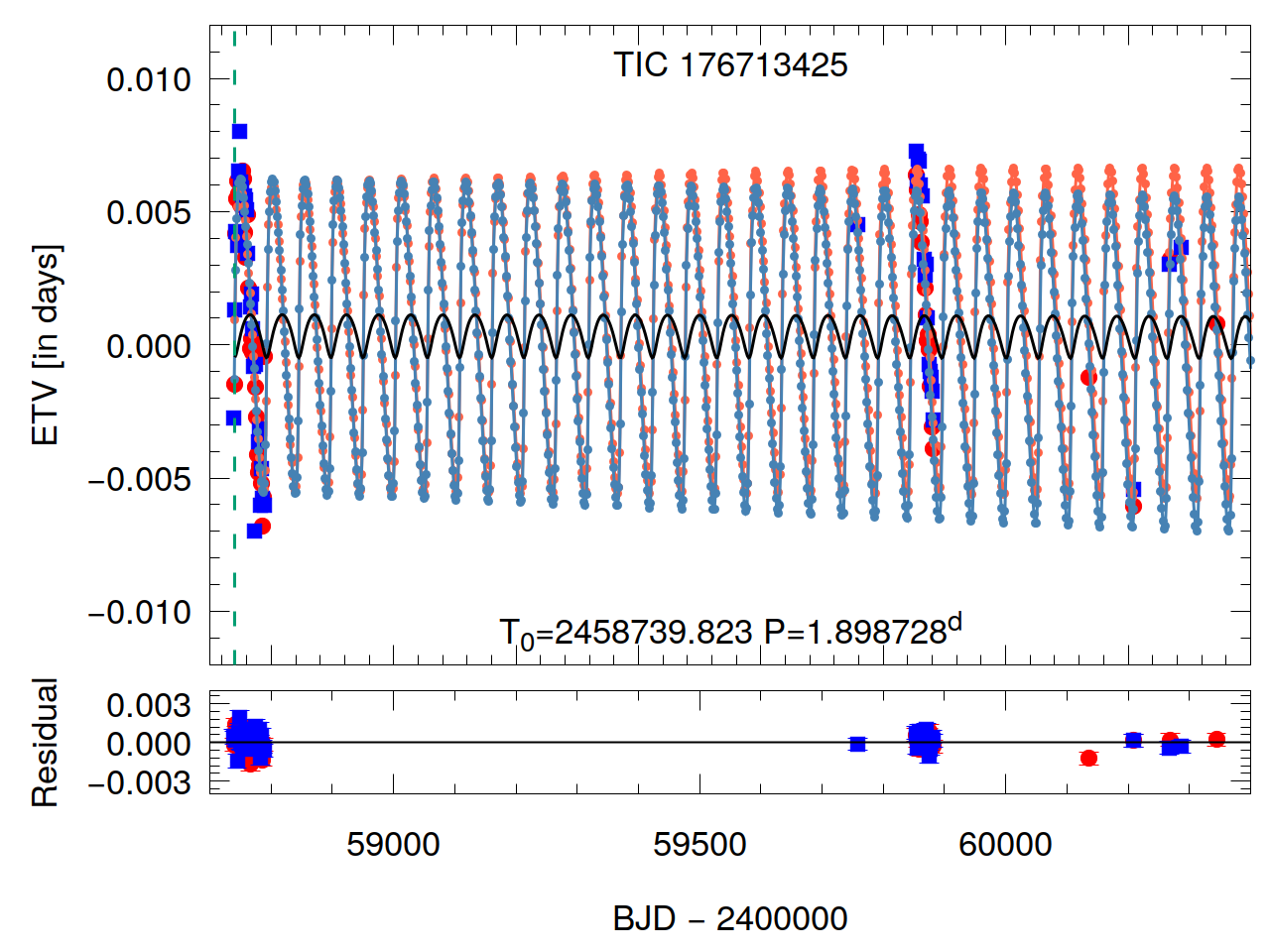}  
     \caption{ETV curves formed from the TESS and ground-based follow up observations with the best-fit photodynamical solution for TIC 176713425. The horizontally centered black curve represents the pure LTTE contribution. Dashed vertical green line marks the position of the only (two-dipped) third-body eclipse event which was observed with TESS.}
\label{fig:176713425ETV}
\end{center}
\end{figure}  
 
\begin{figure}
\centering
     \includegraphics[width=\hsize]{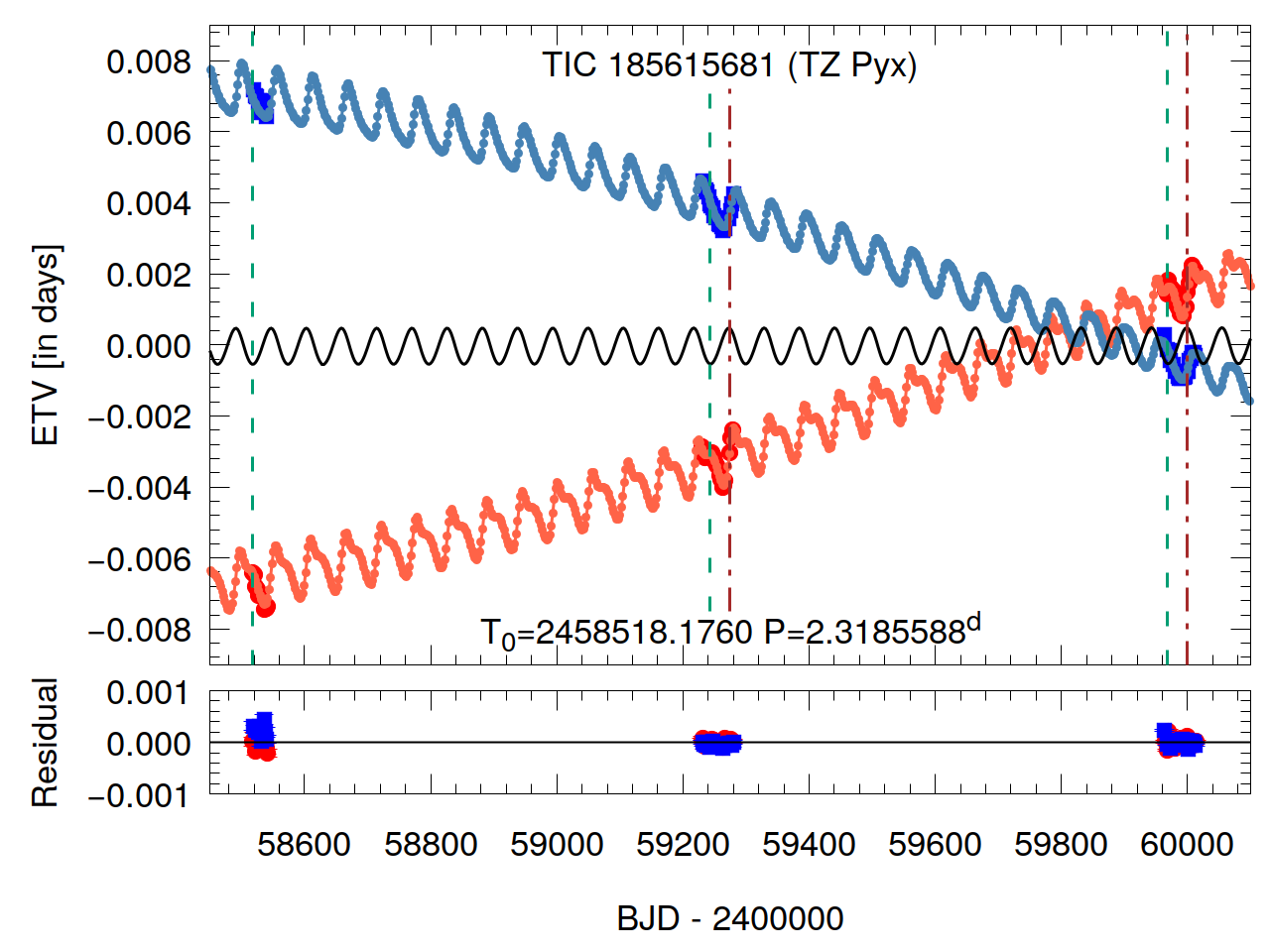}
     \caption{TESS ETV curves with the best-fitted photodynamical solution for TIC 185615681 (TZ Pyx). The horizontally centered black curve represents the pure LTTE contribution. Vertical dashed green lines represent the times of the observed outer eclipses when the binary occults the tertiary star, and vice versa for the dot-dashed red lines.}
\label{fig:185615681ETV}
\end{figure}  

\begin{figure*}
\centering
     \includegraphics[width=1.0\columnwidth]{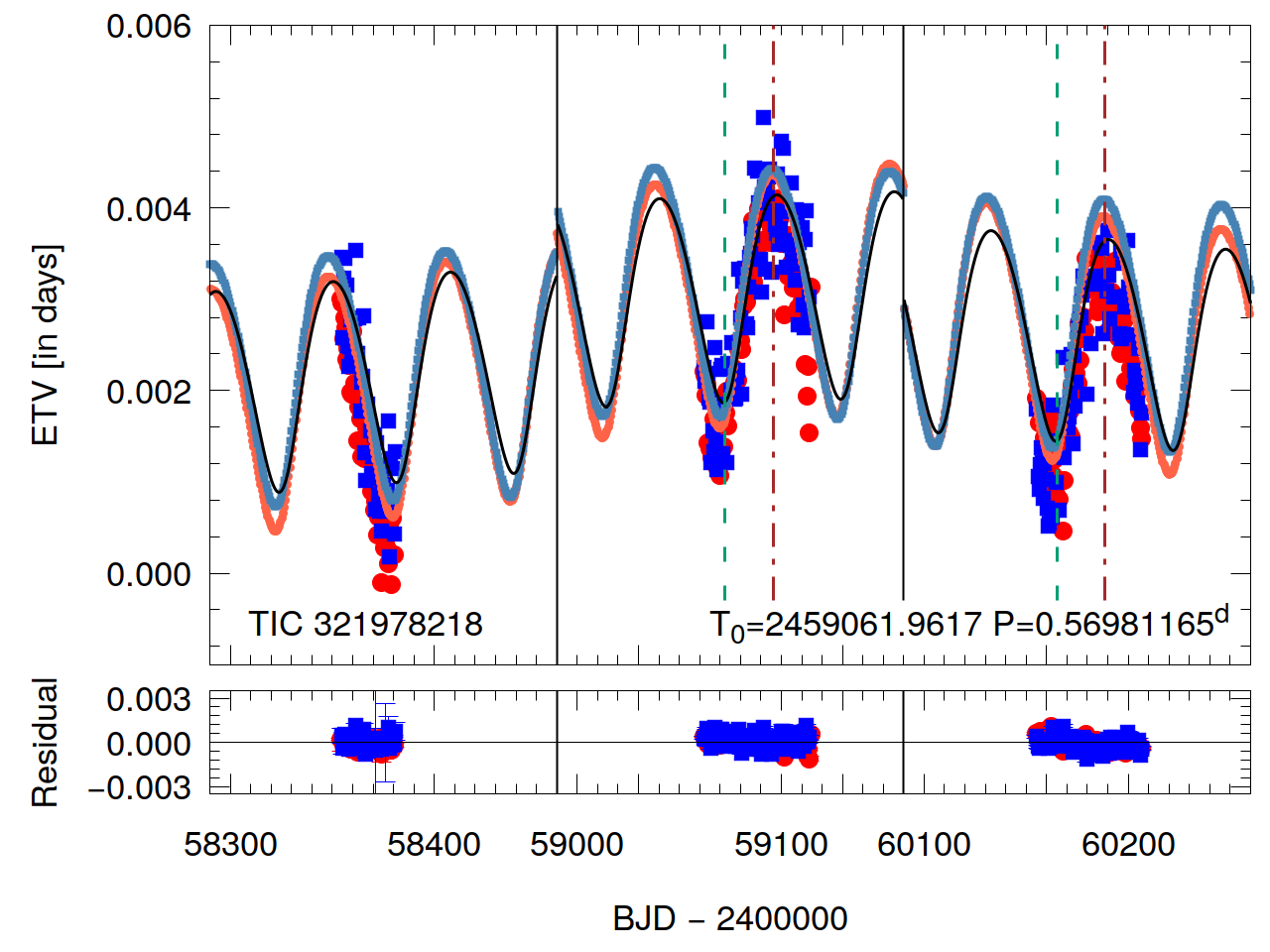}\includegraphics[width=1.0\columnwidth]{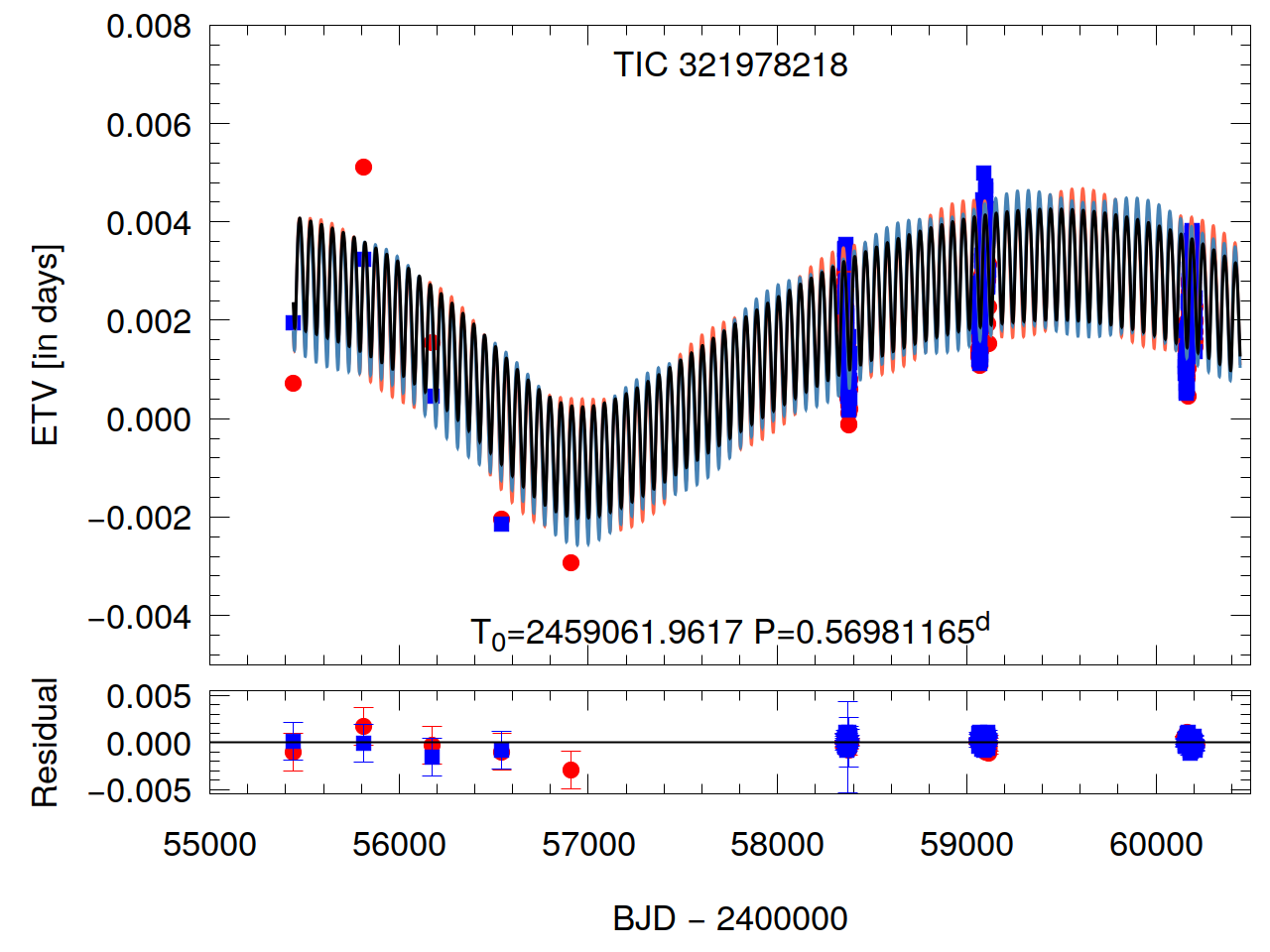}
     \caption{ETV curves for TIC 321978218 with the best-fit photodynamical solution superposed. {\it Left panel:} ETV curve derived from TESS data only. We note that only those sections of the ETV curves, where TESS data are available, are shown. Vertical black lines denote the different sector boundaries; dashed green lines represent the times of the observed outer eclipses when the binary occults the tertiary star, and vice versa for the dot-dashed red lines. {\it Right panel:} Overall ETV curve after adding average `seasonal' ETV points (see text for explanation) calculated from SWASP observations. The longer-term ETV curve also nicely shows an additional variation. We model it with an fourth stellar component, making the system a 2+1+1 quadruple. See text for details. }
\label{fig:321978218ETV}
\end{figure*}  

\begin{figure}
\centering
     \includegraphics[width=\hsize]{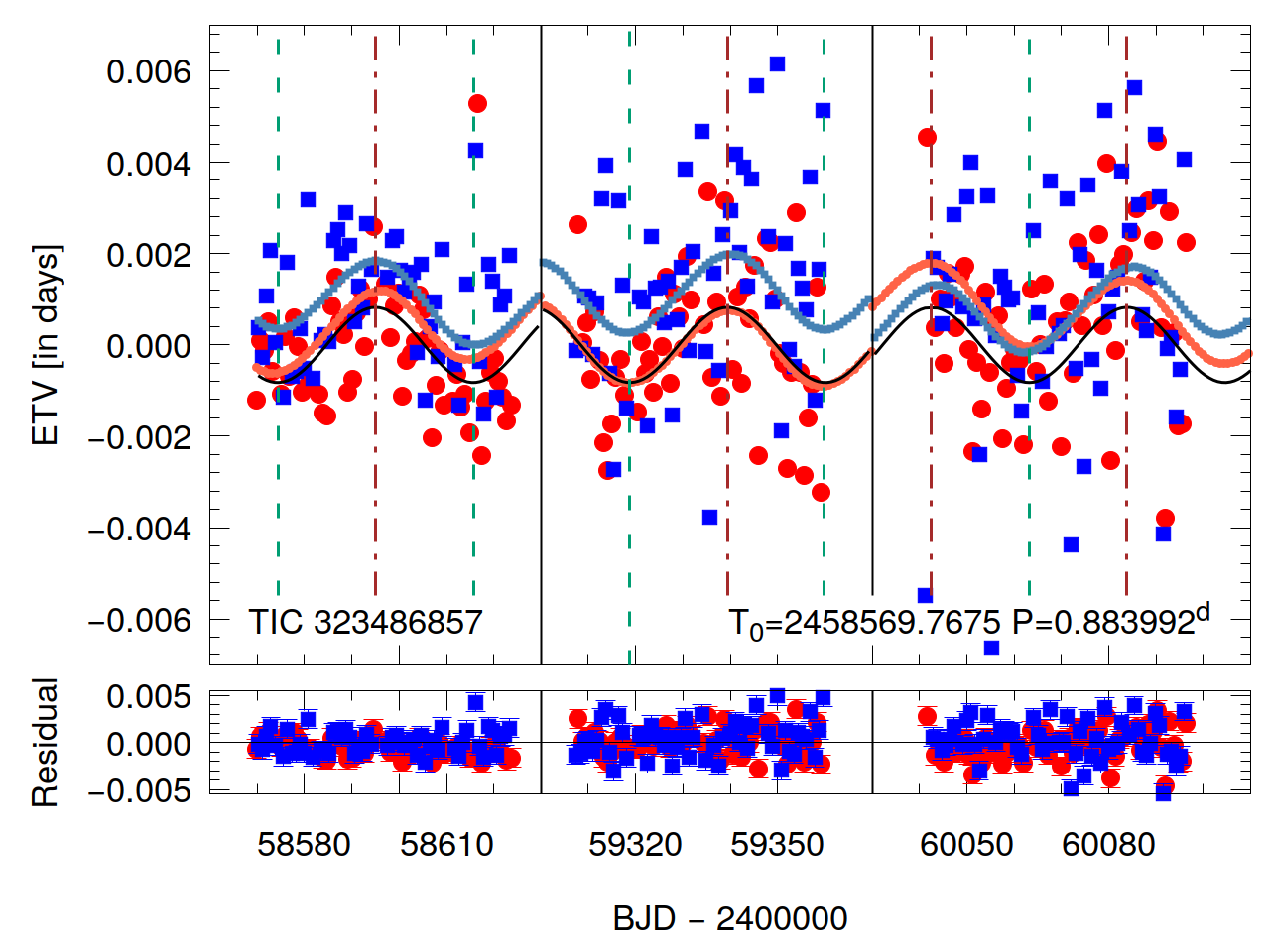}
     \caption{TESS ETV curves with the best-fit photodynamical solution for TIC 323486857. We note that only those sections of the ETV curves, where TESS ETV data are available, are shown. Vertical lines have the same meaning as in Fig.~\ref{fig:321978218ETV}.}
\label{fig:323486857ETV}
\end{figure}  

\begin{figure}[ht]
\begin{center}
     \includegraphics[width=\hsize]{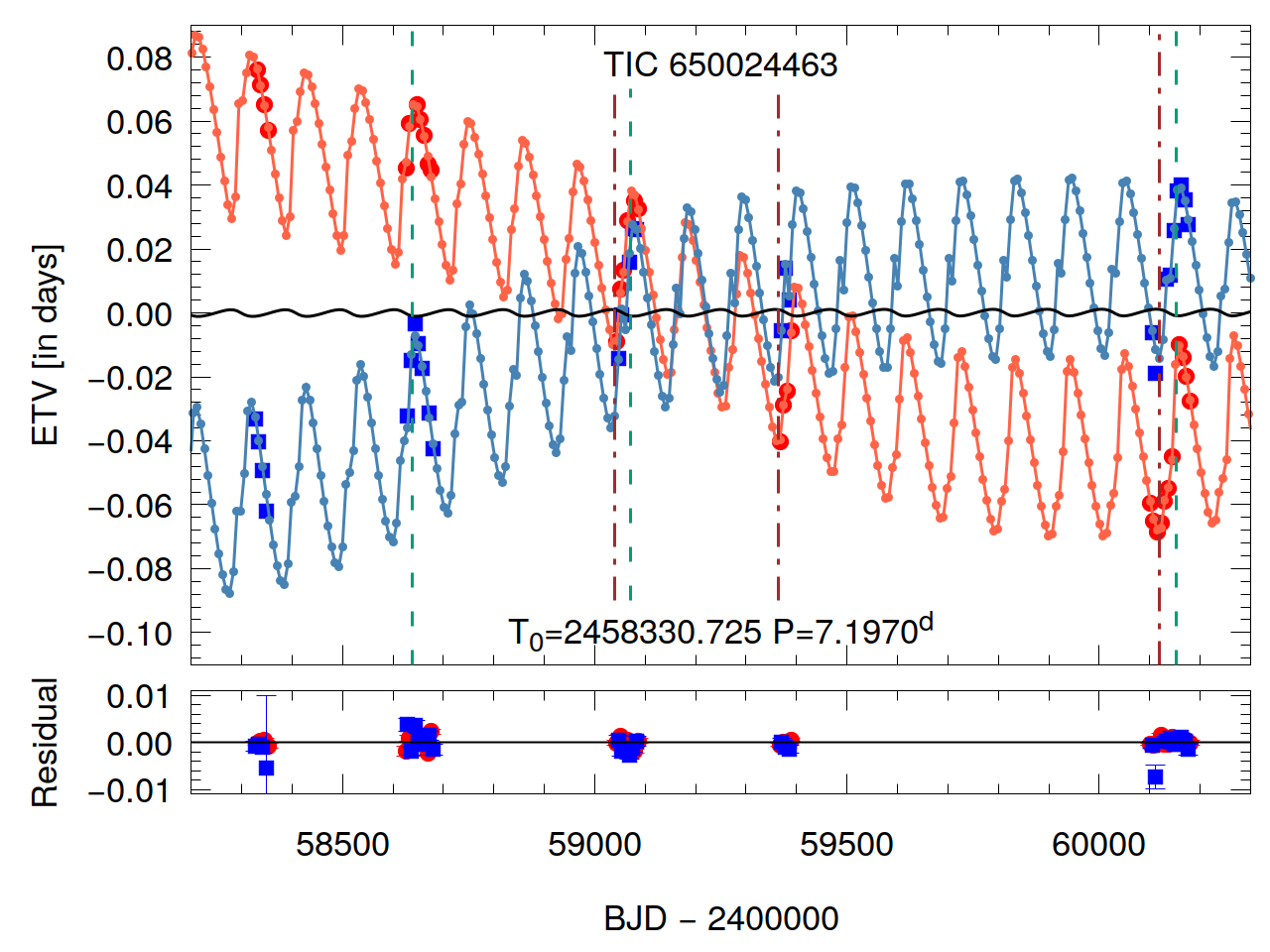}
\caption{TESS ETV curves with the best-fitted photodynamical solution for TIC 650024463. The horizontally centered black curve represents the pure LTTE contribution. Vertical lines have the same meaning as in Fig.~\ref{fig:321978218ETV}.}
\label{fig:650024463ETV}
\end{center}
\end{figure}   

We see all three of these effects, namely, LTTE, dynamical delays, and apsidal motion, in the ETV curves of the seven triple systems reported in this work.  The curves themselves are shown in Figures \ref{fig:133771812ETV} -- \ref{fig:650024463ETV}, while the mid-eclipse times used for the derivation of these ETV curves are tabulated in seven separate tables in Appendix~\ref{app:ToMs}.

{\bf TIC 133771812}: The ETV curve for TIC 133771812 with a 12-day EB and 244-day outer orbit (see Fig.~\ref{fig:133771812ETV}) shows the sparse TESS sampling--essentially two sectors of data every two years.  The smaller red and blue dots, connected with straight lines are the model fits (described in detail in Sect.~\ref{sec:photodynamics}) showing the periodic dynamical delays on the timescale of the outer orbit, and the convergence between the primary and secondary eclipse times, due to the apsidal motion, which in this case is being driven by the tertiary star.  The variations in the ETV curve with the outer period of 244 d are $\sim90$\% dynamical delays and $\sim10$\% LTTE effects. As another way of illustrating the ratio of the LTTE contribution to the overall ETV curve, we superpose the pure LTTE contribution on the ETV curve for TIC 133771812.  We then also show the LTTE contribution to the ETV curves for each of the other sources we investigate.

{\bf TIC 176713425}: Figure \ref{fig:176713425ETV} shows the ETV curve for TIC 176713425.  Here there are 2+1 sectors of data separated by 3 years and, moreover, six ground-based eclipse times obtained within the framework of our photometric follow up campaigns.  Fortunately we know the outer orbital period rather well from the archival data (see Fig.~\ref{fig:outer_orbit_folds}), namely 52.58 days.  Thus, the Sector 16, 17, and 57 data are sufficient to determine satisfactorily the outer orbital parameters.  The variations in the ETV curve with the outer orbital period of 52.58 d are $\sim90$\% dynamical delays and $\sim10$\% LTTE effects.

{\bf TIC 185615681}: The ETV curve for TIC 185615681 shown in Fig.~\ref{fig:185615681ETV} has sparse points taken at roughly two-year intervals. Fortunately, the outer orbital period is also well measured both from the five sets of TESS-observed third-body eclipses and the archival data to be 55.9 d.  The ETV variations with the period of the outer orbit are quite evident, as is the rapid apsidal motion forced by the third body. The former variations in the ETV curve are $\sim$$65$\% dynamical delays and $\sim$$35$\% LTTE effects. We also note that this is the only triple in our sample where ETV data prior to the TESS mission are also available in the literature (see Sect.~\ref{Sect:discussion_TIC185615681}).

\begin{figure}
   \centering
  \includegraphics[width=\hsize]{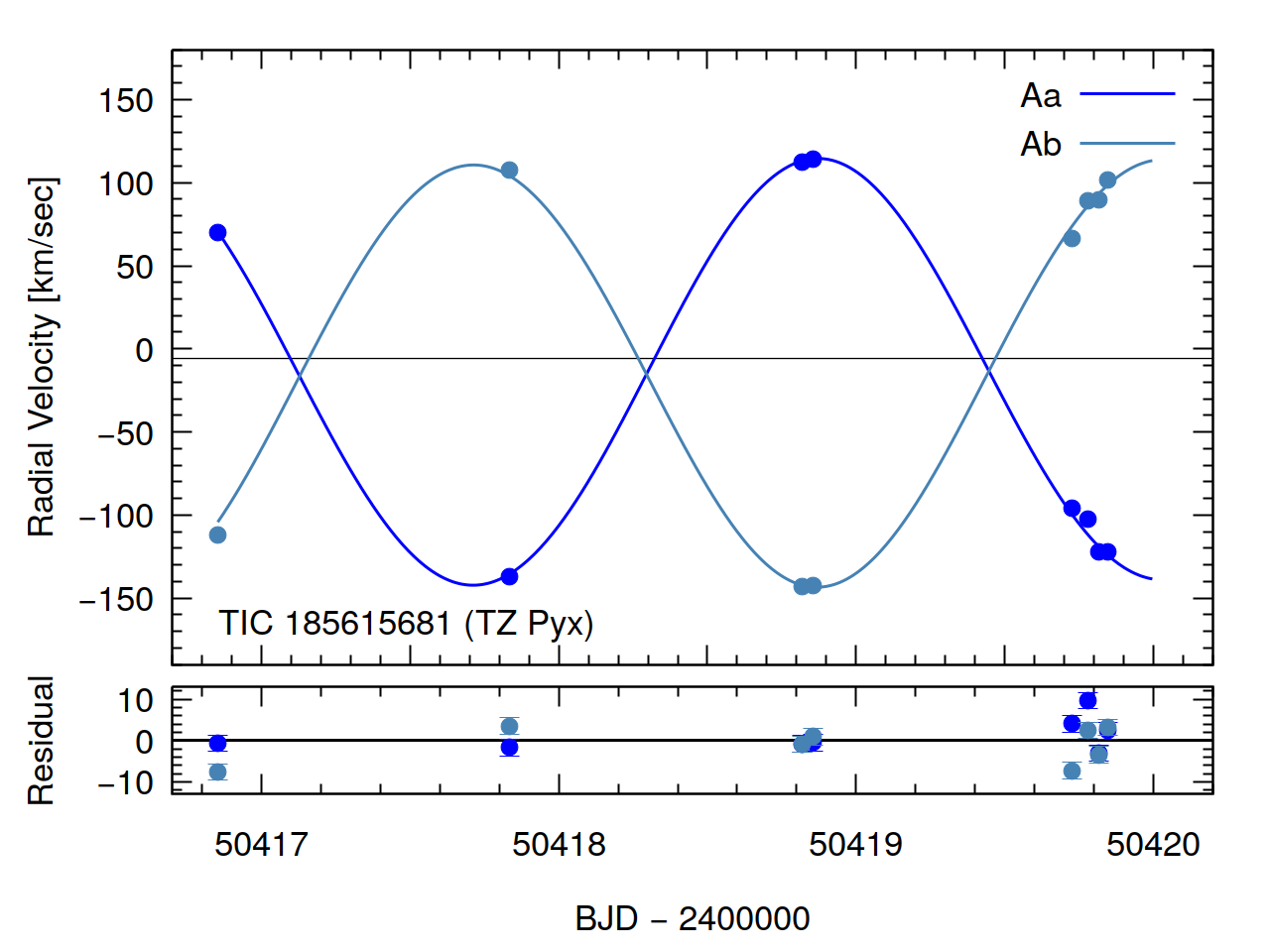}
      \caption{RV curves and model fits for TIC 185615681. The thin, horizontal line at $V=-5.95\,\mathrm{km s}^{-1}$ represents the systemic radial velocity of the triple star system. If TIC 185615681 were a single binary, the RVs of the two components would intersect each other exactly on that line. The different offsets of consecutive intersections of the RV curves from this line is due to the revolution of the inner pair around the center of mass of the whole triple system. The RV points were taken from \citet{duerbeckrucinski07}.}
         \label{fig:185615681rv}
   \end{figure}

This is the only system in our sample where the inner EB is a known double-lined spectroscopic binary (SB2) and, hence, RV data for the EB members are available (see Fig.~\ref{fig:185615681rv}) from the literature \citep{duerbeckrucinski07} and, we used these data for our analysis. 

\begin{figure}
   \centering
  \includegraphics[width=\hsize]{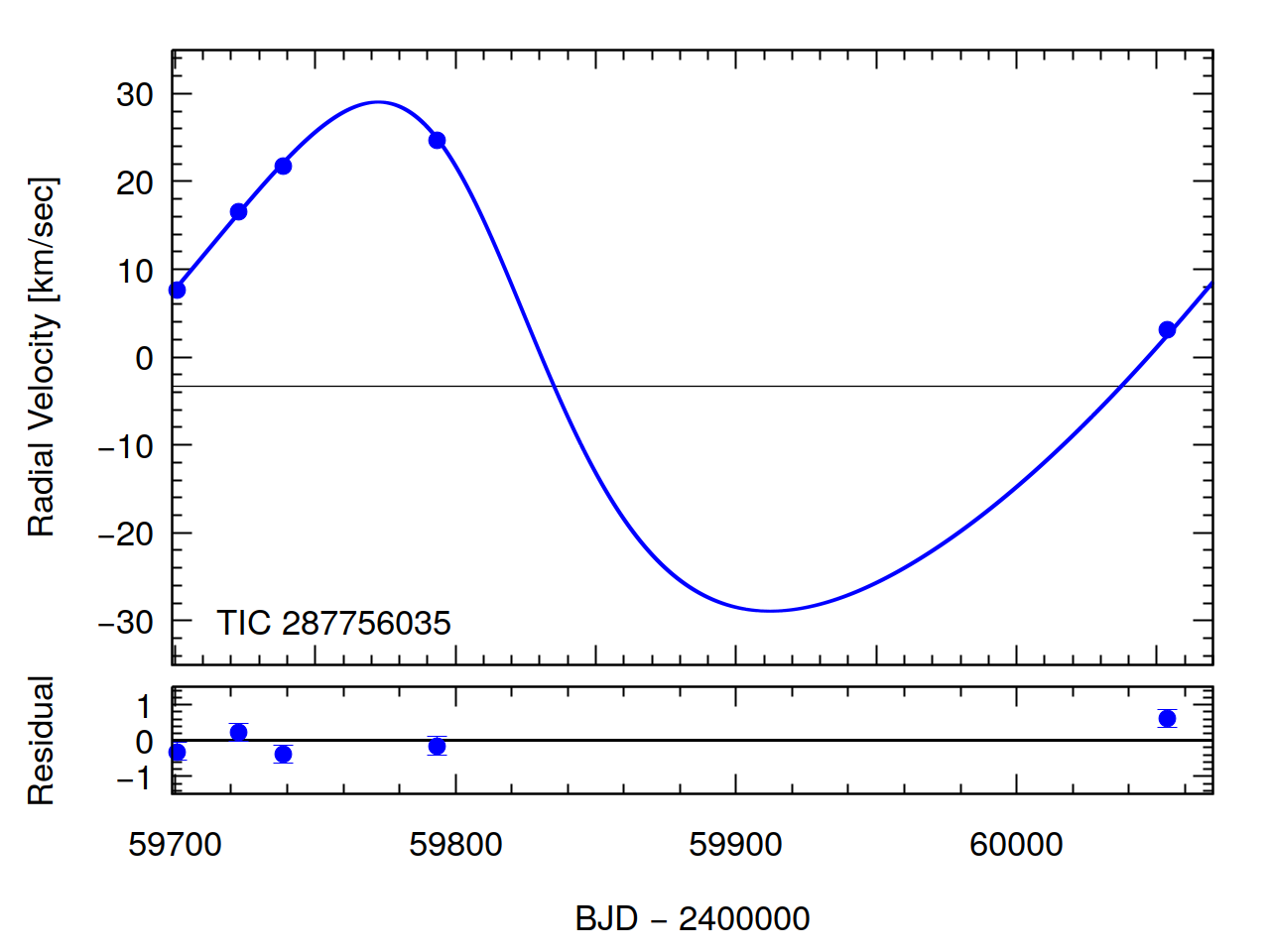}
      \caption{RV curve and model fit for the third component of TIC 287756035. The thin, horizontal line at $V=-3.9\,\mathrm{km s}^{-1}$ represents the systemic radial velocity of the triple star system.}
         \label{fig:287756035rv}
   \end{figure}

{\bf TIC 287756035}: This is the only system in our sample where the ETV curve, derived from the TESS observations, is of no value.  This is the case because the scatter of the measured mid-eclipse times exceeds the full amplitude of the model-ETV by more than an order of magnitude. On the other hand, however, we have obtained five RV points for the tertiary star which provides us with valuable information about the outer orbit in this system (see Fig.~\ref{fig:287756035rv} and Table~\ref{tab:TIC287756035_RVdata} as well). 

\begin{figure}
   \centering
  \includegraphics[width=\hsize]{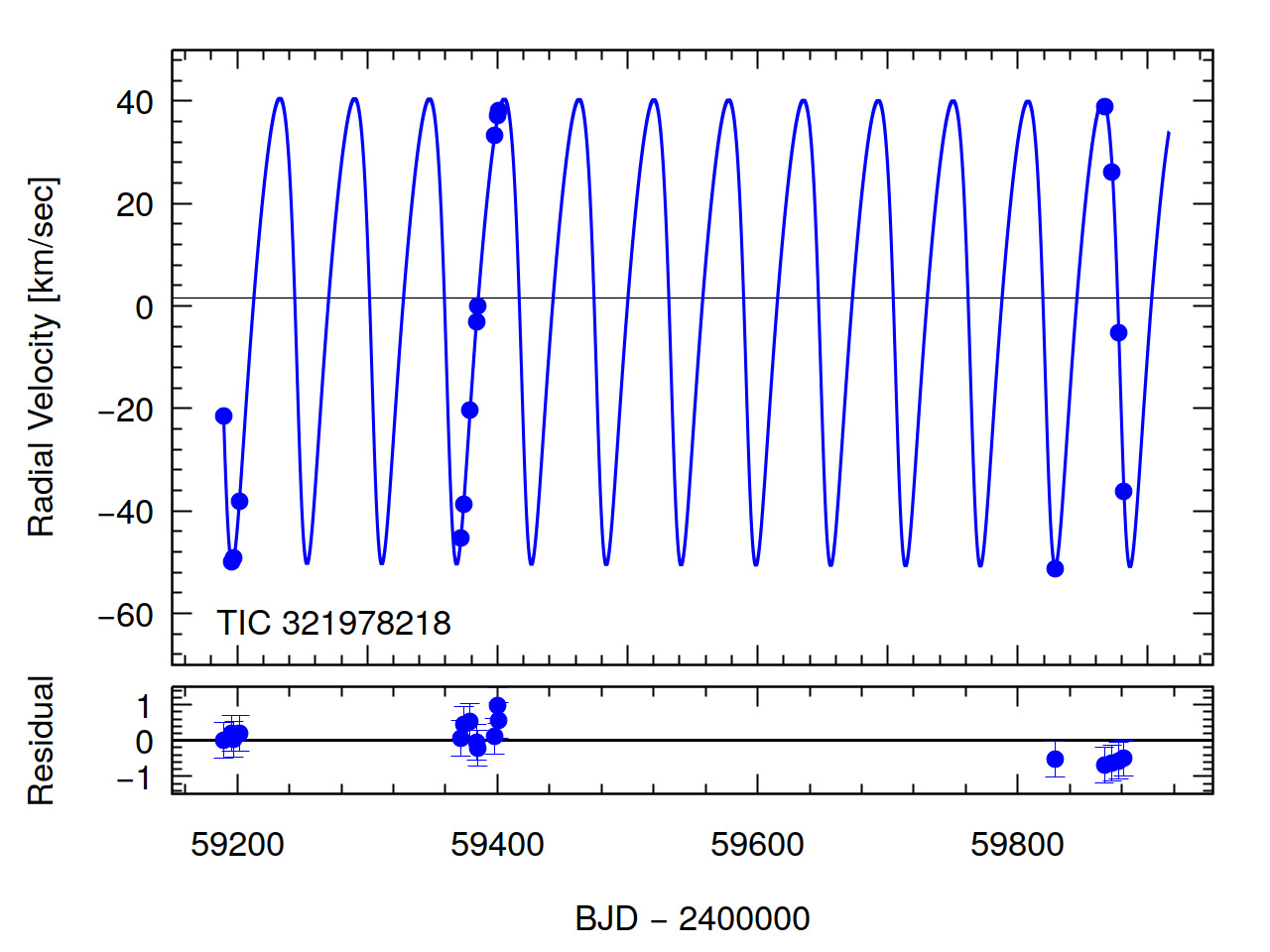}
      \caption{RV curve and model fit for the third component of TIC 321978218. The thin, horizontal line at $V=1.6\,\mathrm{km s}^{-1}$ represents the systemic radial velocity of the quadruple star system.}
         \label{fig:321978218rv}
   \end{figure}

{\bf TIC\,321978218}: The left panel of Fig.\,\ref{fig:321978218ETV} displays the ETV curve of TIC\,321978218 as derived exclusively from the TESS observations. As one can see, the 57.5-day period outer orbit is completely covered both during the Year 3 (Sectors 28 and 29) and Year 5 (Sectors 68 and 69) TESS observations.  This fact makes the determination of the outer orbital period and  the other orbital elements quite robust without the use of the extra information carried by the third-body eclipses.  Our analytic ETV fit also revealed the fact that the ETV is dominated by the LTTE effect ($\sim$$73\%$) over the dynamical delays ($\sim$$27\%$). Besides these data, we also obtained 17 RV points of the dominant tertiary star with the CHIRON instrument.  These RV data also nicely cover the outer orbit (see Fig.~\ref{fig:321978218rv} and Table~\ref{tab:TIC321978218_RVdata}). A combination of the LTTE-dominated ETV curve, which describes the motion of the inner EB around the centre of mass of the triple system, and the RV curve, which reflects the tertiary's motion, makes the outer pair a quasi-SB2 binary.  In turn, one can determine both the outer mass ratio ($q_{\rm out}$) and the total (projected) mass of the inner EB ($M_\mathrm{A}\sin^3i_\mathrm{out}=M_\mathrm{Aa}\sin^3i_\mathrm{out}+M_\mathrm{Ab}\sin^3i_\mathrm{out}$) and the individual (projected) mass of the tertiary ($M_\mathrm{B}\sin^3i_\mathrm{out}$) component purely from the amplitudes of the ETV and RV curves. However, the actual system turned out to be not so simple. The reason is that, after receiving the Sector 68 observations and then deriving ETV points from these data, we realized immediately that the ETV curve exhibits a further, longer timescale, non-linear variation. Sector 69 data confirmed these findings. 

Then, we were able to extend our ETV study thanks to the archival SWASP observations.  As was mentioned above, five seasons of SWASP observations are available for TIC 321978218. The lesser quality of these observations do not allow us to determine individual times of eclipses with the accuracies we achieve with TESS which would be necessary to study the low-amplitude ($\lesssim10^{-3}\,d$) ETV variations.  Hence, instead of using individual eclipse times, we formed seasonal folded, binned light curves from each of the five seasons of SWASP observations (using an EB period of $P_{\rm in} = 0.5698127$ d). With the use of these seasonal average light curves, we determined five times of seasonal `eclipse points' (both for the primary and secondary eclipses). Each of these results in a representative ETV value which we locate near to the mid-time of the observations used to make the average light curve. Moreover, since these seasonal folded EB light curves essentially average out the 57.5-day outer period ETV variations, we chose a location for the mean ETV point that is not only near the SWASP seasonal midtime, but one that is also near the time when the outer 57.5-day outer ETV value would be near zero. (In practice, this means that the tertiary's orbital phase, measured from the middle of first TESS-observed third-body eclipse, was around $0\fp17$ or $0\fp73$.) These nominal seasonal ETV points extend the length of the available ETV data by almost 5\,000 days and confirm that TIC\,321978218 exhibits additional period variation(s) (see right panel of Fig.\,\ref{fig:321978218ETV}). 

At this time we have insufficient information to say anything certain about the nature of these variations either qualitatively or quantitatively. In this study, however, we decided to model this extra ETV variation under the assumption that TIC\,321978218 is a 2+1+1 type quadruple star system. We emphasize, however, that, though this assumption was necessary for a consistent modeling of the available datasets, there is no guarantee that the parameters of the outermost orbit and the fourth star mean much unless the outermost orbit is ultimately confirmed.

{\bf TIC 323486857}: The very noisy ETV curve of TIC~323486857 is shown in Fig.~\ref{fig:323486857ETV}. The large uncertainties of the points, which make the ETV curve nearly unusable, are due to the highly distorted light curve. Fortunately, in addition to the Gaia spectroscopic solution for the outer orbit, nine individual third body eclipses were detected during the six TESS sectors and, hence, the outer period, which is the shortest among our sample, is well characterized. As can be seen in Fig.~\ref{fig:323486857ETV}, the ETV curve is mainly dominated by the LTTE, which is expected for such a nearly flat, doubly circular triple star system.

{\bf TIC 650024463}: Figure \ref{fig:650024463ETV} shows an ETV curve for TIC 650024463 that is fairly well populated with data points from 9 sectors of data spanning more than four years of TESS observations.  In addition to the well mapped out variations with the outer orbital period of 108 days, the better part of 2/3 of a complete apsidal motion cycle is observed. The ETV variations with the outer period of 108 d are $\sim$$90$\% dynamical delays and $\sim$$10$\% LTTE effects.
   
\section{Photodynamical models}
\label{sec:photodynamics}

All seven of the triple systems in this work have been subjected to a detailed photodynamical analysis\footnote{To our knowledge, the first photodynamical analysis of a triply eclipsing triple system was carried out by \citet{carter11} for the system KOI-126.}  in order to extract all of the system parameters.  For the photodynamical analysis we utilize the software package {\sc Lightcurvefactory} \citep[see, e.g.][and references therein]{borkovitsetal19a,borkovitsetal20a}. This code has been developed over the past decade and has been described in a number of previous papers.  The code contains three basic features: (i) emulators for multi-passband light curve(s), the corresponding ETVs, and radial velocity curve (this feature is present whether or not we have actual RV data); (ii) a built-in numerical integrator (a seventh-order Runge-Kutta-Nystr\"om algorithm) to calculate the perturbed three-, or multiple-body orbits, specifically, the coordinates and velocities of the three or more stars in the system;  and (iii) a Markov Chain Monte Carlo (MCMC)-based search routine for determining the best-fit system parameters, and the statistical uncertainties, as well.  The latter feature uses our own implementation of the generic Metropolis-Hastings algorithm \citep[see, e.g., ][]{ford05}. 

{\sc Lightcurvefactory} was developed to analyze multiple star systems, including binaries, triples, and quadruple stars of both the 2+2 and 2+1+1 architecture. The workings of the code, the steps involved in the analysis procedure, and its application to a wide range of multi-stellar systems have been explained in detail \citep{borkovitsetal18,borkovitsetal19a,borkovitsetal19b,borkovitsetal20a,borkovitsetal20b,borkovitsetal21,mitnyanetal20}.  Specifically {\sc Lightcurvefactory} has been used successfully to analyse compact and wider triple systems, both with and without outer eclipses.  

\begin{table}
\centering
\caption{Definitions of triple system parameters in Tables~\ref{tab:syntheticfit_TIC133771812176713425}-- \ref{tab:syntheticfit_TIC323486857650024463}}
\label{tbl:definitions}
\small
\begin{tabular}{lc}
\hline
\hline
Parameter$^a$ & Definition   \\
\hline
$t_0$ & Epoch time for osculating elements    \\
$P$ & Orbital period  \\ 
$a$ & Orbital semimajor axis  \\
$e$ & Orbital eccentricity \\
$\omega$ & Argument of periastron (of secondary) \\
$i$ & Orbital inclination angle \\
$\mathcal{T}_0^\mathrm{inf/sup}$ & Time of conjunction of the secondary$^b$ \\
$\tau$ & Time of periastron passage  \\
$\Omega$ & Longitude of the node relative to \\
& the node of the inner orbit \\
$i_{\rm mut}$ & Mutual inclination angle$^c$   \\
$q$ & Mass ratio (secondary/primary)  \\ 
$K_\mathrm{pri}$ & ``K'' velocity amplitude of primary \\
$K_\mathrm{sec}$ & ``K'' velocity amplitude of secondary \\
$R/a$ & Stellar radius divided by semimajor axis \\
$T_{\rm eff}/T_{\rm eff,Aa}$ & Temperature relative to EB primary \\
fractional flux  & Stellar contribution in the given band \\
$M$ & Stellar mass  \\
$R$ & Stellar radius   \\
$T_\mathrm{eff}$ & Stellar effective temperature  \\ 
$L_\mathrm{bol}$ & Stellar bolometric luminosity  \\
$M_\mathrm{bol}$ & Stellar absolute bolometric magnitude \\
$M_V$ & Stellar absolute visual magnitude \\
$\log g$ & log surface gravity (cgs units) \\
$[M/H]$ & log metallicity abundance to H, by mass \\
$E(B-V)$ & Color excess in B-V bands  \\
extra light, $\ell_4$  & Contaminating flux in the given band   \\
$(M_V)_\mathrm{tot}$ & System absolute visual magnitude   \\ 
distance & Distance to the source  \\
\hline   
\end{tabular}

\textit{Notes}. (a) The units for the parameters are given in Tables~\ref{tab:syntheticfit_TIC133771812176713425}-- \ref{tab:syntheticfit_TIC323486857650024463}. (b) The superscript of ``inf/sup'' indicates inferior vs.~superior conjunctions. (By default we give inferior conjunctions. Superior conjunctions are indicated by asteriks.) (c) More explicitly, this is the angle between the orbital planes of the inner binary and the outer triple orbit.

\end{table} 

Essentially all the details of how this code was used to analyze the compact triply eclipsing triple systems that were found with TESS, were described in \citet{rappaport22}.  Therefore, here we will provide only a high-level overview of the inputs to the code and the parameters that are either fitted or constrained by the MCMC fit.  Altogether, in the case of a hierarchical triple configuration, there are 25 -- 27 system parameters that result directly from the analysis.  Specifically, these are all the stellar parameters (9 = 3 $\times$ mass, radius, $T_{\rm eff}$), all 12 of the elements of the inner and outer orbits, to which, in the case of radial velocity data, one can add the systemic radial velocity as a thirteenth orbital parameter, as well as the 4 system parameters: distance to the source and the interstellar extinction, as well as the system metallicity and age.  Finally, one may optionally adjust the amount of any passband-dependent contaminated (extra) light $\ell_4$ if it is necessary.  

The {\sc Lightcurvefactory} code obviously requires a substantial amount of `input' information in order to allow for the extraction of 26 independent system parameters.  This input information can be divided into two basic categories.  First, there are the `data'.  These include: the (i) EB eclipse profiles, (ii) third-body eclipse profile(s), (iii) times of the EB eclipses which are distinct from the shape and depths of eclipses, (iv) archival SED values, and (v) radial velocities (available only for three of our systems).  Second, we utilize \texttt{PARSEC} model stellar evolution tracks and isochrones as well as model stellar atmospheres \citep{PARSEC}.  The evolution tracks enable us to find the stellar radius and $T_{\rm eff}$ for a given stellar mass, age and metallicity, while the isochrones allow us to compute stellar magnitudes in different passbands in order to fit the SED curve.  The available input information for each of the seven triples is summarized in Table \ref{tbl:input}.

The EB and third-body eclipse profiles and the ETV curves used in the photodynamical analysis were taken from the TESS full-frame images (`FFI').  The photometry on the FFIs was done with Andras Pal's  FITSH package \citep{pal12}. In order to save on computational time we  binned the 10-min cadence data to 30-min cadence\footnote{While {\sc Lightcurvefactory} is able to handle directly finite exposure times, we found that finite exposure (more, strictly speaking cadence-) time corrections were unnecessary to apply even for the 30-min cadence light curves, due to the relatively long durations of both the inner and outer eclipses compared to the cadence times.}, and dropped the out-of-eclipse regions of these light curves, keeping only the $\pm0\fp15$ phase-domain regions around the EB eclipses themselves.  However, whenever a data segment contains a third-body (i.e., `outer') eclipse, we keep the data for an entire binary period before and after the first and last contacts of that particular third-body eclipse. 


As noted in Table \ref{tbl:noticed}, we have RV data for only three of the seven sources.  This may raise the question of how we are able to derive absolute stellar masses, temperatures, and radii.  Here we provide a qualitative answer to this question.  There are in fact several pieces of information that involve combinations of the masses.  First, the ETV curve contains information about the light travel time effect and that, in turn, is equivalent to an SB1 RV measurement of the outer orbit.  We have useable ETV data for six of the seven sources (see Table 3).  And, for that one case where we have no  ETV curve, there are RV data (for the tertiary).  We note that the same information is also encoded into the light curves, as the light curves are fitted in the time-domain. Finally in this regard, the ETV curve (and the light curve, as well) also contain signatures of the dynamical delays which, in turn, carry information about primarily $q_{\rm out}$  and, in a less certain way (through the higher order perturbations), about $q_{\rm in}$ \citep[see, e.g.,][]{borkovitsetal15}.

Additionally, the geometry and timing of the outer eclipses carry significant further information about the ratio of $q_{\rm out}/q_{\rm in}$, as was elaborated on, for example, in Appendix A of \citet{borkovitsetal13}. It is even intuitive that, while the inner eclipses provide information about $R_{\rm Aa,Ab}/a_{\rm in}$, the outer eclipses provide information about $R_{\rm Aa,Ab,B}/a_{\rm out}$ and, hence, their combination leads to the above mentioned ratio of mass ratios.

We did not use spectroscopically determined temperatures and metallicities for most of the systems (as they are unavailable) but, instead, we use SED fits to determine or constrain absolute temperatures. This information is also combined with the results of the simultaneous light curve fits, the latter of which also give combinations of the ratios of the surface brightnesses and, hence, indirectly the absolute temperature of the three stars. Primarily these may be obtained from combinations of the eclipse depths of the inner and outer eclipses. Here one should keep in mind that the ratio of the surface brightnesses of the inner EB components can be obtained not only from the mutual eclipses of the inner components, but also from those events where the inner EB members eclipse (or are eclipsed by) the tertiary star separately. Hence from a light curve containing both inner and outer eclipses, the information which can be mined out is not simply the sum of the parameters that can be determined from two independent eclipsing light curves, but much more. The use of SED information to obtain temperatures is also employed in several other recent papers, for example in \citet{miller20} and \citet{stassun16}.

Finally, we combine this SED information with theoretical, coeval stellar isochrones which provide information not only on the radii and $T_{\rm eff}$ of the stars, but also the masses for a given age (but of course, such masses are no longer independent of the astrophysical assumptions and, hence, they may be somewhat inferior to those dynamical masses which can be directly inferred from high quality RV data). We discuss this question in detail in \citet{borkovitsetal20a,borkovitsetal22a}. All the stars are typically tied together via a ``coeval'' assumption.  And, of course, knowledge of the masses sets the size scales of the system, which then provides for absolute determinations of semi-major axes and stellar radii.



\section{System parameters}
\label{sec:results}

\subsection{Tables of fitted parameters}

With the use of the 25--27 directly fitted system parameters, one can also determine a number of other astrophysically and/or dynamically relevant additional parameters for each star in the triple.  Thus, we can include all the basic stellar parameters in our tabulated results, as well as the inner and outer orbital parameters discussed above, including the relative orientations of the two orbits.  We also give several relative values (e.g., $R/a$, $T_B/T_{Aa}$), some global system parameters such as distance, $[M/H]$, and $E(B-V)$, and a number of parameters that are derived from the fitted parameters.  Each of the results tables contains a few dozen different parameters which are, for the most part, written as symbols only.  These symbols were defined in \citet{rappaport23}, but for the sake of completeness we repeat these definitions here in Table \ref{tbl:definitions}. Then, the system parameters that are derived from the photodynamical analyses are listed in Tables \ref{tab:syntheticfit_TIC133771812176713425} through \ref{tab:syntheticfit_TIC323486857650024463}.

Before discussing the results individually for each system, we make one other important point. The orbital parameters (including also the inner and outer periods) tabulated below are instantaneous, osculating orbital elements, which are valid strictly only for the specific epoch $t_0$ which is given in the very first row of each table. Therefore, the given periods and conjunction times are not applicable for observational predictions (i.e. for determining times of future inner or outer eclipses). For these latter purposes we provide Table~\ref{tab:ephemerides} which contains, among other things, the inner and outer eclipsing periods of the sources, and should be used for planning occasional future observations.

\begin{table*}
 \centering
\caption{Orbital and astrophysical parameters of TICs\,133771812 and 176713425 from the joint photodynamical light curve, ETV, SED and \texttt{PARSEC} isochrone solution. }
 \label{tab:syntheticfit_TIC133771812176713425}
\scalebox{0.91}{\begin{tabular}{@{}lllllll}
\hline
 & \multicolumn{3}{c}{TIC\,133771812} &  \multicolumn{3}{c}{TIC\,176713425}\\
\hline
\multicolumn{7}{c}{Orbital elements} \\
\hline
   & \multicolumn{6}{c}{Subsystem}  \\
   & \multicolumn{2}{c}{Aa--Ab} & A--B & \multicolumn{2}{c}{Aa--Ab} & A--B  \\
  \hline
  $t_0$ [BJD - 2400000]& \multicolumn{3}{c}{$58491.0$} & \multicolumn{3}{c}{$58738.5$} \\
  $P$ [days] & \multicolumn{2}{c}{$12.33390_{-0.00051}^{+0.00045}$} & $243.8900_{-0.0059}^{+0.0048}$ &  \multicolumn{2}{c}{$1.895107_{-0.000090}^{+0.000101}$} & $52.943_{-0.017}^{+0.019}$ \\
  $a$ [R$_\odot$] & \multicolumn{2}{c}{$31.66_{-0.30}^{+0.20}$} & $262.0_{-2.2}^{+1.6}$ & \multicolumn{2}{c}{$8.456_{-0.069}^{+0.082}$} & $89.42_{-0.74}^{+0.77}$ \\
  $e$ & \multicolumn{2}{c}{$0.03368_{-0.00044}^{+0.00043}$} & $0.2174_{-0.0058}^{+0.0058}$ & \multicolumn{2}{c}{$0.00592_{-0.00056}^{+0.00058}$} & $0.4118_{-0.0070}^{+0.0072}$ \\
  $\omega$ [deg]& \multicolumn{2}{c}{$230.39_{-0.61}^{+0.55}$} & $194.91_{-0.81}^{+0.78}$ & \multicolumn{2}{c}{$105.1_{-2.2}^{+2.5}$} & $95.22_{-0.58}^{+0.87}$ \\ 
  $i$ [deg] & \multicolumn{2}{c}{$89.98_{-0.13}^{+0.11}$} & $89.770_{-0.020}^{+0.019}$ & \multicolumn{2}{c}{$88.56_{-0.73}^{+0.66}$} & $88.714_{-0.089}^{+0.084}$ \\
  $\mathcal{T}_0^\mathrm{inf/sup}$ [BJD - 2400000]& \multicolumn{2}{c}{$58501.1638_{-0.0005}^{+0.0004}$} & ${58509.928_{-0.010}^{+0.009}}^*$ & \multicolumn{2}{c}{$58739.8183_{-0.0001}^{+0.0001}$} & ${58740.0972_{-0.0041}^{+0.0040}}^*$ \\
  $\tau$ [BJD - 2400000]& \multicolumn{2}{c}{$58499.8871_{-0.0209}^{+0.0189}$} & $58320.200_{-0.554}^{+0.583}$ & \multicolumn{2}{c}{$58738.951_{-0.012}^{+0.013}$} & $58687.448_{-0.042}^{+0.054}$ \\
  $\Omega$ [deg] & \multicolumn{2}{c}{$0.0$} & $-0.22_{-0.11}^{+0.13}$ & \multicolumn{2}{c}{$0.0$} & $0.62_{-0.77}^{+0.96}$ \\
  $i_\mathrm{mut}$ [deg] & \multicolumn{3}{c}{$0.31_{-0.14}^{+0.14}$} & \multicolumn{3}{c}{$1.00_{-0.52}^{+0.83}$} \\
  $\varpi^\mathrm{dyn}$ [deg]& \multicolumn{2}{c}{$50.39_{-0.61}^{+0.55}$} & $14.91_{-0.81}^{+0.78}$ & \multicolumn{2}{c}{$285.1_{-2.2}^{+2.5}$} & $275.22_{-0.58}^{+0.87}$ \\
  $i^\mathrm{dyn}$ [deg] & \multicolumn{2}{c}{$0.25_{-0.12}^{+0.11}$} & $0.06_{-0.03}^{+0.03}$ & \multicolumn{2}{c}{$0.83_{-0.43}^{+0.69}$} & $0.17_{-0.09}^{+0.14}$ \\
  $\Omega^\mathrm{dyn}$ [deg] & \multicolumn{2}{c}{$228_{-14}^{+15}$} & $48_{-14}^{+15}$ & \multicolumn{2}{c}{$72_{-29}^{+108}$} & $252_{-29}^{+108}$ \\
  $i_\mathrm{inv}$ [deg] & \multicolumn{3}{c}{$89.81_{-0.04}^{+0.03}$} & \multicolumn{3}{c}{$88.68_{-0.11}^{+0.14}$} \\
  $\Omega_\mathrm{inv}$ [deg] & \multicolumn{3}{c}{$-0.18_{-0.09}^{+0.10}$} & \multicolumn{3}{c}{$0.52_{-0.64}^{+0.80}$} \\
  \hline
  mass ratio $[q=M_\mathrm{sec}/M_\mathrm{pri}]$ & \multicolumn{2}{c}{$0.7627_{-0.0057}^{+0.0054}$} & $0.4488_{-0.0092}^{+0.0099}$ & \multicolumn{2}{c}{$0.9334_{-0.0048}^{+0.0052}$} & $0.5150_{-0.0041}^{+0.0042}$ \\
  $K_\mathrm{pri}$ [km\,s$^{-1}$] & \multicolumn{2}{c}{$56.23_{-0.57}^{+0.50}$} & $17.26_{-0.31}^{+0.32}$ & \multicolumn{2}{c}{$109.01_{-0.83}^{+0.87}$} & $31.92_{-0.35}^{+0.24}$ \\ 
  $K_\mathrm{sec}$ [km\,s$^{-1}$] & \multicolumn{2}{c}{$73.76_{-0.61}^{+0.45}$} & $38.44_{-0.34}^{+0.33}$ & \multicolumn{2}{c}{$116.67_{-0.98}^{+1.31}$} & $61.86_{-0.53}^{+0.70}$ \\ 
  \hline
  \multicolumn{7}{c}{Apsidal and nodal motion related parameters} \\
  \hline
$P_\mathrm{apse}$ [year] & \multicolumn{2}{c}{$52.81_{-0.81}^{+0.78}$} & $225.0_{-1.6}^{+1.8}$ & \multicolumn{2}{c}{$7.09_{-0.12}^{+0.13}$} & $59.71_{-0.85}^{+0.82}$ \\ 
$P_\mathrm{apse}^\mathrm{dyn}$ [year] & \multicolumn{2}{c}{$23.64_{-0.32}^{+0.31}$} & $35.96_{-0.34}^{+0.35}$ & \multicolumn{2}{c}{$4.15_{-0.05}^{+0.05}$} & $8.57_{-0.10}^{+0.09}$ \\ 
$P_\mathrm{node}^\mathrm{dyn}$ [year] & \multicolumn{3}{c}{$42.81_{-0.51}^{+0.51}$} & \multicolumn{3}{c}{$10.01_{-0.11}^{+0.11}$} \\
$\Delta\omega_\mathrm{3b}$ [arcsec/cycle] & \multicolumn{2}{c}{$1850_{-24}^{+25}$} & $24064_{-233}^{+231}$ & \multicolumn{2}{c}{$1231_{-13}^{+14}$} & $21915_{-239}^{+255}$ \\ 
$\Delta\omega_\mathrm{GR}$ [arcsec/cycle] & \multicolumn{2}{c}{$0.731_{-0.014}^{+0.009}$} & $0.134_{-0.002}^{+0.002}$ & \multicolumn{2}{c}{$2.207_{-0.036}^{+0.043}$} & $0.381_{-0.007}^{+0.007}$ \\ 
$\Delta\omega_\mathrm{tide}$ [arcsec/cycle] & \multicolumn{2}{c}{$0.179_{-0.009}^{+0.009}$} & $0.00051_{-0.00003}^{+0.00003}$ & \multicolumn{2}{c}{$387_{-17}^{+15}$} & $0.638_{-0.028}^{+0.027}$  \\ 
  \hline  
\multicolumn{7}{c}{Stellar parameters} \\
\hline
   & Aa & Ab &  B & Aa & Ab &  B \\
  \hline
 \multicolumn{7}{c}{Relative quantities} \\
  \hline
 fractional radius [$R/a$]  & $0.0491_{-0.0005}^{+0.0004}$ & $0.0368_{-0.0008}^{+0.0008}$ & $0.0047_{-0.0001}^{+0.0002}$ & $0.2060_{-0.0018}^{+0.0018}$ & $0.1525_{-0.0025}^{+0.0024}$ & $0.0189_{-0.0008}^{+0.0011}$ \\
 temperature relative to $(T_\mathrm{eff})_\mathrm{Aa}$ & $1$ & $0.8345_{-0.0058}^{+0.0049}$ & $0.8486_{-0.0095}^{+0.0139}$ & $1$ & $1.0039_{-0.0022}^{+0.0041}$ & $1.0040_{-0.0095}^{+0.0052}$ \\
 fractional flux [in TESS-band] & $0.3476_{-0.0121}^{+0.0141}$ & $0.1141_{-0.0015}^{+0.0016}$ & $0.1341_{-0.0120}^{+0.0137}$ & $0.3657_{-0.0107}^{+0.0110}$ & $0.2037_{-0.0024}^{+0.0025}$ & $0.3552_{-0.0256}^{+0.0260}$ \\
  fractional flux [in Cousins $R_C$-band] & $-$ & $-$ & $-$ & $0.3808_{-0.0122}^{+0.0123}$ & $0.2134_{-0.0052}^{+0.0063}$ & $0.3770_{-0.0321}^{+0.0171}$ \\
 \hline
 \multicolumn{7}{c}{Physical quantities} \\
  \hline 
 $M$ [M$_\odot$] & $1.588_{-0.044}^{+0.028}$ & $1.209_{-0.033}^{+0.026}$ & $1.254_{-0.035}^{+0.034}$ & $1.166_{-0.029}^{+0.037}$ & $1.091_{-0.026}^{+0.028}$ & $1.164_{-0.031}^{+0.025}$ \\
 $R$ [R$_\odot$] & $1.554_{-0.017}^{+0.017}$ & $1.165_{-0.034}^{+0.031}$ & $1.225_{-0.041}^{+0.046}$ & $1.743_{-0.027}^{+0.024}$ & $1.290_{-0.027}^{+0.027}$ & $1.692_{-0.072}^{+0.105}$ \\
 $T_\mathrm{eff}$ [K]& $7706_{-144}^{+168}$ & $6426_{-89}^{+112}$ & $6561_{-114}^{+103}$ & $6033_{-192}^{+78}$ & $6051_{-153}^{+77}$ & $6033_{-160}^{+97}$ \\
 $L_\mathrm{bol}$ [L$_\odot$] & $7.66_{-0.61}^{+0.67}$ & $2.08_{-0.22}^{+0.23}$ & $2.49_{-0.29}^{+0.35}$ & $3.63_{-0.55}^{+0.20}$ & $2.01_{-0.23}^{+0.13}$ & $3.43_{-0.43}^{+0.42}$ \\
 $M_\mathrm{bol}$ & $2.56_{-0.09}^{+0.09}$ & $3.97_{-0.11}^{+0.12}$ & $3.78_{-0.14}^{+0.13}$ & $3.37_{-0.06}^{+0.18}$ & $4.01_{-0.07}^{+0.13}$ & $3.43_{-0.12}^{+0.15}$ \\
 $M_V           $ & $2.52_{-0.09}^{+0.09}$ & $3.97_{-0.12}^{+0.12}$ & $3.77_{-0.15}^{+0.14}$ & $3.39_{-0.06}^{+0.20}$ & $4.03_{-0.07}^{+0.15}$ & $3.44_{-0.12}^{+0.16}$ \\
 $\log g$ [dex] & $4.253_{-0.008}^{+0.009}$ & $4.388_{-0.015}^{+0.016}$ & $4.358_{-0.021}^{+0.019}$ & $4.022_{-0.007}^{+0.007}$ & $4.254_{-0.012}^{+0.012}$ & $4.043_{-0.047}^{+0.040}$ \\
 \hline
\multicolumn{7}{c}{Global system parameters} \\
  \hline
$\log$(age) [dex] &\multicolumn{3}{c}{$8.760_{-0.053}^{+0.142}$} & \multicolumn{3}{c}{$9.720_{-0.021}^{+0.097}$} \\
$[M/H]$  [dex]    &\multicolumn{3}{c}{$-0.009_{-0.094}^{+0.049}$} & \multicolumn{3}{c}{$0.076_{-0.126}^{+0.080}$} \\
$E(B-V)$ [mag]    &\multicolumn{3}{c}{$0.425_{-0.023}^{+0.024}$} & \multicolumn{3}{c}{$0.158_{-0.045}^{+0.022}$} \\
extra light $\ell_4$ [in TESS-band] & \multicolumn{3}{c}{$0.407_{-0.026}^{+0.016}$} & \multicolumn{3}{c}{$0.075_{-0.022}^{+0.025}$} \\
extra light $\ell_4$ [in Cousins $R_C$-band] & \multicolumn{3}{c}{$-$} & \multicolumn{3}{c}{$0.026_{-0.015}^{+0.031}$} \\
$(M_V)_\mathrm{tot}$  &\multicolumn{3}{c}{$2.03_{-0.10}^{+0.08}$} & \multicolumn{3}{c}{$2.39_{-0.06}^{+0.18}$} \\
distance [pc]           & \multicolumn{3}{c}{$1428_{-27}^{+32}$} & \multicolumn{3}{c}{$2097_{-51}^{+51}$} \\  
\hline
\end{tabular}}
\end{table*}

\begin{table*}
 \centering
\caption{Orbital and astrophysical parameters of TICs\,185615681 and 287756035 from the joint photodynamical light curve, ETV, SED and \texttt{PARSEC} isochrone solution. }
 \label{tab:syntheticfit_TIC185615681287756035}
\scalebox{0.91}{\begin{tabular}{@{}lllllll}
\hline
 & \multicolumn{3}{c}{TIC\,185615681} &  \multicolumn{3}{c}{TIC\,287756035}\\
\hline
\multicolumn{7}{c}{Orbital elements} \\
\hline
   & \multicolumn{6}{c}{Subsystem}  \\
   & \multicolumn{2}{c}{Aa--Ab} & A--B & \multicolumn{2}{c}{Aa--Ab} & A--B  \\
  \hline
  $t_0$ [BJD - 2400000]& \multicolumn{3}{c}{$58516.0$} & \multicolumn{3}{c}{58596.5} \\
  $P$ [days] & \multicolumn{2}{c}{$2.3180401_{-0.0000066}^{+0.0000064}$} & $56.06661_{-0.00066}^{+0.00071}$ &  \multicolumn{2}{c}{$2.082184_{-0.000082}^{+0.000102}$} & $367.923_{-0.023}^{+0.022}$ \\
  $a$ [R$_\odot$] & \multicolumn{2}{c}{$11.718_{-0.057}^{+0.073}$} & $104.78_{-0.52}^{+0.57}$ & \multicolumn{2}{c}{$8.725_{-0.112}^{+0.049}$} & $318.1_{-4.5}^{+2.1}$ \\
  $e$ & \multicolumn{2}{c}{$0.010669_{-0.000071}^{+0.000087}$} & $0.09817_{-0.00084}^{+0.00087}$ & \multicolumn{2}{c}{$0.0036_{-0.0012}^{+0.0012}$} & $0.235_{-0.025}^{+0.019}$ \\
  $\omega$ [deg]& \multicolumn{2}{c}{$207.87_{-0.41}^{+0.58}$} & $308.74_{-0.51}^{+0.54}$ & \multicolumn{2}{c}{$175_{-31}^{+27}$} & $53.70_{-0.66}^{+0.63}$ \\ 
  $i$ [deg] & \multicolumn{2}{c}{$88.06_{-0.12}^{+0.11}$} & $89.826_{-0.097}^{+0.071}$ & \multicolumn{2}{c}{$89.27_{-0.83}^{+0.59}$} & $89.095_{-0.028}^{+0.049}$ \\
  $\mathcal{T}_0^\mathrm{inf/sup}$ [BJD - 2400000]& \multicolumn{2}{c}{$58518.17078_{-0.00004}^{+0.00004}$} & ${58517.269_{-0.015}^{+0.015}}^*$ & \multicolumn{2}{c}{$58599.3048_{-0.0008}^{+0.0009}$} & $59311.672_{-0.021}^{+0.039}$ \\
  $\tau$ [BJD - 2400000]& \multicolumn{2}{c}{$58517.777_{-0.003}^{+0.004}$} & $58496.431_{-0.079}^{+0.082}$ & \multicolumn{2}{c}{$58598.76_{-0.18}^{+0.16}$} & $59071.8_{-1.1}^{+1.4}$ \\
  $\Omega$ [deg] & \multicolumn{2}{c}{$0.0$} & $2.17_{-0.14}^{+0.13}$ & \multicolumn{2}{c}{$0.0$} & $1.22_{-0.54}^{+0.64}$ \\
  $i_\mathrm{mut}$ [deg] & \multicolumn{3}{c}{$2.79_{-0.16}^{+0.17}$} & \multicolumn{3}{c}{$1.39_{-0.49}^{+0.60}$} \\
  $\varpi^\mathrm{dyn}$ [deg]& \multicolumn{2}{c}{$27.83_{-0.41}^{+0.57}$} & $128.74_{-0.51}^{+0.54}$ & \multicolumn{2}{c}{$355_{-31}^{+27}$} & $233.71_{-0.66}^{+0.63}$ \\
  $i^\mathrm{dyn}$ [deg] & \multicolumn{2}{c}{$1.97_{-0.12}^{+0.12}$} & $0.82_{-0.05}^{+0.05}$ & \multicolumn{2}{c}{$1.27_{-0.44}^{+0.55}$} & $0.12_{-0.04}^{+0.05}$ \\
  $\Omega^\mathrm{dyn}$ [deg] & \multicolumn{2}{c}{$50.8_{-1.8}^{+1.9}$} & $230.8_{-1.8}^{+1.9}$ & \multicolumn{2}{c}{$98_{-37}^{+27}$} & $278_{-37}^{+27}$ \\
  $i_\mathrm{inv}$ [deg] & \multicolumn{3}{c}{$89.30_{-0.08}^{+0.08}$} & \multicolumn{3}{c}{$89.12_{-0.06}^{+0.04}$} \\
  $\Omega_\mathrm{inv}$ [deg] & \multicolumn{3}{c}{$1.53_{-0.10}^{+0.10}$} & \multicolumn{3}{c}{$1.12_{-0.50}^{+0.59}$} \\
  \hline
  mass ratio $[q=M_\mathrm{sec}/M_\mathrm{pri}]$ & \multicolumn{2}{c}{$0.998_{-0.002}^{+0.002}$} & $0.221_{-0.003}^{+0.002}$ & \multicolumn{2}{c}{$1.193_{-0.018}^{+0.019}$} & $0.552_{-0.005}^{+0.004}$ \\
  $K_\mathrm{pri}$ [km\,s$^{-1}$] & \multicolumn{2}{c}{$127.76_{-0.63}^{+0.77}$} & $17.21_{-0.16}^{+0.15}$ & \multicolumn{2}{c}{$115.44_{-2.27}^{+1.30}$} & $16.01_{-0.27}^{+0.19}$ \\ 
  $K_\mathrm{sec}$ [km\,s$^{-1}$] & \multicolumn{2}{c}{$127.99_{-0.65}^{+0.80}$} & $77.81_{-0.38}^{+0.57}$ & \multicolumn{2}{c}{$96.58_{-0.53}^{+0.49}$} & $28.95_{-0.27}^{+0.24}$ \\ 
  $V_\gamma$ [km\,s$^{-1}$] & \multicolumn{3}{c}{$-5.97_{-0.16}^{+0.15}$} & \multicolumn{3}{c}{$-3.90_{-0.20}^{+0.27}$} \\
  \hline
  \multicolumn{7}{c}{Apsidal and nodal motion related parameters} \\
  \hline
$P_\mathrm{apse}$ [year] & \multicolumn{2}{c}{$18.17_{-0.12}^{+0.15}$} & $64.38_{-0.11}^{+0.09}$ & \multicolumn{2}{c}{$6.56_{-0.69}^{+0.68}$} & $6429_{-117}^{+142}$ \\ 
$P_\mathrm{apse}^\mathrm{dyn}$ [year] & \multicolumn{2}{c}{$9.28_{-0.05}^{+0.07}$} & $14.67_{-0.05}^{+0.07}$ & \multicolumn{2}{c}{$6.49_{-0.68}^{+0.66}$} & $515_{-9}^{+10}$ \\ 
$P_\mathrm{node}^\mathrm{dyn}$ [year] & \multicolumn{3}{c}{$18.99_{-0.13}^{+0.09}$} & \multicolumn{3}{c}{$560_{-11}^{+10}$} \\
$\Delta\omega_\mathrm{3b}$ [arcsec/cycle] & \multicolumn{2}{c}{$738.6_{-5.9}^{+4.3}$} & $13564_{-65}^{+47}$ & \multicolumn{2}{c}{$25.22_{-0.49}^{+0.45}$} & $2535_{-50}^{+45}$ \\ 
$\Delta\omega_\mathrm{GR}$ [arcsec/cycle] & \multicolumn{2}{c}{$2.832_{-0.028}^{+0.035}$} & $0.391_{-0.004}^{+0.004}$ & \multicolumn{2}{c}{$1.946_{-0.050}^{+0.022}$} & $0.087_{-0.002}^{+0.002}$ \\ 
$\Delta\omega_\mathrm{tide}$ [arcsec/cycle] & \multicolumn{2}{c}{$144.4_{-1.6}^{+1.7}$} & $0.227_{-0.003}^{+0.003}$ & \multicolumn{2}{c}{$1112_{-106}^{+132}$} & $0.131_{-0.012}^{+0.014}$  \\ 
  \hline  
\multicolumn{7}{c}{Stellar parameters} \\
\hline
   & Aa & Ab &  B & Aa & Ab &  B \\
  \hline
 \multicolumn{7}{c}{Relative quantities} \\
  \hline
 fractional radius [$R/a$]  & $0.1749_{-0.0005}^{+0.0005}$ & $0.1744_{-0.0005}^{+0.0005}$ & $0.00749_{-0.00008}^{+0.00008}$ & $0.1129_{-0.0017}^{+0.0018}$ & $0.2707_{-0.0054}^{+0.0061}$ & $0.0212_{-0.0008}^{+0.0006}$ \\
 temperature relative to $(T_\mathrm{eff})_\mathrm{Aa}$ & $1$ & $0.9986_{-0.0014}^{+0.0014}$ & $0.5633_{-0.0057}^{+0.0042}$ & $1$ & $0.9126_{-0.0117}^{+0.0095}$ & $0.8043_{-0.0053}^{+0.0115}$ \\
 fractional flux [in TESS-band] & $0.4840_{-0.0035}^{+0.0039}$ & $0.4806_{-0.0036}^{+0.0039}$ & $0.0126_{-0.0007}^{+0.0007}$ & $0.0372_{-0.0015}^{+0.0016}$ & $0.1537_{-0.0065}^{+0.0065}$ & $0.7708_{-0.0174}^{+0.0170}$ \\
 \hline
 \multicolumn{7}{c}{Physical quantities} \\
  \hline 
 $M$ [M$_\odot$] & $2.009_{-0.030}^{+0.038}$ & $2.005_{-0.029}^{+0.037}$ & $0.888_{-0.013}^{+0.014}$ & $0.935_{-0.023}^{+0.013}$ & $1.116_{-0.053}^{+0.027}$ & $1.133_{-0.056}^{+0.029}$ \\
 $R$ [R$_\odot$] & $2.051_{-0.013}^{+0.012}$ & $2.045_{-0.012}^{+0.012}$ & $0.785_{-0.011}^{+0.011}$ & $0.986_{-0.020}^{+0.017}$ & $2.361_{-0.055}^{+0.060}$ & $6.753_{-0.318}^{+0.205}$ \\
 $T_\mathrm{eff}$ [K]& $9122_{-110}^{+140}$ & $9108_{-109}^{+137}$ & $5136_{-62}^{+64}$ & $6025_{-166}^{+37}$ & $5501_{-207}^{+54}$ & $4833_{-50}^{+53}$ \\
 $L_\mathrm{bol}$ [L$_\odot$] & $23.11_{-2.45}^{+2.30}$ & $22.92_{-2.49}^{+2.29}$ & $0.449_{-0.045}^{+0.050}$ & $1.14_{-0.11}^{+0.06}$ & $4.59_{-0.80}^{+0.26}$ & $22.59_{-2.87}^{+1.04}$ \\
 $M_\mathrm{bol}$ & $1.36_{-0.10}^{+0.12}$ & $1.37_{-0.10}^{+0.12}$ & $5.64_{-0.12}^{+0.11}$ & $4.63_{-0.06}^{+0.11}$ & $3.12_{-0.06}^{+0.21}$ & $1.39_{-0.05}^{+0.15}$ \\
 $M_V           $ & $1.39_{-0.06}^{+0.08}$ & $1.40_{-0.06}^{+0.08}$ & $5.79_{-0.15}^{+0.16}$ & $4.67_{-0.06}^{+0.12}$ & $3.25_{-0.07}^{+0.26}$ & $1.72_{-0.05}^{+0.17}$ \\
 $\log g$ [dex] & $4.116_{-0.003}^{+0.003}$ & $4.118_{-0.003}^{+0.003}$ & $4.595_{-0.006}^{+0.006}$ & $4.421_{-0.012}^{+0.011}$ & $3.738_{-0.021}^{+0.020}$ & $2.835_{-0.025}^{+0.029}$ \\
 \hline
\multicolumn{7}{c}{Global system parameters} \\
  \hline
$\log$(age) [dex] &\multicolumn{3}{c}{$8.768_{-0.059}^{+0.030}$} & \multicolumn{3}{c}{$9.756_{-0.041}^{+0.081}$} \\
$[M/H]$  [dex]    &\multicolumn{3}{c}{$-0.021_{-0.112}^{+0.175}$} & \multicolumn{3}{c}{$-0.262_{-0.122}^{+0.128}$} \\
$E(B-V)$ [mag]    &\multicolumn{3}{c}{$0.093_{-0.027}^{+0.022}$} & \multicolumn{3}{c}{$0.110_{-0.049}^{+0.032}$} \\
extra light $\ell_4$ [in TESS-band] & \multicolumn{3}{c}{$0.023_{-0.007}^{+0.006}$} & \multicolumn{3}{c}{$0.038_{-0.016}^{+0.017}$} \\
$(M_V)_\mathrm{tot}$  &\multicolumn{3}{c}{$0.63_{-0.06}^{+0.08}$} & \multicolumn{3}{c}{$1.42_{-0.04}^{+0.20}$} \\
distance [pc]           & \multicolumn{3}{c}{$942_{-13}^{+12}$} & \multicolumn{3}{c}{$2561_{-102}^{+77}$} \\  
\hline
\end{tabular}}
\end{table*}

\begin{table*}
 \centering
\caption{Orbital and astrophysical parameters of TIC\,321978218 from the joint photodynamical light curve, ETV, SED and \texttt{PARSEC} isochrone solution. }
 \label{tab:syntheticfit_TIC321978218}
\scalebox{0.91}{\begin{tabular}{@{}lllll}
\hline
  \multicolumn{5}{c}{TIC\,321978218} \\
\hline
\multicolumn{5}{c}{Orbital elements} \\
\hline
   & \multicolumn{4}{c}{Subsystem}  \\
   & \multicolumn{2}{c}{Aa--Ab} & A--B & AB--C  \\
  \hline
  $t_0$ [BJD - 2400000]& \multicolumn{4}{c}{$58354.0$} \\
  $P$ [days] & \multicolumn{2}{c}{$0.570267_{-0.000025}^{+0.000041}$} & $57.53549_{-0.00020}^{+0.00020}$ &  $4395_{-84}^{+52}$ \\
  $a$ [R$_\odot$] & \multicolumn{2}{c}{$3.473_{-0.011}^{+0.009}$} & $92.46_{-0.37}^{+0.36}$ & $1691_{-19}^{+18}$ \\
  $e$ & \multicolumn{2}{c}{$0.00100_{-0.00027}^{+0.00043}$} & $0.2583_{-0.0016}^{+0.0016}$ & $0.374_{-0.033}^{+0.049}$ \\
  $\omega$ [deg]& \multicolumn{2}{c}{$140_{-53}^{+22}$} & $122.56_{-0.27}^{+0.28}$ & $68.9_{-4.3}^{+3.1}$ \\ 
  $i$ [deg] & \multicolumn{2}{c}{$87.96_{-1.02}^{+0.37}$} & $90.000_{-0.069}^{+0.089}$ & $95.5_{-6.6}^{+9.2}$ \\
  $\mathcal{T}_0^\mathrm{inf/sup}$ [BJD - 2400000]& \multicolumn{2}{c}{$58354.2542_{-0.0003}^{+0.0002}$} & ${59072.6568_{-0.0031}^{+0.0033}}^*$ & ... \\
  $\tau$ [BJD - 2400000]& \multicolumn{2}{c}{$58354.061_{-0.077}^{+0.040}$} & $59018.153_{-0.071}^{+0.037}$ & $56799_{-24}^{+13}$ \\
  $\Omega$ [deg] & \multicolumn{2}{c}{$0.0$} & $0.06_{-0.57}^{+0.79}$ & $-2.2_{-7.2}^{+7.3}$ \\
  $(i_\mathrm{mut})_\mathrm{A-AB,A-C}$ [deg] & \multicolumn{3}{c}{$2.17_{-0.40}^{+1.08}$} &$11.3_{-5.4}^{+7.0}$ \\
  $(i_\mathrm{mut})_\mathrm{AB-C}$ [deg] & \multicolumn{2}{c}{...} &\multicolumn{2}{c}{$9.9_{-4.9}^{+6.5}$} \\
  $\varpi^\mathrm{dyn}$ [deg]& \multicolumn{2}{c}{$320_{-53}^{+22}$} & $302.53_{-0.30}^{+0.29}$ & $249.2_{-4.4}^{+3.1}$ \\
  $i^\mathrm{dyn}$ [deg] & \multicolumn{2}{c}{$5.7_{-2.4}^{+3.3}$} & $4.2_{-2.1}^{+2.8}$ & $5.7_{-2.9}^{+3.7}$ \\
  $\Omega^\mathrm{dyn}$ [deg] & \multicolumn{2}{c}{$350_{-33}^{+54}$} & $338_{-62}^{+68}$ & $159_{-57}^{+68}$ \\
  $i_\mathrm{inv}$ [deg] & \multicolumn{4}{c}{$92.3_{-2.8}^{+4.0}$} \\
  $\Omega_\mathrm{inv}$ [deg] & \multicolumn{4}{c}{$-0.9_{-3.2}^{+3.3}$} \\
  \hline
  mass ratio $[q=M_\mathrm{sec}/M_\mathrm{pri}]$ & \multicolumn{2}{c}{$0.782_{-0.015}^{+0.018}$} & $0.853_{-0.007}^{+0.008}$ & $0.052_{-0.005}^{+0.006}$ \\
  $K_\mathrm{pri}$ [km\,s$^{-1}$] & \multicolumn{2}{c}{$135.10_{-1.21}^{+1.54}$} & $38.76_{-0.31}^{+0.34}$ & $1.02_{-0.10}^{+0.11}$ \\ 
  $K_\mathrm{sec}$ [km\,s$^{-1}$] & \multicolumn{2}{c}{$172.85_{-2.15}^{+1.79}$} & $45.42_{-0.10}^{+0.09}$ & $19.80_{-0.53}^{+0.49}$ \\ 
  $V_\gamma$ [km\,s$^{-1}$] & \multicolumn{4}{c}{$1.707_{-0.045}^{+0.044}$} \\
  \hline
  \multicolumn{5}{c}{Apsidal and nodal motion related parameters} \\
  \hline
$P_\mathrm{apse}$ [year] & \multicolumn{2}{c}{$0.88_{-0.07}^{+0.05}$} & $526.9_{-2.6}^{+2.4}$ & $1004_{-84}^{+1138}$ \\ 
$P_\mathrm{apse}^\mathrm{dyn}$ [year] & \multicolumn{2}{c}{$0.86_{-0.07}^{+0.05}$} & $35.81_{-0.15}^{+0.14}$& $38.43_{-0.18}^{+0.19}$ \\ 
$P_\mathrm{node}^\mathrm{dyn}$ [year] & \multicolumn{4}{c}{$38.28_{-0.22}^{+0.23}$} \\
$\Delta\omega_\mathrm{3b}$ [arcsec/cycle] & \multicolumn{2}{c}{$101.25_{-0.54}^{+0.49}$} & $5700_{-22}^{+25}$ & $405544_{-7670}^{+6085}$ \\ 
$\Delta\omega_\mathrm{GR}$ [arcsec/cycle] & \multicolumn{2}{c}{$4.111_{-0.026}^{+0.022}$} & $0.307_{-0.003}^{+0.002}$ & $0.0192_{-0.0006}^{+0.0008}$ \\ 
$\Delta\omega_\mathrm{tide}$ [arcsec/cycle] & \multicolumn{2}{c}{$2251_{-119}^{+197}$} & $0.450_{-0.025}^{+0.038}$ & $0.0016_{-0.0001}^{+0.0001}$  \\ 
  \hline  
\multicolumn{5}{c}{Stellar parameters} \\
\hline
   & Aa & Ab &  B & C \\
  \hline
 \multicolumn{5}{c}{Relative quantities} \\
  \hline
 fractional radius [$R/a$]  & $0.2888_{-0.0035}^{+0.0041}$ & $0.2370_{-0.0027}^{+0.0055}$ & $0.0211_{-0.0003}^{+0.0003}$ & $0.00029_{-0.00001}^{+0.00001}$\\
 temperature relative to $(T_\mathrm{eff})_\mathrm{Aa}$ & $1$ & $0.8522_{-0.0073}^{+0.0075}$ & $1.2755_{-0.0093}^{+0.0095}$ & $0.5596_{-0.0156}^{+0.0148}$ \\
 fractional flux [in TESS-band] & $0.0764_{-0.0019}^{+0.0019}$ & $0.0252_{-0.0011}^{+0.0012}$ & $0.6998_{-0.0199}^{+0.0169}$ & $0.0009_{-0.0002}^{+0.0003}$ \\
 fractional flux [in \textit{SWASP}-band] & $0.0672_{-0.0032}^{+0.0030}$ & $0.0169_{-0.0013}^{+0.0013}$ & $0.8795_{-0.0403}^{+0.0254}$ & $0.0004_{-0.0001}^{+0.0001}$ \\
 \hline
 \multicolumn{5}{c}{Physical quantities} \\
  \hline 
 $M$ [M$_\odot$] & $0.970_{-0.018}^{+0.013}$ & $0.757_{-0.006}^{+0.009}$ & $1.473_{-0.023}^{+0.025}$ & $0.165_{-0.016}^{+0.018}$ \\
 $R$ [R$_\odot$] & $1.003_{-0.013}^{+0.014}$ & $0.824_{-0.009}^{+0.017}$ & $1.949_{-0.031}^{+0.027}$ & $0.492_{-0.025}^{+0.024}$ \\
 $T_\mathrm{eff}$ [K]& $4969_{-48}^{+83}$ & $4242_{-53}^{+63}$ & $6326_{-59}^{+147}$ & $2791_{-95}^{+78}$ \\
 $L_\mathrm{bol}$ [L$_\odot$] & $0.553_{-0.025}^{+0.035}$ & $0.198_{-0.014}^{+0.017}$ & $5.46_{-0.24}^{+0.45}$ & $0.013_{-0.002}^{+0.003}$ \\
 $M_\mathrm{bol}$ & $5.41_{-0.07}^{+0.05}$ & $6.53_{-0.09}^{+0.08}$ & $2.93_{-0.09}^{+0.05}$ & $9.47_{-0.20}^{+0.23}$ \\
 $M_V           $ & $5.71_{-0.11}^{+0.07}$ & $7.36_{-0.15}^{+0.15}$ & $2.92_{-0.09}^{+0.05}$ & $13.08_{-0.53}^{+0.65}$ \\
 $\log g$ [dex] & $4.421_{-0.013}^{+0.009}$ & $4.485_{-0.019}^{+0.011}$ & $4.026_{-0.011}^{+0.011}$ & $4.269_{-0.016}^{+0.018}$ \\
 \hline
\multicolumn{5}{c}{Global system parameters} \\
  \hline
$\log$(age) [dex] &\multicolumn{2}{c}{$7.326_{-0.049}^{+0.030}$} & $7.046_{-0.033}^{+0.024}$ & $7.326_{-0.049}^{+0.030}$ \\
$[M/H]$  [dex]    &\multicolumn{4}{c}{$-0.003_{-0.102}^{+0.118}$} \\
$E(B-V)$ [mag]    &\multicolumn{4}{c}{$0.015_{-0.011}^{+0.032}$} \\
extra light $\ell_4$ [in TESS-band] & \multicolumn{4}{c}{$0.197_{-0.017}^{+0.021}$} \\
extra light $\ell_4$ [in \textit{SWASP}-band] & \multicolumn{4}{c}{$0.035_{-0.028}^{+0.045}$} \\
$(M_V)_\mathrm{tot}$  &\multicolumn{4}{c}{$2.82_{-0.09}^{+0.05}$} \\
distance [pc]           & \multicolumn{4}{c}{$636_{-9}^{+10}$} \\  
\hline
\end{tabular}}
\end{table*}

\begin{table*}
 \centering
\caption{Orbital and astrophysical parameters of TICs\,323486857 and 650024463 from the joint photodynamical light curve, ETV, SED and \texttt{PARSEC} isochrone solution. }
 \label{tab:syntheticfit_TIC323486857650024463}
\scalebox{0.91}{\begin{tabular}{@{}lllllll}
\hline
& \multicolumn{3}{c}{TIC\,323486857} & \multicolumn{3}{c}{TIC\,650024463} \\
\hline
\multicolumn{7}{c}{Orbital elements} \\
\hline
 & \multicolumn{3}{c}{Subsystem}  & \multicolumn{3}{c}{Subsystem}  \\
 & \multicolumn{2}{c}{Aa--Ab} & A--B  & \multicolumn{2}{c}{Aa--Ab} & A--B  \\
  \hline
  $t_0$ [BJD - 2400000]& \multicolumn{3}{c}{58569.0} & \multicolumn{3}{c}{$58320.0$} \\
  $P$ [days] &  \multicolumn{2}{c}{$0.88502_{-0.00036}^{+0.00015}$} & $41.4268_{-0.0021}^{+0.0016}$ & \multicolumn{2}{c}{$7.197817_{-0.000028}^{+0.000026}$} & $108.7251_{-0.0013}^{+0.0013}$ \\
  $a$ [R$_\odot$] & \multicolumn{2}{c}{$5.362_{-0.053}^{+0.035}$} & $81.51_{-0.60}^{+0.61}$ & \multicolumn{2}{c}{$18.21_{-0.12}^{+0.07}$} & $127.09_{-0.87}^{+0.47}$ \\
  $e$ & \multicolumn{2}{c}{$0.0021_{-0.0009}^{+0.0012}$} & $0.0066_{-0.0037}^{+0.0042}$ & \multicolumn{2}{c}{$0.02570_{-0.00026}^{+0.00024}$} & $0.32317_{-0.00011}^{+0.00011}$ \\
  $\omega$ [deg] & \multicolumn{2}{c}{$151_{-66}^{+77}$} & $241_{-139}^{+48}$ & \multicolumn{2}{c}{$28.83_{-0.49}^{+0.53}$} & $351.28_{-0.10}^{+0.09}$ \\ 
  $i$ [deg] & \multicolumn{2}{c}{$84.9_{-0.4}^{+2.3}$} & $84.59_{-0.45}^{+0.28}$ & \multicolumn{2}{c}{$90.233_{-0.055}^{+0.059}$} & $90.099_{-0.090}^{+0.069}$ \\
  $\mathcal{T}_0^\mathrm{inf/sup}$ [BJD - 2400000] & \multicolumn{2}{c}{$58569.7694_{-0.0003}^{+0.0003}$} & $58595.020_{-0.028}^{+0.044}$ & \multicolumn{2}{c}{$58330.8185_{-0.0002}^{+0.0002}$} & ${58642.2864_{-0.0086}^{+0.0085}}^*$ \\
  $\tau$ [BJD - 2400000] & \multicolumn{2}{c}{$58569.47_{-0.16}^{+0.20}$} & $58577_{-20}^{+17}$ & \multicolumn{2}{c}{$58325.9422_{-0.0096}^{+0.0100}$} & $58623.745_{-0.024}^{+0.025}$ \\
  $\Omega$ [deg] & \multicolumn{2}{c}{$0.0$} & $1.9_{-2.5}^{+1.2}$ & \multicolumn{2}{c}{$0.0$} & $0.33_{-0.26}^{+0.30}$ \\
  $i_\mathrm{mut}$ [deg] & \multicolumn{3}{c}{$2.3_{-0.7}^{+1.2}$} & \multicolumn{3}{c}{$0.37_{-0.22}^{+0.31}$} \\
  $\varpi^\mathrm{dyn}$ [deg] & \multicolumn{2}{c}{$339_{-70}^{+73}$} & $76_{-142}^{+47}$ & \multicolumn{2}{c}{$208.83_{-0.49}^{+0.53}$} & $171.28_{-0.10}^{+0.09}$ \\
  $i^\mathrm{dyn}$ [deg] & \multicolumn{2}{c}{$2.05_{-0.60}^{+1.07}$} & $0.28_{-0.08}^{+0.14}$ & \multicolumn{2}{c}{$0.30_{-0.18}^{+0.24}$} & $0.07_{-0.04}^{+0.06}$ \\
  $\Omega^\mathrm{dyn}$ [deg] & \multicolumn{2}{c}{$98_{-37}^{+27}$} & $285_{-20}^{+90}$ & \multicolumn{2}{c}{$112_{-12}^{+25}$} & $292_{-12}^{+25}$ \\
  $i_\mathrm{inv}$ [deg] & \multicolumn{3}{c}{$84.68_{-0.47}^{+0.41}$} & \multicolumn{3}{c}{$90.125_{-0.068}^{+0.057}$} \\
  $\Omega_\mathrm{inv}$ [deg] & \multicolumn{3}{c}{$1.7_{-2.2}^{+1.0}$} & \multicolumn{3}{c}{$0.27_{-0.21}^{+0.24}$} \\
  \hline
  mass ratio $[q=M_\mathrm{sec}/M_\mathrm{pri}]$ & \multicolumn{2}{c}{$0.877_{-0.021}^{+0.028}$} & $0.606_{-0.014}^{+0.008}$ & \multicolumn{2}{c}{$0.780_{-0.003}^{+0.004}$} & $0.490_{-0.004}^{+0.004}$ \\
  $K_\mathrm{pri}$ [km\,s$^{-1}$] & \multicolumn{2}{c}{$142.41_{-2.07}^{+3.21}$} & $37.44_{-0.56}^{+0.29}$ & \multicolumn{2}{c}{$56.12_{-0.39}^{+0.26}$} & $20.55_{-0.16}^{+0.14}$ \\ 
  $K_\mathrm{sec}$ [km\,s$^{-1}$] & \multicolumn{2}{c}{$162.74_{-2.52}^{+1.82}$} & $61.80_{-0.75}^{+0.59}$ & \multicolumn{2}{c}{$71.93_{-0.50}^{+0.40}$} & $41.94_{-0.28}^{+0.23}$ \\ 
  \hline
  \multicolumn{7}{c}{Apsidal and nodal motion related parameters} \\
  \hline
$P_\mathrm{apse}$ [year] & \multicolumn{2}{c}{$0.58_{-0.05}^{+0.15}$} & $139.0_{-1.0}^{+1.5}$ & \multicolumn{2}{c}{$15.45_{-0.08}^{+0.08}$} & $62.93_{-0.09}^{+0.08}$ \\ 
$P_\mathrm{apse}^\mathrm{dyn}$ [year] & \multicolumn{2}{c}{$0.56_{-0.04}^{+0.14}$} & $14.80_{-0.07}^{+0.14}$ & \multicolumn{2}{c}{$6.87_{-0.03}^{+0.03}$} & $10.34_{-0.03}^{+0.03}$ \\ 
$P_\mathrm{node}^\mathrm{dyn}$ [year] & \multicolumn{3}{c}{$16.56_{-0.19}^{+0.10}$} & \multicolumn{3}{c}{$12.38_{-0.05}^{+0.04}$} \\
$\Delta\omega_\mathrm{3b}$ [arcsec/cycle] & \multicolumn{2}{c}{$356.5_{-4.2}^{+2.2}$} & $9917_{-94}^{+50}$ & \multicolumn{2}{c}{$3716_{-17}^{+15}$} & $37300_{-109}^{+102}$ \\ 
$\Delta\omega_\mathrm{GR}$ [arcsec/cycle] & \multicolumn{2}{c}{$4.068_{-0.077}^{+0.053}$} & $0.429_{-0.006}^{+0.007}$ & \multicolumn{2}{c}{$0.710_{-0.009}^{+0.006}$} & $0.169_{-0.002}^{+0.001}$ \\ 
$\Delta\omega_\mathrm{tide}$ [arcsec/cycle] & \multicolumn{2}{c}{$5243_{-1121}^{+481}$} & $12.6_{-3.2}^{+5.0}$ & \multicolumn{2}{c}{$0.378_{-0.024}^{+0.026}$} & $0.0014_{-0.0001}^{+0.0001}$  \\ 
  \hline  
\multicolumn{7}{c}{Stellar parameters} \\
\hline
   & Aa & Ab &  B   & Aa & Ab &  B\\
  \hline
 \multicolumn{7}{c}{Relative quantities} \\
  \hline
 fractional radius [$R/a$] & $0.3479_{-0.0155}^{+0.0092}$ & $0.2514_{-0.0098}^{+0.0181}$ & $0.0925_{-0.0049}^{+0.0083}$ & $0.0531_{-0.0008}^{+0.0009}$ & $0.0369_{-0.0003}^{+0.0002}$ & $0.00593_{-0.00006}^{+0.00006}$ \\
 temperature relative to $(T_\mathrm{eff})_\mathrm{Aa}$ & $1$ & $0.9816_{-0.0066}^{+0.0109}$ & $0.7741_{-0.0128}^{+0.0145}$ & $1$ & $0.8104_{-0.0038}^{+0.0041}$ & $0.8998_{-0.0067}^{+0.0063}$ \\
 fractional flux [in TESS-band] & $0.0972_{-0.0080}^{+0.0071}$ & $0.0496_{-0.0035}^{+0.0050}$ & $0.6820_{-0.0609}^{+0.0569}$ & $0.1013_{-0.0028}^{+0.0029}$ & $0.0207_{-0.0007}^{+0.0008}$ & $0.0411_{-0.0020}^{+0.0021}$ \\
 \hline
 \multicolumn{7}{c}{Physical quantities} \\
  \hline 
 $M$ [M$_\odot$] & $1.408_{-0.043}^{+0.021}$ & $1.223_{-0.028}^{+0.052}$ & $1.588_{-0.038}^{+0.050}$ & $0.877_{-0.018}^{+0.012}$ & $0.684_{-0.014}^{+0.008}$ & $0.765_{-0.017}^{+0.010}$ \\
 $R$ [R$_\odot$] & $1.873_{-0.105}^{+0.046}$ & $1.344_{-0.057}^{+0.110}$ & $7.541_{-0.455}^{+0.750}$ & $0.967_{-0.015}^{+0.015}$ & $0.673_{-0.009}^{+0.005}$ & $0.754_{-0.011}^{+0.008}$ \\
 $T_\mathrm{eff}$ [K]& $6550_{-44}^{+61}$ & $6434_{-60}^{+86}$ & $5068_{-71}^{+76}$ & $5682_{-21}^{+28}$ & $4608_{-31}^{+31}$ & $5114_{-39}^{+43}$ \\
 $L_\mathrm{bol}$ [L$_\odot$] & $5.80_{-0.69}^{+0.34}$ & $2.78_{-0.32}^{+0.64}$ & $33.51_{-2.44}^{+5.16}$ & $0.877_{-0.020}^{+0.021}$ & $0.182_{-0.005}^{+0.006}$ & $0.349_{-0.015}^{+0.015}$ \\
 $M_\mathrm{bol}$ & $2.86_{-0.06}^{+0.14}$ & $3.66_{-0.23}^{+0.13}$ & $0.96_{-0.16}^{+0.08}$ & $4.91_{-0.03}^{+0.03}$ & $6.62_{-0.03}^{+0.03}$ & $5.91_{-0.04}^{+0.05}$ \\
 $M_V           $ & $2.84_{-0.06}^{+0.14}$ & $3.65_{-0.23}^{+0.13}$ & $1.22_{-0.15}^{+0.10}$ & $4.99_{-0.03}^{+0.03}$ & $7.14_{-0.05}^{+0.05}$ & $6.15_{-0.06}^{+0.06}$ \\
 $\log g$ [dex] & $4.043_{-0.019}^{+0.036}$ & $4.267_{-0.050}^{+0.028}$ & $2.882_{-0.067}^{+0.049}$ & $4.409_{-0.015}^{+0.014}$ & $4.617_{-0.003}^{+0.004}$ & $4.566_{-0.006}^{+0.006}$ \\
 \hline
\multicolumn{7}{c}{Global system parameters} \\
  \hline
$\log$(age) [dex] & \multicolumn{3}{c}{$9.374_{-0.038}^{+0.013}$} &\multicolumn{3}{c}{$10.001_{-0.033}^{+0.039}$} \\
$[M/H]$  [dex]    & \multicolumn{3}{c}{$0.015_{-0.058}^{+0.018}$} &\multicolumn{3}{c}{$-0.107_{-0.055}^{+0.036}$} \\
$E(B-V)$ [mag]    & \multicolumn{3}{c}{$0.340_{-0.020}^{+0.020}$} &\multicolumn{3}{c}{$0.088$} \\
extra light $\ell_4$ [in TESS-band] & \multicolumn{3}{c}{$0.172_{-0.063}^{+0.064}$} & \multicolumn{3}{c}{$0.837_{-0.003}^{+0.003}$} \\
$(M_V)_\mathrm{tot}$  & \multicolumn{3}{c}{$0.93_{-0.15}^{+0.07}$} &\multicolumn{3}{c}{$4.56_{-0.01}^{+0.01}$} \\
distance [pc]         & \multicolumn{3}{c}{$1434_{-74}^{+123}$} &\multicolumn{3}{c}{$1126$} \\  
\hline
\end{tabular}}
\end{table*}

\begin{table*}
\centering 
\caption{Derived ephemerides for the seven triple systems to be used for planning future observations.}
 \label{tab:ephemerides}
 \begin{tabular}{llllllll}
\hline 
TIC ID               & 133771812  & 176713425  & 185615681 & 287756035  & 321978218  & 323486857 & 650024463 \\
\hline
&\multicolumn{7}{c}{Inner binary} \\
\hline
$P$  & 12.36949 &  1.898728   & 2.3185738 & 2.0814225 & 0.5698104  & 0.883992 & 7.1971 \\
     & 12.36844 &             & 2.3185448 &           &            &          & 7.1969 \\
$\mathcal{T}_0$ & 60\,035.027 & 60\,317.664 & 60\,055.382 & 60\,127.070 & 60\,434.640 & 60\,150.345 & 60\,252.278\\
                & 60\,020.719 &             & 60\,056.539 &             &             &            & 60\,248.733 \\
$\mathcal{A}_\mathrm{ETV}$  & 0.030 & 0.006  & 0.0007 & 0.0025 & 0.0013 & 0.0008 & 0.028 \\
$D$  & 0.322 & 0.228 & 0.269  & 0.283 & 0.085 & 0.168 & 0.307 \\
\hline
&\multicolumn{7}{c}{Wide binary (third body eclipses)} \\
\hline
$P$ & 242.3 & 52.55 & 55.84 & 367.64 & 57.52 & 41.356 & 107.95 \\
$\mathcal{T}_0^\mathrm{inf}$  & 58\,600.1: & (58762.2:) & 58\,547.5 & 59\,311.55 & 59\,095.85 & 58\,595.02  &  59\,579.5 \\
$D^\mathrm{inf}$ & 6.5 & 6.3: & 1.45 & 4.21 & 1.01 & 1.29 & 2.70 \\
$\mathcal{T}_0^\mathrm{sup}$ & 58\,510.4 & 58\,740.1 & (58\,572.9) & 59\,470.28: & 59\,072.45 & (58\,615.71) & 59\,503.8 \\
$D^\mathrm{sup}$ & 6.3 & 0.89 & 1.90 & 3.90: & 0.68 & 1.27 & 3.18 \\
\hline
\end{tabular}

\textit{Notes.} (a) For the inner binaries: $P$, $\mathcal{T}_0$, $\mathcal{A}_\mathrm{ETV}$, $D$ are the period, reference time of a primary minimum, half-amplitude of the ETV curve, and the full duration of an eclipse, respectively. $\mathcal{T}_0$ is given in BJD -- 2\,400\,000, while the other quantities are in days. For all those triples where the inner eccentricities are very small and, hence, the shifts of the secondary eclipses relative to phase 0.5 are negligible (quantitatively, they are much smaller than the full durations of the individual eclipses), the same reference times and periods can be used to predict the times of the secondary eclipses. In the case of the three eccentric EBs we give a separate period and reference time for the secondary eclipses, listing them below the primary ephemerides. (b) For the outer orbits we give separate reference times for the third body eclipses around the inferior and superior conjunctions of the tertiary component. The eclipse durations, $D$, of the third-body eclipses do not give the extent of any specific third body events.  Rather $D$ represents the time difference corresponding to the very first and last moments around a given third-body conjunction when the first/last contact of a third-body event may occur). Double dots (:) call attention to the less certain superior/inferior conjunction times at those types of third-body events (i.e., primary vs.~secondary outer eclipses) because they were not observed by TESS. Conjunction data, in parentheses, indicate that only very shallow third-body eclipses may occur which can hardly be observed with ground-based instruments.
\end{table*} 

\subsection{Results for the individual systems}

\subsubsection{TIC 133771812}
\label{Sect:discussion_TIC133771812}

This is the only target in our sample for which we were unable to obtain a reasonable third-body orbit fold from the archival data. Thus, in order to determine the outer period we had to restrict ourselves to just the three sets of third-body eclipses observed with TESS in years 1, 3 and 5, and to the high quality ETV data obtained from these observations.   Our analytic ETV solutions preferred an outer period of $P_\mathrm{out}\sim244$\,d, but allowed also for half that period, namely, $P_\mathrm{out}\sim122$\,d. We produced preliminary photodynamical solutions for both outer periods, and we found that the true period should actually be about $P_\mathrm{out}\sim244$\,days. In other words, we were unable to find both dynamically and astrophysically consistent and reliable solutions with the shorter outer period. We have tabulated the median posteriors of the adjusted parameters and several further derived quantities of the complex photodynamical analysis in the first three columns of Table~\ref{tab:syntheticfit_TIC133771812176713425}.

According to our results, all three components of TIC~133771812 are somewhat more massive than our Sun. The inner EB consists of the most and least massive ($M_\mathrm{Aa}=1.58\pm0.04\,\mathrm{M}_\sun$ and $M_\mathrm{Ab}=1.20\pm0.03\,\mathrm{M}_\sun$, respectively) and, correspondingly, the hottest and coolest ($T_\mathrm{Aa}=7700\pm160$\,K and $T_\mathrm{Ab}=6410\pm100$\,K) stars of the triple.  The third component lies in between the two EB members both in its mass and temperature, but is closer to the secondary of the inner EB with $M_\mathrm{B}=1.25\pm0.03\,\mathrm{M}_\sun$ and $T_\mathrm{B}=6550\pm110$\,K. The inferred masses, in the absence of RV data, have relatively large uncertainties of $\sim$$3-4$\%. Relative quantities, such as the mass ratios, however, can be deduced with better accuracies. These are found to be $q_\mathrm{in}=0.763\pm0.006$ and $q_\mathrm{out}=0.449\pm0.010$.

Amongst all the triply eclipsing triple stars which were analyzed photodynamically in our present and former papers, TIC~133771812 has by far the longest inner EB period with $P_\mathrm{in}=12\fd33$. Thus, despite the fact that the inferred outer period ($P_\mathrm{out}=243\fd9$) is the second longest in the current set, this target is still the second tightest ($P_\mathrm{out}/P_\mathrm{in}\simeq19.8$) among  these seven triply eclipsing triples. Due the relative large size of the inner EB and, hence, the small fractional radii ($r_\mathrm{Aa}=R_\mathrm{Aa}/a_\mathrm{in}=0.0491\pm0.0005$ and $r_\mathrm{Ab}=0.0368\pm0.0008$), tidal effects are certainly negligible. Hence, not surprisingly, this inner EB has the largest eccentricity with $e_\mathrm{in}=0.0337\pm0.0004$ among our present sample. The combination of inner and outer ($e_\mathrm{out}=0.217\pm0.006$) eccentricities, combined with the tightness of the system, still leaves it safely within the dynamically stable region according to the \citet{mardlingaarseth01} stability criterion.

\subsubsection{TIC 176713425}
\label{Sect:discussion_TIC176713425}

This is the only northern (ecliptic) hemisphere target amongst the currently investigated triples. Its original FITSH light curves were partially blended with the nearby overcontact binary ATO J353.9138+42.3630 (TIC 176713436). Hence, to eliminate the signal of the latter object from the light curve of our target, we applied a principal component analysis (PCA) method implemented in the lightkurve \citep{2018ascl.soft12013L} Python package and its dependencies: astropy \citep{2018AJ....156..123A}, astroquery \citep{2019AJ....157...98G} and TESScut \citep{2019ascl.soft05007B}.  

We achieved this by creating 60$\times$60 pixel size cutouts around the target and choosing an aperture such that the contaminating star also has some contribution in the image region outside the aperture. After that, we created a design matrix containing regressors from all pixels surrounding the aperture and calculated the principal component vectors using lightkurve's built-in PCA method. These principal components are a combination of signals from scattered light, spacecraft motion and the nearby contaminating star. As a final step, we used the built-in RegressionCorrector method of lightkurve that uses linear algebra to find a combination of these principal component vectors that will transform the input light curve closest to zero. As the mean flux level of the light curve is different from zero, an additional offset term has to be used by appending a constant term to the design matrix. In this way, all of the previously mentioned signals were modeled and eliminated from the light curve of the target.

Due to the northerly position of the target, we were able to obtain photometric follow up observations in Hungarian observatories, and thereby observed eight additional regular eclipses (four primary and four secondary minima)\footnote{We note, however, that the last four eclipse times were obtained only after the photodynamical analysis was finalized. Despite this, they are plotted in Fig.~\ref{fig:176713425ETV} and also tabulated in Table~\ref{Tab:TIC_176713425_ToM}. These additional measurements are in perfect accord with the photodynamical model.}  with the two identical 80 cm RC telescopes of Baja Astronomical Observatory and Gothard Astrophysical Observatory, Szombathely \citep[for short descriptions of the instruments used see][]{borkovitsetal22a}. Despite the fact that TESS observed only one set of third-body events, the parameters of the outer orbit can be determined robustly not only due to the archival observations which provide the outer orbital period ($P_\mathrm{out}$), but also because of the very high quality ETV data (Fig.~\ref{fig:176713425ETV}). In this regard, we note that the ground-based follow up eclipse times were found to be very helpful in excluding some possible alternative outer periods.

Turning to the astrophysical implications of the photodynamical results (see last columns of Table~\ref{tab:syntheticfit_TIC133771812176713425}), we found that the system consists of nearly identical triplet stars. The masses of the EB primary and the tertiary agree to within their 1-$\sigma$ statistical uncertainties ($M_\mathrm{Aa}=1.17\pm0.03\,\mathrm{M}_\sun$ vs. $M_\mathrm{B}=1.16\pm0.03\,\mathrm{M}_\sun$), and the secondary of the inner EB has a lower mass by only $6-7\%$ ($M_\mathrm{Ab}=1.09\pm0.03\,\mathrm{M}_\sun$). (Referring to the more precise relative quantities, we find: $q_\mathrm{in}=0.933\pm0.005$ and $q_\mathrm{out}=0.515\pm0.004$, respectively.) What makes this system more interesting from an astrophysical and, especially, evolutionary point of view is that the two more massive components (stars Aa and B) are very likely at the end of their TAMS or, just at the beginning of their evolution from the MS toward the giant branch. This means that, although these stars are more massive, the slightly less massive secondary actually looks a bit hotter, though, the small difference is well within the 1-$\sigma$ posterior uncertainties and, thus, looks statistically insignificant ($T_\mathrm{Aa}=6033_{-192}^{+78}$\,K and $T_\mathrm{B}=6033_{-160}^{+97}$\,K vs. $T_\mathrm{Ab}=6051_{-153}^{+77}$\,K). At the same time, despite the quite similar masses of the inner EB members, the radii of the EB components differ substantially, being $R_\mathrm{Aa}=1.555\pm0.015\,\mathrm{R}_\sun$ vs. $R_\mathrm{Ab}=1.137\pm0.045\,\mathrm{R}_\sun$ (or, again citing the precise fractional radii, we find $r_\mathrm{Aa}=R_\mathrm{Aa}/a_\mathrm{in}=0.2060\pm0.0018$ vs. $r_\mathrm{Ab}=0.1525\pm0.0025$). 

These results emphasize that a more robust and precise determination of the stellar masses with the addition of RV observations would be very important, since the inner EB of this triple may be used to strongly constrain stellar evolutionary models. Moreover, another aspect that should be considered is that, because of the nearly equal surface brightnesses of the primary and secondary components of the inner circular EB, one can assume that due to the total eclipses, the secondary-to-primary eclipse depth ratio is dominated by the limb darkening law of the binary components. Therefore, in order to obtain a better characterization of limb darkening law(s) -- a recently intensively investigated area, having crucial importance in the precision analysis of transiting exoplanets, or even of eclipsing binaries \citep[see~e.g.][]{csizmadiaetal13,maxted23} --  high-precision photometric observations with good time resolution of future regular (binary) eclipses would also be very useful.

Turning to the dynamical properties of the triple, the period ratio $P_\mathrm{out}/P_\mathrm{in}=27.9$ indicates that TIC~176713425 is a moderately tight system, where one can expect the effects of significant third-body perturbations. This can be seen nicely in the ETV curve, where the dynamical delays strongly dominate over the pure, geometric LTTE. On the other hand, however, due to the small physical size of the inner binary (and, hence, the significant fractional radii of stars Aa, Ab) the tidal effects are non-negligible and the tidally forced nominal\footnote{We say 'nominal' because, due to the nearly circular shape of the inner orbit, one cannot really expect observable apsidal motion.} apsidal advance rate in the inner EB is of the same order as the dynamically forced one ($\Delta\omega_\mathrm{in}^\mathrm{tide}=387''\pm15''\,\mathrm{cycle}^{-1}$ vs. $\Delta\omega_\mathrm{in}^\mathrm{3b}=1230''\pm15''\,\mathrm{cycle}^{-1}$).

\subsubsection{TIC 185615681 = TZ Pyx}
\label{Sect:discussion_TIC185615681}

This is the only system in our current sample of triples where the inner EB has been known already for a long time. The variability of the system was discovered by \citet{strohmeier966}. \citet{duerbeckrucinski07} obtained and analyzed 8 RV points for each of the two stellar components of the EB. They found that the inner EB consists of two very similar stars (quantitatively, they found a spectroscopic mass ratio of $q^\mathrm{spec}=0.996\pm0.020$) that are hot and likely of late A spectral type.  These authors, however, did not take note of the slight eccentricity of the binary's orbit. It was \citet{otero07} who reported for the first time not only the binary eccentricity, but also evidence for the apsidal motion in the EB. This effect was then studied first by \citet{zasche12} who, with the use of 16 badly scattered ground-based eclipse times (spanning an interval of $\sim$7000\,days), determined an apsidal motion period of $P_\mathrm{apse}=157\pm37$\,yr.  But, as they note: ``The apsidal motion fit is still not very convincing and would have even shorter period [...], but only further observations would confirm this hypothesis.'' From this apsidal motion period they inferred that the (mean) apsidal motion or, internal structure constant, $k_2$, is in accord with the theoretically expected value. Naturally, they were not aware of the presence of the third stellar component in the system, whose perturbations, as our analysis clearly reveals, completely dominates over the tidal effects. Hence, even if their apsidal motion period had been correct, it would not be conclusive in regard to the internal structure constants of the stars. Most recently \citet{kimetal18} repeated the ETV analysis, using the then available 20 data points, and got a much shorter apsidal motion period of $P_\mathrm{apse}=22.89\pm0.30$\,yr, which is much closer to our findings.

Turning to our analysis of the available former ground-based observational material, we used only the RV data of \citet{duerbeckrucinski07}.  Neither ground-based light curves, nor ETV data were used for our photodynamical runs. While neglecting the ground-based light curve data is well justifiable given the availability of the far superior TESS light curves, the ETV points, in theory, might be usable to obtain a more precise detection of the apsidal motion period. Unfortunately, however, the historical ETV data have such bad scatter that, from our perspective, these are unusable to mine out any additional information.

The twin components of TZ Pyx ($M_\mathrm{Aa}=2.013\pm0.034\,\mathrm{M}_\sun$; $M_\mathrm{Aa}=2.008\pm0.033\,\mathrm{M}_\sun$) are the hottest, fully radiative stars in our sample. Moreover, this is the only triple in our sample, where the third, outer component is substantially less massive ($M_\mathrm{B}=0.888\pm0.014\,\mathrm{M}_\sun$) than either of the binary members.  (Referring again to the more precise relative quantities, we find: $q_\mathrm{in}=0.998\pm0.002$ and $q_\mathrm{out}=0.221\pm0.003$, respectively.)  In such a way, the contribution of the third component to the total system flux is negligibly small, only $\sim$$1\%$. On the other hand, the depths of the primary third-body eclipses may reach $\sim$$6-8\%$, where, for such a relatively bright ($V=10.7$) variable star, such extra dips would be serendipitously detectable even with sub-metre size ground-based telescopes.  Despite this fact, no indications of the triple nature of this system have been previously  reported. This fact emphasizes that one should not be surprised to occasionally find a triply eclipsing tertiary hidden even amongst the best observed and brightest EBs.

Regarding the orbital properties, we found a small, but significant, eccentricity for the inner orbit, being $e_\mathrm{in}=0.01067\pm0.00009$. This value agrees well with the results of \citet{kimetal18} who obtained $e_\mathrm{in}=0.01068\pm0.00008$.\footnote{One should keep in mind, however, that the two values cannot be perfectly compared, as our orbital elements are instantaneous osculating elements, while the ones listed in the \citet{kimetal18} paper are some kind of average elements, deduced from the observations.} The outer orbit was also found to be slightly eccentric with $e_\mathrm{out}=0.0982\pm0.0008$.  This is the second lowest outer eccentricity in our current sample.

Despite the moderately tight nature of the triple system ($P_\mathrm{out}/P_\mathrm{in}=24.2$) the full-amplitudes of the primary and secondary ETVs, on the $P_\mathrm{out}$-timescale, remain under $1.5$\,min ($\sim0.001$\,d). This effect is hardly detectable with incidental, highly inhomogeneous eclipse times measured with ground-based photometry. The ETVs, determined from the precise TESS photometry, however, clearly show the presence of typical, third-body induced dynamical ETVs (see Fig.~\ref{fig:185615681ETV}). Our solution reveals that the ETVs on this timescale are nearly equally contributed by the third-body perturbations and the geometric LTTE. 

On a longer timescale, however, the main effect seen in the ETV curves is due to apsidal motion. In turn, the third-body-driven apsidal motion dominates over the classic tidal component, but the latter contribution is non-negligible ($\Delta\omega_\mathrm{in}^\mathrm{3b}=738''\pm5''\,\mathrm{cycle}^{-1}$ vs. $\Delta\omega_\mathrm{in}^\mathrm{tide}=144''\pm2''\,\mathrm{cycle}^{-1}$). In this regard, we note that even though three inner eccentric EBs of the current sample of seven systems exhibit well observed apsidal motion, TZ Pyx is the only one where the classic tidal effects are important.  Hence, this is the only system in our sample where the proper settings of the internal structure constants have real significance. Therefore, we set the first two apsidal motion constants to $k_2=0.005$ and $k_3=0.001$ for both members of the inner EB, in accord with the recent tables of \citet{claret23}. Taking into account all three different contributions of the apsidal motion (though the relativistic contribution was found to be much smaller than the other two, with $\Delta\omega_\mathrm{in}^\mathrm{GR}=2\farcs83\pm0\farcs04\,\mathrm{cycle}^{-1}$) our solution gives a theoretical apsidal motion period of $P_\mathrm{apse}^\mathrm{theo}=18.2\pm0.2$\,yr.

\subsubsection{TIC 287756035}
\label{Sect:discussion_TIC287756035}

This system has, by far, the longest of the outer periods among this current set of seven triply eclipsing triples at 368 days. In fact, this outer period is so close to that of an Earth year, that we were able to see only $\sim$3/4 of the phases in the outer orbital fold of the archival data (see left panel, middle row of Fig.~\ref{fig:outer_orbit_folds}).  The system is quite flat with the inner and outer orbital inclinations near 89$^\circ$ and a mutual inclination angle of $i_{\rm mut} = 1.4^\circ$. 

The outer orbit has a quite typical eccentricity for these systems of $e_\mathrm{out}=0.23\pm0.02$, while the observable argument of the outer periastron was found to be $\omega_\mathrm{out}=53\fdg7\pm0\fdg7$. One should keep in mind, however, that both TESS and the formerly mentioned ground-based surveys observed only one type of third-body eclipse, namely, those where the tertiary star occulted the inner binary.  Therefore we are not able to constrain the quantity $e_\mathrm{out}\cos\omega_\mathrm{out}$ through the offset of the other type of third-body eclipse from an orbital phase of $0\fp5$.  In addition to this fact, due the lack of any usable ETV curve (which would offer additional constraints on $e_\mathrm{out}$ and $\omega_\mathrm{out}$), our primary source for information about $e_\mathrm{out}$ and $\omega_\mathrm{out}$ is only the poorly covered RV curve of the tertiary (see Fig.~\ref{fig:287756035rv}). Thus, the obtained $e_\mathrm{out}$ and $\omega_\mathrm{out}$ values are less robust than in the case of the other systems studied in this work.

One can, however, make, an indirect, inverse check on the reliability of the obtained outer eccentricity and argument of periastron values. Our photodynamical solution (of which $e_\mathrm{out}\cos\omega_\mathrm{out}$ was a direct, adjustable parameter) resulted in a median posterior of $e_\mathrm{out}\cos\omega_\mathrm{out}=0.14\pm0.01$. Taking into account that the phase of the superior conjunction relative to the inferior conjunction can be approximated as $\phi_\mathrm{sup}\simeq \frac{1}{2}-\frac{2}{\pi}e\cos\omega$ \citep[see, e.g.,][]{sterne39}, one can calculate that the other type of third-body eclipse should occur around phase $0\fp411\pm0\fp007$. Taking a look at the archival folded light curve of TIC~287756035 (Fig.~\ref{fig:outer_orbit_folds}), one can see that this phase value is very close to the left edge of the seasonal gap of the observations, but the target was definitely observable during this orbital phase. Despite this, no dip can be noticed in the fold around that phase. One should keep in mind, however, that this phase is very close to the beginning of the seasonal gaps, and one may expect that the target was observable only for a short time just after sunset and hence, only observations reduced in number and quality are available during these intervals. Another possibility might be that, due to the weaker constraints on $e_{\rm out}\cos\omega_{\rm out}$, the obtained posteriors might have larger statistical uncertainties than we found. Thus, further RV observations would be very important to clarify this question.

Apart from this issue discussed above, what makes this system very interesting from an astrophysical point of view is that the two more massive stellar components, with very similar masses of $M_\mathrm{B}=1.133\pm0.04\,\mathrm{M}_\sun$ and $M_\mathrm{Ab}=1.116\pm0.04\,\mathrm{M}_\sun$, are evolved to very different levels\footnote{$M_B$ is only 0.017 M$_\odot$ greater than $M_{\rm Ab}$, but is within 1-$\sigma$ of being equal.  However, $M_B$ could be as much as 0.14 M$_\odot$ more massive at the 1.5-$\sigma$ level. Also, using the more accurate mass ratios, we find that $M_\mathrm{Ab}/M_\mathrm{B}=0.986\pm0.009$, which is less than unity at the 1.5-$\sigma$ level.}. The slightly more massive tertiary is quite evolved at $R_\mathrm{B}=6.7\pm0.3\,\mathrm{R}_\sun$, and dominates the light of the system. On the other hand, the secondary component of the inner EB is just at the very beginning of its evolution toward the giant branch with $R_\mathrm{B}=2.36\pm0.06\,\mathrm{R}_\sun$. For such near solar mass stars, they must be fairly old at 5.7 Gyr in order to be ascending the giant branch. Interestingly, this shows what will become of our Sun within a short (astronomical) time.  We also note that even the third, less massive star with $M_\mathrm{Aa}=0.93\pm0.02\,\mathrm{M}_\sun$ and $R_\mathrm{Aa}=0.98\pm0.02\,\mathrm{R}_\sun$ is currently located on the TAMS just before its evolution toward the giant branch. Thus, at the current epoch, this star is the hotter one and, hence, it is occulted by its more massive companion in the inner binary during the primary eclipses. Thus, we call star $Aa$ the primary of the inner EB, even though it has the lower mass of the pair.  It is somewhat surprising that Gaia has not yet measured the spectroscopic orbit of the tertiary given that it dominates the system light, is reasonably bright at $G$ = 13.5, and has relatively narrow lines with $T_{\rm eff} \simeq 4830$ K. Only the tertiary in TIC 323486857 is somewhat more evolved ($R \simeq 7.5$ R$_\odot$) among our current set of triples.
		
The system is the opposite of `tight' with $P_{\rm out}/P_{\rm in} \simeq  368/2.1 \simeq 175$, and therefore, given the flatness of the system and small-to-modest eccentricities for the inner and outer orbits, respectively, no large dynamical effects are expected.

Finally, we note that star $Ab$, at $R = 2.36$ R$_\odot$, is only modest underfilling its Roche lobe which is currently about 3.45 R$_\odot$. Thus, since $Ab$ is already well evolved off the main sequence it will overflow its Roche lobe relatively soon.  By contrast, the tertiary star, at $R = 6.75$ R$_\odot$, underfills its Roche of 104 R$_\odot$ by a large margin.  However, since the tertiary is more evolved than star $Ab$, and therefore has a big evolutionary lead over $Ab$, it is an interesting question as to which star will first overflow its Roche lobe.  It appears that the tertiary will actually grow to 104 R$_\odot$ within 100 Myr and overflow its Roche lobe, while star $Ab$ will require about 40\% more time to overflow its Roche lobe in the EB.

\subsubsection{TIC 321978218}
\label{Sect:discussion_TIC321978218}

TIC~321978218 has the largest archival observational material among our current sample of triples. Besides the above mentioned four seasons of SWASP observations, this target was also spectroscopically surveyed by the Radial Velocity Experiment (RAVE), and detailed spectroscopic parameters ($T_\mathrm{eff}$, $\log g$, different abundances, etc.) were published in their data release 5 \citep{kunderetal17}. Our own spectroscopic follow up has also resulted in 17 high-quality RV points. The system was found to be single lined in the RAVE analysis, and we also found that only the lines of the tertiary component are observable.  These are in nice accord with the findings of the photodynamical analysis, which shows that the system's light is largely dominated by the tertiary component. Hence, we can conclude that the quantitative spectroscopic results of RAVE basically describe the tertiary component. Consequently, the effective temperature of $T_\mathrm{eff}=5974\pm150$\,K given in RAVE DR5 yields the effective temperature of the dominant tertiary.  This result becomes especially interesting when we combine the preliminary analytic RV and ETV solutions to find an estimate for the mass of this tertiary star.  A first analytic solution for the high-quality RV curve (see Fig.~\ref{fig:321978218rv}) yields a spectroscopic mass function of $f_\mathrm{sp}(M_\mathrm{A})=0.506\pm0.003\,\mathrm{M}_\sun$. On the other hand, the preliminary analytic LTTE ETV solution provides a mass function of $f_\mathrm{ETV}(M_\mathrm{B})=0.290\pm0.030\,\mathrm{M}_\sun$.  Writing these out explicitly
\begin{equation}
f_\mathrm{ETV}(M_\mathrm{B})=M_\mathrm{B}\sin^3i_\mathrm{out}\left(\frac{q_\mathrm{out}}{1+q_\mathrm{out}}\right)^2
\end{equation}
and,
\begin{equation}
f_\mathrm{sp}(M_\mathrm{A})=M_\mathrm{A}\sin^3i_\mathrm{out}\left(\frac{q^{-1}_\mathrm{out}}{1+q^{-1}_\mathrm{out}}\right)^2=M_\mathrm{A}\sin^3i_\mathrm{out}\left(\frac{1}{1+q_\mathrm{out}}\right)^2,
\end{equation}
one can readily see that
\begin{equation}
M_\mathrm{B}\sin^3i_\mathrm{out}=f^{1/3}_\mathrm{ETV}\left(f^{1/3}_\mathrm{ETV}+f^{1/3}_\mathrm{sp}\right)^2=1.41\pm0.10\,\mathrm{M}_\sun.
\end{equation}

Thus, one may immediately realize that a star with an effective temperature of $T_\mathrm{eff}\simeq 5900$\,K and stellar mass of $m_\mathrm{B}\approx1.4\,\mathrm{M}_\sun$ cannot be a main sequence object (we would expect a range of $T_{\rm eff} \approx 6500-7000$ K for this mass). This suspicion became even stronger when we made efforts to model the light curve without taking into account any astrophysical constraints, that is, by directly adjusting the fractional radii of the stars and their temperature ratios. In such a way we found a quite good fit, but the resultant $\log g$'s for the inner EB stars were found to be $\sim$$4.4$ together temperatures of less than $5000$\,K. In other words, the fractional radii of the EB members\footnote{For the connection between the fractional radii and the surface gravity see, e.g., \citealt{southworthetal04,hajduetal17}.} (this latter quantity is mainly dictated by the phase-durations of the regular eclipses) were found to be too large.

After an initial consideration, these findings call to mind three plausible explanations; however, two of them can be readily refuted. First, for a main-sequence tertiary with a mass of $M_B\approx1.8\,\mathrm{M}_\sun$ and corresponding effective temperature of $T_\mathrm{eff}\approx8000$\,K, a consistent model can be obtained with three coeval MS stellar components, where the total mass of the inner EB is well constrained by the spectroscopic mass function. In this case, however, (i) the ETV fit becomes considerably weaker, (ii) the model cumulative SED clearly lies far above the measured catalogued SED points and, (iii) the observed spectrum is clearly inconsistent with such a hot source.  Second, one can assume that the EB members are, again, main sequence stars, but they have radii that are inflated by 10-15\% due to tidal interactions with each other (see, e.g., \citealt{spada13}). The problem in this case, however, is that if one assumes the tertiary is a quite evolved\footnote{This qualifier is necessary for $T_\mathrm{eff}\approx6000$\,K in the case of such a massive star.} $\sim$$1.4\,\mathrm{M}_\sun$ star, the EB components with the proper masses would be far too bright for the observed eclipse depths. This means that for relatively brighter EB components, the deeper third-body eclipses, which occur when the EB stars transit in front of the tertiary would be too shallow, while the other third-body eclipses, when the EB stars are occulted by the third tertiary star, would be too deep.  Moreover, the regular EB eclipses would also be deeper than are observed.

After the above arguments, we found only one really plausible explanation for the system properties.  According to our interpretation TIC~321978218 is a very young, pre-main sequence system. This assumption would explain both the low temperature of the third star relative to its mass and the large fractional radii of the less massive EB members.  An additional problem, however, arises even in this pre-MS scenario. We were unable to find strictly coeval stars which would satisfy all the observational constraints. Instead we allowed a different age for the tertiary component than that of the EB members, the latter two of which  were considered to be strictly coeval.  As one can see in Table~\ref{tab:syntheticfit_TIC321978218}, we found satisfactory solutions with a slightly younger tertiary $t_\mathrm{B}\approx11$\,Myr vs.~$t_\mathrm{Aa,Ab,C}\approx21$\,Myr. Such a small difference of $\approx10$\,Myr might be comparable with even the life time of a circumbinary disk, and also may imply the operation of a sequential disk instability mechanism as a possible explanation of multiple stellar system formation \citep[see, e.g.][and further references therein]{tokovinin21}. It may also amount to only some minor inconsistencies in the pre-main sequence stellar evolution tracks.

Having established that TIC 321978218 is a young system, we return to the interesting system structure initially introduced in Sect.~\ref{sec:etvs_rvs}.  The securely understood part of the orbital architecture is that of a flat edge-on triple with a 0.57-day inner EB and a tertiary in a 57.5 day outer period. The mutual inclination angle is only $\sim$2$^\circ$. The tertiary contains some 90\% of the system bolometric luminosity.

In addition, as we found from the ETV curves in Fig.~\ref{fig:321978218rv} there is a fairly clear additional long-term structure superposed on the ETV of the triple system itself. If we assume that this additional perturbation to the light curve is due to another body orbiting the triple farther out, we find evidence for a quadruple system with a 2+1+1 hierarchy. If this is the correct interpretation, the outer orbital period is $\approx 4400$ d, the outer eccentricity is $\sim$0.37, and the mutual inclination with the inner triple is $\sim$10$^\circ$.  

The only astrophysical parameter associated with  the fourth star that is actually constrained by the available observational material is its (very low) mass of $M_\mathrm{C}=0.16\,\mathrm{M}_\odot$. Because of its negligibly small contribution to the system's total flux, as well as the absence of any other effects on the light curve, all other parameters such as stellar radius and effective temperature remain basically undetermined.  Due the consistency of the model, however, {\sc Lightcurvefactory} calculates all the astrophysical parameters of this fourth star in the same manner as for the other three components, namely, via the use of \texttt{PARSEC} isochrones. During all the runs, for technical reasons, the age of this fourth star was considered to be equal to the age of the inner EB system (21 Myr). The resultant effective temperature ($T_\mathrm{C}=2790\pm100$\,K) follows from its mass and age. The highly bloated radius ($R_\mathrm{C}=0.49\pm0.02\,\mathrm{R}_\sun$) can be understood in the long time it takes for a star this low in mass to contract onto the main sequence.  However, none of these parameters for the fourth star should be taken too seriously until the existence of the fourth star is confirmed by further observations.  

\subsubsection{TIC 323486857}
\label{Sect:discussion_TIC323486857}

This is another target where the TESS light curve was strongly contaminated by nearby, and likely variable, stars. Thus, we had to apply our PCA analysis again in order to separate the signal of the triply eclipsing triple system from other stars. Due to the strongly variable nature of the contaminating light, this was not a simple task and, during the process, we probably lost a significant fraction of the flux coming from our target system. Moreover, even after the PCA process, we obtained a substantially distorted light curve. We assume that this remaining signal is real, and comes from our system, revealing strong chromospheric activity in the late type stars. We made some further efforts to remove, or at least reduce, the effects of these extra distortions to the geometric triple star signal (the latter of which we are investigating), but these efforts met with only partial success. Finally, we carried out our usual complex photodynamical analysis, but simultaneous to the triple-star light-curve fitting, we also fit the extra distortions together with harmonic functions with four frequencies, as we have done previously in some other cases \citep[see][for a detailed description of this process]{borkovitsetal18}. In such a way we were able to obtain a quite reliable solution, but the parameter uncertainties are somewhat larger than in the case of the other triple systems.

TIC 323486857 is comprised of three comparable-mass stars in highly circular inner and outer orbits that are nearly coplanar.  The mutual inclination angle between these two orbits is only $2\fdg5 \pm 1\degr$. The tertiary star has a mass of 1.6 M$_\odot$ compared to 1.4 M$_\odot$ for the primary star of the inner binary.  Just this small difference in mass, however, is sufficient for the tertiary star to be substantially evolved ($R_{\rm B} = 7.5\,\mathrm{R}_\odot$), while the primary EB star has not evolved much. 

\begin{figure}
\begin{center}
\includegraphics[width=1.000 \columnwidth]{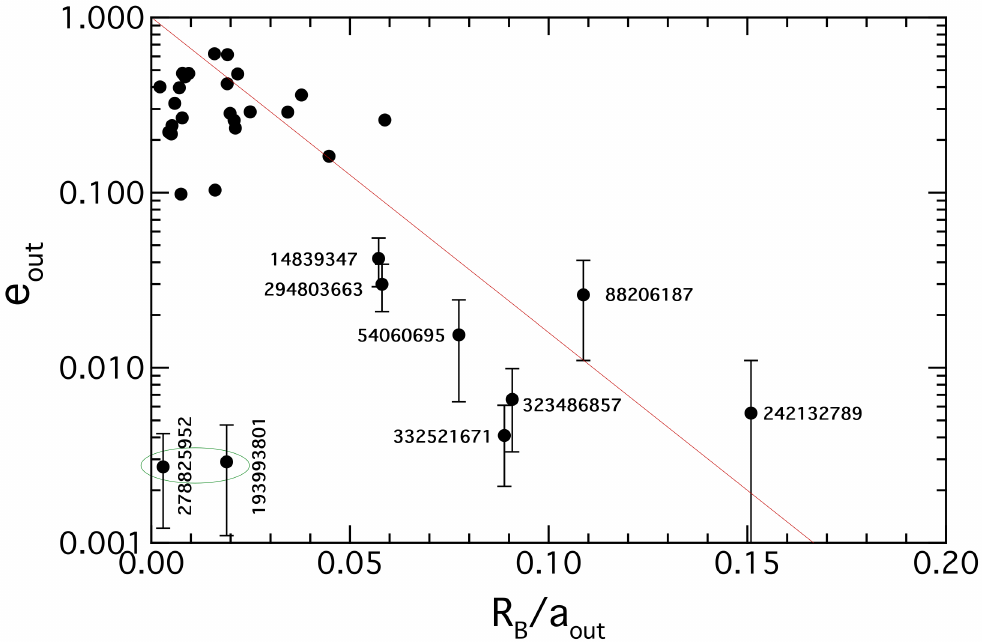} 
\caption{Parameter $e_{\rm out}$ plotted against the fractional radius of the tertiary component $r_\mathrm{B}=R_{\rm B}/a_\mathrm{out}$ for 33 triply eclipsing triple systems that we have studied recently (see Sect.\,\ref{sec:discuss} for details).  The red line sketches out the inverse correlation between these two parameters, with a functional form of $e_{\rm out} \simeq \exp(-42 r_{\rm B})$.  Exceptions are the two systems marked with a faint green ellipse with very low eccentricities, but not large values of $r_{\rm B}$. Numbers next to individual sources are the TIC numbers.  We omit error bars on the systems with $e_{\rm out} \gtrsim 0.1$ for increased clarity. }
\label{fig:evsRB}
\end{center}
\end{figure} 

This system has three related properties that make it rather interesting.  It has a combination of (i) a very short outer orbital period (41 d), (ii) a quite low outer eccentricity ($0.0066 \pm 0.0040$), and (iii) a fairly large tertiary fractional radius $r_\mathrm{B}=R_{\rm B}/a_{\rm out}= 0.093\pm0.007$.  In Fig.~\ref{fig:evsRB} we plot the essence of this information in the plane of $e_{\rm out}$--$r_{\rm B}$ for all 33 triply eclipsing triples that we have studied recently (see Sect.\,\ref{sec:discuss} for more information), including TIC 323486857.  With the exception of the two sources at the bottom left of the plot (marked with a faint green ellipse) there is a strong inverse correlation between the outer eccentricity and the tertiary's fractional radius.  We believe this is caused by dissipation in the tides raised in the evolving giants by the companion EB.  The most circularized systems (with the two aforementioned exceptions) have outer orbital separations of only $\sim$7-13 times the radius of the tertiary, while the tidally induced decay of the eccentricity goes as a high power of $R_{\rm B}/a_{\rm out}$ \citep{zahn977}. This is completely analogous to the circularization process that operates in ordinary binary stars.  The other two systems with low eccentricity, but relatively small values of $r_{\rm B}$, seem otherwise perfectly normal systems with well measured parameters--we presume these were born circular.

\subsubsection{TIC 650024463 = Gaia DR3 4618836249918572544}

The stellar identification of this system is somewhat problematic. This is because, according to Gaia DR3, there are three potential candidate stars within 10$''$ of each other (see Fig.~\ref{fig:650024463ID}). These are Gaia DR3 4618836249918572928 = TIC 302965294 with G=13.28 mag; 4618836249918572544 = TIC 650024463 with G=14.84 mag at 2$''$ from the former target; and 4618836249918735488 = TIC 302965293 with G=15.13 mag, a bit farther at 9$''$. The three targets have completely different DR3 parallaxes and proper motions, hence, there are no physical relations amongst them, at least not with any certainty. On the other hand, it is also clear that the images of the three stars cannot be disentangled due to the large pixel size of TESS and, naturally, the blended light curve of the three stars is  totally dominated by the brightest target star. 

Thus, our first assumption was that the triply eclipsing triple stellar system is hosted by the brightest stellar image, namely that it belongs to TIC 302965294. The very first light curve fitting trials, however,  revealed that a consistent EB and triply eclipsing model can be obtained only with the assumption of $80-85\%$ contaminated light. {However, for the brightest star, TIC 302965294, it is clearly not possible that this much contaminated light comes from one of the fainter neighboring stars.  On the other hand, if the triple resides in one of the fainter stars (i.e., TICs 650024463 or 302965293), then the extra light is nicely accounted for by the brightest star which does not host the triple.  These two fainter stars have $G_{\rm RP}$ band fluxes of $\approx0.26$ and $\approx0.19$, respectively, relative to the brightest star.

\begin{figure}[ht]
\begin{center}
    \includegraphics[width=0.75\columnwidth]{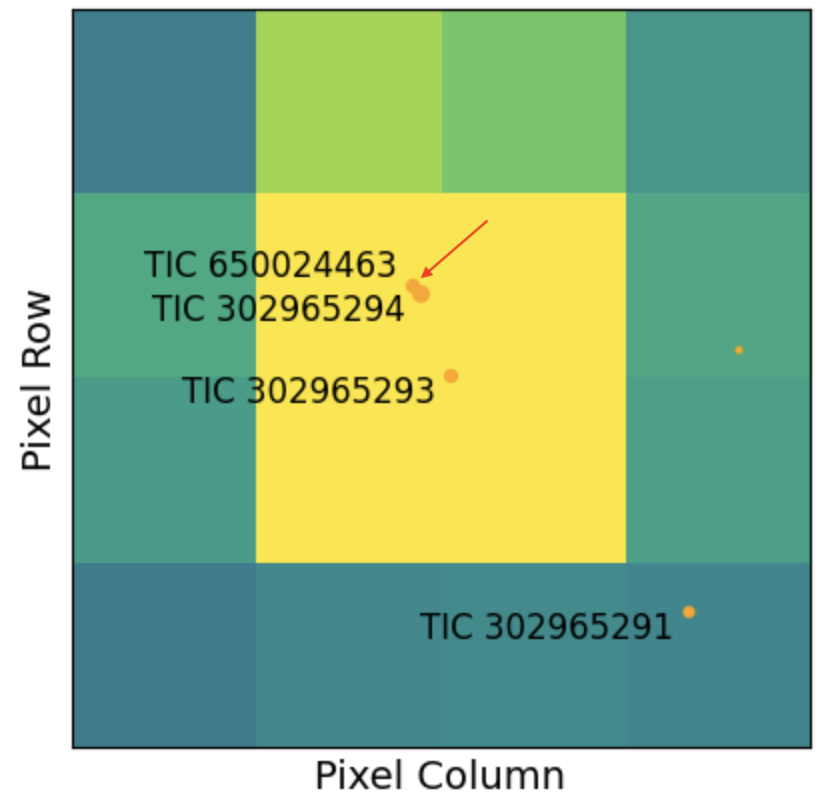}  \vglue0.1cm \hglue-0.5cm
   \includegraphics[width=0.85\columnwidth]{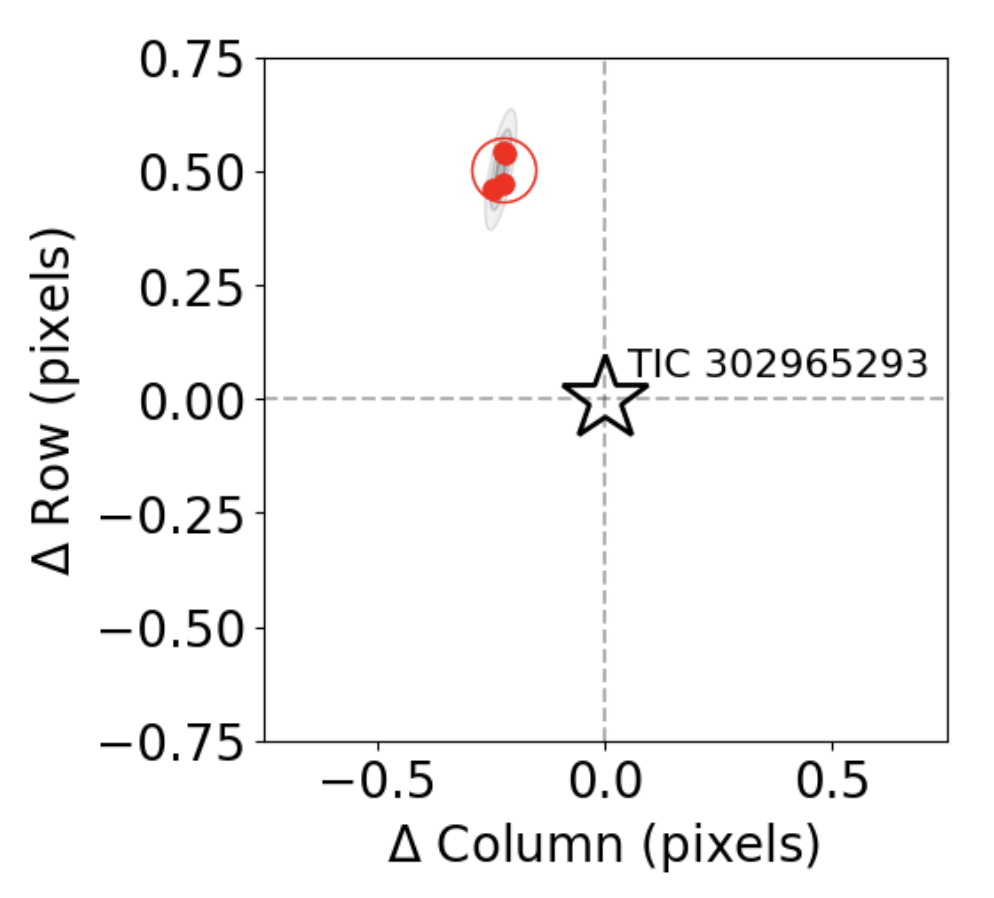}
   \caption{Positional information for TIC 650024463 and its two neighbors. {\it Upper panel}: A $4 \times 4$ pixels TESS image near the position of TIC 650024463 (marked by the red arrow) from Sector 1, highlighting nearby field stars (orange dots). {\it Lower panel}: Photocenter measurements for the primary eclipses detected in Sector 1 compared to the relative pixel position of TIC 302965293. The small red dots represent the measurements for each eclipse (four in Sector 1); the large red circle represents the average photocenter; and the grey contours represent the corresponding 1-, 2- and 3-$\sigma$ confidence intervals. As seen from the panel, the measurements clearly rule out TIC 302965293 as a potential source of the detected eclipses.}
\label{fig:650024463ID}
\end{center}
\end{figure}   

Because TIC 302965293 is 9$''$ from the brightest star TIC 302965294, we carried out a photo-center analysis using the eclipses observed in the source (see, e.g., \citealt{kostov24}). The top panel of Fig.~\ref{fig:650024463ID} shows the layout of the neighboring stars at the TESS pixel level.  The bottom panel indicates the relative location of TIC 302965293 (G=15.13), while the red points are the photocenter locations of the primary EB eclipses in comparison.  The grey contours are the confidence levels for the eclipse locations.  Thus we see that TIC 302965293 is robustly ruled out as the source of the eclipses. This photo-center analysis is not quite sufficient to distinguish TIC 302965294 from TIC 650024463 because they are only 2$''$ apart. However, TIC 302965294 has already been ruled out via the shallow depth of the eclipses (see above).

Thus, even though we now know that TIC 650024463 is the correct host of the triple system, due its extreme proximity to the brighter star TIC 302965294, there are no available cataloged passband magnitudes for this target --  with the exception of the three Gaia DR3 magnitudes.  Hence, in this case, we departed from our usual SED + \texttt{PARSEC} fitting procedure and simply fixed the Gaia DR3 parallax (and, hence, distance). We also set the interstellar reddening $E(B-V)$ parameter to the value given in TIC v8.2, and kept it fixed.

Apart from the light contamination issue, the complex photodynamical analysis resulted in quite robust results for the system parameters.  In particular, the orbital dynamics and 3D configuration of the entire stellar system were well determined and, moreover, the relative (dimensionless) stellar quantities (i.e. the mass, temperature and radius ratios of the stars) are also determined with high accuracies.  This success is due to the fact that TESS observed the target during nine sectors (the most amongst the currently investigated seven systems). These observations recorded not only six third-body eclipses (Fig.~\ref{fig:650024463lcs}), but the derived ETV curves (see in Fig.~\ref{fig:650024463ETV}) show marked, large amplitude, dynamically dominated $P_\mathrm{out}$-period cycles, and cover more than half of a full apsidal revolution cycle (the latter of which is also clearly dominated by third-body effects).

\begin{figure*}[h!]
\begin{center}
     \includegraphics[width=0.75\textwidth]{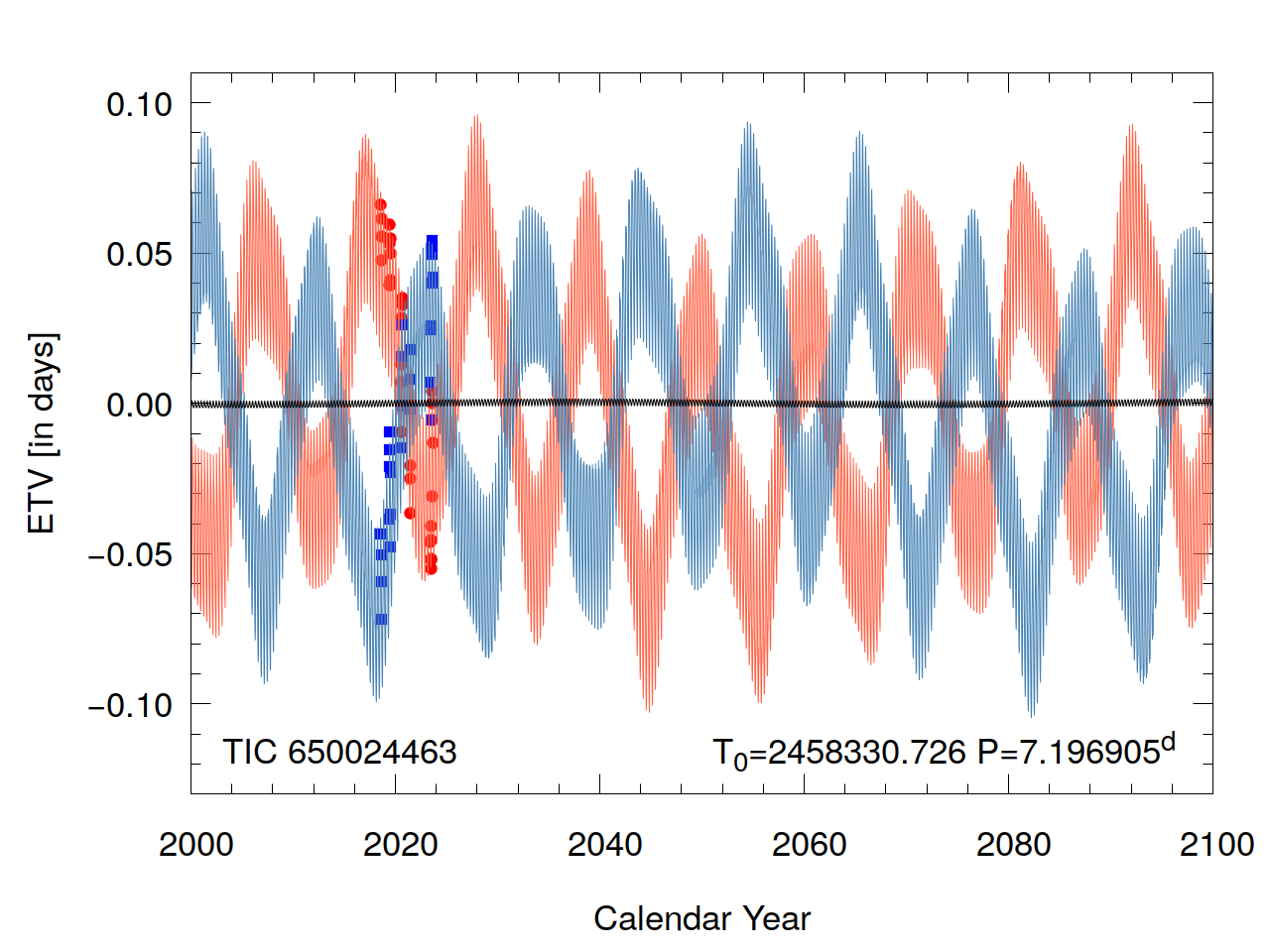}
    \includegraphics[width=0.42\textwidth]{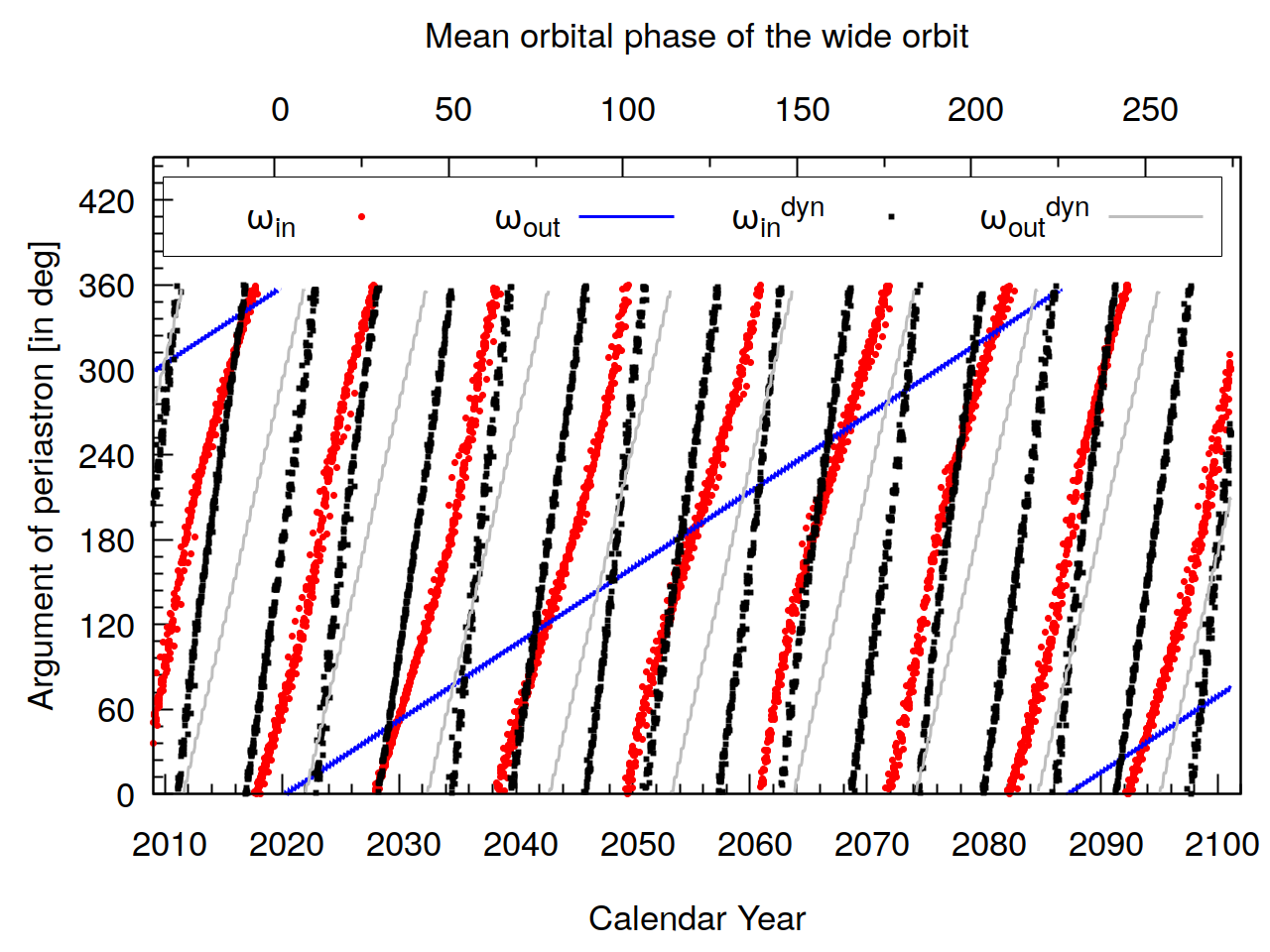}  
   \includegraphics[width=0.42\textwidth]{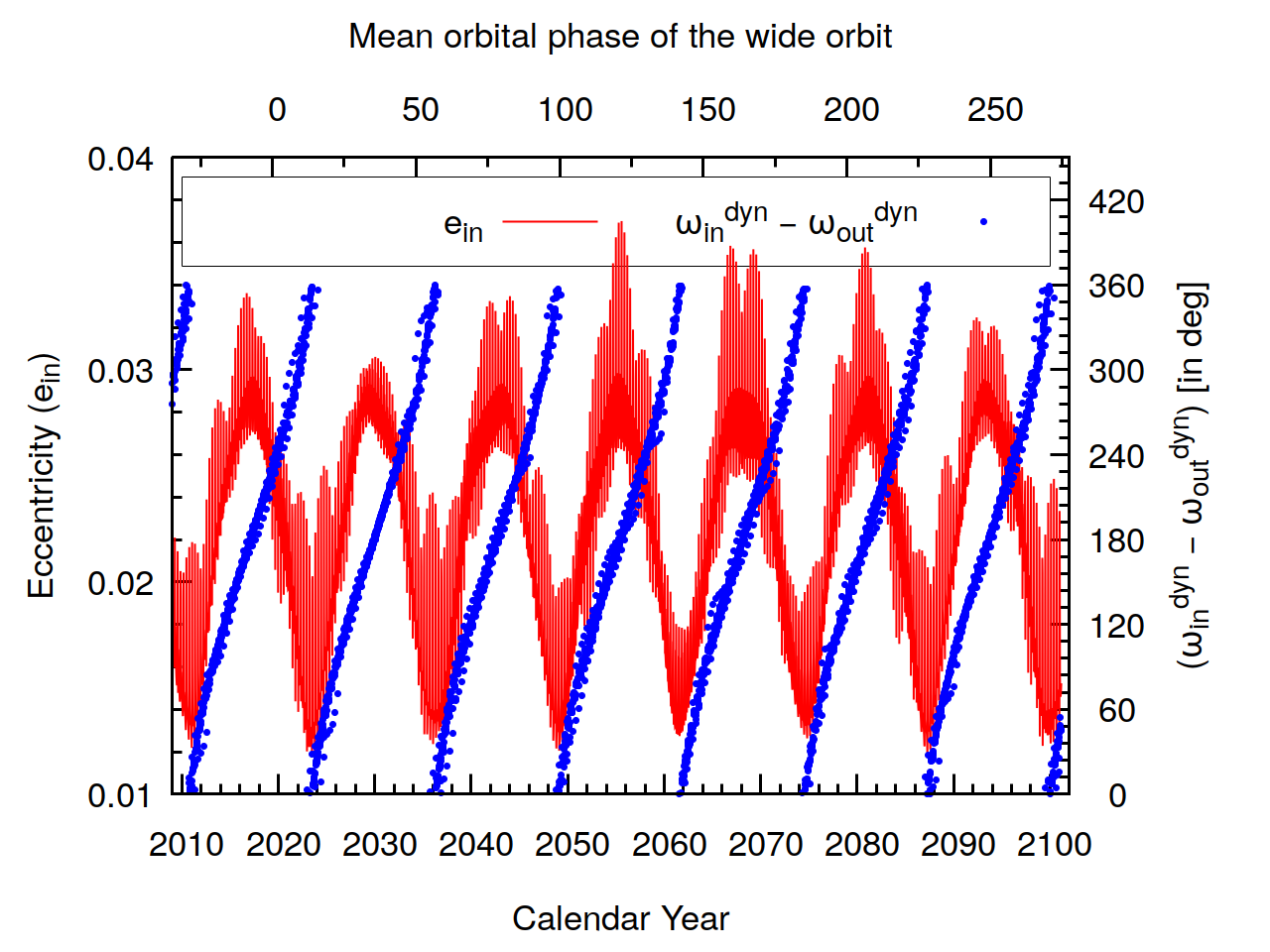}
   \caption{Numerically integrated models for TIC 650024463 for the entire $21^\mathrm{st}$ century.  {\it Upper panel:} Model ETV curve: red and blue lines represent the primary and secondary curves, respectively, while red circles and blue squares stand for the TESS-observed ETV points. Besides the rapidly oscillating vertical `spikes', which come from the $P_\mathrm{out}$ period of 108 d, namely, the so-called medium term third-body perturbations, the 10.9 year third-body forced apsidal motion is also clearly visible. The amplitudes of consecutive half-cycles vary substantially which is the consequence of the octupole-order eccentricity perturbations. {\it Lower left panel:} Variations of the observable and dynamical arguments of periastron (i.e., the apsidal motion of the triple) until year 2100. {\it Lower right panel:} Cyclic variations of the inner eccentricity ($e_\mathrm{in}$ -- red) and the corresponding, similar-period variations of the differences of the dynamical inner and outer arguments of periastrons ($\omega_\mathrm{in}^\mathrm{dyn}-\omega_\mathrm{out}^\mathrm{dyn}$ -- blue) during the present century. See text for further details.}
\label{fig:650024463numint}
\end{center}
\end{figure*}   

These ETV properties are consequences of the fact that this system is by far the tightest triple in our sample with $P_\mathrm{in}=7\fd20$, $P_\mathrm{out}=108\fd73$ and, hence, $P_\mathrm{out}/P_\mathrm{in}\approx15.1$.  From a dynamical perspective, this triple star system is very similar to the three triples (KIC~9714358 -- $P_\mathrm{in}=6\fd47$, $P_\mathrm{out}=104\fd08$; KIC~5771589 -- $10\fd68$, $113\fd87$; TIC~219885468 -- $7\fd51$, $111\fd51$) which were recently analyzed in detail by \citet{borkovitsmitnyan23}.  Besides the very similar inner and outer periods, further similarities are that all of the four systems are (i) quite flat (well within $i_\mathrm{mut}=0\fdg5$), and have (ii) small, but significantly non-zero inner eccentricities ($0.0036\lesssim e_\mathrm{in}\lesssim0.042$), and (iii) moderate outer ($0.16\lesssim e_\mathrm{out}\lesssim0.39$) eccentricities.

As it was pointed out by \citet{borkovitsmitnyan23}, such tight, eccentric and flat systems are ideal for observational detections of higher order third-body perturbations on timescales as short as a few years. The different apsidal motion amplitudes of the ETV curves in the first and the second half of the TESS observations of TIC 650024463 (Fig.~\ref{fig:650024463ETV}) suggest  the additional presence of observable effects of higher, octupole order third-body perturbations in the current system.  

To further study this effect, we numerically integrated the motion of the triple system for the forthcoming few centuries. From these integrations we show and discuss some representative results for the current century.  In the upper panel of Fig.~\ref{fig:650024463numint} we plot the numerically generated ETV curve for the entire $21^{st}$ century. The $P_\mathrm{apse}^\mathrm{obs}=10.9$\,yr-period, dynamically forced apsidal motion is readily visible from the anticorrelated nature of the primary (red) and secondary (blue) ETV curves. These variations correspond nicely to the period of revolution of the inner orbital ellipse in the observational frame of reference, that is, to the variation of the observational argument of periastron ($\omega_\mathrm{in}$) plotted with red dots in the lower left panel of Fig.~\ref{fig:650024463numint}.  This period of 10.9\,yr, however, is substantially shorter than the theoretical value of $P_\mathrm{apse}^\mathrm{theo}=15.45\pm0.08$\,yr (see in Table~\ref{tab:syntheticfit_TIC323486857650024463}) which was calculated from the simple, quadrupole-order theory.  This discrepancy indicates that the quadrupole-order perturbation theory may well be insufficient in the present situation.

Besides this weak, indirect indication for the presence of the higher-order perturbation effects, however, the very same ETV curve carries more clear and well observable evidence about the presence of higher order effects.  As one can see in the upper panel of Fig.~\ref{fig:650024463numint}, the amplitudes of the consecutive half-apsidal motion cycles in the ETV curve are really highly variable during the entire $21^{st}$ century.  These amplitude variations do not correspond strictly to the apsidal motion cycles, (and hence, the amplitude varies cycle-by-cycle), but instead, correspond to the $\sim$$12.9$\,yr-period (inner) eccentricity cycles. The lower right panel of Fig.~\ref{fig:650024463numint} illustrates that the inner eccentricity (again, apart from the $P_\mathrm{out}$-period mid-timescale variations) oscillates between $(e_\mathrm{in})_\mathrm{min}\sim 0.015$ and $(e_\mathrm{in})_\mathrm{max}\sim0.028$ with the period mentioned above. As is known from the theory of the secular hierarchical three-body problem, for such an almost perfectly flat ($i_\mathrm{m}=0\fdg41\pm0\fdg27$) triple, the lowest order (quadrupole) perturbations in the inner eccentricity vanish \citep[see, e.g.][]{naoz16} and, consequently, these $e$-cycles cannot be explained with the usual, quadrupole-order perturbations.  On the other hand, these oscillations are in phase with the variations of the differences of the inner and outer dynamical arguments of periastron ($\Delta\equiv\omega^\mathrm{dyn}_\mathrm{in}-\omega^\mathrm{dyn}_\mathrm{out}$) and reach their extrema (i.e., minima and maxima of $e_\mathrm{in}$'s) when $\Delta=0\degr$ or $180\degr$ and, hence, the octupole-order dominant perturbation term in $e_\mathrm{in}$, which is proportional to $\sin\Delta$ \citep[see, e.g.,][Eq.~(12)]{borkovitsmitnyan23}, disappears. Thus, one can readily conclude that these $e$-cycles, and the concomitant variations in the shape of the apsidal ETV curves are direct, observable manifestations of the higher, octupole order third-body perturbation effects. 


\begin{figure*}
\begin{center}
\includegraphics[width=0.319 \textwidth]{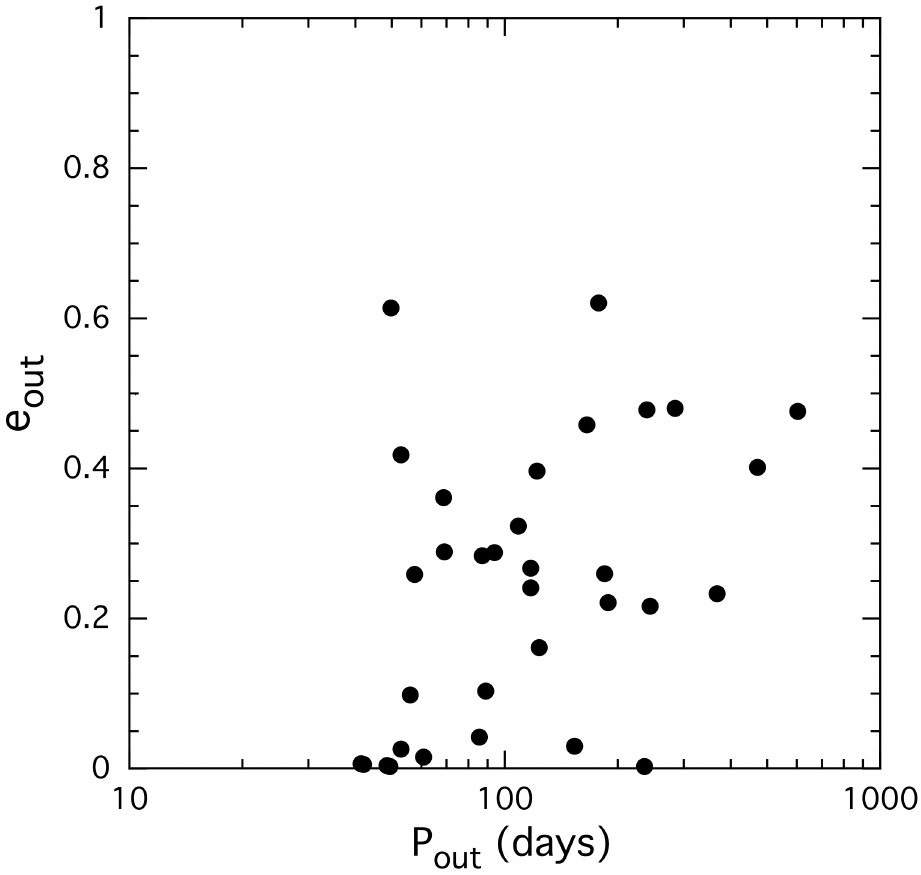} \hglue-0.10cm
\includegraphics[width=0.308 \textwidth]{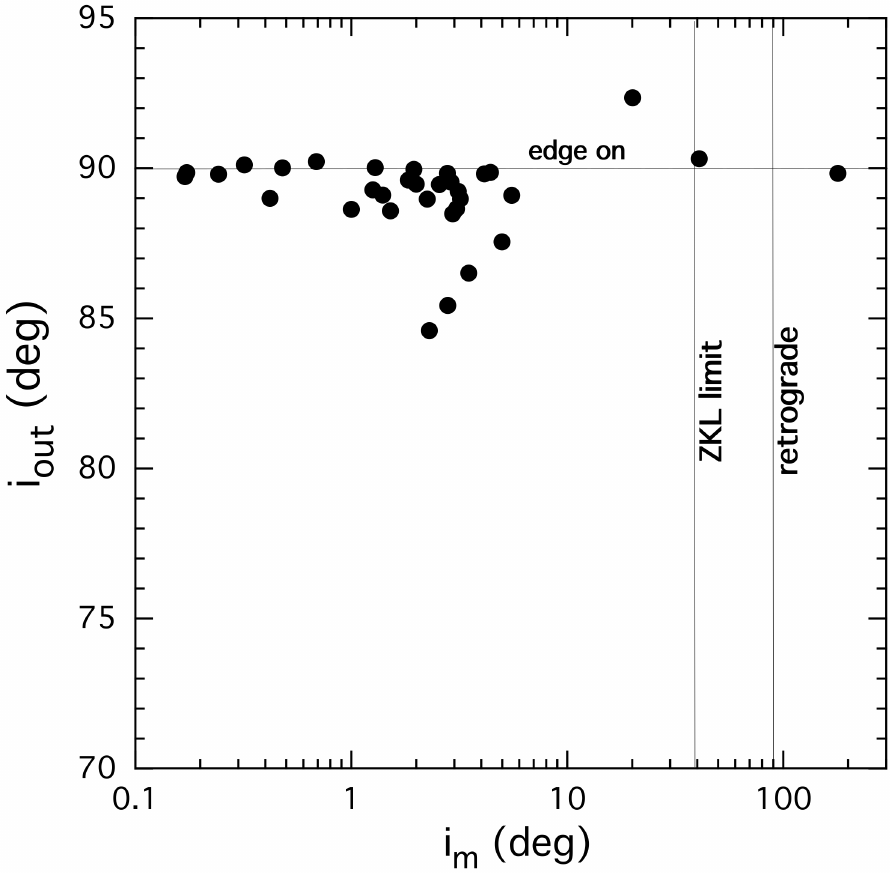} \hglue0.1cm   \vglue0.1cm \hglue0.20cm
\includegraphics[width=0.315 \textwidth]{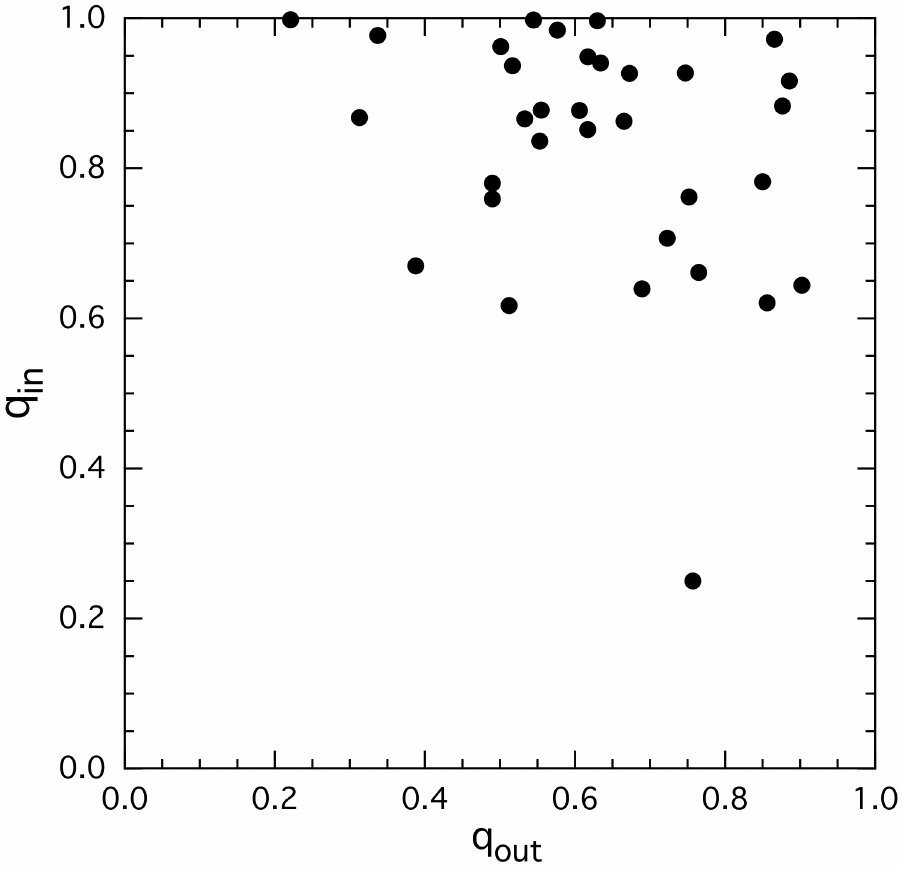} \hglue0.15cm \hglue0.0cm
\includegraphics[width=0.310 \textwidth]{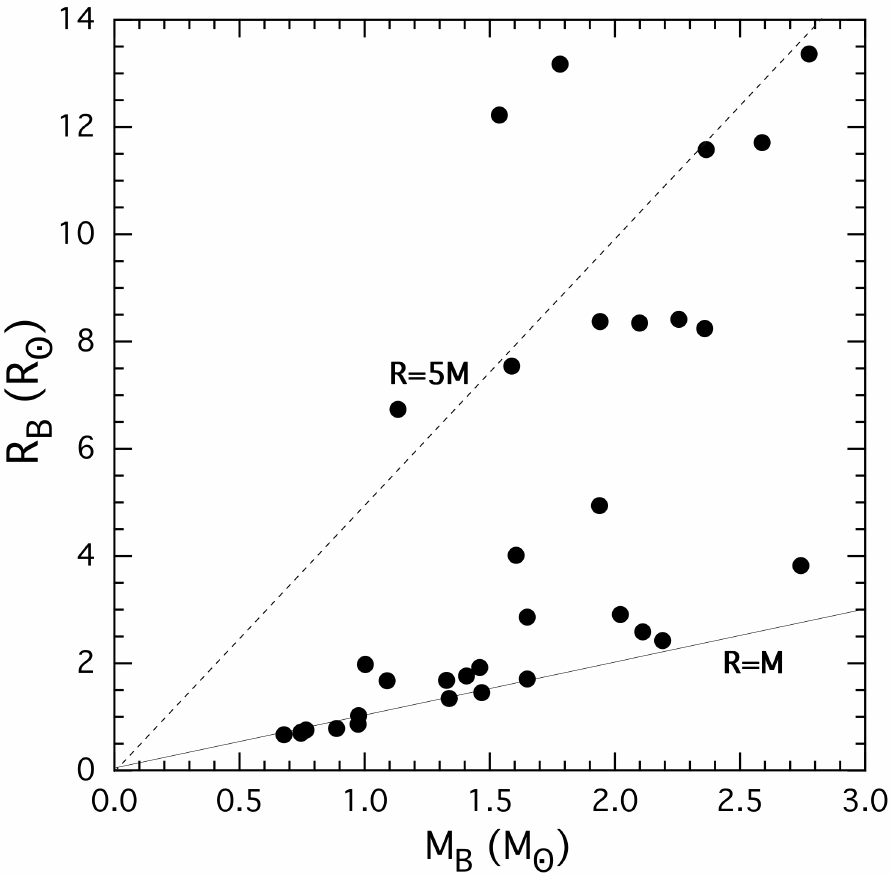} \hglue0.2cm
\caption{Statistical plots for properties of 33 triply eclipsing triples uniformly analyzed (see text for references).  {\it Top-row panels:} $e_{\rm out}$ vs.~$P_{\rm out}$ and $i_{\rm out}$ vs.~ $i_{\rm mut}$. In the upper right panel the vertical lines denote the transition to the ZLK cycles, and to retrograde orbits, respectively. {\it Bottom-row panels:} $q_{\rm in}$ vs.~$q_{\rm out}$ and $R_{\rm B}$ vs.~$M_{\rm B}$. The sloped dashed lines in the latter plot are for $R_{\rm B}$ =1\,M$_{\rm B}$ and = 5\,M$_{\rm B}$ (both expressed in solar units), as rough guides of unevolved and quite evolved stars, respectively.}
\label{fig:statistics}
\end{center}
\end{figure*} 

\section{Summary, discussion, and conclusions}
\label{sec:discuss}

In this work we have presented seven new compact triply eclipsing triples with a full solution to their stellar and orbital parameters.  All of these systems were found by searching through millions of TESS photometric light curves looking for third body eclipses in binary systems (see e.g., \citealt{kristiansen22,rappaport22}).  

We utilized the TESS photometric light curves, the ETV points derived from the light curves, archival SED data, archival photometry from ASAS-SN and ATLAS, and in some cases, ground-based follow up RV observations as well as eclipse photometry.  These were combined in a complex photodynamical analysis where we solve for all the system parameters, as well as the distance to the source.  Typical uncertainties on the masses and radii are in the range of a couple of per cent to not much more than $\sim$5 per cent.  Uncertainties on the angles associated with the orbital planes (e.g., $i_{\rm out}$ and $i_{\rm mut}$) range from a fraction of a degree to about a degree.  See Tables  \ref{tab:syntheticfit_TIC133771812176713425} through  \ref{tab:syntheticfit_TIC323486857650024463}. 


In Fig.~\ref{fig:statistics} we present a set of four correlation plots among some of the physically interesting parameters associated with our collection of 33 recently analyzed triply eclipsing triples.  To the seven sources presented in this work, we have added the 15 triples from our previous closely related papers \citep{rappaport22} and \citep{rappaport23}, as well as eleven triples studied in \citet{borkovitsetal19a,borkovitsetal20b,borkovitsetal22a,borkovitsetal22b}, \citet{czavalinga23b}, and \citet{mitnyanetal20} using largely the same selection criteria and methods of analysis.  

The top left panel in Fig.~\ref{fig:statistics} shows how $e_{\rm out}$ varies with $P_{\rm out}$.  In general, there is little correlation, except for the fact that the most circular outer orbits tend to have the relatively shorter outer periods, where the circularization may be brought about by tidal damping with an evolved tertiary. See also Fig.~\ref{fig:evsRB} and the associated discussion.

\begin{figure*}[ht]
\begin{center}
     \includegraphics[width=0.6\textwidth]{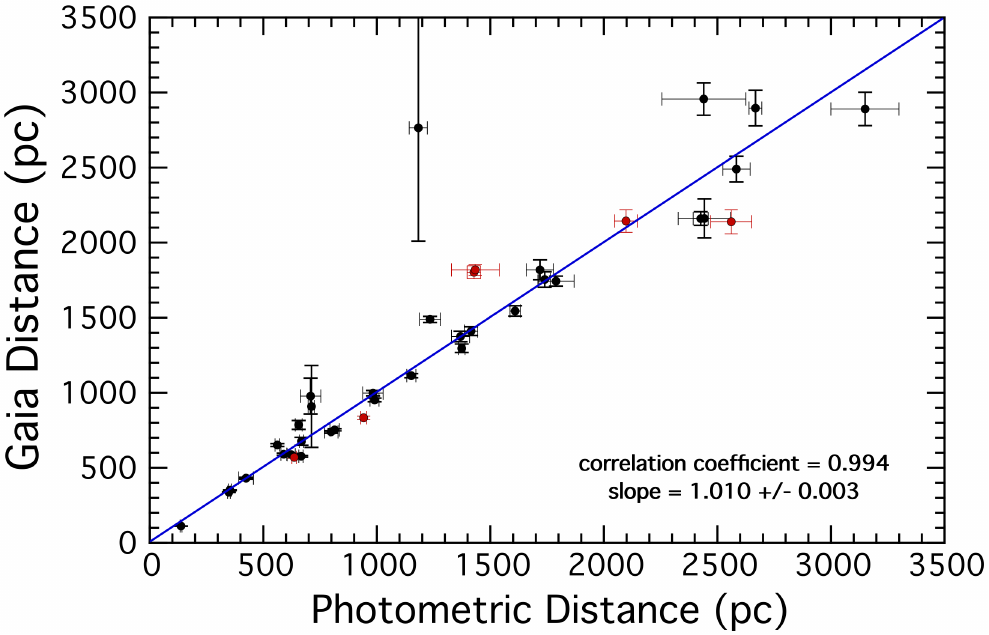}
   \caption{Comparison of Gaia distances \citep{bailer-jonesetal21} to 40 triple systems with distances found from our photodynamical fits to the system parameters. The systems marked in red are 6 from the current work with fitted distances. The blue curve is the line where the Gaia and our photometric distances would match. The formal correlation coefficient between the two sets of data is 0.994, and the fitted slope is $1.010 \pm 0.003$.  There is one system, TIC 52041148, at a Gaia distance of $5931 \pm 400$ pc and photometric distance of  $1357 \pm 30$, that is off the plot.}
\label{fig:distances}
\end{center}
\end{figure*}   

The top right panel of Fig.~\ref{fig:statistics} shows the inclination angle of the outer orbit vs. the mutual inclination angle of the inner binary with respect to the outer orbit.  The fact that most of the values of $i_{\rm out}$ lie near $90^\circ$ is a selection effect since these triples were, in fact, discovered from the presence of third-body eclipses. To a lesser extent, the same selection also holds for the small values of $i_{\rm mut}$, otherwise third body eclipses would be somewhat more difficult to detect.  Two of the systems have large enough $i_{\rm mut}$ (20$^\circ$ and 40$^\circ$) to undergo substantial precession of their orbital planes. Finally, there is one triple (TIC 276162169) in a nearly flat system, but where the outer orbit is retrograde with respect to the inner EB.  These are rare systems. 

The $q_{\rm in}$ vs.~$q_{\rm out}$ plot (lower left panel in Fig.~\ref{fig:statistics}) shows an interesting feature, namely that the ratio of $q_{\rm out} \equiv M_{\rm B}/M_{\rm EB}$ never exceeds unity (and not just by definition) and has a median value of 0.62. This result is not obviously caused by any selection effects.  For purposes of this plot only we define $q_{\rm in}$ as the ratio of the lower mass to the higher mass star in the EB.  

This result for $\bar{q}_{\rm out}$ having a mean of $\sim$0.62 and never exceeding unity, favors a formation scenario for tight triples involving sequential disk fragmentation and subsequent accretion proposed by \citet{tokovinin_moe} \citep[see also][for recent reviews]{tokovinin21,offneretal23},  In this scenario, the accretion rate onto the protostellar cores has an inverse relation with their mass-ratio. Hence, the less massive core of the progenitor EB secondary accretes faster while it has lower mass. Similarly, the tertiary, with the youngest core, will accrete faster while it has lower mass than the sum of the elder, inner cores. Consequently, under the assumption that there is enough matter, the triple tends toward a `double twin' scenario, which means that both $q_{\rm in}$ and $q_{\rm out}$ would be about unity. Naturally, the accretion may stop earlier, resulting in $q_{\rm out} <1$, but it does not allow for $q_{\rm out} >1$.

The bottom right panel of Fig.~\ref{fig:statistics} shows the relation between the tertiary radius and its mass.  About a dozen of the stars lie fairly near the main sequence, roughly represented by the $R=M$ line in the figure.  However, the other $\sim$20 of the tertiary stars have evolved distinctly away from the main sequence.  That is due both to an observational selection effect involving the detectability of third-body eclipses, and the fact that the EB stars cannot grow as large as the tertiaries without overflowing their Roche lobes. 

A couple of the statistical plots rely on the following assumption made in our photodynamical analyses.  Here we have generally adopted the assumption that the inner EB stars are coeval with the tertiary stars and, specifically, that no mass has been transferred among the stars or lost from the system.  In turn, this includes the assumption that the tertiary star itself was always a single star, and not the merger product of another binary star. 

Another item worthy of a brief discussion is the photometric distance we derive vs.~the parallactic distance found by Gaia.  Our distances are plotted in  Fig.~\ref{fig:distances} where they are compared to Gaia's distances.  These are the same 33 triply eclipsing systems discussed above (of which we have photometric distances for 31), and augmented by 9 other triple systems analyzed in the same manner, but which are not triply eclipsing.  These latter sources are taken from \citet{borkovitsetal20a,borkovitsetal22b}, \citet{gaulme22}, and \citet{borkovitsmitnyan23}.  


The overall general agreement between the Gaia distances and our photometric distances is quite striking, and with generally comparable error bars.  The formal correlation coefficient between the two sets of distances is 0.993 and the fitted slope between them is $1.028 \pm 0.004$.  Aside from this general agreement, there are a fair number of points where the two distances differ by more than a few statistical error bars.  In these cases, it is not obvious from an inspection of this figure which distance is the more accurate one. This may be a case where one or both sets have underestimated uncertainties.  In the case of Gaia, the outer orbit, which can be of the order of a 1/2-1 year, may be slightly distorting the parallax measurement\footnote{In this context we note that it was pointed out in Sect.~5.3 of \citet{borkovitsetal22a} that, in the case of TIC~52041148, the extreme outlier Gaia EDR3 parallax-derived distance of \citet{bailer-jonesetal21} cannot be physically realistic.}.  For our photometric distances, there may be issues with inhomogeneous SED data, some of which may be taken during eclipses, the PARSEC isochrones are based on isolated and non-rotating stars, and there are uncertainties in the wavelength-dependent interstellar extinction that we use.

Now that the 33 triply eclipsing triples discussed above (see Fig.~\ref{fig:statistics}) have their basic parameters determined, and all but three have outer periods shorter than a year, this would make for an interesting follow-up ground-based eclipse timing project.  Most of these systems are accessible with modest-size amateur telescopes since the  majority of the 33 objects have G magnitudes of $\lesssim 13$.  The ETV data from TESS itself was typically quite fundamental in the determination of the systems parameters via the photodynamical analyses. Therefore, future timing observations of the regular EB eclipses in these systems could significantly improve the parameter determinations.  The ETV delays are typically in the minute range, so readily within the reach of amateur observations.  Since there are a large number of these triply eclipsing triples, there should be a system to observe nearly every night.

\begin{acknowledgements}
This project has received funding from the HUN-REN Hungarian Research Network.

T.\,B., T.\,M., I.\,B.\,B. and A.\,P. acknowledge the financial support of the Hungarian National Research, Development and Innovation Office -- NKFIH Grants K-147131 and K-138962.

This paper makes extensive use of data collected by the TESS mission. Funding for the TESS mission is provided by the NASA Science Mission directorate. Some of the data presented in this paper were obtained from the Mikulski Archive for Space Telescopes (MAST). STScI is operated by the Association of Universities for Research in Astronomy, Inc., under NASA contract NAS5-26555. Support for MAST for non-HST data is provided by the NASA Office of Space Science via grant NNX09AF08G and by other grants and contracts.

Distances and other astrometric properties for all targets, and outer orbits for two of our sources,  were taken from the archives of European  Space Agency (ESA)  mission {\it Gaia}\footnote{\url{https://www.cosmos.esa.int/gaia}},  processed  by  the {\it   Gaia}   Data   Processing   and  Analysis   Consortium   (DPAC)\footnote{\url{https://www.cosmos.esa.int/web/gaia/dpac/consortium}}.  Funding for the DPAC  has been provided  by national  institutions, in  particular the institutions participating in the {\it Gaia} Multilateral Agreement.

Some of the SED fluxes and magnitudes were obtained with the Wide-field Infrared Survey Explorer, which is a joint project of the University of California, Los Angeles, and the Jet Propulsion Laboratory/California Institute of Technology, funded by the National Aeronautics and Space Administration. 

Additionally, some of the SED fluxes and magnitudes were obtained with the Two Micron All Sky Survey, which is a joint project of the University of Massachusetts and the Infrared Processing and Analysis Center/California Institute of Technology, funded by the National Aeronautics and Space Administration and the National Science Foundation.

We  used the  Simbad  service  operated by  the  Centre des  Donn\'ees Stellaires (Strasbourg,  France) and the ESO  Science Archive Facility services (data  obtained under request number 396301).  

This research has also made use of the VizieR catalogue access tool, CDS, Strasbourg, France (DOI : 10.26093/cds/vizier). The original description of the VizieR service was published in \citet{ochsenbein00}.

Our searches for outer orbital periods from photometry utilized archival data from the Asteroid Terrestrial-impact Last Alert System (ATLAS) project. The Asteroid Terrestrial-impact Last Alert System (ATLAS) project is primarily funded to search for near earth asteroids through NASA grants NN12AR55G, 80NSSC18K0284, and 80NSSC18K1575; byproducts of the NEO search include images and catalogs from the survey area. This work was partially funded by Kepler/K2 grant J1944/80NSSC19K0112 and HST GO-15889, and STFC grants ST/T000198/1 and ST/S006109/1. The ATLAS science products have been made possible through the contributions of the University of Hawaii Institute for Astronomy, the Queen's University Belfast, the Space Telescope Science Institute, the South African Astronomical Observatory, and The Millennium Institute of Astrophysics (MAS), Chile.

Our searches for outer orbital periods also made use of archival data from ASAS-SN data for this work. The ASAS-SN Sky Patrol (All-Sky Automated Survey for Supernovae) is an automated program to search for supernovae and other astronomical transients, headed by astronomers from Ohio State University (US). It has 20 robotic telescopes in both the northern and southern hemispheres and surveys the entire sky approximately once every day.

\end{acknowledgements}

%
%

\begin{appendix} 

\onecolumn

\section{Eclipse times of the inner EBs of the seven triples}
\label{app:ToMs}

In this appendix we tabulate the times of the individual primary and secondary eclipses of the inner EBs of the triples considered in this study.  These naturally include mostly eclipses from TESS, plus a few that were observed from the ground (Tables~\ref{Tab:TIC_133771812_ToM}--\ref{Tab:TIC_650024463_ToM}).

\begin{table*}[h!]
\caption{Eclipse times for TIC~133771812}
 \label{Tab:TIC_133771812_ToM}
\scalebox{0.88}{\begin{tabular}{@{}lrllrllrllrl}
\hline
BJD & Cycle  & std. dev. & BJD & Cycle  & std. dev. & BJD & Cycle  & std. dev. & BJD & Cycle  & std. dev. \\ 
$-2\,400\,000$ & no. &   \multicolumn{1}{c}{$(d)$} & $-2\,400\,000$ & no. &   \multicolumn{1}{c}{$(d)$} & $-2\,400\,000$ & no. &   \multicolumn{1}{c}{$(d)$} & $-2\,400\,000$ & no. &   \multicolumn{1}{c}{$(d)$} \\ 
\hline
58495.15572 &    -0.5 & 0.00129 & 58538.30974 &     3.0 & 0.00107 & 59261.99860 &    61.5 & 0.00484 & 59985.53480 &   120.0 & 0.00044  \\ 
58501.18645 &     0.0 & 0.00051 & 59230.98511 &    59.0 & 0.00059 & 59274.37006 &    62.5 & 0.00490 & 59991.73798 &   120.5 & 0.00114  \\ 
58507.52038 &     0.5 & 0.00597 & 59243.35923 &    60.0 & 0.00052 & 59966.99857 &   118.5 & 0.00099 & 59997.91101 &   121.0 & 0.00029  \\ 
58513.56058 &     1.0 & 0.00052 & 59249.63045 &    60.5 & 0.00090 & 59973.15918 &   119.0 & 0.00038 & 60004.10785 &   121.5 & 0.00073  \\ 
58519.89039 &     1.5 & 0.00158 & 59255.73720 &    61.0 & 0.00181 & 59979.36605 &   119.5 & 0.00101 & 60010.28418 &   122.0 & 0.00041  \\ 
58525.93650 &     2.0 & 0.00051  \\ 
\hline
\end{tabular}}

\textit{Notes}. Integer and half-integer cycle numbers (here and the following tables) denote primary and secondary eclipses, respectively.
\end{table*}

\begin{table*}[h!]
\caption{Eclipse times for TIC~176713425}
 \label{Tab:TIC_176713425_ToM}
\scalebox{0.88}{\begin{tabular}{@{}lrllrllrllrl}
\hline
BJD & Cycle  & std. dev. & BJD & Cycle  & std. dev. & BJD & Cycle  & std. dev. & BJD & Cycle  & std. dev. \\ 
$-2\,400\,000$ & no. &   \multicolumn{1}{c}{$(d)$} & $-2\,400\,000$ & no. &   \multicolumn{1}{c}{$(d)$} & $-2\,400\,000$ & no. &   \multicolumn{1}{c}{$(d)$} & $-2\,400\,000$ & no. &   \multicolumn{1}{c}{$(d)$} \\ 
\hline
58738.87091 &    -0.5 & 0.00016 & 58760.71389 &    11.0 & 0.00015 & 58784.43845 &    23.5 & 0.00015 & 59869.56727 &   595.0 & 0.00007  \\ 
58739.82153 &     0.0 & 0.00026 & 58761.66181 &    11.5 & 0.00020 & 58785.38567 &    24.0 & 0.00015 & 59870.51853 &   595.5 & 0.00010  \\ 
58740.77369 &     0.5 & 0.00023 & 58762.60987 &    12.0 & 0.00014 & 58786.33606 &    24.5 & 0.00015 & 59871.46504 &   596.0 & 0.00007  \\ 
58741.72591 &     1.0 & 0.00018 & 58765.45725 &    13.5 & 0.00024 & 58787.28548 &    25.0 & 0.00016 & 59872.41525 &   596.5 & 0.00009  \\ 
58742.67534 &     1.5 & 0.00014 & 58766.40505 &    14.0 & 0.00019 & 58788.23455 &    25.5 & 0.00038 & 59873.36401 &   597.0 & 0.00008  \\ 
58743.62592 &     2.0 & 0.00024 & 58767.35647 &    14.5 & 0.00011 & 58789.18950 &    26.0 & 0.00086 & 59874.31226 &   597.5 & 0.00009  \\ 
58744.57356 &     2.5 & 0.00019 & 58768.30415 &    15.0 & 0.00013 & 59758.49509$^a$ &   536.5 & 0.00004 & 59875.26216 &   598.0 & 0.00010  \\ 
58745.52534 &     3.0 & 0.00010 & 58769.25387 &    15.5 & 0.00022 & 59853.43423 &   586.5 & 0.00009 & 59876.21071 &   598.5 & 0.00009  \\ 
58746.47509 &     3.5 & 0.00019 & 58770.20236 &    16.0 & 0.00013 & 59854.38270 &   587.0 & 0.00009 & 59877.15953 &   599.0 & 0.00009  \\ 
58747.42357 &     4.0 & 0.00014 & 58771.15121 &    16.5 & 0.00020 & 59855.33184 &   587.5 & 0.00009 & 59878.10921 &   599.5 & 0.00010  \\ 
58748.37530 &     4.5 & 0.00022 & 58772.10136 &    17.0 & 0.00011 & 59856.28088 &   588.0 & 0.00011 & 59879.05671 &   600.0 & 0.00009  \\ 
58749.32190 &     5.0 & 0.00015 & 58773.04374 &    17.5 & 0.00196 & 59857.23138 &   588.5 & 0.00008 & 59880.00742 &   600.5 & 0.00011  \\ 
58750.27213 &     5.5 & 0.00024 & 58773.99852 &    18.0 & 0.00022 & 59858.17868 &   589.0 & 0.00010 & 59880.95465 &   601.0 & 0.00008  \\ 
58752.17061 &     6.5 & 0.00019 & 58774.94872 &    18.5 & 0.00023 & 59859.13005 &   589.5 & 0.00011 & 59881.90507 &   601.5 & 0.00010  \\ 
58753.12064 &     7.0 & 0.00015 & 58775.89613 &    19.0 & 0.00031 & 59860.07715 &   590.0 & 0.00009 & 60135.38686$^b$ &   735.0 & 0.00010  \\ 
58754.06873 &     7.5 & 0.00016 & 58777.79345 &    20.0 & 0.00017 & 59861.02787 &   590.5 & 0.00013 & 60208.48369$^b$ &   773.5 & 0.00003  \\ 
58755.01904 &     8.0 & 0.00012 & 58778.74329 &    20.5 & 0.00025 & 59861.97509 &   591.0 & 0.00008 & 60209.43240$^b$ &   774.0 & 0.00004  \\ 
58755.96778 &     8.5 & 0.00016 & 58779.69151 &    21.0 & 0.00018 & 59862.92622 &   591.5 & 0.00009 & 60267.35271$^b$ &   804.5 & 0.00003  \\ 
58756.91576 &     9.0 & 0.00027 & 58780.64248 &    21.5 & 0.00015 & 59866.72126 &   593.5 & 0.00009 & 60268.30225$^b$ &   805.0 & 0.00007  \\ 
58757.86637 &     9.5 & 0.00019 & 58781.59047 &    22.0 & 0.00013 & 59867.66957 &   594.0 & 0.00008 & 60286.34061$^b$ &   814.5 & 0.00003  \\ 
58758.81356 &    10.0 & 0.00014 & 58782.53835 &    22.5 & 0.00020 & 59868.61941 &   594.5 & 0.00010 & 60344.24895$^b$ &   845.0 & 0.00004  \\ 
58759.76472 &    10.5 & 0.00019 & 58783.48854 &    23.0 & 0.00010  \\ 
\hline
\end{tabular}}

\textit{Notes}. $^a$: Observed at BAO; $^b$: Observed at GAO; the other eclipses observed with \text{TESS}. 
\end{table*}

\newpage

\begin{table*}[h!]
\caption{Eclipse times for TZ~Pyx (TIC~185615681)}
 \label{Tab:TZ_Pyx_(TIC_185615681)_ToM}
\scalebox{0.88}{\begin{tabular}{@{}lrllrllrllrl}
\hline
BJD & Cycle  & std. dev. & BJD & Cycle  & std. dev. & BJD & Cycle  & std. dev. & BJD & Cycle  & std. dev. \\ 
$-2\,400\,000$ & no. &   \multicolumn{1}{c}{$(d)$} & $-2\,400\,000$ & no. &   \multicolumn{1}{c}{$(d)$} & $-2\,400\,000$ & no. &   \multicolumn{1}{c}{$(d)$} & $-2\,400\,000$ & no. &   \multicolumn{1}{c}{$(d)$} \\ 
\hline
58518.16959 &     0.0 & 0.00009 & 59238.09281 &   310.5 & 0.00003 & 59274.02312 &   326.0 & 0.00003 & 59986.98206 &   633.5 & 0.00002  \\ 
58519.34246 &     0.5 & 0.00009 & 59239.24465 &   311.0 & 0.00004 & 59275.18922 &   326.5 & 0.00004 & 59988.14316 &   634.0 & 0.00002  \\ 
58520.48810 &     1.0 & 0.00013 & 59240.41124 &   311.5 & 0.00003 & 59276.34210 &   327.0 & 0.00003 & 59989.30061 &   634.5 & 0.00002  \\ 
58521.66086 &     1.5 & 0.00007 & 59242.72960 &   312.5 & 0.00003 & 59277.50798 &   327.5 & 0.00003 & 59990.46173 &   635.0 & 0.00002  \\ 
58522.80632 &     2.0 & 0.00009 & 59243.88184 &   313.0 & 0.00003 & 59278.66089 &   328.0 & 0.00003 & 59991.61920 &   635.5 & 0.00002  \\ 
58523.97938 &     2.5 & 0.00008 & 59245.04810 &   313.5 & 0.00003 & 59279.82680 &   328.5 & 0.00004 & 59992.78021 &   636.0 & 0.00002  \\ 
58525.12478 &     3.0 & 0.00006 & 59246.20042 &   314.0 & 0.00005 & 59963.79769 &   623.5 & 0.00002 & 59993.93777 &   636.5 & 0.00002  \\ 
58526.29777 &     3.5 & 0.00010 & 59247.36660 &   314.5 & 0.00004 & 59964.95824 &   624.0 & 0.00003 & 59996.25634 &   637.5 & 0.00002  \\ 
58527.44319 &     4.0 & 0.00011 & 59248.51888 &   315.0 & 0.00004 & 59967.27669 &   625.0 & 0.00002 & 59997.41758 &   638.0 & 0.00002  \\ 
58528.61638 &     4.5 & 0.00008 & 59249.68494 &   315.5 & 0.00004 & 59968.43440 &   625.5 & 0.00002 & 59998.57505 &   638.5 & 0.00002  \\ 
58530.93462 &     5.5 & 0.00009 & 59250.83729 &   316.0 & 0.00004 & 59969.59564 &   626.0 & 0.00004 & 59999.73656 &   639.0 & 0.00005  \\ 
58535.57206 &     7.5 & 0.00009 & 59252.00343 &   316.5 & 0.00004 & 59970.75292 &   626.5 & 0.00002 & 60000.89359 &   639.5 & 0.00005  \\ 
58536.71703 &     8.0 & 0.00010 & 59253.15574 &   317.0 & 0.00004 & 59971.91399 &   627.0 & 0.00002 & 60002.05537 &   640.0 & 0.00002  \\ 
58537.89034 &     8.5 & 0.00010 & 59256.64039 &   318.5 & 0.00003 & 59973.07121 &   627.5 & 0.00002 & 60003.21247 &   640.5 & 0.00002  \\ 
58539.03562 &     9.0 & 0.00011 & 59257.79257 &   319.0 & 0.00005 & 59974.23254 &   628.0 & 0.00002 & 60004.37418 &   641.0 & 0.00002  \\ 
58540.20874 &     9.5 & 0.00008 & 59258.95891 &   319.5 & 0.00004 & 59976.55100 &   629.0 & 0.00002 & 60005.53109 &   641.5 & 0.00002  \\ 
58541.35423 &    10.0 & 0.00008 & 59260.11101 &   320.0 & 0.00004 & 59977.70821 &   629.5 & 0.00002 & 60006.69298 &   642.0 & 0.00002  \\ 
59229.97068 &   307.0 & 0.00004 & 59261.27743 &   320.5 & 0.00004 & 59978.86955 &   630.0 & 0.00002 & 60007.84981 &   642.5 & 0.00002  \\ 
59231.13745 &   307.5 & 0.00004 & 59262.42937 &   321.0 & 0.00004 & 59980.02659 &   630.5 & 0.00002 & 60009.01147 &   643.0 & 0.00002  \\ 
59232.28914 &   308.0 & 0.00004 & 59263.59588 &   321.5 & 0.00004 & 59981.18779 &   631.0 & 0.00002 & 60010.16837 &   643.5 & 0.00002  \\ 
59233.45587 &   308.5 & 0.00004 & 59264.74799 &   322.0 & 0.00004 & 59983.50631 &   632.0 & 0.00002 & 60011.32996 &   644.0 & 0.00002  \\ 
59234.60749 &   309.0 & 0.00004 & 59265.91448 &   322.5 & 0.00004 & 59984.66356 &   632.5 & 0.00002 & 60012.48686 &   644.5 & 0.00002  \\ 
59235.77432 &   309.5 & 0.00003 & 59267.06667 &   323.0 & 0.00004 & 59985.82477 &   633.0 & 0.00002 & 60013.64839 &   645.0 & 0.00002  \\ 
59236.92609 &   310.0 & 0.00004 & 59272.87044 &   325.5 & 0.00003  \\ 
\hline
\end{tabular}}
\end{table*}

\begin{table*}[h!]
\caption{Eclipse times for TIC~287756035}
 \label{Tab:TIC_287756035_ToM}
\scalebox{0.88}{\begin{tabular}{@{}lrllrllrllrl}
\hline
BJD & Cycle  & std. dev. & BJD & Cycle  & std. dev. & BJD & Cycle  & std. dev. & BJD & Cycle  & std. dev. \\ 
$-2\,400\,000$ & no. &   \multicolumn{1}{c}{$(d)$} & $-2\,400\,000$ & no. &   \multicolumn{1}{c}{$(d)$} & $-2\,400\,000$ & no. &   \multicolumn{1}{c}{$(d)$} & $-2\,400\,000$ & no. &   \multicolumn{1}{c}{$(d)$} \\ 
\hline
58597.24221 &    -1.0 & 0.09984 & 58623.26018 &    11.5 & 0.00368 & 59336.13166 &   354.0 & 0.00281 & 60041.73549 &   693.0 & 0.00109  \\ 
58598.27373 &    -0.5 & 0.00665 & 59308.03345 &   340.5 & 0.00214 & 59337.16701 &   354.5 & 0.00323 & 60042.78350 &   693.5 & 0.00173  \\ 
58599.31567 &     0.0 & 0.00452 & 59309.07420 &   341.0 & 0.00187 & 59338.20811 &   355.0 & 0.00230 & 60043.81491 &   694.0 & 0.00149  \\ 
58600.35967 &     0.5 & 0.00178 & 59313.29341 &   343.0 & 0.00326 & 59339.24448 &   355.5 & 0.00283 & 60044.85216 &   694.5 & 0.00216  \\ 
58601.40239 &     1.0 & 0.00181 & 59314.28649 &   343.5 & 0.00230 & 59340.30039 &   356.0 & 0.00243 & 60049.01666 &   696.5 & 0.00268  \\ 
58602.43427 &     1.5 & 0.00197 & 59315.31438 &   344.0 & 0.00171 & 59341.33582 &   356.5 & 0.00191 & 60050.05709 &   697.0 & 0.00119  \\ 
58603.47656 &     2.0 & 0.00287 & 59316.35173 &   344.5 & 0.00173 & 59342.37244 &   357.0 & 0.00175 & 60051.09339 &   697.5 & 0.00169  \\ 
58604.52050 &     2.5 & 0.00177 & 59317.39805 &   345.0 & 0.00126 & 59343.41287 &   357.5 & 0.00139 & 60052.13384 &   698.0 & 0.00146  \\ 
58605.55465 &     3.0 & 0.00206 & 59318.45364 &   345.5 & 0.00198 & 59344.46241 &   358.0 & 0.00121 & 60053.17517 &   698.5 & 0.00140  \\ 
58606.59093 &     3.5 & 0.00339 & 59320.52522 &   346.5 & 0.00240 & 59345.50382 &   358.5 & 0.00176 & 60054.22572 &   699.0 & 0.00123  \\ 
58607.63434 &     4.0 & 0.00276 & 59321.56601 &   347.0 & 0.00180 & 59347.57886 &   359.5 & 0.00250 & 60056.29312 &   700.0 & 0.00144  \\ 
58608.67791 &     4.5 & 0.00162 & 59322.59519 &   347.5 & 0.00209 & 59348.61854 &   360.0 & 0.00194 & 60057.33468 &   700.5 & 0.00128  \\ 
58611.79733 &     6.0 & 0.14594 & 59323.64770 &   348.0 & 0.00162 & 59349.65894 &   360.5 & 0.00157 & 60058.38237 &   701.0 & 0.00136  \\ 
58612.82854 &     6.5 & 0.01613 & 59324.69195 &   348.5 & 0.00193 & 59350.70147 &   361.0 & 0.00163 & 60059.42664 &   701.5 & 0.00177  \\ 
58613.88105 &     7.0 & 0.02453 & 59325.71101 &   349.0 & 0.00163 & 59351.74403 &   361.5 & 0.00250 & 60060.45615 &   702.0 & 0.00129  \\ 
58614.92035 &     7.5 & 0.00293 & 59326.74897 &   349.5 & 0.00241 & 59352.78385 &   362.0 & 0.00221 & 60061.50146 &   702.5 & 0.00194  \\ 
58615.96147 &     8.0 & 0.00275 & 59327.80051 &   350.0 & 0.00170 & 59353.81992 &   362.5 & 0.00290 & 60062.55115 &   703.0 & 0.00127  \\ 
58617.00661 &     8.5 & 0.00457 & 59328.83824 &   350.5 & 0.00197 & 59354.86158 &   363.0 & 0.00164 & 60063.59200 &   703.5 & 0.00167  \\ 
58618.04633 &     9.0 & 0.00338 & 59329.88489 &   351.0 & 0.00171 & 59355.90671 &   363.5 & 0.00237 & 60064.63478 &   704.0 & 0.00149  \\ 
58619.08382 &     9.5 & 0.00349 & 59330.92391 &   351.5 & 0.00194 & 59356.94344 &   364.0 & 0.00228 & 60065.67316 &   704.5 & 0.00227  \\ 
58620.12676 &    10.0 & 0.18136 & 59331.95671 &   352.0 & 0.00185 & 59357.98665 &   364.5 & 0.00497 & 60066.70700 &   705.0 & 0.00271  \\ 
58621.16943 &    10.5 & 0.00432 & 59334.05750 &   353.0 & 0.00395 & 59359.02178 &   365.0 & 0.00192 & 60067.74751 &   705.5 & 0.00188  \\ 
58622.21359 &    11.0 & 0.00359 & 59335.09428 &   353.5 & 0.00328 & 59360.06104 &   365.5 & 0.00397  \\ 
\hline
\end{tabular}}
\end{table*}

\newpage

\begin{table*}[h!]
\caption{Eclipse times for TIC~321978218}
 \label{Tab:TIC_321978218_ToM}
\scalebox{0.88}{\begin{tabular}{@{}lrllrllrllrl}
\hline
BJD & Cycle  & std. dev. & BJD & Cycle  & std. dev. & BJD & Cycle  & std. dev. & BJD & Cycle  & std. dev. \\ 
$-2\,400\,000$ & no. &   \multicolumn{1}{c}{$(d)$} & $-2\,400\,000$ & no. &   \multicolumn{1}{c}{$(d)$} & $-2\,400\,000$ & no. &   \multicolumn{1}{c}{$(d)$} & $-2\,400\,000$ & no. &   \multicolumn{1}{c}{$(d)$} \\ 
\hline
55432.54833 & -6369.5 & 0.00031 & 59064.81212 &     5.0 & 0.00006 & 59097.29401 &    62.0 & 0.00005 & 60170.53185 &  1945.5 & 0.00009  \\ 
55432.83201 & -6369.0 & 0.00015 & 59065.09753 &     5.5 & 0.00013 & 59097.57864 &    62.5 & 0.00009 & 60170.81671 &  1946.0 & 0.00004  \\ 
55803.49700 & -5718.5 & 0.00034 & 59065.38192 &     6.0 & 0.00005 & 59097.86390 &    63.0 & 0.00007 & 60171.10179 &  1946.5 & 0.00010  \\ 
56180.70955 & -5056.5 & 0.00033 & 59065.66669 &     6.5 & 0.00012 & 59098.14911 &    63.5 & 0.00011 & 60171.38676 &  1947.0 & 0.00007  \\ 
56180.99554 & -5056.0 & 0.00029 & 59065.95165 &     7.0 & 0.00006 & 59098.43374 &    64.0 & 0.00006 & 60171.67182 &  1947.5 & 0.00007  \\ 
56533.99015 & -4436.5 & 0.00044 & 59066.23684 &     7.5 & 0.00012 & 59098.71815 &    64.5 & 0.00015 & 60171.95628 &  1948.0 & 0.00005  \\ 
56534.27516 & -4436.0 & 0.00017 & 59066.52152 &     8.0 & 0.00006 & 59099.00333 &    65.0 & 0.00012 & 60172.24210 &  1948.5 & 0.00007  \\ 
56896.10467 & -3801.0 & 0.00031 & 59066.80643 &     8.5 & 0.00010 & 59099.28814 &    65.5 & 0.00010 & 60172.52667 &  1949.0 & 0.00005  \\ 
58354.25863 & -1242.0 & 0.00005 & 59067.09170 &     9.0 & 0.00012 & 59099.57334 &    66.0 & 0.00007 & 60172.81093 &  1949.5 & 0.00008  \\ 
58354.54400 & -1241.5 & 0.00053 & 59067.37738 &     9.5 & 0.00006 & 59099.85891 &    66.5 & 0.00014 & 60173.09665 &  1950.0 & 0.00005  \\ 
58354.82840 & -1241.0 & 0.00077 & 59067.66102 &    10.0 & 0.00005 & 59100.14269 &    67.0 & 0.00017 & 60173.38171 &  1950.5 & 0.00009  \\ 
58355.11293 & -1240.5 & 0.00041 & 59067.94616 &    10.5 & 0.00010 & 59100.99845 &    68.5 & 0.00045 & 60173.66615 &  1951.0 & 0.00005  \\ 
58355.39782 & -1240.0 & 0.00005 & 59068.23123 &    11.0 & 0.00006 & 59101.28153 &    69.0 & 0.00013 & 60173.95097 &  1951.5 & 0.00008  \\ 
58355.68338 & -1239.5 & 0.00037 & 59068.51567 &    11.5 & 0.00006 & 59102.42163 &    71.0 & 0.00010 & 60174.23623 &  1952.0 & 0.00005  \\ 
58355.96786 & -1239.0 & 0.00006 & 59068.80059 &    12.0 & 0.00007 & 59102.70658 &    71.5 & 0.00010 & 60174.52103 &  1952.5 & 0.00011  \\ 
58356.25342 & -1238.5 & 0.00029 & 59069.08609 &    12.5 & 0.00014 & 59102.99178 &    72.0 & 0.00007 & 60174.80594 &  1953.0 & 0.00006  \\ 
58356.53756 & -1238.0 & 0.00023 & 59069.37046 &    13.0 & 0.00005 & 59103.27701 &    72.5 & 0.00008 & 60175.37574 &  1954.0 & 0.00005  \\ 
58356.82293 & -1237.5 & 0.00009 & 59069.65543 &    13.5 & 0.00009 & 59103.56120 &    73.0 & 0.00007 & 60175.66128 &  1954.5 & 0.00006  \\ 
58357.10703 & -1237.0 & 0.00026 & 59069.94014 &    14.0 & 0.00007 & 59103.84650 &    73.5 & 0.00010 & 60175.94620 &  1955.0 & 0.00006  \\ 
58357.39208 & -1236.5 & 0.00008 & 59070.22620 &    14.5 & 0.00014 & 59104.13146 &    74.0 & 0.00003 & 60176.23119 &  1955.5 & 0.00009  \\ 
58357.67696 & -1236.0 & 0.00004 & 59070.51025 &    15.0 & 0.00010 & 59104.41625 &    74.5 & 0.00012 & 60176.51561 &  1956.0 & 0.00004  \\ 
58357.96167 & -1235.5 & 0.00007 & 59070.79605 &    15.5 & 0.00007 & 59104.70096 &    75.0 & 0.00004 & 60176.80099 &  1956.5 & 0.00009  \\ 
58358.24659 & -1235.0 & 0.00039 & 59071.08036 &    16.0 & 0.00009 & 59104.98676 &    75.5 & 0.00013 & 60177.08567 &  1957.0 & 0.00005  \\ 
58358.53170 & -1234.5 & 0.00053 & 59071.36490 &    16.5 & 0.00014 & 59105.27098 &    76.0 & 0.00004 & 60177.37105 &  1957.5 & 0.00010  \\ 
58358.81611 & -1234.0 & 0.00005 & 59071.64989 &    17.0 & 0.00008 & 59105.55572 &    76.5 & 0.00012 & 60177.65572 &  1958.0 & 0.00006  \\ 
58359.10180 & -1233.5 & 0.00008 & 59071.93526 &    17.5 & 0.00024 & 59105.84032 &    77.0 & 0.00006 & 60177.94107 &  1958.5 & 0.00012  \\ 
58359.38589 & -1233.0 & 0.00012 & 59072.78988 &    19.0 & 0.00012 & 59106.12549 &    77.5 & 0.00013 & 60178.22531 &  1959.0 & 0.00005  \\ 
58359.67124 & -1232.5 & 0.00033 & 59073.07423 &    19.5 & 0.00016 & 59106.41026 &    78.0 & 0.00009 & 60178.51088 &  1959.5 & 0.00010  \\ 
58359.95639 & -1232.0 & 0.00016 & 59073.35993 &    20.0 & 0.00008 & 59106.69512 &    78.5 & 0.00011 & 60178.79519 &  1960.0 & 0.00006  \\ 
58360.24095 & -1231.5 & 0.00013 & 59073.64513 &    20.5 & 0.00013 & 59106.98023 &    79.0 & 0.00006 & 60179.08063 &  1960.5 & 0.00012  \\ 
58360.52551 & -1231.0 & 0.00017 & 59073.92936 &    21.0 & 0.00013 & 59107.26571 &    79.5 & 0.00011 & 60179.36579 &  1961.0 & 0.00006  \\ 
58360.81125 & -1230.5 & 0.00048 & 59075.35427 &    23.5 & 0.00014 & 59107.54988 &    80.0 & 0.00005 & 60179.64921 &  1961.5 & 0.00017  \\ 
58361.09545 & -1230.0 & 0.00011 & 59075.63916 &    24.0 & 0.00007 & 59107.83547 &    80.5 & 0.00010 & 60179.93525 &  1962.0 & 0.00011  \\ 
58361.95049 & -1228.5 & 0.00009 & 59075.92461 &    24.5 & 0.00006 & 59108.11935 &    81.0 & 0.00008 & 60181.92901 &  1965.5 & 0.00018  \\ 
58362.23444 & -1228.0 & 0.00007 & 59076.20896 &    25.0 & 0.00007 & 59108.40407 &    81.5 & 0.00012 & 60184.49411 &  1970.0 & 0.00006  \\ 
58362.52030 & -1227.5 & 0.00006 & 59076.49390 &    25.5 & 0.00010 & 59108.68935 &    82.0 & 0.00006 & 60184.77872 &  1970.5 & 0.00009  \\ 
58362.80462 & -1227.0 & 0.00043 & 59076.77904 &    26.0 & 0.00006 & 59108.97479 &    82.5 & 0.00011 & 60185.06332 &  1971.0 & 0.00004  \\ 
\hline
\end{tabular}}

\textit{Note}. The first eight eclipse times are SWASP seasonal normal minima.
\end{table*}

\newpage

\addtocounter{table}{-1}

\begin{table*}[h!]
\caption{Eclipse times for TIC~321978218 (continued)}
\scalebox{0.88}{\begin{tabular}{@{}lrllrllrllrl}
\hline
BJD & Cycle  & std. dev. & BJD & Cycle  & std. dev. & BJD & Cycle  & std. dev. & BJD & Cycle  & std. dev. \\ 
$-2\,400\,000$ & no. &   \multicolumn{1}{c}{$(d)$} & $-2\,400\,000$ & no. &   \multicolumn{1}{c}{$(d)$} & $-2\,400\,000$ & no. &   \multicolumn{1}{c}{$(d)$} & $-2\,400\,000$ & no. &   \multicolumn{1}{c}{$(d)$} \\ 
\hline
58363.09051 & -1226.5 & 0.00044 & 59077.06362 &    26.5 & 0.00012 & 59109.25888 &    83.0 & 0.00004 & 60185.34860 &  1971.5 & 0.00009  \\ 
58363.37465 & -1226.0 & 0.00018 & 59077.34871 &    27.0 & 0.00004 & 59109.54467 &    83.5 & 0.00012 & 60185.63322 &  1972.0 & 0.00003  \\ 
58363.65930 & -1225.5 & 0.00036 & 59077.63344 &    27.5 & 0.00010 & 59109.82896 &    84.0 & 0.00006 & 60185.91880 &  1972.5 & 0.00008  \\ 
58363.94412 & -1225.0 & 0.00011 & 59077.91895 &    28.0 & 0.00007 & 59110.11457 &    84.5 & 0.00012 & 60186.20319 &  1973.0 & 0.00006  \\ 
58364.22942 & -1224.5 & 0.00073 & 59078.20347 &    28.5 & 0.00008 & 59110.39867 &    85.0 & 0.00005 & 60186.48824 &  1973.5 & 0.00008  \\ 
58364.51351 & -1224.0 & 0.00022 & 59078.48835 &    29.0 & 0.00004 & 59110.68353 &    85.5 & 0.00010 & 60186.77318 &  1974.0 & 0.00005  \\ 
58364.79930 & -1223.5 & 0.00028 & 59078.77336 &    29.5 & 0.00010 & 59110.96839 &    86.0 & 0.00007 & 60187.05835 &  1974.5 & 0.00009  \\ 
58365.08375 & -1223.0 & 0.00072 & 59079.05850 &    30.0 & 0.00006 & 59111.25309 &    86.5 & 0.00016 & 60187.34295 &  1975.0 & 0.00004  \\ 
58365.65311 & -1222.0 & 0.00008 & 59079.34428 &    30.5 & 0.00011 & 59111.53759 &    87.0 & 0.00005 & 60187.62724 &  1975.5 & 0.00010  \\ 
58365.93808 & -1221.5 & 0.00027 & 59079.62838 &    31.0 & 0.00007 & 59111.82388 &    87.5 & 0.00011 & 60189.33714 &  1978.5 & 0.00010  \\ 
58366.22315 & -1221.0 & 0.00005 & 59079.91356 &    31.5 & 0.00010 & 59112.10796 &    88.0 & 0.00005 & 60189.62196 &  1979.0 & 0.00004  \\ 
58366.50759 & -1220.5 & 0.00026 & 59080.19822 &    32.0 & 0.00007 & 59112.39400 &    88.5 & 0.00011 & 60189.90762 &  1979.5 & 0.00009  \\ 
58366.79308 & -1220.0 & 0.00006 & 59080.48254 &    32.5 & 0.00010 & 59112.67687 &    89.0 & 0.00006 & 60190.19141 &  1980.0 & 0.00004  \\ 
58367.07824 & -1219.5 & 0.00025 & 59080.76794 &    33.0 & 0.00007 & 59112.96266 &    89.5 & 0.00013 & 60190.47693 &  1980.5 & 0.00007  \\ 
58368.78739 & -1216.5 & 0.00014 & 59081.05359 &    33.5 & 0.00011 & 59113.24702 &    90.0 & 0.00008 & 60190.76149 &  1981.0 & 0.00005  \\ 
58369.07163 & -1216.0 & 0.00020 & 59081.33822 &    34.0 & 0.00006 & 59113.53268 &    90.5 & 0.00017 & 60191.04627 &  1981.5 & 0.00008  \\ 
58369.35674 & -1215.5 & 0.00014 & 59081.62342 &    34.5 & 0.00013 & 59113.81609 &    91.0 & 0.00017 & 60191.33106 &  1982.0 & 0.00006  \\ 
58369.64203 & -1215.0 & 0.00012 & 59081.90811 &    35.0 & 0.00008 & 59114.38750 &    92.0 & 0.00016 & 60191.61713 &  1982.5 & 0.00011  \\ 
58369.92672 & -1214.5 & 0.00012 & 59082.19335 &    35.5 & 0.00009 & 60154.86236 &  1918.0 & 0.00005 & 60191.90127 &  1983.0 & 0.00006  \\ 
58370.21158 & -1214.0 & 0.00004 & 59082.47818 &    36.0 & 0.00006 & 60155.14715 &  1918.5 & 0.00010 & 60192.47097 &  1984.0 & 0.00005  \\ 
58370.49678 & -1213.5 & 0.00073 & 59082.76255 &    36.5 & 0.00010 & 60155.43215 &  1919.0 & 0.00004 & 60192.75595 &  1984.5 & 0.00011  \\ 
58370.78086 & -1213.0 & 0.00010 & 59083.04770 &    37.0 & 0.00006 & 60155.71622 &  1919.5 & 0.00011 & 60193.04077 &  1985.0 & 0.00008  \\ 
58371.06578 & -1212.5 & 0.00485 & 59083.33232 &    37.5 & 0.00010 & 60156.00186 &  1920.0 & 0.00006 & 60193.32650 &  1985.5 & 0.00008  \\ 
58371.35077 & -1212.0 & 0.00021 & 59083.61766 &    38.0 & 0.00008 & 60156.28589 &  1920.5 & 0.00012 & 60193.61105 &  1986.0 & 0.00009  \\ 
58371.63632 & -1211.5 & 0.00038 & 59083.90258 &    38.5 & 0.00015 & 60156.57152 &  1921.0 & 0.00006 & 60195.31965 &  1989.0 & 0.00007  \\ 
58371.92021 & -1211.0 & 0.00017 & 59084.18769 &    39.0 & 0.00005 & 60156.85597 &  1921.5 & 0.00007 & 60195.60460 &  1989.5 & 0.00009  \\ 
58372.20573 & -1210.5 & 0.00022 & 59084.47272 &    39.5 & 0.00010 & 60157.14150 &  1922.0 & 0.00005 & 60195.88963 &  1990.0 & 0.00004  \\ 
58372.49022 & -1210.0 & 0.00012 & 59084.75741 &    40.0 & 0.00006 & 60157.42597 &  1922.5 & 0.00009 & 60196.17532 &  1990.5 & 0.00008  \\ 
58372.77538 & -1209.5 & 0.00015 & 59085.04258 &    40.5 & 0.00013 & 60157.71122 &  1923.0 & 0.00005 & 60196.45910 &  1991.0 & 0.00005  \\ 
58373.06046 & -1209.0 & 0.00006 & 59085.32708 &    41.0 & 0.00005 & 60157.99562 &  1923.5 & 0.00012 & 60196.74417 &  1991.5 & 0.00009  \\ 
58373.34500 & -1208.5 & 0.00020 & 59085.61266 &    41.5 & 0.00016 & 60158.28095 &  1924.0 & 0.00008 & 60197.02894 &  1992.0 & 0.00005  \\ 
58373.91459 & -1207.5 & 0.00045 & 59085.89765 &    42.0 & 0.00009 & 60158.56504 &  1924.5 & 0.00009 & 60197.31481 &  1992.5 & 0.00006  \\ 
58374.19894 & -1207.0 & 0.00056 & 59086.18239 &    42.5 & 0.00021 & 60158.85060 &  1925.0 & 0.00006 & 60197.59918 &  1993.0 & 0.00005  \\ 
58374.48490 & -1206.5 & 0.00020 & 59086.46679 &    43.0 & 0.00005 & 60159.13523 &  1925.5 & 0.00009 & 60197.88383 &  1993.5 & 0.00009  \\ 
58374.76950 & -1206.0 & 0.00017 & 59086.75295 &    43.5 & 0.00014 & 60159.42018 &  1926.0 & 0.00005 & 60198.16890 &  1994.0 & 0.00005  \\ 
58375.05473 & -1205.5 & 0.00054 & 59087.03672 &    44.0 & 0.00025 & 60159.70487 &  1926.5 & 0.00009 & 60198.45395 &  1994.5 & 0.00012  \\ 
58375.33894 & -1205.0 & 0.00015 & 59088.46235 &    46.5 & 0.00013 & 60159.99034 &  1927.0 & 0.00006 & 60198.73804 &  1995.0 & 0.00005  \\ 
58375.62430 & -1204.5 & 0.00019 & 59088.74683 &    47.0 & 0.00008 & 60160.27437 &  1927.5 & 0.00009 & 60199.02389 &  1995.5 & 0.00010  \\ 
58375.90879 & -1204.0 & 0.00004 & 59089.03181 &    47.5 & 0.00005 & 60160.55997 &  1928.0 & 0.00005 & 60199.30817 &  1996.0 & 0.00004  \\ 
58376.19406 & -1203.5 & 0.00264 & 59089.31654 &    48.0 & 0.00005 & 60160.84398 &  1928.5 & 0.00009 & 60199.59430 &  1996.5 & 0.00008  \\ 
58376.47909 & -1203.0 & 0.00014 & 59089.60100 &    48.5 & 0.00010 & 60161.12967 &  1929.0 & 0.00004 & 60199.87805 &  1997.0 & 0.00006  \\ 
58376.76409 & -1202.5 & 0.00013 & 59089.88607 &    49.0 & 0.00008 & 60161.41383 &  1929.5 & 0.00009 & 60200.16360 &  1997.5 & 0.00009  \\ 
58377.04837 & -1202.0 & 0.00010 & 59090.17046 &    49.5 & 0.00008 & 60161.69909 &  1930.0 & 0.00007 & 60200.44787 &  1998.0 & 0.00007  \\ 
58377.61802 & -1201.0 & 0.00004 & 59090.45626 &    50.0 & 0.00007 & 60162.26977 &  1931.0 & 0.00006 & 60200.73325 &  1998.5 & 0.00008  \\ 
58377.90299 & -1200.5 & 0.00026 & 59090.74131 &    50.5 & 0.00009 & 60162.55352 &  1931.5 & 0.00007 & 60201.01743 &  1999.0 & 0.00005  \\ 
58378.18822 & -1200.0 & 0.00021 & 59091.02573 &    51.0 & 0.00006 & 60162.83907 &  1932.0 & 0.00005 & 60201.30270 &  1999.5 & 0.00008  \\ 
58378.47388 & -1199.5 & 0.00006 & 59091.31199 &    51.5 & 0.00009 & 60163.12455 &  1932.5 & 0.00009 & 60201.87283 &  2000.5 & 0.00008  \\ 
58378.75741 & -1199.0 & 0.00018 & 59091.59603 &    52.0 & 0.00010 & 60163.40869 &  1933.0 & 0.00007 & 60202.15700 &  2001.0 & 0.00007  \\ 
58379.04321 & -1198.5 & 0.00044 & 59091.88104 &    52.5 & 0.00013 & 60163.69397 &  1933.5 & 0.00008 & 60202.44170 &  2001.5 & 0.00012  \\ 
58379.32794 & -1198.0 & 0.00013 & 59092.16592 &    53.0 & 0.00006 & 60163.97868 &  1934.0 & 0.00003 & 60202.72691 &  2002.0 & 0.00004  \\ 
58379.61339 & -1197.5 & 0.00020 & 59092.44995 &    53.5 & 0.00015 & 60164.54811 &  1935.0 & 0.00006 & 60203.01175 &  2002.5 & 0.00007  \\ 
58379.89794 & -1197.0 & 0.00005 & 59092.73559 &    54.0 & 0.00008 & 60164.83315 &  1935.5 & 0.00012 & 60203.29637 &  2003.0 & 0.00006  \\ 
58380.18249 & -1196.5 & 0.00030 & 59093.02059 &    54.5 & 0.00010 & 60165.68847 &  1937.0 & 0.00016 & 60203.58142 &  2003.5 & 0.00011  \\ 
58380.46717 & -1196.0 & 0.00014 & 59093.30530 &    55.0 & 0.00008 & 60165.97246 &  1937.5 & 0.00011 & 60203.86651 &  2004.0 & 0.00005  \\ 
58380.75321 & -1195.5 & 0.00061 & 59093.59059 &    55.5 & 0.00008 & 60166.25749 &  1938.0 & 0.00013 & 60204.15125 &  2004.5 & 0.00008  \\ 
59061.96391 &     0.0 & 0.00004 & 59093.87482 &    56.0 & 0.00004 & 60167.96658 &  1941.0 & 0.00011 & 60204.43604 &  2005.0 & 0.00005  \\ 
59062.24885 &     0.5 & 0.00012 & 59094.16049 &    56.5 & 0.00010 & 60168.25338 &  1941.5 & 0.00011 & 60204.72083 &  2005.5 & 0.00012  \\ 
59062.53368 &     1.0 & 0.00005 & 59094.44458 &    57.0 & 0.00006 & 60168.53695 &  1942.0 & 0.00005 & 60205.00588 &  2006.0 & 0.00005  \\ 
59062.81852 &     1.5 & 0.00009 & 59094.72981 &    57.5 & 0.00008 & 60168.82209 &  1942.5 & 0.00009 & 60205.29123 &  2006.5 & 0.00009  \\ 
59063.10327 &     2.0 & 0.00005 & 59095.01455 &    58.0 & 0.00008 & 60169.10719 &  1943.0 & 0.00004 & 60205.57546 &  2007.0 & 0.00006  \\ 
59063.38898 &     2.5 & 0.00009 & 59095.29898 &    58.5 & 0.00006 & 60169.39239 &  1943.5 & 0.00009 & 60205.85994 &  2007.5 & 0.00008  \\ 
59063.67320 &     3.0 & 0.00005 & 59096.43927 &    60.5 & 0.00013 & 60169.67701 &  1944.0 & 0.00004 & 60206.14508 &  2008.0 & 0.00008  \\ 
59063.95823 &     3.5 & 0.00008 & 59096.72412 &    61.0 & 0.00005 & 60169.96218 &  1944.5 & 0.00009 & 60206.43015 &  2008.5 & 0.00013  \\ 
59064.24238 &     4.0 & 0.00004 & 59097.00893 &    61.5 & 0.00018 & 60170.24681 &  1945.0 & 0.00005 & 60206.71478 &  2009.0 & 0.00007  \\ 
59064.52777 &     4.5 & 0.00010  \\ 
\hline
\end{tabular}}
\end{table*}

\newpage

\begin{table*}[h!]
\caption{Eclipse times for TIC~323486857}
 \label{Tab:TIC_323486857_ToM}
\scalebox{0.88}{\begin{tabular}{@{}lrllrllrllrl}
\hline
BJD & Cycle  & std. dev. & BJD & Cycle  & std. dev. & BJD & Cycle  & std. dev. & BJD & Cycle  & std. dev. \\ 
$-2\,400\,000$ & no. &   \multicolumn{1}{c}{$(d)$} & $-2\,400\,000$ & no. &   \multicolumn{1}{c}{$(d)$} & $-2\,400\,000$ & no. &   \multicolumn{1}{c}{$(d)$} & $-2\,400\,000$ & no. &   \multicolumn{1}{c}{$(d)$} \\ 
\hline
58569.76629 &     0.0 & 0.00043 & 58611.31389 &    47.0 & 0.00030 & 59335.30793 &   866.0 & 0.00042 & 60056.64267 &  1682.0 & 0.00029  \\ 
58570.20988 &     0.5 & 0.00031 & 58611.75668 &    47.5 & 0.00037 & 59335.74280 &   866.5 & 0.00071 & 60057.08554 &  1682.5 & 0.00023  \\ 
58570.65159 &     1.0 & 0.00051 & 58612.19847 &    48.0 & 0.00033 & 59336.18785 &   867.0 & 0.00061 & 60057.52397 &  1683.0 & 0.00031  \\ 
58571.09320 &     1.5 & 0.00072 & 58612.63979 &    48.5 & 0.00029 & 59336.63213 &   867.5 & 0.00056 & 60057.96929 &  1683.5 & 0.00035  \\ 
58571.53535 &     2.0 & 0.00022 & 58613.08175 &    49.0 & 0.00029 & 59337.07349 &   868.0 & 0.00046 & 60058.40908 &  1684.0 & 0.00029  \\ 
58571.97855 &     2.5 & 0.00027 & 58613.52515 &    49.5 & 0.00041 & 59337.51399 &   868.5 & 0.00074 & 60058.85300 &  1684.5 & 0.00030  \\ 
58572.41999 &     3.0 & 0.00032 & 58613.96601 &    50.0 & 0.00043 & 59337.95543 &   869.0 & 0.00056 & 60059.29364 &  1685.0 & 0.00022  \\ 
58572.86354 &     3.5 & 0.00069 & 58614.41042 &    50.5 & 0.00046 & 59338.40096 &   869.5 & 0.00050 & 60059.73704 &  1685.5 & 0.00033  \\ 
58573.30288 &     4.0 & 0.00026 & 58614.84916 &    51.0 & 0.00023 & 59338.84369 &   870.0 & 0.00068 & 60060.17791 &  1686.0 & 0.00023  \\ 
58573.74551 &     4.5 & 0.00052 & 58616.18135 &    52.5 & 0.00022 & 59340.16948 &   871.5 & 0.00046 & 60060.61934 &  1686.5 & 0.00035  \\ 
58575.07036 &     6.0 & 0.00033 & 58616.62435 &    53.0 & 0.00031 & 59340.60799 &   872.0 & 0.00044 & 60061.06166 &  1687.0 & 0.00034  \\ 
58575.51230 &     6.5 & 0.00024 & 58617.06071 &    53.5 & 0.00038 & 59341.05470 &   872.5 & 0.00045 & 60061.50254 &  1687.5 & 0.00033  \\ 
58575.95563 &     7.0 & 0.00039 & 58617.50064 &    54.0 & 0.00022 & 59341.49356 &   873.0 & 0.00046 & 60061.94381 &  1688.0 & 0.00033  \\ 
58576.39925 &     7.5 & 0.00036 & 58617.94355 &    54.5 & 0.00017 & 59341.93654 &   873.5 & 0.00055 & 60063.71521 &  1690.0 & 0.00034  \\ 
58576.83874 &     8.0 & 0.00042 & 58618.38582 &    55.0 & 0.00040 & 59342.37567 &   874.0 & 0.00067 & 60064.15847 &  1690.5 & 0.00037  \\ 
58577.28074 &     8.5 & 0.00012 & 58618.83083 &    55.5 & 0.00044 & 59342.82239 &   874.5 & 0.00041 & 60064.59739 &  1691.0 & 0.00030  \\ 
58577.72402 &     9.0 & 0.00037 & 58619.27046 &    56.0 & 0.00021 & 59343.26177 &   875.0 & 0.00053 & 60065.04066 &  1691.5 & 0.00030  \\ 
58578.16485 &     9.5 & 0.00026 & 58619.71444 &    56.5 & 0.00034 & 59343.70379 &   875.5 & 0.00071 & 60065.48196 &  1692.0 & 0.00029  \\ 
58578.60739 &    10.0 & 0.00046 & 58620.15474 &    57.0 & 0.00056 & 59344.14506 &   876.0 & 0.00042 & 60065.92317 &  1692.5 & 0.00027  \\ 
58579.04978 &    10.5 & 0.00045 & 58620.59589 &    57.5 & 0.00040 & 59344.59012 &   876.5 & 0.00054 & 60066.36728 &  1693.0 & 0.00030  \\ 
58579.49038 &    11.0 & 0.00030 & 58621.03823 &    58.0 & 0.00018 & 59345.03023 &   877.0 & 0.00045 & 60066.80792 &  1693.5 & 0.00038  \\ 
58579.93289 &    11.5 & 0.00048 & 58621.48190 &    58.5 & 0.00024 & 59345.47615 &   877.5 & 0.00066 & 60067.24872 &  1694.0 & 0.00028  \\ 
58580.37447 &    12.0 & 0.00036 & 58621.92189 &    59.0 & 0.00022 & 59345.91006 &   878.0 & 0.00076 & 60067.69553 &  1694.5 & 0.00024  \\ 
58580.82057 &    12.5 & 0.00034 & 58622.36610 &    59.5 & 0.00048 & 59347.68278 &   880.0 & 0.00051 & 60069.01844 &  1696.0 & 0.00058  \\ 
58581.25860 &    13.0 & 0.00033 & 58622.80536 &    60.0 & 0.00021 & 59348.12482 &   880.5 & 0.00066 & 60069.46017 &  1696.5 & 0.00046  \\ 
58581.70065 &    13.5 & 0.00011 & 58623.25098 &    60.5 & 0.00022 & 59348.56668 &   881.0 & 0.00067 & 60069.89970 &  1697.0 & 0.00037  \\ 
58583.02629 &    15.0 & 0.00032 & 58623.68970 &    61.0 & 0.00019 & 59349.00738 &   881.5 & 0.00062 & 60070.34433 &  1697.5 & 0.00050  \\ 
58583.46947 &    15.5 & 0.00064 & 59307.45870 &   834.5 & 0.00048 & 59349.44945 &   882.0 & 0.00046 & 60070.78644 &  1698.0 & 0.00052  \\ 
58583.90988 &    16.0 & 0.00024 & 59307.90346 &   835.0 & 0.00030 & 59349.89660 &   882.5 & 0.00049 & 60071.23112 &  1698.5 & 0.00037  \\ 
58584.35359 &    16.5 & 0.00033 & 59308.34274 &   835.5 & 0.00033 & 59350.33224 &   883.0 & 0.00062 & 60071.67084 &  1699.0 & 0.00061  \\ 
58584.79381 &    17.0 & 0.00033 & 59308.78486 &   836.0 & 0.00031 & 59350.77254 &   883.5 & 0.00048 & 60072.10753 &  1699.5 & 0.00066  \\ 
58585.23743 &    17.5 & 0.00030 & 59309.22788 &   836.5 & 0.00047 & 59351.21601 &   884.0 & 0.00048 & 60072.55328 &  1700.0 & 0.00048  \\ 
58585.68021 &    18.0 & 0.00032 & 59309.66930 &   837.0 & 0.00037 & 59351.66064 &   884.5 & 0.00057 & 60072.99538 &  1700.5 & 0.00052  \\ 
58586.12364 &    18.5 & 0.00031 & 59310.11184 &   837.5 & 0.00023 & 59352.09770 &   885.0 & 0.00052 & 60073.44014 &  1701.0 & 0.00043  \\ 
58586.56482 &    19.0 & 0.00039 & 59310.55204 &   838.0 & 0.00030 & 59352.54232 &   885.5 & 0.00055 & 60073.88186 &  1701.5 & 0.00064  \\ 
58587.00787 &    19.5 & 0.00027 & 59310.99457 &   838.5 & 0.00035 & 59352.98380 &   886.0 & 0.00055 & 60074.32231 &  1702.0 & 0.00066  \\ 
58587.44784 &    20.0 & 0.00022 & 59311.43751 &   839.0 & 0.00025 & 59353.42594 &   886.5 & 0.00042 & 60074.76122 &  1702.5 & 0.00050  \\ 
58587.89134 &    20.5 & 0.00022 & 59311.87970 &   839.5 & 0.00039 & 59353.87131 &   887.0 & 0.00059 & 60075.20774 &  1703.0 & 0.00046  \\ 
58588.33155 &    21.0 & 0.00015 & 59312.32043 &   840.0 & 0.00022 & 59354.31209 &   887.5 & 0.00053 & 60075.65137 &  1703.5 & 0.00068  \\ 
58588.77622 &    21.5 & 0.00014 & 59312.76598 &   840.5 & 0.00031 & 59354.75179 &   888.0 & 0.00047 & 60076.53155 &  1704.5 & 0.00061  \\ 
58589.21430 &    22.0 & 0.00139 & 59313.20262 &   841.0 & 0.00037 & 59355.19564 &   888.5 & 0.00047 & 60076.97495 &  1705.0 & 0.00061  \\ 
58589.65949 &    22.5 & 0.00025 & 59313.65069 &   841.5 & 0.00024 & 59355.63353 &   889.0 & 0.00043 & 60077.41750 &  1705.5 & 0.00052  \\ 
58590.09857 &    23.0 & 0.00030 & 59314.08602 &   842.0 & 0.00032 & 59356.07916 &   889.5 & 0.00050 & 60077.86026 &  1706.0 & 0.00052  \\ 
58590.54181 &    23.5 & 0.00017 & 59314.53014 &   842.5 & 0.00041 & 59356.51878 &   890.0 & 0.00062 & 60078.29890 &  1706.5 & 0.00051  \\ 
58590.98400 &    24.0 & 0.00035 & 59314.97102 &   843.0 & 0.00031 & 59356.96604 &   890.5 & 0.00082 & 60078.74227 &  1707.0 & 0.00048  \\ 
58591.42658 &    24.5 & 0.00036 & 59315.41203 &   843.5 & 0.00034 & 59357.40351 &   891.0 & 0.00055 & 60079.18897 &  1707.5 & 0.00049  \\ 
58591.86852 &    25.0 & 0.00022 & 59315.85605 &   844.0 & 0.00026 & 59357.84515 &   891.5 & 0.00039 & 60079.62981 &  1708.0 & 0.00044  \\ 
58592.31010 &    25.5 & 0.00038 & 59316.30190 &   844.5 & 0.00031 & 59358.28964 &   892.0 & 0.00051 & 60080.06855 &  1708.5 & 0.00063  \\ 
58592.75126 &    26.0 & 0.00039 & 59316.74043 &   845.0 & 0.00026 & 59358.73201 &   892.5 & 0.00048 & 60080.50730 &  1709.0 & 0.00053  \\ 
58593.19595 &    26.5 & 0.00026 & 59317.18404 &   845.5 & 0.00038 & 59359.16913 &   893.0 & 0.00052 & 60080.95305 &  1709.5 & 0.00051  \\ 
58593.63629 &    27.0 & 0.00045 & 59317.62363 &   846.0 & 0.00031 & 59359.61948 &   893.5 & 0.00051 & 60081.39369 &  1710.0 & 0.00047  \\ 
58594.07894 &    27.5 & 0.00024 & 59318.06535 &   846.5 & 0.00044 & 60041.16669 &  1664.5 & 0.00030 & 60081.83729 &  1710.5 & 0.00058  \\ 
58594.52186 &    28.0 & 0.00039 & 59320.27523 &   849.0 & 0.00029 & 60041.61872 &  1665.0 & 0.00025 & 60082.27957 &  1711.0 & 0.00046  \\ 
58597.17258 &    31.0 & 0.00026 & 59320.71976 &   849.5 & 0.00023 & 60042.94206 &  1666.5 & 0.00029 & 60082.72362 &  1711.5 & 0.00060  \\ 
58597.61472 &    31.5 & 0.00038 & 59321.16077 &   850.0 & 0.00034 & 60043.38255 &  1667.0 & 0.00034 & 60083.16379 &  1712.0 & 0.00058  \\ 
58598.05540 &    32.0 & 0.00032 & 59321.60364 &   850.5 & 0.00031 & 60043.82585 &  1667.5 & 0.00023 & 60084.49029 &  1713.5 & 0.00049  \\ 
58598.49952 &    32.5 & 0.00033 & 59322.04407 &   851.0 & 0.00035 & 60044.26715 &  1668.0 & 0.00025 & 60084.93224 &  1714.0 & 0.00038  \\ 
58598.94008 &    33.0 & 0.00027 & 59322.48491 &   851.5 & 0.00023 & 60044.70862 &  1668.5 & 0.00028 & 60085.37741 &  1714.5 & 0.00048  \\ 
58599.38361 &    33.5 & 0.00029 & 59322.92837 &   852.0 & 0.00030 & 60045.14974 &  1669.0 & 0.00029 & 60085.81676 &  1715.0 & 0.00037  \\ 
58599.82437 &    34.0 & 0.00023 & 59323.37305 &   852.5 & 0.00034 & 60045.59309 &  1669.5 & 0.00039 & 60086.25885 &  1715.5 & 0.00039  \\ 
58600.26685 &    34.5 & 0.00036 & 59323.81164 &   853.0 & 0.00028 & 60046.03571 &  1670.0 & 0.00035 & 60086.69829 &  1716.0 & 0.00044  \\ 
58600.70610 &    35.0 & 0.00037 & 59324.25592 &   853.5 & 0.00025 & 60046.47768 &  1670.5 & 0.00026 & 60087.14042 &  1716.5 & 0.00049  \\ 
58601.15038 &    35.5 & 0.00036 & 59324.69728 &   854.0 & 0.00028 & 60046.91919 &  1671.0 & 0.00020 & 60087.58317 &  1717.0 & 0.00044  \\ 
58601.59086 &    36.0 & 0.00032 & 59325.13993 &   854.5 & 0.00043 & 60047.36298 &  1671.5 & 0.00025 & 60088.02407 &  1717.5 & 0.00049  \\ 
58602.03462 &    36.5 & 0.00023 & 59325.58062 &   855.0 & 0.00024 & 60047.80250 &  1672.0 & 0.00026 & 60088.46892 &  1718.0 & 0.00046  \\ 
58602.47502 &    37.0 & 0.00022 & 59326.02314 &   855.5 & 0.00047 & 60048.68766 &  1673.0 & 0.00028 & 60088.90923 &  1718.5 & 0.00048  \\ 
58602.91879 &    37.5 & 0.00026 & 59326.46614 &   856.0 & 0.00023 & 60049.12895 &  1673.5 & 0.00026 & 60089.35204 &  1719.0 & 0.00038  \\ 
\hline
\end{tabular}}
\end{table*}

\newpage

\addtocounter{table}{-1}

\begin{table*}[h!]
\caption{Eclipse times for TIC~323486857 (continued)}
\scalebox{0.88}{\begin{tabular}{@{}lrllrllrllrl}
\hline
BJD & Cycle  & std. dev. & BJD & Cycle  & std. dev. & BJD & Cycle  & std. dev. & BJD & Cycle  & std. dev. \\ 
$-2\,400\,000$ & no. &   \multicolumn{1}{c}{$(d)$} & $-2\,400\,000$ & no. &   \multicolumn{1}{c}{$(d)$} & $-2\,400\,000$ & no. &   \multicolumn{1}{c}{$(d)$} & $-2\,400\,000$ & no. &   \multicolumn{1}{c}{$(d)$} \\ 
\hline
58603.35927 &    38.0 & 0.00027 & 59326.90804 &   856.5 & 0.00025 & 60049.57184 &  1674.0 & 0.00029 & 60089.79635 &  1719.5 & 0.00086  \\ 
58603.80102 &    38.5 & 0.00023 & 59327.34780 &   857.0 & 0.00034 & 60050.01534 &  1674.5 & 0.00022 & 60090.23819 &  1720.0 & 0.00098  \\ 
58604.24427 &    39.0 & 0.00032 & 59327.78911 &   857.5 & 0.00036 & 60050.45400 &  1675.0 & 0.00024 & 60090.67897 &  1720.5 & 0.00063  \\ 
58604.68696 &    39.5 & 0.00034 & 59328.23375 &   858.0 & 0.00034 & 60050.90010 &  1675.5 & 0.00029 & 60091.11812 &  1721.0 & 0.00049  \\ 
58605.12795 &    40.0 & 0.00024 & 59328.67518 &   858.5 & 0.00027 & 60051.33576 &  1676.0 & 0.00024 & 60091.55560 &  1721.5 & 0.00064  \\ 
58605.56796 &    40.5 & 0.00028 & 59329.11724 &   859.0 & 0.00033 & 60051.78066 &  1676.5 & 0.00030 & 60091.99793 &  1722.0 & 0.00051  \\ 
58606.01130 &    41.0 & 0.00028 & 59329.56032 &   859.5 & 0.00033 & 60052.22169 &  1677.0 & 0.00027 & 60092.44365 &  1722.5 & 0.00061  \\ 
58606.45355 &    41.5 & 0.00027 & 59330.00054 &   860.0 & 0.00031 & 60052.66168 &  1677.5 & 0.00026 & 60092.88863 &  1723.0 & 0.00042  \\ 
58606.89314 &    42.0 & 0.00024 & 59330.44647 &   860.5 & 0.00034 & 60053.10468 &  1678.0 & 0.00031 & 60093.32787 &  1723.5 & 0.00063  \\ 
58607.33810 &    42.5 & 0.00029 & 59330.88656 &   861.0 & 0.00025 & 60053.54895 &  1678.5 & 0.00027 & 60093.76998 &  1724.0 & 0.00052  \\ 
58607.77827 &    43.0 & 0.00036 & 59331.32648 &   861.5 & 0.00029 & 60053.99122 &  1679.0 & 0.00027 & 60094.21013 &  1724.5 & 0.00050  \\ 
58608.22090 &    43.5 & 0.00030 & 59331.76958 &   862.0 & 0.00041 & 60054.43533 &  1679.5 & 0.00034 & 60094.65193 &  1725.0 & 0.00058  \\ 
58608.66302 &    44.0 & 0.00039 & 59332.21264 &   862.5 & 0.00028 & 60054.87346 &  1680.0 & 0.00025 & 60095.09516 &  1725.5 & 0.00055  \\ 
58609.10723 &    44.5 & 0.00030 & 59333.98326 &   864.5 & 0.00090 & 60055.30942 &  1680.5 & 0.00058 & 60095.53596 &  1726.0 & 0.00048  \\ 
58609.54582 &    45.0 & 0.00071 & 59334.42102 &   865.0 & 0.00033 & 60055.75857 &  1681.0 & 0.00031 & 60095.98375 &  1726.5 & 0.00049  \\ 
58610.88147 &    46.5 & 0.00087 & 59334.86242 &   865.5 & 0.00040 & 60056.20025 &  1681.5 & 0.00027 & 60096.42393 &  1727.0 & 0.00070  \\ 
\hline
\end{tabular}}
\end{table*}

\begin{table*}[h!]
\caption{Eclipse times for TIC~650024463}
 \label{Tab:TIC_650024463_ToM}
\scalebox{0.88}{\begin{tabular}{@{}lrllrllrllrl}
\hline
BJD & Cycle  & std. dev. & BJD & Cycle  & std. dev. & BJD & Cycle  & std. dev. & BJD & Cycle  & std. dev. \\ 
$-2\,400\,000$ & no. &   \multicolumn{1}{c}{$(d)$} & $-2\,400\,000$ & no. &   \multicolumn{1}{c}{$(d)$} & $-2\,400\,000$ & no. &   \multicolumn{1}{c}{$(d)$} & $-2\,400\,000$ & no. &   \multicolumn{1}{c}{$(d)$} \\ 
\hline
58327.08816 &    -0.5 & 0.00110 & 58654.64552 &    45.0 & 0.00049 & 59079.24315 &   104.0 & 0.00037 & 60122.70738 &   249.0 & 0.00042  \\ 
58330.79614 &     0.0 & 0.00050 & 58658.16613 &    45.5 & 0.00115 & 59082.83269 &   104.5 & 0.00108 & 60129.91097 &   250.0 & 0.00033  \\ 
58334.27812 &     0.5 & 0.00069 & 58661.83767 &    46.0 & 0.00048 & 59086.43768 &   105.0 & 0.00050 & 60133.57915 &   250.5 & 0.00093  \\ 
58337.98821 &     1.0 & 0.00030 & 58669.02564 &    47.0 & 0.00051 & 59367.04776 &   144.0 & 0.00040 & 60137.11216 &   251.0 & 0.00043  \\ 
58341.46618 &     1.5 & 0.00114 & 58672.54609 &    47.5 & 0.00100 & 59370.68081 &   144.5 & 0.00100 & 60140.77733 &   251.5 & 0.00086  \\ 
58345.17921 &     2.0 & 0.00042 & 58676.22071 &    48.0 & 0.00053 & 59374.25614 &   145.0 & 0.00042 & 60144.31914 &   252.0 & 0.00029  \\ 
58348.65041 &     2.5 & 0.01546 & 58679.73200 &    48.5 & 0.00111 & 59377.89761 &   145.5 & 0.00084 & 60147.98837 &   252.5 & 0.00108  \\ 
58352.36818 &     3.0 & 0.00053 & 59043.21409 &    99.0 & 0.00048 & 59381.45739 &   146.0 & 0.00036 & 60155.19765 &   253.5 & 0.00099  \\ 
58625.84233 &    41.0 & 0.00112 & 59046.80730 &    99.5 & 0.00102 & 59385.08446 &   146.5 & 0.00079 & 60158.74801 &   254.0 & 0.00030  \\ 
58629.36316 &    41.5 & 0.00144 & 59050.42746 &   100.0 & 0.00036 & 59388.67342 &   147.0 & 0.00037 & 60162.39658 &   254.5 & 0.00094  \\ 
58633.05334 &    42.0 & 0.00067 & 59054.01795 &   100.5 & 0.00108 & 60101.12259 &   246.0 & 0.00029 & 60165.94101 &   255.0 & 0.00044  \\ 
58636.57755 &    42.5 & 0.00082 & 59057.63048 &   101.0 & 0.00045 & 60104.77423 &   246.5 & 0.00074 & 60169.58876 &   255.5 & 0.00076  \\ 
58643.78601 &    43.5 & 0.00103 & 59064.84274 &   102.0 & 0.00041 & 60108.31387 &   247.0 & 0.00037 & 60173.13221 &   256.0 & 0.00033  \\ 
58647.45317 &    44.0 & 0.00055 & 59068.42848 &   102.5 & 0.00101 & 60111.95859 &   247.5 & 0.00251 & 60176.77814 &   256.5 & 0.00111  \\ 
58650.97684 &    44.5 & 0.00119 & 59075.63588 &   103.5 & 0.00079 & 60115.50744 &   248.0 & 0.00028 & 60180.32145 &   257.0 & 0.00053  \\ 
\hline
\end{tabular}}
\end{table*}


\section{Radial velocities of TICs~287756035 and 321978218}
\label{app:RVtables}

In this appendix we list the RV data obtained for two tertiary components out of our seven triples (Tables~\ref{tab:TIC287756035_RVdata}-\ref{tab:TIC321978218_RVdata}).
All of these data were obtained with the CHIRON spectrometer on the SMARTS telescope.

\begin{table}
\centering
\caption{Measured radial velocities of the tertiary component of TIC~287756035. The date is given as BJD -- 2\,400\,000, while the RVs are in km\,s$^{-1}$.}
\label{tab:TIC287756035_RVdata}
\begin{tabular}{lrlrlr}
\hline
\hline
Date & RV$_\mathrm{B}$ & Date & RV$_\mathrm{B}$ & Date & RV$_\mathrm{B}$  \\  
\hline
$59700.608356$ & $  7.628$ & $59738.546993$ & $ 21.727$ & $60053.676004$ & $  3.070$ \\ 
$59722.571059$ & $ 16.556$ & $59793.483242$ & $ 24.726$ \\
\hline
\end{tabular}
\end{table}  


\begin{table}
\centering
\caption{Measured radial velocities of the tertiary component of TIC~321978218. The date is given as BJD -- 2\,400\,000, while the RVs are in km\,s$^{-1}$.}
\label{tab:TIC321978218_RVdata}
\begin{tabular}{lrlrlr}
\hline
\hline
Date & RV$_\mathrm{B}$ & Date & RV$_\mathrm{B}$  \\  
\hline
$59197.534786$ & $ -49.069$ & $59378.910596$ & $ -20.313$ & $59828.655080$ & $ -51.324$ \\ 
$59201.567894$ & $ -38.221$ & $59383.923361$ & $  -3.060$ & $59866.610970$ & $  38.891$ \\
$59189.515319$ & $ -21.500$ & $59384.912542$ & $   0.014$ & $59872.611652$ & $  26.137$ \\
$59195.592569$ & $ -49.890$ & $59397.889185$ & $  33.266$ & $59877.609567$ & $  -5.269$ \\
$59371.888074$ & $ -45.392$ & $59399.767719$ & $  37.061$ & $59881.591328$ & $ -36.279$ \\
$59373.907426$ & $ -38.801$ & $59400.828602$ & $  38.036$ && \\
\hline
\end{tabular}
\end{table}  


\end{appendix}

\end{document}